\documentclass[12pt,a4paper]{book}

\usepackage[latin1]{inputenc}
\usepackage[english]{babel}
\usepackage{frontespizio}
\usepackage[T1]{fontenc}

\usepackage{faktor}
\usepackage{booktabs}
\usepackage{braket}
\usepackage{mathrsfs}
\usepackage{caption}
\usepackage{booktabs}
\usepackage{url}
\usepackage{footnote}
\usepackage{tabu}
\usepackage{setspace}
\usepackage{appendix}
\usepackage{slashed}
\usepackage{tabularx}
\usepackage{latexsym}
\usepackage[nottoc]{tocbibind}
\usepackage{subfig}
\usepackage{multirow}
\usepackage{booktabs}
\usepackage{enumerate}
\usepackage{mathrsfs}
\usepackage[section]{placeins}
\usepackage{float}
\usepackage{midpage}
\def\nn{\nonumber}

\def\0 {\nonumber} 
\def\del{\partial} 
 
\def\bj{\begin{align}r \jmath}
\def\minicent#1#2{
  \begin{minipage}{#1 cm}
    \begin{center}
     #2 
    \end{center}
  \end{minipage}
}

\newcommand{\Kahler}{\ensuremath{\text{K}\ddot{\text{a}}\text{hler}\,}}
\newcommand{\be}{\begin{equation}}
\newcommand{\ee}{\end{equation}}
\newcommand{\bi}{\begin{itemize}}
\newcommand{\ei}{\end{itemize}}
\newcommand{\ben}{\begin{enumerate}}
\newcommand{\een}{\end{enumerate}}
\newcommand{\comments}[1]{}
\def\nn{\nonumber}

\def\del{\partial}
\def\SM{{\scriptscriptstyle \rm SM}}
\def\KK{{\scriptscriptstyle \rm KK}}
\def\dS{{\scriptscriptstyle \rm dS}}
\def\lp{{\scriptscriptstyle \rm loop}}
\def\DM{{\scriptscriptstyle \rm DM}}

\newcommand\La{{\mathcal{L}}}
\newcommand{\ov}{\overline}
\newcommand\vo{{\mathcal{V}}}

\newcommand{\mc}{\mathcal}

\newcommand{\beqa}{\begin{eqnarray}}
\newcommand{\eeqa}{\end{eqnarray}}
\newcommand{\V}{{\cal{V}}}

\usepackage{color}
\usepackage{amsfonts}
\usepackage{multicol}
\usepackage{amsmath,amssymb}
\usepackage{graphicx}

\graphicspath{{Tesi/}}
\usepackage{tensor}

\begin{document}

\frontmatter

\begin{titlepage}

\begin{center}
{{\large{\mbox{ALMA MATER STUDIORUM - UNIVERSITA' DI~BOLOGNA}}}} \\
\vspace{2cm}
\Large{Dottorato di Ricerca in Fisica}\\
\large{Ciclo XXVIII}\\
\end{center}
\vspace{1cm}
Settore concorsuale di afferenza: 02/A2\\
Settore scientifico disciplinare: FIS/02
\vspace{1cm}
\begin{center}
\LARGE
\textbf{Sequestered String Models:\vspace{3mm}\\Supersymmetry Breaking\\and\vspace{3mm}\\Cosmological Applications\vspace{3mm}}
\normalsize
\end{center}
\vspace{2cm}
\par
\noindent
\raggedright
\begin{minipage}[t]{0.47\textwidth}
{\textbf{Presentata da: Francesco Muia\\
\vspace{1cm}}}
\end{minipage}\\
\begin{minipage}[t]{0.47\textwidth}
{\textbf{Coordinatore Dottorato:\\
Prof. Gastone Castellani}}
\end{minipage}
\hfill
\begin{minipage}[t]{0.47\textwidth}\raggedleft
{\textbf{Relatore:\\
Dr. Michele Cicoli}\\
\vspace{2cm}}
\end{minipage}
\centering
\textbf{Esame finale anno 2016}
\end{titlepage}

\cleardoublepage

\begin{midpage}
\begin{large}
\begin{center}
\textit{``...non per un dio, ma nemmeno per gioco...''}
\end{center}
\end{large}
\begin{flushright}
\textit{F. de Andr\'e}
\end{flushright}
\vspace{5cm}
\begin{large}
\begin{flushright}
\textit{Dedicata alle cinque persone importanti:\\
mamma, pap\'a, Fabrizia, Roxana, Miscell.}
\end{flushright}
\end{large}
\end{midpage}

\cleardoublepage

\begin{midpage}
\begin{center}
\huge{Acknowledgements}
\end{center}
\vspace{1cm}
I would like to thank all the collaborators who made the realization of this thesis possible: Rouzbeh Allahverdi, Luis Aparicio, Michele Cicoli, Bhaskar Dutta, Sven Krippendorf, Anshuman Maharana, Francisco Pedro, Fernando Quevedo. I'm particularly grateful to my supervisor, Michele Cicoli, who taught me how to do research, and to Fernando Quevedo, who represents a rare combination of competence and kindness.\\

This thesis has been partially developed during my stays at the DESY Institute in Hamburg, at the ICTP Institute in Trieste and at the IFT Institute in Madrid, where I spent a few months as a visitor.\\
\end{midpage}

\newpage

\begin{midpage}
\begin{center}
\huge Declaration
\end{center}

This thesis is based on the results presented in the following papers:\\
\ben
\item[] \cite{Aparicio:2014wxa} \textbf{2014, November.}\\
L. Aparicio, M. Cicoli, S. Krippendorf, A. Maharana, \textbf{F. Muia}, F. Quevedo,\\
``Sequestered de Sitter String Scenarios: Soft-terms'',\\
JHEP {\bf 1411} (2014) 071, arXiv:1409.1931 [hep-th].\\

\item[] \cite{Aparicio:2015sda} \textbf{2015, May.}\\
L. Aparicio, B. Dutta, M. Cicoli, S. Krippendorf, A. Maharana, \textbf{F. Muia}, F. Quevedo,\\
``Non-thermal CMSSM with a 125 GeV Higgs'',\\
JHEP {\bf 1505} (2015) 098, arXiv:1502.05672 [hep-ph].\\

\item[] \cite{Cicoli:2015wja} \textbf{2015, September.}\\
M. Cicoli, \textbf{F. Muia}, F. G. Pedro,\\
``Microscopic Origin of Volume Inflation'',\\
JCAP {\bf 1512} (2015) 040, arXiv:1509.07748 [hep-th].\\

\item[] \cite{Cicoli:2015bpq} \textbf{2015, November.}\\
M. Cicoli, \textbf{F. Muia},\\
``General Analysis of Dark Radiation in Sequestered String Models'',\\
JHEP {\bf 1512} (2015) 152, arXiv:1511.05447 [hep-th].\\
\een

\end{midpage}

\newpage

\begin{midpage}
\begin{center}
\huge{Abstract}
\end{center}

In the present thesis I studied the particle phenomenology and cosmology arising from a class of string models called \textit{sequestered compactifications}, which were born with the aim of getting low-energy SUSY from strings. This is not an easy task if combined with cosmological constraints, since the mechanism of moduli stabilization fixes both the scale of supersymmetric particles and the scale of moduli, which tend to be of the same order. However, if on the one hand supersymmetric particles with TeV mass are desired in order to address the electroweak hierarchy problem, on the other hand the cosmological moduli problem requires the moduli to be heavier than $100 \,$ TeV. The specific setup of sequestered compactifications makes this hierarchy achievable, at least in principle: as in these models the visible sector is located on a stack of D3-branes at singularities, a physical separation between the visible degrees of freedom and the SUSY-breaking sources takes place. Such decoupling translates into a hierarchy between the scale of SUSY-breaking and the spectrum of supersymmetric particles. Moreover, it is interesting to notice that moduli are the four-dimensional manifestation of the existence of extra-dimensions, and then their presence is a common feature of all string compactifications. Since they are only gravitationally coupled, they could decay late in the history of the universe, affecting in a significant way its cosmological evolution. Possible deviations of the cosmological observables from the values predicted by the standard Hot Big Bang Theory constitute an interesting alternative for the discovery of new physics beyond the Standard Model, which is complementary to the particle physics search. For this reason in addition to SUSY-breaking in sequestered models, I also studied several cosmological scenarios arising from them, such as production of non-thermal dark matter and dark radiation, reheating from moduli decay and inflation.
\end{midpage}

\tableofcontents

\mainmatter

\part{Introduction}

\chapter{State of the Art}
\label{chap:StateOfArt}

\raggedbottom

The higgs boson has been for a long time the last missing block of the \textit{Standard Model of particle physics} (in the following \textit{Standard Model} or SM)~\cite{Agashe:2014kda}. With the announcement of the discovery of a new particle in 2012~\cite{Aad:2012tfa, Chatrchyan:2012xdj}, whose mass is around $125 \,$ GeV and whose quantum numbers are compatible with those of the theorized higgs boson~\cite{Higgs:1964pj}, the scientific community finally celebrated the complete affirmation of this theory. The Standard Model describes all known particles\footnote{Except massive neutrinos.} and their interactions through three out of the four known forces in nature: the \textit{electromagnetic force}, the \textit{weak force} and the \textit{strong force}. Fermionic particles compose the so-called \textit{matter} of the universe, and are called leptons and quarks. They are organized in three different families, of which the first one contains the lightest and stable particles, such as the electron. Forces are mediated by bosonic particles called gauge bosons: photon, gluons and the bosons $W^\pm$ and $Z$.\\

The Standard Model is an extraordinarily successful scientific theory: to the extent that we can compute physical quantities to make predictions which can be experimentally verified, it has passed every single test. To give an idea of the amazing agreement between Standard Model predictions and experimental data, in Tab.~\ref{tab:BosonMasses} we report measured and predicted values for the masses of the $W^\pm$ and $Z$ bosons, which highlight an impressive accord~\cite{Agashe:2014kda}.\\
\begin{table}[h!]
\begin{center}
\begin{tabular}{ccc}
\hline
 & Measured values (GeV) & SM predictions (GeV) \\
\hline
Mass of $W^\pm$ boson & $80.376 \pm 0.033$ & $80.363 \pm 0.006$ \\
\hline
Mass of $Z$ boson & $91.1876 \pm 0.0021$ & $91.1880 \pm 0.0020$ \\
\hline
\end{tabular}
\end{center}
\caption{Predicted and measured values for the masses of $W^\pm$ and $Z$~\cite{Agashe:2014kda}. The measured value of $W^\pm$ is taken from LEP2 results~\cite{Schael:2013ita}.}
\label{tab:BosonMasses}
\end{table}

In order to locate the Standard Model within a wider scientific context, it is worth recalling that it represents a specific example of a general tool, called \textit{quantum field theory}, which allows to describe the physics of relativistic particles at the quantum level. Quantum field theory represents one of the two main pillars on which the whole understanding of the universe is currently based, and it is useful to describe the physics taking place at small distances: in particular the Standard Model describes the physics up to $10^{-17} \,$ cm. On the other hand, the second main pillar is \textit{General Relativity}~\cite{Einstein:1915ca}, which describes at the classical level the fourth force of nature: \textit{gravity}. Given the weakness of the force of gravity, which becomes relevant only when extremely large amounts of mass are taken into account, General Relativity is useful to describe the physics involving the largest scales of the universe. The agreement between predictions of General Relativity and experiments has been tested to a high degree of accuracy~\cite{Will:2001mx}.\\

The Standard Model supplemented by General Relativity is commonly interpreted as an \textit{Effective Field Theory} (EFT) which is valid up to the energies explored at the \textit{Large Hadron Collider} (LHC) outside Geneve ($M_{\rm EW} \sim 100 \,$ GeV). To be precise, the description provided by the Standard Model is expected to break down not far beyond the electroweak scale $M_{\rm EW}$, while the effective description of gravity provided by General Relativity is supposed to be a valid approximation all the way up to the Planck scale $M_{\rm P} \sim 10^{18} \,$ GeV (or some other cut-off scale below $M_{\rm P}$ at which the four-dimensional description of General Relativity ceases to be valid, as it happens for example in theories with extra-dimensions). As we will see in the next sections, there are many good arguments in support of the idea that new physics should appear at energies just above $M_{\rm EW}$.\\

According to the \textit{Hot Big Bang Theory}, in its early stage the universe was composed by a hot and dense thermal bath of relativistic particles, whose temperature decreased following its expansion. As the temperature of the thermal bath can be put in correspondence with the age of the universe, it turns out that the electroweak energy scale $M_{\rm EW}$ corresponds to a tiny age: $t_{\rm EW} = 2 \times 10^{-11} \,$ sec. As a consequence, the EFT composed by General Relativity plus the Standard Model is perfectly suitable to describe the evolution of the entire universe from a tiny fraction of a second after the Big Bang ($t_{\rm EW}$) to the present. The Hot Big Bang Theory is the cosmological counterpart of the Standard Model of particle physics.\\

Despite its outstanding success, the Standard Model (and consequently also the Hot Big Bang Theory) is definitely incomplete, since at least it fails to describe gravity at the quantum level. More in detail, there are two main gaps which have to be filled in order to get a comprehensive understanding of the physics governing the universe. On the one hand, the first and more fundamental gap is the absence of a quantum theory of gravity. Quantum gravitational effects become important either when a big amount of energy density is confined in a small region of space (as in black holes or in the universe at $t_{\rm P} \simeq 10^{-44} \,$ sec), or when the physics taking place at the smallest distances $\ell_{\rm P} \simeq 10^{-34} \,$ cm is considered\footnote{As mentioned before, the description of General Relativity can actually break down at distances larger than $\ell_{\rm P}$.}. The second gap on the other hand, is the lack of comprehension of the physics taking place between $M_{\rm EW}$ and $M_{\rm P}$, which however should be consistently described in the framework of quantum field theory. From a cosmological point of view this second gap translates into a lack of understanding of the evolution of the universe before $t_{\rm EW}$.  Hopefully, very soon new experimental data coming from the second run of the LHC will cut through the fog, driving us towards the right way among the myriads of alternatives available at the moment for the physics beyond the Standard Model.\\

There are a few main principles which have played a role of guidance in the development of the current understanding of the universe, and that are extensively used also in the investigation for new physics. The first one is the principle of \textit{necessity}. As we remarked above, Standard Model is a specific example of a more general tool: quantum field theory, whose birth is exactly due to the \textit{necessity} of unifying \textit{quantum mechanics} and \textit{special relativity} into a single framework. Indeed, quantum mechanics is not able to describe relativistic particles, since it does not provide the possibility of creating and/or annihilating particles, which is instead allowed by the equivalence between energy and mass of special relativity, along with the Heisenberg's uncertainty principle. Similarly, a conceptual obstruction is encountered as soon as one tries to formulate a quantum version of General Relativity, since the prescriptions dictated by quantum field theory to quantize it lead to a non-renormalizable theory~\cite{Nilles:1983ge, Cerdeno:1998hs}. Given that this obstacle appears to be unavoidable, it suggests that probably a completely new tool is needed exactly in the same way as quantum field theory was needed in order to describe relativistic particles. As we will explain in more detail in Sec.~\ref{sec:StringTheory}, the only known candidate theory providing a description of gravity at the quantum level, and containing all the building blocks of the Standard Model, is \textit{string theory}~\cite{Polchinski:1998rq, Polchinski:1998rr, Green:1987sp, Green:1987mn, Ibanez:2012zz}.\\

The second principle comes out from a very intuitive observation: physical phenomena taking place at different length scales (which correspond to different energy scales) do not affect each other. This is a quite obvious point, since clearly it is not necessary to understand quantum mechanics in order to play ping-pong, as well as it is not needed to appreciate the dynamics of quarks within nuclei in order to study the chemical features of a material. Physical phenomena operating at different energy scales simply decouple from each other. This concept is ubiquitous in physics at every level of complexity, and it is often referred to as \textit{naturalness}. The naturalness principle is also incorporated in quantum field theory~\cite{Susskind:1978ms, Veltman:1980mj, Dimopoulos:1979qi}, in which context for instance the authors of~\cite{Gaillard:1974hs} were able to predict the mass of the \textit{charm} quark, only by using naturalness arguments. This idea is strictly related to that of effective field theory: let us assume to know the renormalizable lagrangian of an effective theory, like for instance the Standard Model. It is possible to compute loop corrections to the parameters of the lagrangian (e.g. the masses of particles), as a function of the cut-off $\Lambda$. The energy scale at which the corrections get larger than the measured values of some of the parameters defines $\Lambda_{\rm UV}$: at this scale either some degree of fine-tuning is needed in order to cancel off different loop contributions and keep the values of the parameters equal to the measured ones, or new physics intervenes, altering the low-energy description. The latter option is the \textit{natural} one, and it essentially relies on some underlying symmetry of the complete theory, which keeps the corrections small. This idea works perfectly well in the Standard Model - where for instance the chiral symmetry keeps fermionic masses small - with a single major exception: the higgs mass. Such a breakdown is due to the generic UV sensitivity of scalar particles, and it is a further reason to consider the Standard Model unsatisfactory. Naturalness has probably been the main guidance principle in the search for new physics beyond the electroweak scale $M_{\rm EW}$ in the last decades.\\

Strictly related to the idea of naturalness, there is the problem of hierarchies of the fundamental scales of nature, also known as \textit{Dirac's naturalness}. This issue arises as a consequence of simple dimensional analysis in physics, according to which dimensionless quantities should be of order one. Such a problem was first noticed by Dirac, who was interested in understanding the origin of the hierarchy between the proton mass and the Planck mass: $m_{\rm proton}/M_{\rm P} \sim 10^{-18}$. It is commonly accepted that every time dimensional analysis dramatically fails to predict the ratio between two scales of nature, the hierarchy should be explained in a dynamical way. In the particular case at hand the solution is asymptotic freedom of QCD, which gives
\begin{align}
\label{eq:ProtonVsPlanckMass}
\frac{m_{\rm proton}}{M_{\rm P}} \simeq e^{-\frac{a}{g_s^2}} \sim 10^{-18}\,,
\end{align}
where $a$ is a order one coefficient, while $g_s$ is the strong coupling constant\footnote{In the rest of the thesis $g_s$ will denote the string coupling constant, unless differently specified.}. On the other hand the hierarchy between the electroweak scale and the Planck scale
\begin{align}
\label{eq:EWVsPlanckMass}
\frac{M_{\rm EW}}{M_{\rm P}} \sim 10^{-16} \,,
\end{align}
still needs an explanation.\\

Finally, it is worth mentioning a trend which is manifest in the development of physics in the last century. Starting from the unification of \textit{electricity} and \textit{magnetism} into a single description through Maxwell's equations~\cite{Maxwell:1865zz}, to end with the unification of weak and electromagnetic forces in the Standard Model, it is possible to recognize a general tendency\, pointing towards the \textit{unification} of the interactions described in the quantum field theory framework\footnote{To be distinguished from the previously discussed unification of the description of all forces into a single framework, such for example string theory.}. This observation is also reinforced by noting that the Renormalization Group evolution of the coupling constants of the Standard Model makes them qualitatively converge towards a single value at high energies. This could imply that at high energies, such as in the primordial universe, the forces are indistinguishable and they get differentiated through various symmetry breaking steps, the last one of which takes place at $M_{\rm EW}$ leaving us with the unbroken electromagnetism. This observation has been deeply exploited in the search for new physics, in order to fill the gap between $M_{\rm EW}$ and $M_{\rm P}$.\\

In the present chapter we first give a brief description of the state of the art, both in particle physics and in cosmology, in Sec.~\ref{sec:CurrentUnderstanding}. Afterwards we describe the possible alternatives for the physics beyond the current understanding in Sec.~\ref{sec:BSM}, focusing specially on the role played by supersymmetric theories. Finally we report a brief introduction to the main concepts of string theory in Sec.~\ref{sec:StringTheory}.\\

\section{Current Understanding}
\label{sec:CurrentUnderstanding}

\subsection{Standard Model of Particle Physics}
\label{ssec:StandardModel}

The Standard Model of particle physics~\cite{Agashe:2014kda} is a quantum field theory based on the gauge group
\begin{align}
\label{eq:SMGaugeGroup}
G_{\rm SM} = SU(3)_{\rm col} \times SU(2)_{\rm L} \times U(1)_{Y}\,.
\end{align}
The gauge theory based on $SU(3)_{\rm col}$ describes strong interactions and it is called \textit{Quantum Chromo-Dynamics} (QCD), while the $SU(2)_{\rm L} \times U(1)_{\rm Y}$ component describes electroweak (EW) interactions. Matter particles are organized in three families of quarks and leptons, as summarized in Tab.~\ref{tab:SMFamilies}, transforming under the gauge groups as reported in Tab.~\ref{tab:SMMatter}.\\

\begin{small}
\renewcommand{\arraystretch}{1.6}
\begin{table}[h!]
\begin{center}
\begin{tabular}{cccc}
\hline
Field & $1^{\rm st}\,$ family & $2^{\rm nd}\,$ family & $3^{\rm rd}\,$ family \\
 & mass (in GeV) & mass (in GeV) & mass (in GeV) \\
\hline
$u^i$ quarks & $u$ & $c$ & $t$ \\
 & $(1.5 - 3.3) \times 10^{-3}$ & $1.14-1.34$ & $169.1-173.3$\\
\hline
$d^i$ quarks & $d$ & $s$ & $b$ \\
 & $(3.5-6) \times 10^{-3}$ & $0.07-0.13$ & 4.13-4.37 \\
\hline
leptons & $e$ & $\mu$ & $\tau$ \\
 & $0.51 \times 10^{-3}$ & $1.05 \times 10^{-1}$ & $1.78$\\
\hline
neutrinos & $\nu_e$ & $\nu_\mu$ & $\nu_\tau$ \\
 & $< 2 \times 10^{-9}$ & $< 0.19 \times 10^{-3}$ & $< 18.2 \times 10^{-3}$\\
\hline
\end{tabular}
\end{center}
\caption{Families of matter particles of the SM. We report the respective masses measured in GeV. Uncertainties on quark masses are predominantly theoretical, since they do not exist in free-states. The bounds on neutrino masses can be sharpened by making assumptions on the nature of their mass terms.}
\label{tab:SMFamilies}
\end{table}
\end{small}

\begin{small}
\renewcommand{\arraystretch}{1.6}
\begin{table}[h!]
\begin{center}
\begin{tabular}{ccccc}
\hline
Field & $SU(3)_{\rm col}$ & $SU(2)_{\rm L}$ & $U(1)_{\rm Y}$ & $U(1)_{\rm em}$\\
\hline
$Q^i = \begin{pmatrix} u^i_{\rm L} \\ d^i_{\rm L} \end{pmatrix}$ & $\mathbf{3}$ & $\mathbf{2}$ & $\frac{1}{6}$ & $\begin{pmatrix} +2/3 \\ -1/3 \end{pmatrix}$ \\
\hline
$u^i_{\rm R}$ & $\overline{\mathbf{3}}$ & $\mathbf{1}$ & $- \frac{2}{3}$ & $+2/3$ \\
\hline
$d^i_{\rm R}$ & $\overline{\mathbf{3}}$ & $\mathbf{1}$ & $\frac{1}{3}$ & $-1/3$ \\
\hline
$L^i = \left( \begin{matrix}\nu^i \\ e^i_{\rm L} \end{matrix}\right)$ & $\mathbf{1}$ & $\mathbf{2}$ & $- \frac{1}{2}$& $\begin{pmatrix} 0 \\ -1 \end{pmatrix}$ \\
\hline
$e^i_{\rm R}$ & $\mathbf{1}$ & $\mathbf{1}$ & $1$ & $-1$ \\
\hline
$H = \left( \begin{matrix} H^- \\ H^0 \end{matrix}\right)$ & $\mathbf{1}$ & $\mathbf{2}$ & $- \frac{1}{2}$ & $\begin{pmatrix} -1 \\ 0 \end{pmatrix}$\\
\hline
\end{tabular}
\end{center}
\caption{Matter particles of the SM. The index $i$ runs over the three families in Tab.~\ref{tab:SMFamilies}.}
\label{tab:SMMatter}
\end{table}
\end{small}

The most interesting feature of the strong sector is asymptotic freedom, which explains the hierarchy in eq. \eqref{eq:ProtonVsPlanckMass}. The coupling constant of strong interactions depends on the energy $Q$ as
\begin{align}
\label{eq:StrongCouplingRunning}
\alpha^2_s (Q) \equiv \frac{g_s^2 (Q)}{4 \pi} = \frac{1}{1 + \frac{11 N_{\rm c} - 2 N_{f}}{12 \pi} \log \left(\frac{Q^2}{\Lambda_{\rm QCD}^2}\right)} \,,
\end{align}
where $N_{\rm c} = 3$ is the number of colors, $N_f$ is the number of kinematically accessible quark flavours (e.g. $N_f = 5$ at $Q = M_Z$) and from measurements it turns out that $\Lambda_{\rm QCD} \simeq 200 \,$ MeV \cite{Agashe:2014kda}. In the limit $Q^2 \rightarrow \infty$ the strong coupling constant vanishes, giving rise to asymptotic freedom. At low $Q^2$ on the other hand, QCD is in the strong regime and quarks are constrained to form bound states called \textit{hadrons}, such as the proton, the pions, etc..\\

The dynamics of the electroweak sector is much more involved than that of the strong one. SM is a chiral theory, i.e. left and right components of the fermionic fields transform differently under the gauge groups, as it can be observed in Tab.~\ref{tab:SMMatter}. The chiral nature of the SM interactions is reflected in the fact that mass terms are forbidden for all the fermions of the theory as a consequence of gauge invariance. The higgs field is a complex field transforming as a doublet under $SU(2)_{\rm L}$ and its dynamics is determined by the scalar potential
\begin{align}
\label{eq:HiggsPotential}
V_{\rm Higgs} = \lambda \left(|H|^2 - \frac{v^2}{2}\right)^2 \,,
\end{align}
whose minimum is given by $\langle \left|H\right|^2\rangle = \frac{v^2}{2}$. The higgs VEV triggers a spontaneous symmetry breaking of the group of electroweak interactions down to the group of symmetry of the electromagnetism
\begin{align}
\label{eq:SMSymmetryBreaking}
SU(2)_{\rm L} \times U(1)_{\rm Y} \rightarrow U(1)_{\rm EM}\,.
\end{align}
The \textit{ElectroWeak Symmetry Breaking} mechanism (EWSB) is also called \textit{higgs mechanism}. Using gauge invariance it is possible to show that without loss of generality the higgs field can be written (in the unitary gauge) as
\begin{align}
\label{eq:HiggsFieldUnitaryGauge}
H(x) = \begin{pmatrix} 0 \\ \frac{v}{\sqrt{2}} + \frac{h^0(x)}{\sqrt{2}} \end{pmatrix}\,,
\end{align}
where $h^0(x)$ is the unique, electrically neutral, physical scalar excitation about the higgs vacuum. As a consequence of the higgs mechanism, three higgs degrees of freedom are swallowed by the gauge bosons $W^\pm$ and $Z$, which become massive. Moreover, through the Yukawa couplings
\begin{align}
\label{eq:SMYukawacouplings}
\mathcal{L}_{\rm Yuk} = Y^{ij}_U \overline{Q}^i u^j_R H^* + Y^{ij}_D \overline{Q}^i d^j_R H + Y^{ij}_{\rm L} \overline{L}^i e^j_R H + \text{h.c.} \,,
\end{align}
the higgs mechanism gives mass to all the fermions of the theory, except the neutrinos. The fermion masses are given by
\begin{align}
\label{eq:FermionMasses}
m_f = \frac{y_f}{\sqrt{2}} v\,,
\end{align}
where $v = 170 \,$ GeV, while $y_f$ is an eigenvalue of the proper Yukawa matrix among those appearing in eq. \eqref{eq:SMYukawacouplings}.\\

The Yukawa couplings in eq. \eqref{eq:SMYukawacouplings} are the most general compatibly with gauge invariance and renormalizability, and they feature an accidental invariance under the global symmetries \textit{baryon number} $B$ and \textit{lepton number} $L$. A clear manifestation of the conservation of the baryon number is the proton lifetime, which is constrained by the Kamiokande experiment to be $\tau_{\rm proton} > 10^{32}-10^{33} \,$ years \cite{Hirata:1989kn}, while lepton number conservation\footnote{Lepton number is actually violated by neutrino oscillations.} puts a strict bound on processes such as $\mu \rightarrow e \gamma$, whose branching ratio is currently $Br(\mu \rightarrow e \gamma) < 5.7 \times 10^{-13}$ \cite{Adam:2013mnn}. The hypercharge assignments of Tab.~\ref{tab:SMMatter} are dictated by anomaly cancellation.\\

A crucial point in the phenomenology of the SM is that the fields enumerated in Tab.~\ref{tab:SMMatter} form a basis of gauge eigenstates, which does not diagonalize the mass matrices obtained through the Yukawa couplings in eq. \eqref{eq:SMYukawacouplings}. In order to get the mass eigenstates it is necessary to diagonalize them by unitary transformations $V_{\rm L}^{U, D, L}$, $V_{\rm R}^{U, D, L}$ acting on the left and right handed fermions respectively. Eventually the $W^\pm$ bosons couple to the quarks through the Cabibbo-Kobayashi-Maskawa (CKM) matrix, which is not diagonal
\begin{align}
\label{eq:CKM}
\left|V_{\rm CKM}\right| = \left|V^U_{\rm L} \left(V_{\rm L}^D\right)^\dagger \right| = \begin{pmatrix} 0.9742 & 0.226 & 0.0036 \\ 0.226 & 0.9733 & 0.042 \\ 0.0087 & 0.041 & 0.99913 \end{pmatrix}\,,
\end{align}
and produces a mixing of the quark flavors. On the other hand the CKM matrix does not appear in the couplings of the $Z$ boson with quarks, which implies the absence of \textit{flavour-changing neutral currents}. A similar mixing does not happen in the lepton sector of the SM, as a consequence of the fact that neutrinos are massless. The CKM matrix has complex entries, but many of them can be eliminated by re-parametrization of the fields. Eventually a single phase turns out to be physical, producing CP-violation in the SM. A second source of CP-violation is the $\theta$-term
\begin{align}
\label{eq:ThetaTerm}
\mathcal{L_{\theta}} = \frac{\theta}{32 \pi^2} G_{\mu \nu} \tilde{G}^{\mu \nu} \,,
\end{align}
which can be added to the SM lagrangian, where $G_{\mu \nu}$ is the field strength of QCD. Although this term is a total derivative, it can contribute to physical processes through gauge instantons. Since it would give a contribution to the electric dipole moment of the neutron, which is highly constrained, the value of $\theta$ has to be minuscule: $\theta < 10^{-10}$. This unexplained high degree of fine-tuning required to match the experimental data is known as \textit{Strong CP Problem}.

\subsection{Standard Model of Cosmology}
\label{ssec:StandardModelCosmology}

As already discussed in the introduction to the chapter, the Standard Model of cosmology is based on the Standard Model of particle physics supplemented by General Relativity, which describes gravity at the classical level in terms of the dynamics of the space-time metric $g_{\mu \nu}$. The equation of motion for the metric tensor can be inferred from the \textit{Einstein-Hilbert} action, which reads\footnote{We neglect the cosmological constant here.}
\begin{align}
\label{eq:EinsteinHilbertAction}
S_{\rm EH} = \frac{M_{\rm P}^2}{2} \int d^4x \, \sqrt{-g} \, \mathcal{R} \,,
\end{align}
where $g = \text{det} \left(g_{\mu \nu}\right)$ and $\mathcal{R}$ is the \textit{Ricci scalar} associated to the space-time metric $g_{\mu \nu}$. $M_{\rm P}$ denotes the reduced Planck mass, which in terms of the universal gravitational constant $G$ can be written as
\begin{align}
\label{eq:PlanckMass}
M_{\rm P} = \frac{1}{\sqrt{8 \pi G}} \simeq 2.4 \times 10^{18} \, \rm GeV \,.
\end{align}
From the action in eq. \eqref{eq:EinsteinHilbertAction} it is possible to infer the famous Einstein's equation
\begin{align}
\label{eq:EinsteinEquation}
\mathcal{R}_{\mu \nu} - \frac{1}{2} \mathcal{R} g_{\mu \nu} = 8 \pi G T_{\mu \nu} \,,
\end{align}
where $T_{\mu \nu}$ is the energy-momentum tensor and $\mathcal{R}_{\mu \nu}$ is the \textit{Ricci tensor}.\\

The Standard Model of cosmology~\cite{Gorbunov:2011zz}, also called \textit{Hot Big Bang Theory}, assumes that the universe started in a hot and dense state. As the universe expanded, the thermal bath cooled and progressively became less dense. One of the most exciting discoveries in cosmology dates back to 1998, when two independent projects~\cite{Riess:1998cb, Perlmutter:1998np} observed that the universe is currently expanding at an accelerate rate. The basic observation about the universe is that it is isotropic and homogeneous at large spatial scales, above the size of the largest observed structures i.e. supercluster of galaxies, which can reach a diameter of tens of Megaparsec~\cite{Galaxies}. As the observations suggest that space is essentially flat~\cite{Ade:2013zuv}, isotropy and homogeneity allow us to write the metric of the expanding universe as
\begin{align}
\label{eq:FRWMetric}
ds^2 = dt^2 - a^2(t) d \mathbf{x}^2 \,,
\end{align}
which is called \textit{Friedmann-Robertson-Walker} (FRW) metric\footnote{The FRW in eq. \eqref{eq:FRWMetric} is written in the case of flat universe.}. The entire dynamics of the universe is described by the evolution of the time-dependent \textit{scale factor} $a(t)$. Since the universe expands, $a(t)$ is an increasing monotonic function of the time. The rate of expansion is written in terms of the \textit{Hubble parameter}
\begin{align}
\label{eq:HubbleParameter}
H(t) = \frac{\dot{a}(t)}{a(t)} \,,
\end{align}
whose current value $H_0$\footnote{The subscript $0$ is used to denote current values of physical quantities in cosmology.} defines the Hubble law
\begin{align}
\label{eq:HubbleLaw}
v = H_0 r \,,
\end{align}
where $v$ is the speed of a far galaxy and $r$ is its distance from the Earth. The reciprocal of the current values of the Hubble constant gives an estimate of the age of the universe, which is approximately $t_0 = H_0^{-1} \simeq 14 \times 10^9$ yrs.\\

Depending on the source of energy which dominates the universe in a given epoch, the scale factor has a different time-dependence, as reported in Tab.~\ref{tab:ScaleFactor}.
\begin{small}
\renewcommand{\arraystretch}{1.6}
\begin{table}[h!]
\begin{center}
\begin{tabular}{ccc}
\hline
Source of Energy & $a(t)$ & Hubble parameter  \\
\hline
Radiation & $\sim t^{1/2}$ & $\frac{1}{2 t}$ \\
Matter & $\sim t^{2/3}$ & $\frac{2}{3 t}$ \\
Vacuum Energy & $\sim e^{H t}$ & const.\\
\hline
\end{tabular}
\end{center}
\caption{Scale factor as a function of time, depending on the source of energy dominating the universe in a given epoch.}
\label{tab:ScaleFactor}
\end{table}
\end{small}
The dynamics of the scale factor is dictated by the Einstein's equation. Assuming that the equation of state of the universe takes the perfect fluid form
\begin{align}
\label{eq:EqStatePerfectFluid}
p = w \rho \,,
\end{align}
and using the FRW metric in eq. \eqref{eq:FRWMetric} it is possible to rewrite the Einstein's equation as
\begin{align}
\label{eq:FriedmannEquation}
H^2 =& \, \frac{8 \pi G}{3} \left(\rho_{\rm matter} + \rho_{\rm radiation} + \rho_{\Lambda} + \rho_{\rm curv}\right) \,, \nonumber \\  
&\dot{H} + H^2 = \frac{\ddot{a}}{a} = -\frac{1}{6} \left(\rho + 3 p\right)\,,
\end{align}
where $\rho_i$ denote the contributions to the total energy density of the universe $\rho$ coming from different sources, and the derivatives are taken with respect to $t$. Equations in \eqref{eq:FriedmannEquation} are called \textit{Friedmann's Equations}. In the first Friedmann's equation we also included a contribution $\rho_{\rm curv}$ due to the curvature of the universe, to be as generic as possible, even if observations show that it is negligibly small~\cite{Ade:2013zuv}. In terms of the critical value of the energy density $\rho_{\rm c} = \frac{3}{8 \pi G} H_0^2 = 5 \times 10^{-6} \, \text{GeV}/\text{cm}^3$, it is possible to define the current fractions of energy density
\begin{align}
\label{eq:OmegaParameters}
\Omega_i = \frac{\rho_{i,0}}{\rho_{\rm c}} \,,
\end{align}
in terms of which the first Friedmann's equation takes the nice form
\begin{align}
\label{eq:SmartFriedmannEquation}
H^2 = \frac{8 \pi G}{3} \rho_{\rm c} \left[\Omega_{\rm matter} \left(\frac{a_0}{a}\right)^3 + \Omega_{\rm radiation} \left(\frac{a_0}{a}\right)^4 + \Omega_{\Lambda} + \Omega_{\rm curv} \left(\frac{a_0}{a}\right)^2\right] \,.
\end{align}
A few observations about eq. \eqref{eq:SmartFriedmannEquation} are in order. First, at the current epoch the various components sum up exactly to one. The second observation is that from the scaling of the different sources of energy density, it is possible to understand the sequence of the dominating ones. At the very early stages of the universe, when $a \ll a_0$, radiation dominated. Subsequently matter started to dominate at the age $t_{\rm matter} = 10^5 \,$ yrs. The vacuum energy, also called \textit{Cosmological Constant} $\Lambda$, has started his period of domination very late, at the age of $t_{\Lambda} \simeq 7 \times 10^9$ yrs. The parameters $\Omega_i$ represent the current fractions of the energy density, which are
\begin{align}
\label{eq:CurrentFractions}
\Omega_{\rm radiation} \approx 5 \times 10^{-5} \,, \quad \Omega_{\rm matter} \simeq 0.27 \,, \quad \Omega_{\Lambda} \simeq 0.73 \,, \quad \left|\Omega_{\rm curv}\right| < 0.005 \,,
\end{align}
so that the universe is currently dominated by the Cosmological Constant $\Lambda$, with a minor contribution from matter, divided into \textit{baryonic matter} and \textit{Dark Matter} (DM) in the following proportions
\begin{align}
\label{eq:BaryonAndDM}
\Omega_{\rm baryons} \simeq 0.046 \,, \qquad \Omega_{\rm DM} \simeq 0.23 \,.
\end{align}

DM is presumably composed by stable massive particles not included in the SM, which do not interact electromagnetically and strongly. These compose clumps of energy density which account for most of the mass of galaxies and clusters of galaxies. There are several experimental evidences for DM existence, which also provide estimates for its abundance, namely
\ben
\item Accurate reconstructions of the mass distributions in clusters of galaxies is obtained through the observations of gravitational lensing of the light coming from further galaxies~\cite{Lesgourgues:2007te, Massey:2007gh}.
\item X-ray observations of clusters just after their collisions show that most of baryons are in a hot, ionized intergalactic gas. The total mass of clusters exceeds the mass of baryons in luminous matter by an order of magnitude~\cite{Vikhlinin:2005mp}.
\item The study of the motion of stars at the periphery of galaxies gives further motivations for DM. Assuming circular motion, the dependence of the velocity $v(R)$ on the distance $R$ from galactic center follows the Newton's law
\begin{align}
\label{eq:NewtonLaw}
v(R) = \sqrt{\frac{G M(R)}{R}}\,.
\end{align}
Observationally, $v(R) = \text{const.}$ sufficiently far away from the center, and this deviation from the expectations can be explained by assuming that luminous matter is embedded into dark clouds of larger size~\cite{Begeman:1991iy}.
\item Another possibility is to measure the speed of galaxies in galactic clusters and use the virial theorem to infer the gravitational potential and hence the total mass in the cluster itself. It turns out that the estimated total mass by far exceeds the sum of masses of individual galaxies in clusters. This discrepancy can be explained if most of the mass is due to DM which is distributed smoothly over the cluster.
\een

The presence of DM in the universe is quite important for our own existence, since without it the universe today would still be pretty homogeneous, and as a consequence large scale structures (like galaxies and clusters of galaxies) would have never formed. DM can be of three different types, depending on the mass $m_{\rm DM}$ of the particles which compose it: if $m_{\rm DM} > 30 \, \text{KeV}$ it is called \textit{Cold Dark Matter} (CDM), while if $m_{\rm DM} = 1-10 \, \text{KeV}$ then it is called \textit{Warm Dark Matter} (WDM), if $m_{\rm DM} < 1 \,$ KeV it is called \textit{Hot Dark Matter} (HDM). In the present thesis we will deal with CDM. In this case, one of the best candidates for DM is represented by \textit{Weakly Interacting Massive Particles} (WIMP's). In fact the abundance of DM is roughly given by\footnote{The entropy density $s$ is often used to measure the number of particles $N_i$ of a given particle species in a comoving volume, since in case of entropy conservation $s \sim a^{-3}$, and then $N_i = n_i/s$, where $n_i$ is the number density of particles.}
\begin{align}
\label{eq:DMAbundance}
\frac{\rho_{\rm DM}}{s_0} \approx \langle \sigma v \rangle^{-1} \,,
\end{align}
where $\rho_{\rm DM}$ is the energy density of DM, $s_0$ is the current value of the density of entropy and $\langle \sigma v \rangle$ is the annihilation cross section for DM particles. If the annihilation cross section is of the same order of magnitude as that of weakly interacting particles $\langle \sigma v \rangle \sim 10^{-8} \,$ GeV$^{-2}$, the abundance of DM roughly matches the observed value. This observation is usually called the \textit{WIMP miracle}.\\

The estimate in eq. \eqref{eq:DMAbundance} contains a concept which is fundamental in the understanding of the thermal history of the universe: as the expansion of the universe $H$ equals the rate of interaction of a given particle $\Gamma$ then the particle \textit{freezes-out}, decoupling from the thermal bath. This concept is useful to determine the abundance of CDM, as well as for instance the epoch of neutrino decoupling.\\

The standard cosmological model which includes CDM and dark energy with abundances close to those reported in eq. \eqref{eq:CurrentFractions} and eq. \eqref{eq:BaryonAndDM} is called $\Lambda$CDM. We summarize the key events in the thermal history of the the universe, according to the $\Lambda$CDM, (see also Tab.~\ref{tab:ThermalHistory})
\bi
\item \textbf{Baryogenesis -} The universe features an asymmetry between observed matter and antimatter\footnote{We ignore the possible existence of compact objects made of antimatter, such that there is no asymmetry as in eq. \eqref{eq:MatterAntiMatterAsymmetry}. Such compact objects are predicted in some models~\cite{Khlopov:2000as, Dolgov:2008wu} and do not contradict observations~\cite{Khlopov:2002ww, Dolgov:2010cb}.}, which can be quantified through the ratio
\begin{align}
\label{eq:MatterAntiMatterAsymmetry}
\eta = \frac{n_b}{s_0} \simeq 8 \times 10^{-10}\,,
\end{align}
where $n_b$ is the number density of baryonic particles. Since we do not have experimental access to the energies at which baryogenesis presumably took place, we do not know which physical process has produced the asymmetry in eq. \eqref{eq:MatterAntiMatterAsymmetry}, despite many mechanisms have been proposed.
\item \textbf{Electroweak Phase Transition - } Assuming that the universe was reheated at a temperature above $M_{\rm EW}$, when the thermal bath reached a temperature of about $100 \,$ GeV, the electroweak phase transition occurred~\cite{Kirzhnits:1972iw, Kirzhnits:1972ut, Dolan:1973qd, Weinberg:1974hy}, causing the higgs field to get a VEV. As a consequence the electroweak group $SU(2)_{\rm L} \times U(1)_{\rm Y}$ got broken to the electromagnetism one $U(1)_{\rm em}$ and the particles of the SM became massive.
\item \textbf{QCD Phase Transition - } Above $150 \,$ MeV quarks were \textit{asymptotically free}, and the thermal bath was composed by a phase of matter called \textit{Quark Gluon Plasma}. Around the temperature of $150 \,$ MeV strong interactions became relevant and the quarks were suddenly constrained to form hadrons.
\item \textbf{DM Freeze-Out - } Assuming that DM is of the cold type and that it is composed of WIMPs, it decoupled at a temperature around $1 \,$ MeV.
\item \textbf{Neutrino decoupling - } Since neutrinos interacted with the thermal bath only through weak interactions, the same estimate as for WIMP DM holds, with a minor correction, so that the freeze-out took place at a temperature around $0.8 \,$ MeV.
\item \textbf{Electron-Positron annihilation - } Electrons and positrons annihilated soon after neutrino decoupling at a temperature of $0.5 \,$ MeV, slightly reheating the photon thermal bath. As a consequence the energy density of neutrinos $\rho_\nu$ is related to that of photons $\rho_\gamma$ as
\begin{align}
\label{eq:NeutrinoPhotonTemperature}
\rho_\nu = \frac{7}{8} N_{\rm eff} \left(\frac{4}{11}\right)^{4/3} \rho_\gamma \,,
\end{align}
where $N_{\rm eff}$ is the \textit{effective number of neutrinos}, and this expression is valid after neutrino decoupling. The effective number of neutrinos after $e^+ e^-$ annihilation in the SM is $N_{\rm eff, SM} = 3.046$.

\item \textbf{Big Bang Nucleosynthesis (BBN) - } At a temperature of about $300 \,$ KeV, light elements started to be formed. This is the furthest epoch of which we have proper and detailed comprehension~\cite{Alpher:1948ve, Dempster:1948zz}, both from a phenomenological and from a theoretical point of view.
\item \textbf{Recombination - } At a temperature of about $0.3 \,$ KeV the neutral hydrogen formed via the reaction $e^- + p^+ \rightarrow H + \gamma$, since its reverse reaction had become disfavored.
\item \textbf{Photon Decoupling - } Before recombination the photons were coupled to the rest of the plasma essentially through Thomson scattering $e^- + \gamma \rightarrow e^- + \gamma$. The sharp drop in the free electron density after recombination made Thomson scattering inefficient and the photons decoupled at $0.25 \,$ eV. As a consequence, they started their free-stream through the universe, and are today observed as the Cosmic Microwave Background (CMB).
\ei

The CMB consists of free-streaming photons which last-scattered about $400000 \,$ years after the ``birth'' of the universe. They have an excellent black-body spectrum with temperature~\cite{Gawiser:2000az}
\begin{align}
\label{eq:CMBTemperature}
T_0 = 2.725 \pm 0.001 \, \text{K} \,.
\end{align}
This spectrum is almost perfectly uniform in all directions, so that we can infer that the universe at photon decoupling was almost perfectly isotropic and homogeneous. Small perturbations to this uniformity have been measured to be of order $\frac{\delta T}{T_0} \sim 10^{-5}$~\cite{deBernardis:2000sbo, Hanany:2000qf, Spergel:2006hy, Ade:2015xua}.

\begin{small}
\renewcommand{\arraystretch}{1.6}
\begin{table}[h!]
\begin{center}
\begin{tabular}{ccc}
\hline
Event & Time $t$ & Temperature $T$ \\
\hline
Baryogenesis & ? & ? \\
EW Phase Transition & $2 \times 10^{-11}\,$ sec & $100\,$ GeV \\
QCD Phase Transition & $2 \times 10^{-5}$ & $150\,$ MeV \\
DM Freeze-Out & ? & ? \\
Neutrino Decoupling & $1\,$ sec & $1\,$ MeV \\
Electron-Positron annihilation & $6\,$ sec & $500 \,$ KeV \\
Big Bang Nucleosynthesis (BBN) & $3\,$ min & $100\,$ KeV \\
Matter-Radiation Equality & $60 \times 10^3\,$ yrs & $0.75 \,$ eV \\
Recombination & $260-380 \times 10^3\,$ yrs & $0.26-0.33\,$ eV \\
Photon Decoupling & $380 \times 10^3 \,$ yrs & $0.23-0.28\,$ eV \\
Dark energy-Matter equality & $9 \times 10^9\,$ yrs & $0.33 \,$ meV \\
Present & $13.8 \times 10^9 \,$ yrs & $0.24 \,$ meV \\
\hline
\end{tabular}
\end{center}
\caption{Key events in the thermal history of the universe.}
\label{tab:ThermalHistory}
\end{table}
\end{small}

\section{Beyond the Standard Paradigm}
\label{sec:BSM}

\subsection{Beyond the Standard Model of Particle Physics}
\label{ssec:BSM}

We start by listing all the problems featured by the Standard Model of particle physics:
\ben
\item \textbf{Observational issues}.
The Standard Model does not describe
\bi
\item gravity,
\item dark energy,
\item dark matter,
\item neutrino masses.
\ei
\item \textbf{Theoretical issues}.
\bi
\item \textit{Electroweak Hierarchy Problem -} This problem is two-fold: the first issue is to understand the physical origin of the hierarchy between the scales $M_{\rm EW}$ and $M_{\rm P}$, while the second issue is to understand how it is possible to keep this hierarchy stable against quantum corrections. In fact, as we mentioned in the introduction to the chapter, in general scalar fields are not protected by any symmetry, and then they can receive arbitrarily large quantum corrections to their mass. This is the case for the higgs field, which receives quadratic corrections of the form
\begin{align}
\label{eq:HiggsCorrections}
\delta m^2_{\rm Higgs} \simeq \frac{y_{\rm t}}{8 \pi^2} \Lambda_{\rm cutoff}^2 \,,
\end{align}
where $\Lambda_{\rm cutoff}$ is the energy scale at which physics beyond the SM appears, ranging from $M_{\rm EW}$ to $M_{\rm P}$, while $y_{\rm t} \sim 1$ is the Yukawa coupling of the largest correction due to the top quark. If $\Lambda_{\rm cutoff} \simeq M_{\rm P}$ the fine-tuning required in order to keep the higgs mass light is of the same order of the ratio $M_{\rm EW}/M_{\rm P} \sim 10^{-16}$.
\item \textit{Strong CP Problem -} As we mentioned in Sec.~\ref{ssec:StandardModel}, a term as in eq. \eqref{eq:ThetaTerm} can be added to the SM lagrangian, and the angle $\theta$ is severely constrained by the experimental bounds on the electric dipole moment of the neutron: $\theta < 10^{-10}$. The lack of a dynamical mechanism to this huge fine-tuning of $\theta$ is known as Strong CP Problem.
\item \textit{Gauge Coupling Unification - } The running of the Renormalization Group (RG) makes gauge coupling constants evolve in such a way that they tend to meet at a high scale. However, in the Standard Model the gauge coupling unification is not sufficiently precise to imply that it is not simply an accident.
\item \textit{Arbitrariness -} There are many arbitrary parameters in the lagrangian of the SM, whose values have to be inserted by hand, in order for the predictions to match with experimental data. These parameters include the number of families of the SM, the masses of fermions (excluding neutrinos) which span a range of five order of magnitude (from $m_e \simeq 0.5 \,$ MeV to $m_t \simeq 170 \,$ GeV.) and are intimately related to Yukawa couplings, CKM angles, neutrino masses and mixing angles. All these parameters deserve a deeper explanation in terms of dynamical mechanisms which fix their values to the physical ones. Analogously, also the higgs scalar potential is assumed to have the form given in eq. \eqref{eq:HiggsPotential} for no good reason, in order to produce the electroweak symmetry breaking.
\ei
\een

\subsubsection{Supersymmetry}
\label{sssec:SUSY}

Supersymmetry (SUSY) is a space-time symmetry which mixes bosonic and fermionic degrees of freedom of a theory. In its ten-dimensional version, supersymmetry is an essential ingredient of string theory, as we observe in Sec. \ref{sec:StringTheory}. Moreover, two-dimensional worldsheet supersymmetry is crucial in order for string theory to admit space-time fermions.\\

Four-dimensional global supersymmetry has been the main candidate for physics beyond the Standard Model for decades, mainly due to its ability to address the hierarchy problem~\cite{Nilles:1995ci, Martin:1997ns}. The simplest supersymmetric extension of the Standard Model of particle physics is the \textit{Minimal Supersymmetric Standard Model} (MSSM), whose chiral spectrum is summarized in Tab.~\ref{tab:MSSMSpectrum}. Essentially it provides a doubling of the degrees of freedom of the SM: to each SM particle corresponds a superpartner with different spin and equal mass. An exception is represented by the higgs sector, where anomalies cancellation requires an additional higgs doublet along with the superpartners. All the fields in Tab.~\ref{tab:MSSMSpectrum} can be accommodated in $\mathcal{N} = 1$ chiral supermultiplets. In addition to the fields in Tab.~\ref{tab:MSSMSpectrum} there are also vector multiplets composed of the gauge bosons $B, W^0, W^\pm, g^i$ and the respective \textit{gauginos} $\widetilde{B}, \widetilde{W}^0, \widetilde{W}^\pm, \widetilde{g}^i$, where $i = 1, \dots, 8$ since \textit{gluons} $g^i$ and \textit{gluinos} $\widetilde{g}^i$ transform in the adjoint representation $\mathbf{8}$ of $SU(3)_{\rm col}$. The fermionic fields $\widetilde{B}$ and $\widetilde{W}$ are called \textit{bino} and \textit{wino} respectively.\\

\renewcommand{\arraystretch}{1.6}
\begin{table}[tb]
\begin{center}
\begin{tabular}{|c|c|c|c|c|}
\hline
\multicolumn{2}{|c|}{Names} 
& spin 0 & spin 1/2 & $SU(3)_{\rm col} ,\, SU(2)_{\rm L} ,\, U(1)_{\rm Y}$
\\  \hline\hline
squarks, quarks & $Q^i$ & $({\widetilde u}^i_{\rm L}\>\>\>{\widetilde d}^i_{\rm L} )$&
 $(u^i_{\rm L}\>\>\>d^i_{\rm L})$ & $(\>{\bf 3},\>{\bf 2}\>,\>{\frac{1}{6}})$
\\
($\times 3$ families) & $\overline{u}^i$
&${\widetilde u}^{i \, *}_R$ & $u^{i \, \dagger}_R$ & 
$(\>{\bf \overline 3},\> {\bf 1},\> -{\frac{2}{3}})$
\\ & $\overline{d}^i$ &${\widetilde d}^{i \, *}_R$ & $d^{i \, \dagger}_R$ & 
$(\>{\bf \overline 3},\> {\bf 1},\> {\frac{1}{3}})$
\\  \hline
sleptons, leptons & $L^i$ &$({\widetilde \nu}^i\>\>{\widetilde e}^i_{\rm L} )$&
 $(\nu^i\>\>\>e^i_{\rm L})$ & $(\>{\bf 1},\>{\bf 2}\>,\>-{1\over 2})$
\\
($\times 3$ families) & $\overline{e}^i$
&${\widetilde e}^{i \, *}_R$ & $e^{i \, \dagger}_R$ & $(\>{\bf 1},\> {\bf 1},\>1)$
\\  \hline
higgs, higgsinos &$H_u$ &$(H_u^+\>\>\>H_u^0 )$&
$(\widetilde H_u^+ \>\>\> \widetilde H_u^0)$& 
$(\>{\bf 1},\>{\bf 2}\>,\>+{\frac{1}{2}})$
\\ &$H_d$ & $(H_d^0 \>\>\> H_d^-)$ & $(\widetilde H_d^0 \>\>\> \widetilde H_d^-)$& 
$(\>{\bf 1},\>{\bf 2}\>,\>-{\frac{1}{2}})$
\\  \hline
\end{tabular}
\caption{Chiral supermultiplets in the Minimal Supersymmetric Standard Model. The spin-$0$ fields are complex scalars, and the spin-$1/2$ fields are left-handed two-component Weyl fermions. The index $i$ runs over the three families. \label{tab:MSSMSpectrum}}
\vspace{-0.6cm}
\end{center}
\end{table}

The Yukawa couplings between matter fermions and the higgs field are given by the minimal phenomenologically viable superpotential
\begin{align}
\label{eq:MSSMSuperpotential}
W_{\rm MSSM} = \overline{u}^i Y^{ij}_U Q^j H_u - \overline{d}^i Y^{ij}_D Q^j H_d + \overline{e}^i Y^{ij}_{\rm L} L^j H_d + \mu H_u H_d \,,
\end{align}
so that the masses of leptons and quarks are determined as usual by the higgs mechanism. Eq. \eqref{eq:MSSMSuperpotential} contains also an explicit mass term for the higgs, which is called $\mu$-term and determines the mass of higgsinos. The $\mu$-term is required in order to get a phenomenologically satisfactory EW symmetry breaking. The superpotential in eq. \eqref{eq:MSSMSuperpotential} is not the most generic renormalizable and gauge invariant one: terms violating $B$ and $L$ have been excluded in order to avoid rapid proton decay. All the excluded terms violate an additional symmetry of the MSSM, called $R$-parity, under which SM particles are even and their superpartners are odd. $R$-parity can be interpreted as a discrete $\mathbb{Z}_2$ subgroup of the symmetry $U(1)_{B-L}$, which is spontaneously broken to $\mathbb{Z}_2$ by the VEV of a scalar with charge $2$ under $B-L$. $R$-parity has important consequences from a phenomenological point of view, since it implies that the \textit{Lightest Supersymmetric Particle} (LSP) is stable and consequently it could be a viable candidate for DM.\\

Supersymmetry must be spontaneously broken in nature, since a selectron with mass $m_{\tilde{e}} \simeq 0.5 \,$ MeV has not been observed. Regardless of the explicit mechanism of SUSY-breaking, its effects can be parametrized in terms of the \textit{soft-terms}, i.e. additional mass terms and cubic interactions~\cite{Chung:2003fi}
\begin{align}
\label{eq:MSSMSoftTerms}
\mathcal{L}_{\rm soft} =&\, \frac{1}{2} \left(\sum_a M_a \lambda_a \lambda_a + \text{c.c.}\right) - m_{ij}^2 C^i \overline{C}^j - \nonumber \\
& - \left(A_{ijk} Y_{ijk} C^i C^j C^k + B H_d H_u + \text{c.c.}\right) \,,
\end{align}
where $M_a$ are the masses of the gauginos of the MSSM, $m_{ij}$ are the masses of the scalar fields $C^i$ of the MSSM, $A_{ijk}$ are called $A$-terms and $B$ is called $B$-term. Moreover we defined $Y_{Kij} \equiv Y_K^{ij}$, where $K = U, D, L$. The lagrangian in eq. \eqref{eq:MSSMSoftTerms} explicitly breaks supersymmetry, since it introduces mass terms only for the superpartners of the SM particles. The minimal model of SUSY-breaking with universal soft-terms: $M_1 = M_2 = M_3 = M$, $m_{ij} \propto \delta_{ij}$, $A_{ijk} = A_i \delta_{jk}$, is called \textit{Constrained Minimal Supersymmetric Standard Model} (CMSSM) or also \textit{Minimal Supergravity Model} (mSUGRA).\\

There are many theoretical motivations to look at supersymmetry as a good candidate for the physics beyond the SM.
\bi
\item It potentially solves the hierarchy problem, since the divergent diagrams are exactly canceled by the analogous diagrams involving the superpartners, in the case of unbroken supersymmetry. When supersymmetry is spontaneously broken by soft-terms, the quadratic divergences are still eliminated, but there are non-vanishing logarithmic divergences. The hierarchy problem is addressed provided that the mass of squarks is not far beyond $1 \,$ TeV.
\item If $R$-parity is present, the LSP constitutes a good candidate for DM.
\item Provided that the masses of gauginos and higgsinos lie around the TeV scale, the MSSM provides a unification of the coupling constants at the GUT scale $M_{\rm GUT} = 2 \times 10^{16} \,$ GeV. Scalars do not contribute to gauge coupling unification since they enter the equations of the renormalization group for gauge couplings only at two-loops order.
\ei

It is worth recalling that the higgs potential in the MSSM takes the form
\begin{align}
\label{eq:MSSMHiggsPotential}
V_{\rm Higgs} =&\, \frac{1}{8} \left(g_1^2 + g_2^2\right) \left(\left|H_d^0\right|^2-\left|H_u^0\right|^2\right)^2 + \left(B H_d^0 H_u^0 + \text{h.c.}\right) + \nonumber \\
&+ \left(m_{H_d}^2 + |\mu|^2\right) \left|H_d^0\right|^2 + \left(m_{H_u}^2 + |\mu|^2\right) \left|H_u^0\right|^2  \,.
\end{align}
EWSB requires that a linear combination of $H_u^0$ and $H_d^0$ has negative squared mass in the origin $H_u^0 = H_d^0 = 0$. Requiring also that the scalar potential is bounded from below translates into the EWSB conditions\footnote{From the first condition in eq. \eqref{eq:MSSMEWSB} it is possible to infer the so-called \textit{$\mu$-problem}: $\mu$ is a supersymmetric quantity but in order to get a satisfactory SUSY-breaking it has to be of the same order of $m_{H_u}$, $m_{H_d}$, which are non-supersymmetric quantities.}
\begin{align}
\label{eq:MSSMEWSB}
\mu^2 = \frac{m_{H_d}^2 - m_{H_u}^2 \tan^2 \beta}{\tan^2 \beta - 1} - \frac{m_Z^2}{2} \,, \nonumber \\
\sin\left(2 \beta\right) = \frac{2 \left|B \mu\right|}{m_{H_d}^2 - m_{H_u}^2 + 2 \mu^2} \,,
\end{align}
where $\tan \beta$ is determined by the ratio between the VEVs of the two higgs doublets
\begin{align}
\label{eq:TanBeta}
\tan \beta = \frac{\langle H_u^0 \rangle}{\langle H_d^0 \rangle}\,.
\end{align}
The conditions in eq. \eqref{eq:MSSMEWSB} are not trivially satisfied by the scalar potential in eq. \eqref{eq:MSSMHiggsPotential}. However EWSB can be triggered by radiative corrections to $m_{H_u}^2$, which receives negative contributions from one-loop diagrams involving the top squark. Consequently also the higgs mechanism finds a dynamical explanation in the MSSM.\\

The EWSB can produce a mixing among the gauge eigenstates of the MSSM, whose superpartner mass spectrum is given by the
\bi
\item \textit{Gluinos -} $\tilde{g}^i$ transform in the adjoint representation $\mathbf{8}$ of $SU(3)_{\rm col}$, so they do not mix with any other particle of the MSSM.
\item \textit{Neutralinos -} Neutral higgsinos $\left(\widetilde{H}_u^0, \widetilde{H}_d^0\right)$ and neutral gauginos $\left(\widetilde{B}, \widetilde{Z}^0\right)$ combine to form neutral mass eigenstates called \textit{neutralinos} $\widetilde{N}_i$, where $i = 1, \dots, 4$. If the lightest neutralino is also the LSP, it can play the role of DM in supersymmetric theories which preserve $R$-parity.
\item \textit{Charginos -} Charged higgsinos $\left(\widetilde{H}_u^+, \widetilde{H}_d^-\right)$ and charged gauginos $\left(\widetilde{W}^+, \widetilde{W}^-\right)$ combine to form two mass eigenstates with charge $\pm 1$, called \textit{charginos} $\widetilde{C}^\pm_i$, where $i = 1, 2$.
\item \textit{Squarks and sleptons - } In principle all the scalars with the same electric charge, $R$-parity and color quantum numbers can mix with each other. Fortunately only squarks of the third generation actually mix, due to the large Yukawa coupling $y_t \sim 1$. It turns out that the mass eigenstates effectively coincide with gauge eigenstates for all scalars except the stop, whose left and right components $(\widetilde{t}_{\rm L}, \widetilde{t}_{\rm R})$ mix and give rise to the physical stops $(\widetilde{t}_1, \widetilde{t}_2)$.
\item \textit{Higgs fields -} Once EWSB has taken place, three out of the four higgs degrees of freedom become the longitudinal polarizations of the electroweak bosons $Z, W^\pm$. The remaining five degrees of freedom are: two CP-even neutral scalars $h^0, H^0$, one CP-odd neutral scalar $A^0$, two charged scalars $H^\pm$. $h^0$ turns out to be the lightest one, and then it plays the role of the SM higgs field. The tree-level masses are given by
\begin{align}
m^2_{A^0} &= 2 |\mu|^2 + m_{H_u}^2 + m_{H_d}^2 \,, \\
m_{h^0, H^0}^2 &= \frac{1}{2} \left(m_{A^0}^2 + m_Z^2 \mp \sqrt{\left(m_{A^0}^2 - m_Z^2\right)^2 + 4 m_Z^2 m_{A^0}^2 \sin^2 (2 \beta)}\right) \,,\\ 
m_{H^\pm}^2 &= m_{A^0}^2 + m_W^2 \,,
\end{align}
According to these expressions $m_{h^0} < m_{Z^0}$, but loop-corrections to the lightest higgs mass can raise its value up to $125 \,$ GeV.
\ei

\subsubsection{Supergravity}
\label{sssec:SUGRA}

Supersymmetry, as described in the previous section, is a global symmetry. Interestingly, as soon as we promote the parameters of a supersymmetry transformation to a space-time-dependent function, general coordinate transformations are included in the theory, which becomes a theory of gravity~\cite{Nilles:1983ge, Cerdeno:1998hs, Wess:1992cp}, and it is called ($\mathcal{N} = 1$) \textit{supergravity}. From a field theoretical point of view, the inclusion of gravity can be described by an additional multiplet, which is called \textit{gravity multiplet}. It contains the mediator of the gravitational interactions: a spin-two massless particle called graviton $g_{\mu \nu}$ and its superpartner, a spin-$3/2$ massless particle called \textit{gravitino} $\psi^\mu_\alpha$ (where $\alpha = 1, 2$ is a spinorial index). The gravitino plays the role of the gauge field of local supersymmetry.\\

A supergravity theory can be completely determined starting from the following three functions of the chiral superfields $\Phi^i$:
\ben
\item The \Kahler potential $K(\Phi^i, \overline{\Phi}^i)$: it is a real function which determines the kinetic terms of the chiral fields.
\item The holomorphic superpotential $W(\Phi^i)$: along with the \Kahler potential it determines the scalar potential of the supergravity theory.
\item The gauge kinetic functions $f_a(\Phi^i)$: they are holomorphic functions of the chiral superfields $\Phi^i$, which determine the kinetic terms of the gauge bosons and their couplings to axions.
\een

As for global supersymmetry, also the local one has to be spontaneously broken. Local supersymmetry breaking produces the analogous of the higgs mechanism: a scalar field in the hidden sector acquires a VEV which breaks the local supersymmetry. The fermionic superpartner (goldstino) of this scalar field is swallowed by the massless gravitino, which acquires mass. This local SUSY-breaking is called \textit{Super-Higgs mechanism}. From a phenomenological point of view it is interesting to notice that local SUSY-breaking automatically generates soft-terms in the global supersymmetric lagrangian, realizing the so-called \textit{gravity mediated SUSY-breaking}. Usually, the scale of SUSY-breaking is roughly given by the gravitino mass.\\

Unfortunately, even if local supersymmetry automatically includes gravity, it is neither a finite theory nor a renormalizable one, meaning that it does not provide an UV completion of General Relativity.

\subsubsection{GUT Theories}
\label{sssec:GUTs}

As we have seen in Sec.~\ref{sssec:SUSY}, in supersymmetric models gauge coupling unification is achieved at the scale $M_{\rm GUT}$. Another kind of models which provide gauge coupling unification are \textit{Grand Unified Theories} (GUT theories)\footnote{From which the subscript in the unification scale $M_{\rm GUT}$.}~\cite{Ross:1985ai}. They assume that the SM gauge group is actually contained in a larger simple group $G_{\rm GUT}$ which is spontaneously broken to $G_{\rm SM}$ around the GUT scale by a higgs-like mechanism. In this picture, at energies above $M_{\rm GUT}$ gauge couplings are unified into a single value: the gauge coupling associated to the GUT group. In the last decades different alternatives have been studied in the context of GUT theories, the most relevant ones being those with gauge groups $SU(5)$, $SO(10)$ or $E(6)$.\\

For example in the $SU(5)$ GUT, each SM fermionic generation fits into a reducible $SU(5)$ representation $\mathbf{5} + \overline{\mathbf{10}}$. It provides the existence of $24$ bosons, of which $12$ are the SM ones, while the remaining $12$ transform as $SU(2)_{\rm L}$ doublets and $SU(3)_{\rm col}$ triplets. The higgs sector is composed by a $24$ scalar fields transforming in the representation $\mathbf{24}$ of $SU(5)$. The VEVs acquired by some of these fields trigger the symmetry breaking which leads to the SM gauge group. An interesting point about GUT theories is that, since quarks and leptons fit into the same GUT multiplets, they can transform into each other by emission/absorption of a heavy gauge boson, giving rise to lepton/baryon number violation. These interactions are clearly suppressed by the mass of the heavy bosons, leading to a proton lifetime of about $\tau_{\text{proton}, SU(5)} \simeq 4 \times 10^{29} \,$ yrs, which is quite below the super-Kamiokande lower bound $\tau_{\rm proton} \simeq 10^{32} - 10^{33}$, so that the simplest $SU(5)$ GUT is ruled out.\\

GUT theories with different gauge groups have similar features, with increasingly more involved higgs sectors and symmetry breaking patterns. For example the breaking of $SO(10)$ to $G_{\rm SM}$ requires a ``GUT-Higgs'' transforming in the adjoint representation $\mathbf{45}$ of $SO(10)$, plus additional scalar fields transforming in the representation $\mathbf{16}$. Analogously, the lowest-dimensional non-trivial representation of $E(6)$ is the $\mathbf{27}$, which contains the representation $\mathbf{16}$ of $SO(10)$, since it decomposes as
\begin{align}
\label{eq:27GroupDecomposition}
\mathbf{27} = \mathbf{16} + \mathbf{10} + \mathbf{1} \,.
\end{align}
Symmetry breaking in $E(6)$ models is triggered by the VEVs of the ``GUT-Higgs'' field in the adjoint representation of $E(6)$ and of the scalar fields in the $\mathbf{27}$. A possible symmetry breaking pattern is given by
\begin{align}
\label{eq:E6Symmetrybreaking}
E(6) \quad \longrightarrow \quad SU(3)_{\rm col} \times SU(3)_{\rm L} \times SU(3)_{\rm R} \quad \longrightarrow \quad G_{\rm SM}\,,
\end{align}
where the intermediate gauge group, in which the QCD symmetry is treated on equal footing with left-right symmetries, is called \textit{trinification model}.\\

The most interesting GUT models are the supersymmetric versions, since they provide a much better agreement with the experimental values for the Weinberg angle and the proton lifetime.

\subsubsection{Further Alternatives}
\label{sssec:Alternatives}

We mention a couple of alternatives to supersymmetric theories:
\ben
\item \textit{Brane-world and large extra-dimensions - } The possibility that there exist more dimensions than those that we naively observe is a very old idea. In the '20s in fact, T. Kaluza and O. Klein attempted to unify gravity with electromagnetism by postulating the existence of a circular fifth dimension~\cite{Kaluza:1921tu, Klein:1926tv}, whose radius is small enough to escape the detection. The idea of extra-dimensions turns out to be completely natural in string theory, where there are six curled up extra-dimensions, which are usually assumed to be very small. There is a further alternative, according to which extra-dimensions are large enough to be in the detection range of the LHC, called \textit{Brane-World Scenario}. In this scenario the SM particles are confined in a four-dimensional subspace of a higher-dimensional space-time. Such a subspace is called \textit{brane}. On the other hand, gravitons can move freely in the full higher dimensional space-time, so that in this picture the weakness of the gravitational interaction with respect to the other forces can be explained by its dilution \cite{Witten:1996mz}. For example, assuming that there exist $n$ extra-dimensions compactified to circles with radii $R$, then the Planck scale can be written as
\begin{align}
\label{eq:PlanckScaleBraneWorld}
M_{\rm P}^2 = 8 \pi M^2 (2 \pi R)^n\,,
\end{align}
where $M$ is the $(n+4)$-dimensional Planck scale. The hierarchy problem can be addressed by assuming a large $R$, so that the value of $M$ is lowered to $1 \,$ TeV. Unfortunately the large value of $R$ has not a dynamical explanation in this scenario.

\item \textit{Strong dynamics - } A possible alternative solution to the hierarchy problem is by assuming that the higgs field is not an elementary particle, but a bound state resulting from some yet-unknown strong dynamics~\cite{Dimopoulos:1979qi}. In this picture the hierarchy between $M_{\rm EW}$ and $M_{\rm P}$ can be explained in the same way as the hierarchy in eq. \eqref{eq:ProtonVsPlanckMass} between the proton mass $m_{\rm p}$ and the Planck scale $M_{\rm P}$. Models of physics beyond the SM built following the idea that some strong dynamics is hiding behind the higgs boson usually go under the name of \textit{Technicolor Models}. Despite their attractiveness, these models do not seem to feature phenomenological viable patterns for fermion masses and mixings.
\een

\subsection{Beyond the Standard Model of Cosmology}
\label{ssec:BSMCosmology}

Despite its undeniable success, the Standard Model of cosmology contains many problems:
\ben
\item \textit{Initial singularity - } Tracing back the scale factor $a(t)$ using the Einstein's equation, in the Standard Model of cosmology it inevitably ends up in a singularity: all the energy of the universe is supposed to be concentrated in a single point at $t = 0$. This is a clear signal of the breakdown of General Relativity in the very early universe, when the energy was concentrated in such a small region that both quantum mechanics and gravity had important effects. This problem can only be cured within a quantum theory of gravity.
\item \textit{Horizon problem - } The high uniformity of the CMB temperature is unexplained within the Hot Big Bang Theory: tracing back the evolution of the universe just using General Relativity and the Standard Model, it is possible to check that at photon decoupling there were many patches that could have never been in causal contact.
\item \textit{CMB anisotropies - } As we discussed in Sec.~\ref{ssec:StandardModelCosmology} the uniformity of the CMB is astonishing: anisotropies in its temperature have a tiny size with respect to the background temperature $T_0$: $\frac{\delta T}{T_0} \sim 10^{-5}$. However, such deviations from uniformity can not be produced within the Standard Model of cosmology.
\item \textit{Flatness problem - } As we mentioned in Sec.~\ref{ssec:StandardModelCosmology}, the universe today appears to be extremely flat, with $\left|\Omega_{\rm curv}\right| < 0.005$. However, even a tiny curvature in the very early universe would have increased to large values at present time, so that a huge fine-tuning of order $\frac{\rho_{\rm curv}}{\rho_c}(t_{\rm Planck}) \sim 10^{-61}$ is required in order satisfy the current bounds.
\item \textit{Monopole problem - } According to GUT theories, topological defects such as monopoles have been created in the early universe. Since they are stable objects, they should be still present to date in such a quantity that they would dominate the energy density of the universe. However, experimental searches for topological defect have failed, setting stringent bounds on their energy density in the universe.
\een

Finally we should emphasize that the thermal cosmological history depicted in Sec.~\ref{ssec:StandardModelCosmology} is completely understood and experimentally tested only from the Big Bang Nucleosynthesis on. In particular
\bi
\item If there is a unification mechanism as in GUT theories, additional phase transitions may have occurred in the universe, provided that it was reheated to a sufficiently high temperature.
\item There are several mechanisms proposed for baryogenesis, but none of them has found an experimental validation. It is commonly accepted that baryogenesis happened at some stage after inflation, since otherwise the baryon asymmetry would have been smoothed out by the exponential expansion of the universe. In order to get baryon asymmetry the \textit{Sakharov conditions} have to be met~\cite{Sakharov:1967dj}: \textit{i)} baryon number violation, \textit{ii)} $C$ and $CP$ violation, \textit{iii)} thermal non-equilibrium (or CPT violation). Among the several mechanisms for baryogenesis we mention the \textit{electroweak mechanism}~\cite{Rubakov:2002fi, Kuzmin:1985mm, Kolb:1990vq}, which makes use of the non-perturbative baryon and lepton number violation in the SM, the \textit{GUT mechanism}, which exploits the baryon and lepton number violation intrinsic in GUT theories, and the \textit{Affleck-Dine mechanism}~\cite{Affleck:1984fy}, which utilizes the dynamics of MSSM flat directions to develop a baryon asymmetry. Interestingly, in \cite{ADB} the authors realized a string theory embedding of the Affleck-Dine mechanism. All the proposed mechanisms (including the electroweak one) make use of physics beyond the Standard Model.
\item It is commonly accepted that the decoupling of neutrinos at a temperature of about $1 \,$ MeV gave rise to a neutrino background similar to the CMB\footnote{Assuming that the universe was reheated at a temperature larger than $1\,$ MeV.}, but with a smaller temperature of about $T_{\nu, 0} \simeq 1.95 \,$ K. Unfortunately, it is very hard to observe the cosmic neutrino background, due to the weakness of neutrino interactions.
\ei

\subsubsection{Inflation}
\label{sssec:Inflation}

Problems 2., 3., 4., 5. in the previous list can all be cured at once by assuming that the universe underwent a period of exponential expansion at very early stages, when the age of the universe was about $t_{\rm inf} \sim 10^{-34} \,$ sec~\cite{Guth:1980zm, Linde:1981mu, Albrecht:1982wi}. Such period is called \textit{inflation}. Inflation explains the uniformity of CMB temperature since, in this picture, regions of the universe which apparently have never talked to each other, were in causal contact in the very early stages before the accelerated expansion. Inflation provides also a fascinating explanation to the origin of the anisotropies observed in the CMB~\cite{Mukhanov:1990me, Gorbunov:2011zzc}: according to the inflationary paradigm quantum fluctuations taking place in the universe during inflation have been stretched out by the accelerated expansion, resulting in the tiny anisotropies observed in the CMB temperature~\cite{Ade:2015xua, Ade:2015lrj}. Interestingly, also the presence of large scale structures in the universe can be explained as the result of the evolution of quantum fluctuations in the early universe under the action of inflation: a perfectly homogeneous and isotropic universe would not give rise to galaxies and stars. Finally, inflation addresses also the flatness and monopole problems, since the accelerated expansion would drive the universe towards flatness, diluting any possible unwanted relics like topological defects.\\

The simplest model of inflation requires the presence of a scalar field $\phi$, called \textit{inflaton}, minimally coupled to gravity, so that the action reads
\begin{align}
\label{eq:InflationLagrangian}
S_{\rm inf} = \int d^4x\, \sqrt{-g} \left[\frac{M_{\rm P}^2}{2} \, \mathcal{R} + \frac{1}{2} g^{\mu \nu} \partial_\mu \phi \partial_\nu \phi - V(\phi)\right]\,,
\end{align}
where $V(\phi)$ is the scalar potential. Assuming that the scalar field is homogeneous $\phi(\mathbf{x},t) \equiv \phi(t)$, it behaves like a perfect fluid whose equation of state can be written as
\begin{align}
\label{eq:EqStateInflaton}
w_\phi = \frac{p_\phi}{\rho_\phi} = \frac{\dot{\phi}/2 - V}{\dot{\phi}/2 + V}\,.
\end{align}
The equations of motion for the inflaton and for the scale factor read
\begin{align}
\label{eq:InflatonEqMotion}
&\ddot\phi + 3 H \dot\phi + \partial_\phi V = 0 \,, \nonumber \\
&H^2 = \frac{1}{3} \left(\frac{1}{2} \dot{\phi}^2 + V(\phi) \right) \,, \nonumber \\
& \frac{\ddot{a}}{a} = H^2 \left(1 - \frac{\dot{\phi}^2}{H^2}\right) \,.
\end{align}
From eq. \eqref{eq:EqStateInflaton} and eq. \eqref{eq:InflatonEqMotion} it can be inferred that a period of accelerated expansion can be sustained for a quite long time if
\begin{align}
\label{eq:SlowRollCondition}
\epsilon = \frac{M_{\rm P}^2}{2} \left(\frac{\partial_{\phi}V}{V}\right)^2 \ll 1 \,, \qquad \eta = M_{\rm P}^2 \frac{\partial^2_{\phi, \phi} V}{V} \ll 1 \,.
\end{align}
The conditions in eq. \eqref{eq:SlowRollCondition} are called \textit{slow-roll conditions}. In the slow-roll regime the Hubble parameter is almost constant, and the space-time is approximately de Sitter:
\begin{align}
\label{eq:deSitterSlowRoll}
H^2 \approx \frac{V}{3} \sim \text{const.} \,, \qquad a(t) \sim e^{H t} \,,
\end{align}
which corresponds to an exponential expansion.\\

As already mentioned, at the classical level inflation would produce a completely uniform universe, which is not what we observe. The formation of large scale structures and of the anisotropies of the CMB are consequences of the quantum fluctuations of the inflaton $\delta \phi$ and of the metric $\delta g_{\mu \nu}$ around the homogeneous background during inflation (labeled with a bar)
\begin{align}
\label{eq:FieldsFluctuations}
\phi(\mathbf{x}, t) = \overline{\phi}(t) + \delta \phi(\mathbf{x},t) \,, \qquad g_{\mu \nu}(\mathbf{x}, t) = \overline{g}_{\mu \nu} + \delta g_{\mu \nu}(\mathbf{x}, t) \,.
\end{align}
For practical purposes, after having eliminated redundant degrees of freedom, it turns out that there are three relevant perturbations: one scalar perturbation $\mathfrak{R}$ and two tensor perturbations $h = h^+, h^\times$\footnote{+ and $\times$ refer to the two polarizations of a gravitational wave.}. Two important statistical measures of the primordial fluctuations are the power spectra of $\mathfrak{R}$ and $h$, which can be defined as
\begin{align}
\label{eq:PowerSpectraDefinition}
&\Delta_s^2 = \frac{k^3}{2 \pi^2} P_{\mathfrak{R}}(k) \,, \qquad &\langle \mathfrak{R}_{\mathbf{k}} \mathfrak{R}_{\mathbf{k'}} \rangle = (2 \pi)^3 \delta \left(\mathbf{k} + \mathbf{k'}\right) P_{\mathfrak{R}}(k) \,, \\
&\Delta_t^2 = \frac{k^3}{\pi^2} P_h(k) \,, \qquad &\langle h_{\mathbf{k}} h_{\mathbf{k'}} \rangle = (2 \pi)^3 \delta \left(\mathbf{k} + \mathbf{k'}\right) P_h(k) \,,
\end{align}
where the subscript $\mathbf{k}$ denotes a mode of the Fourier expansion of $\mathfrak{R}$ and $h$, and $k = |\mathbf{k}|$. In the slow-roll approximation, the power spectra turn out to be~\cite{Maldacena:2002vr, Weinberg:2005vy}
\begin{align}
\label{eq:PowerSpectra}
\Delta_s^2(k) = \frac{H_*^2}{(2 \pi)^2} \frac{H_*^2}{\dot\phi_*^2} \,, \qquad \Delta_t^2(k) = \frac{8}{M_{\rm P}^2} \left(\frac{H_*}{2 \pi}\right)^2 \,,
\end{align}
where the subscript $*$ denotes that the quantities have been computed at the horizon exit of a given mode $k$: $k = a_* H_*$. Given that the amplitude of scalar perturbations has been measured to be $\Delta_s^2 \sim 10^{-9}$, it is possible to define the \textit{tensor-to-scalar ratio} $r$
\begin{align}
r = \frac{\Delta_t^2}{\Delta_s^2} \,,
\end{align}
in terms of which the energy scale of inflation reads
\begin{align}
\label{eq:EnergyScaleInflation}
V^{1/4} \simeq \left(\frac{r}{0.01}\right)^{1/4} \, 10^{16} \, \rm GeV \,.
\end{align}
For a detailed survey of the observed values of the inflationary parameters we refer to the Planck 2015 analysis~\cite{Ade:2015lrj}, where it is shown that the tensor-to-scalar ratio is bounded to be $r \lesssim 0.1$, while the scalar spectrum is nearly scale invariant, with a value for the spectral index $n_s$ close to unity
\begin{align}
n_s = 1 + \frac{d \ln \Delta_s^2}{d \ln k} \simeq 0.96\,.
\end{align}

We shall emphasize that inflation is still an hypothesis, whose predictions have been repeatedly confronted and found to be in agreement with cosmological observations. Among these we would like to recall the impressive agreement between the predicted angular power spectrum of CMB temperature fluctuations and the observed one. The most striking correspondence between observations and inflationary predictions is probably the fact that all the Fourier modes of perturbations generated during inflation have the same phase~\cite{Dodelson:2003ft, Baumann:2009ds}. Such observation is sufficient to rule out many competing alternatives to inflation.

\section{String Theory}
\label{sec:StringTheory}

All the alternatives for the physics beyond the Standard Model described so far have been inspired by the principle of naturalness and by the search for the unification of forces, but they do not provide a new conceptual framework in which gravity can be conciliated with quantum mechanics. To date, the most promising candidate as a theory of quantum gravity is \textit{string theory}. In a first approach string theory can be defined as a theory in which the elementary objects are not point-like as in particle physics, but rather they are one-dimensional objects with typical length $\ell_s$, sweeping out a two-dimensional surface, called the \textit{worldsheet} $\Sigma$. Since strings have not been observed in particle accelerators, it is necessary to assume that $\ell_s$ is smaller than the scales currently explored in experiments. Different patterns of vibration of the strings effectively produce the crowd of particles of the SM or extensions thereof.\\

Interestingly, the quantization of string theory produces the unavoidable presence of a spin-two particle in the massless spectrum~\cite{Scherk:1974ca}, which behaves like a graviton: gravity is naturally included in string theory at the quantum level. The reason why string theory works nicely as a theory of quantum gravity is intuitive: the typical length $\ell_s$ acts as an intrinsic cut-off for the theory, eliminating the divergences encountered in the straightforward quantization of General Relativity. Furthermore, although not completely clear at a first glance, string theory potentially contains also all the building blocks necessary to reproduce the SM in the low-energy limit, such as non-abelian gauge interactions, chiral fermions in three families, and so on. For this reason string theory is supposed to be a framework in which all the interactions can be finally unified into a single description. A further nice feature of string theory is that it does not contain any unknown external parameter except the string length $\ell_s$, so that all the SM parameters (like the number of generations, the Yukawa couplings, etc.) have to be determined by the dynamics of the theory itself. On the cosmological side, inflation can be naturally accommodated in the low-energy limit of string theory, since it contains many fundamental scalar fields which can play the role of the inflaton. More in detail, all the alternative scenarios for physics beyond the Standard Models of particle physics and cosmology described in Sec.~\ref{sec:BSM} can be potentially embedded in string theory as low-energy effective field theories. However, a honest statement about the state of the art is that a completely consistent model which reproduces all the experimentally established observations has not been found to date, although we much progress has been made in the last years.

\subsection{Basic Facts About Strings}
\label{ssec:BasicsOfStrings}

Strings naturally come in two types: open strings, with two free end-points, and closed strings, with no end-points. They respectively sweep out worldsheets with and without boundaries. The two-dimensional worldsheet $\Sigma$ swept out by strings can be locally parametrized by the coordinates $(\tau, \sigma)$. $\Sigma$ embeds in a generically $D$-dimensional space-time through a set of functions $X^M(\tau, \sigma)$, where $M = 0, \dots, D-1$. The dynamics of strings is described by the Polyakov action~\cite{Polchinski:1998rq, Green:1987sp}
\begin{align}
\label{eq:PolyakovAction}
S_{\rm P} = -\frac{1}{4 \pi \alpha'} \int_\Sigma d^2x \, \sqrt{- \det{g}}\, g^{ab}(\tau, \sigma) \partial_a X^M \partial_b X^N \eta_{MN} \,, \qquad a = 1,2 \,,
\end{align}
where $g_{ab}$ is a metric on $\Sigma$ and $\eta_{MN}$ is the Minkovski metric on the $D$-dimensional space-time, while $\ell_s = 2 \pi \sqrt{\alpha'}$. The Polyakov action in eq. \eqref{eq:PolyakovAction} features several symmetries:
\bi
\item global $D$-dimensional space-time Poincar\'e invariance,
\item invariance under two-dimensional local worldsheet coordinate re-parametriza-\\tion,
\item invariance under local rescaling of the two-dimensional metric (Weyl invariance).
\ei
Fixing the invariance under the local coordinate re-parametrization (light-cone gau-\\ge), it is possible to see that only oscillations orthogonal to the worldsheet are physical, so that the dynamics described by the Polyakov action reduces to that of $D-2$ two-dimensional free massless scalar fields. Solving the equations of motion for this system gives rise to a tower of infinite decoupled harmonic oscillators, which correspond to the oscillations of the string. Finally, the quantization of the system leads to the spectrum of space-time fields in string theory.\\

In the perturbative regime, space-time string interactions can be treated in a way similar to that used in particle physics. Scattering amplitudes between asymptotic states can be computed as a quantum path integral, summing over all the possible topologies of the worldsheet which interpolate between the asymptotic states:
\begin{align}
\label{eq:ScatteringAmplitude}
\bra{\text{out}} \text{evolution} \ket{\text{in}} = \sum_{\text{worldsheets}} \int [D X] e^{-S_{\rm P}[X]} \mathcal{O}_{\rm in} \mathcal{O}_{\rm out} \,,
\end{align}
where the \textit{vertex operators} $\mathcal{O}_{\rm in/out}$ encode the information about in and out asymptotic states. Similarly to what happens in particle physics, each ``Feynman diagram'' can be built by combining a set of basic string interactions vertices, each of which is weighted with a suitable power of the string coupling $g_s \ll 1$. Since string theory does not contain external unknown arbitrary parameters except $\ell_s$, $g_s$ has to be dynamically determined by the theory itself. Indeed, it is given by the VEV of a scalar field which is always present in the massless string spectrum, namely the \textit{dilaton} $\phi$:
\begin{align}	
\label{eq:StringCouplingConstant}
g_s = e^{\langle\phi\rangle}\,.
\end{align}
The perturbative expansion for closed strings is obtained by adding handles to the worldsheet, while the expansion for open strings is obtained by adding boundaries. Each ``Feynman diagram'' is weighted in the perturbative expansion by $g_s^{-\chi}$, where $\chi$ is the \textit{genus} of $\chi$ of $\Sigma$, namely
\begin{align}
\label{eq:Euler}
\chi = 2 - 2 g - n_b\,,
\end{align}
for a worldsheet with $g$ handles and $n_b$ boundaries. Given the structure of the basic interaction vertices, closed strings are always present in an interacting theory, while the presence of open strings is not mandatory.\\

Weyl invariance plays a central role in the dynamics of string theory. In fact, the requirement that this symmetry is not an anomalous one in the quantum theory determines the dimension of the space-time $D$. As soon as a non-trivial background is considered (as an example a metric different from Minkovski for the $D$-dimensional space-time), the theory described by the Polyakov action becomes an interacting one and it is not exactly solvable. However, it can be studied perturbatively around the free theory, and the parameter governing the perturbative expansion is $\alpha'/R^2$, where $R$ denotes the typical length-scale of variation of the space-time fields\footnote{Eventually $R$ boils down to be the typical size of the compact space made up of the extra-dimensions.}. As a consequence, string theory in a general background features a double expansion: the \textit{genus-expansion} controlled by $g_s$ and the $\alpha'$\textit{-expansion} governed by $\alpha'/R^2$.

\subsubsection{Spectrum}

In order to get fermions in the spectrum of string theory, it is necessary to consider a generalization of the Polyakov action, such that the two-dimensional theory is supersymmetric. Consequently new fermionic degrees of freedom $\psi^M$ are included as the superpartners of the bosonic fields $X^M$. In this picture, the requirement of Weyl anomaly cancellation fixes the dimension of the space-time to ten. Depending on the way of quantizing it, string theory leads to five different possibilities for the spectrum, whose massless components are summarized in Tab.s~\ref{tab:TypeIIspectrum} and~\ref{tab:HeteroticTypeISpectrum}. Typically, the spectrum is given by a tower of string states, whose spacing is determined by the unique external unknown parameter of the theory
\begin{align}
\label{eq:StringTower}
M_n^2 \simeq n M^2_s \equiv \frac{n}{\ell_s^2} \,,
\end{align}
where $M_s = 1/\ell_s$ is the \textit{string scale}. Since we are interested in the physics taking place at distances larger than the string length, we have to consider the effective field theory of massless string states, corresponding to energies $E \ll M_s$. The acronyms NS and R in Tab.s~\ref{tab:TypeIIspectrum} and~\ref{tab:HeteroticTypeISpectrum} refer to different possibilities for the choice of fermionic boundary conditions:
\ben
\item Neveu-Schwarz (NS): $\psi^i_{\rm L/R}(\tau + \sigma + \ell_s) = - \psi^i_{\rm L/R}(\tau + \sigma)$,
\item Ramond (R): $\psi^i_{\rm L/R}(\tau + \sigma + \ell_s) = \psi^i_{\rm L/R}(\tau + \sigma)$,
\een
where $\psi^i_{\rm L/R}$ denote respectively the the \textit{left-movers} and \textit{right-movers} fermionic degrees of freedom. Boundary conditions can be chosen independently for left- and right-movers, and the two sectors have to be glued together in a way that preserves the modular invariance of the partition function (this step is called GSO projection). Modular invariance is a crucial property of string theory, since it is the underlying mathematical structure which allows it to behave properly as a quantum theory of gravity. More in detail:
\bi
\item \textit{Type II string theory -} There are two Type II string theories: Type IIA and Type IIB. Both of them contain only closed strings, and they differ in the way of performing the GSO projection. Left and right sectors are treated on equal footing in Type II string theories: both sectors contain bosonic and fermionic degrees of freedom. The massless spectrum is divided into four sectors: NSNS, NSR, RNS and RR. The NSNS is common to Type IIA and Type IIB and contains the ten-dimensional metric $g_{MN}$, the dilaton $\phi$ and the two-form field $B_{MN}$. The RR sector contains form-fields, whose degrees are different in Type IIA and Type IIB. Both theories contain two gravitinos and dilatinos, which have opposite chiralities in Type IIA and the same chiralities in Type IIB. In the former case the massless spectrum features a ten-dimensional $\mathcal{N} = (1,1)$ supersymmetry, while in the latter case the massless spectrum features a ten-dimensional $\mathcal{N} = (2,0)$ supersymmetry. Massless spectra coincide with those of the ten-dimensional Type IIA and Type IIB supergravity theories.
\item \textit{Heterotic theories - } Heterotic theories contain only closed strings as well, but in this case left and right sectors are not treated on equal footing: the left sector contains only bosonic degrees of freedom, while the right one contains both bosonic and fermionic degrees of freedom. Due to the GSO projection, it turns out that there are two consistent ways to get a modular invariant partition function: in the first case the field $A^a_M$ in Tab.~\ref{tab:HeteroticTypeISpectrum} is the vector potential of a $SO(32)$ gauge theory, while in the second case it is the vector potential of a $E_8 \times E_8$ gauge theory. $\lambda^a_{\dot{\alpha}}$ is the corresponding gaugino. Both $A^a_M$ and $\lambda^a_{\dot{\alpha}}$ transform in the adjoint representation of the gauge group. The massless spectrum contains a single gravitino $\psi_{M\, \alpha}$ and dilatino $\lambda_\alpha$, corresponding to ten-dimensional $\mathcal{N} = (1,0)$ supersymmetry.
\item \textit{Type I theory - } Type I theory is a theory of unoriented open and closed strings. Unorientedness is crucial in order to get a consistent theory, since otherwise Type I strings would suffer from RR tadpoles. The field $A_M$ reported in Tab.~\ref{tab:HeteroticTypeISpectrum} is the vector potential of a $SO(32)$ gauge theory, coming from the open sector. $\lambda^a_{\dot \alpha}$ is the corresponding gaugino. Both $A^a_M$ and $\lambda^a_{\dot{\alpha}}$ transform in the adjoint representation of the gauge group. The massless spectrum contains a single gravitino $\psi_{M\, \alpha}$ and dilatino $\lambda_\alpha$, and eventually the theory features a ten-dimensional $\mathcal{N} = (1,0)$ supersymmetry, corresponding to Type I supergravity.
\ei

\renewcommand{\arraystretch}{1.4}
\begin{table}[tb]
\begin{center}
\begin{tabular}{cc}
\multicolumn{2}{c}{Type IIA} \\
\hline
Sector & $10$-dimensional fields \\
\hline
NSNS & $\phi$, $B_{MN}$, $g_{MN}$ \\
NSR & $\lambda^1_{\dot{\alpha}}$, $\psi^1_{M \, \dot{\alpha}}$ \\
RNS & $\lambda^2_{\alpha}$, $\psi^2_{M \, \alpha}$ \\
RR & $C_M$, $C_{MNP}$ \\
\hline
\end{tabular} \qquad 
\begin{tabular}{cc}
\multicolumn{2}{c}{Type IIB} \\
\hline
Sector & $10$-dimensional fields \\
\hline
NSNS & $\phi$, $B_{MN}$, $g_{MN}$ \\
NSR & $\lambda^1_{\alpha}$, $\psi^1_{M \, \alpha}$ \\
RNS & $\lambda^2_{\alpha}$, $\psi^2_{M \, \alpha}$ \\
RR & $C_0$, $C_{MN}$, $C_{MNPQ}$ \\
\hline
\end{tabular}
\caption{Type IIA and Type IIB massless spectrum.\label{tab:TypeIIspectrum}}
\vspace{-0.6cm}
\end{center}
\end{table}

\renewcommand{\arraystretch}{1.4}
\begin{table}[tb]
\begin{center}
\begin{tabular}{cc}
\multicolumn{2}{c}{Heterotic} \\
\hline
Sector & $10$-dimensional fields \\
\hline
NS & $\phi$, $B_{MN}$, $g_{MN}$ \\
R & $\lambda_{\alpha}$, $\psi_{M \, \alpha}$ \\
NS & $A_M$ \\
R & $\lambda_{\dot\alpha}$\\
\hline
\end{tabular} \qquad 
\begin{tabular}{cc}
\multicolumn{2}{c}{Type I} \\
\hline
Sector & $10$-dimensional fields \\
\hline
Closed spectrum &  \\
NSNS & $\phi$, $g_{MN}$ \\
NSR + RNS & $\lambda_{\alpha}$, $\psi_{M \, \alpha}$ \\
RR & $C_{MN}$ \\
\hline
Open spectrum & \\
NS & $A_M$ \\
R & $\lambda_{\dot\alpha}$\\
\hline
\end{tabular}
\caption{Heterotic and Type I massless spectrum.\label{tab:HeteroticTypeISpectrum}}
\vspace{-0.6cm}
\end{center}
\end{table}

\subsection{String Phenomenology}
\label{ssec:StringPhenomenology}

In spite of the fact that string theory is supposed to be a fundamental theory of nature providing a framework in which all the interactions are unified, during the last forty years its study has led to many developments in several areas of physics, and its application goes well beyond that expected from a theory of quantum gravity. The birth of AdS/CFT in 1997 for instance~\cite{Maldacena:1997re, Aharony:1999ti} has made string theory a useful tool to explore corners of physics which apparently are completely disconnected from quantum gravity. For example the AdS/CFT correspondence has applications in condensed matter physics~\cite{Sachdev:2010ch} and in the physics of strong interactions at high temperatures and densities~\cite{Sadeghi:2013zma}. Furthermore, the study of string theory has led to important developments also in mathematics, the most important of which is the discovery of mirror symmetry~\cite{Candelas:1989hd, Greene:1990ud}. Remarkably, using string theory, Strominger and Vafa were able to exactly reproduce the known result for the Bekenstein-Hawking black-hole entropy~\cite{Strominger:1996sh}, giving support to the idea that it is the correct microscopic description of gravity. Consequently, string theory can be fairly described using prof. M. Greene's words: to date string theory is not simply a ``theory of string-like elementary particles'', but rather a

\begin{center}
\textit{magnificent theoretical framework that interrelates a very wide range of topics in physics and mathematics\footnote{Citation taken from~\cite{ConlonBook}.}.}
\end{center}

Nevertheless, string theory is primarily supposed to be a fundamental theory of nature, which is able to unify all the interactions at the quantum level. In order to study the truthfulness of this claim, it is necessary to connect the ten-dimensional picture to the four-dimensional one, which is the goal of the branch of string theory usually referred to as \textit{string phenomenology}~\cite{Ibanez:2012zz}. This is done by assuming that six out of the ten dimensions are curled up to form a compact space, whose typical size is smaller than the distances currently explored in the particle accelerators, so that extra-dimensions escaped the detection so far. As it is intuitive, the four-dimensional physics depends on the details of the geometry and of the topology of the compact space, but unfortunately there are an infinite number of possible choices for it. Given that a selection principle for the compact space is missing so far, it is very difficult to get predictions from string theory. Phenomenological requirements like four-dimensional supersymmetry can help to restrict the choice to a certain class of compact spaces, called \textit{Calabi-Yau} manifolds~\cite{Greene:1996cy, Hubsch:1992nu}, but still there is a huge number of spaces of this kind. Connecting string theory to the real world then seems a quite difficult task.\\

Historically, Heterotic theories have been the first ones to be studied with the intention of connecting the ten-dimensional picture to the real world~\cite{Candelas:1985en, Donagi:2004ia, Greene:1986bm}. This is because they contain non-abelian gauge symmetries in their massless spectrum, and upon compactification they can give rise to a four-dimensional chiral theory. On the other hand Type II theories did not appear that promising at first, since they neither feature a non-abelian sector in their massless spectrum, nor they give rise to chiral theories in four dimensions. Nevertheless, the search for the right compactification space is quite hard in Heterotic theories, since SM fields arise directly from the closed massless string spectrum, whose fields inhabit the whole compact space and depend on its details~\cite{Derendinger:1985kk, Dine:1985rz, Burgess:1995aa, Gukov:2003cy, deWit:1986mwo}. This way of dealing with string phenomenology is usually called \textit{top-down approach}.\\

With the advent of D-branes in 1995~\cite{Polchinski:1995mt}, it was realized that four-dimensional chiral theories can also be obtained in Type II string theories. Indeed, SM fields can arise from the open string sector which is supported by D-branes~\cite{Angelantonj:2002ct}. D-branes also constitute a source for RR fluxes, so that they allow the field strengths for the form-fields to acquire non-vanishing VEVs. Furthermore, with the introduction of D-branes, there is also a major advantage from a technical point of view: since they can be located in small regions of the compact space, interesting physical quantities of the SM arising from them should be independent of the details of the whole compact space. It is then possible to locally build a D-branes configuration which gives rise to a visible sector with the desired phenomenological properties, and eventually embed it into a global compact space in a consistent way~\cite{Blumenhagen:2005mu, Giddings:2001yu, Blumenhagen:2008zz, Verlinde:2005jr, Aldazabal:2000sa}. This is the approach to string phenomenology used in the present thesis, and it is usually referred to as \textit{bottom-up approach}.\\

Given the difficulty of choosing a precise compact space, the best approach to string phenomenology is to look for physical effects which are shared by large classes of compactifications. For instance, regardless of the specific choice of the compact space, all string compactifications come with a large number of gravitationally coupled scalar fields, called \textit{moduli}, which are singlets under the SM gauge group. Such fields are the four-dimensional manifestation of the existence of extra-dimensions. Depending on their masses moduli can play a significant role in the cosmological evolution of the universe, and can give rise to observable effects. It is important to remark that the observation of any of these effects \textit{would not prove the correctness of string theory}. This in fact can only be established either by the direct observation of strings or by the observation of a physical effect which can arise exclusively by string theory. Nevertheless, the (non-)observation of physical effects due to moduli can help us to understand whether a given class of models is representative of the physical reality or not, and to drive us towards the right way in the never-ending search for a fundamental theory of nature. On the theoretical side, string theory provides the unique available framework in which one can explicitly test if a given model describing some aspects of nature can be consistently embedded in a quantum theory of gravity.

\subsubsection{Effective Action for Type II Closed Superstrings}
\label{sssec:EffectiveActionsTypeII}

As we discussed in the previous section, in the present thesis we use the bottom-up approach to string phenomenology, hence we are interested in Type II string theories. In this section we review some basic facts about their low-energy limit, showing explicitly the effective actions for the closed massless fields in Tab.~\ref{tab:TypeIIspectrum}.\\

In the string frame the bosonic effective action for massless states of Type II string theories takes the form
\begin{align}
\label{eq:EffectiveAction}
S_{\rm IIB} = \frac{1}{2 \kappa_{10}^2} \int d^{10}x\,\sqrt{-g} \left[\La_{\rm NSNS} + \La_{\rm RR} + \La_{CS}\right] + S_{\rm loc}\,,
\end{align}
where the term $S_{\rm loc}$ takes into account possible local sources, and $\kappa_{10}^2 = \frac{\ell_s^8}{4 \pi}$. The NSNS term is given by
\begin{align}
\label{eq:EffectiveActionNSNS}
\La_{\rm NSNS} = e^{-2 \phi} \left[R + 4 \partial_M \phi \partial^M \phi - \frac{1}{12} |H_3|^2\right]\,,
\end{align}
where the field strength of the Kalb-Ramond field
\begin{align}
\label{eq:H3Definition}
H_3 = dB_2 \,,
\end{align}
is a three-form. The RR piece takes the form
\begin{align}
\label{eq:EffectiveActionRR}
\La_{\rm RR} = - \frac{1}{2} \sum_p \frac{1}{p!} |F_p|^2 \,,
\end{align}
where $p = 0, 2, 4$ in Type IIA, while $p = 1, 3, 5$ in Type IIB. $F_p$ are the field strengths (also called \textit{fluxes}) of the RR forms, defined as
\begin{align}
\label{eq:FluxesFirstConvention}
F_p = \hat{F}_p - H_3 \wedge C_{p-3} \,,
\end{align}
where $\hat{F}_p = dC_{p-1}$. In Type IIB $F_5$ is constrained to be self-dual:
\begin{align}
\label{eq:F5SelfDuality}
F_5 = *_{10} F_5 \,.
\end{align}
Finally the Chern-Simons term can be written as
\begin{align}
\label{eq:CSAction}
\La_{\rm CS, IIA} = -\frac{1}{2} B_2 \wedge F_4 \wedge F_4 \,,\\
\La_{\rm CS, IIB} = -\frac{1}{2} C_4 \wedge H_3 \wedge F_3 \,.
\end{align}

It is useful to introduce an alternative convention for the RR fields, called the \textit{democratic formulation}~\cite{Bergshoeff:2001pv}. This convention consists in replacing eq. \eqref{eq:EffectiveActionRR} with
\begin{align}
\label{eq:EffectiveActionRRDemocratic}
\La_{\rm RR} = \frac{1}{2} \sum_p |F_p|^2\,,
\end{align}
where $p = 0, 2, 4, 6, 8, 10$ in Type IIA, and $p = 1, 3, 5, 7, 9$ in Type IIB. In order to halve the propagating degrees of freedom, it is necessary to impose a self-duality constraint:
\begin{align}
\label{eq:RRSelfDuality}
F_p = \left(-1\right)^{\rm{Int}\left(\frac{p}{2}\right)} *_{10} F_{10-p} \,,
\end{align}
which automatically includes the constraint on $F_5$ in eq. \eqref{eq:F5SelfDuality}.\\

Gauge transformations of form-fields, which leave the fluxes invariant, can be written as
\begin{align}
\label{eq:GaugeTransformations}
B_2 \rightarrow B'_2 = B_2 + d \lambda_1 \,, \quad C_p \rightarrow C^{'}_p = C_p + d \lambda_{p-1} - H_3 \wedge \lambda_{p-3} \,,
\end{align}
where $\lambda_k$ denotes a $k$-form. Equations of motion and Bianchi identities for fluxes are reported in Sec.~\ref{ssec:BasicFluxes}.

\chapter{Flux Compactifications and Model Building}
\label{chap:Compactifications&ModelBuilding}

As we have pointed out in Sec.~\ref{ssec:BasicsOfStrings}, in the present thesis we use the bottom-up approach to string phenomenology. We consider Type II String Theories, in which SM gauge interactions take place on D-branes localized in a small region of the compact space~\cite{Ibanez:2012zz, Aldazabal:2000sa}. Roughly, in these constructions it is possible to separate global issues (i.e. moduli stabilization) from local ones (i.e. the search for a D-branes configuration reproducing the MSSM or proper extensions thereof) as we will describe more accurately in Sec.~\ref{ssec:VisibleSector}.\\

We have already observed in Sec.~\ref{ssec:BasicsOfStrings} that the quantization of string theory gives rise naturally to the string scale, defined as
\begin{align}
M_s = \frac{1}{\ell_s} \,,
\end{align}
which is the only external unknown parameter of string theory and determines the spacing of the tower of string states. In order to connect string theory with the real world, we are interested in the physics taking place at energies below the string scale (or equivalently, at distances larger than the string length $\ell_s$)
\begin{align}
\label{eq:SUGRAApproximation}
E \ll M_s\,,
\end{align}
so that only massless modes of the string spectrum are excited. As pointed out in Sec.~\ref{ssec:BasicsOfStrings}, massless fields of Type II string spectra correspond to Type II Supergravity spectra, so that eq.~\eqref{eq:SUGRAApproximation} is called \textit{supergravity approximation}. Supergravity spectra and the corresponding effective actions of Sec.~\ref{ssec:StringPhenomenology} are the starting point for the compactification procedure.\\

Given the nice features of $\mathcal{N} = 1$ four-dimensional supersymmetry that we reviewed in Sec.~\ref{sssec:SUSY}, we look for a class of compactifications which preserve $\mathcal{N} = 1$ in four dimensions. In the first part of the present chapter, Sec.~\ref{sec:FluxCompactifications}, we briefly review the properties that such a compact space has to satisfy, starting from the most generic possibility and showing how phenomenological arguments along with our poor computational skills restrict the choice of the internal space to a very specific kind of manifolds. We conclude the section by reporting the effective action for the four-dimensional effective theory which arises from the compactification. In the second part of the chapter, Sec.~\ref{sec:ModelBuilding}, we recall basic notions about D3-branes at singularities, which represent the D-branes setup used in the present thesis to embed the visible sector into string compactifications. We conclude the section with an explicit example of a consistent string compactification.

\section{Flux Compactifications}
\label{sec:FluxCompactifications}

The first phenomenological requirement that we ask for is the maximal symmetry of the four-dimensional space, which allows us to write the most general expression for the ten-dimensional metric in the form
\begin{align}
\label{eq:GeneralMetric}
ds^2 = e^{2 A(y)} \tilde{g}_{\mu \nu} dx^\mu dx^\nu + g_{mn} dx^m dx^n \,,
\end{align}
where $e^{2 A(y)}$ is the \textit{warp-factor}, which depends only on the internal coordinates collectively denoted by $y$. $\tilde{g}_{\mu \nu}$ is a maximally symmetric four-dimensional metric and $g_{mn}$ is a six-dimensional metric on the internal space. Maximal symmetry of the four-dimensional space restricts the choice of $\tilde{g}_{\mu \nu}$ to a Minkovski, dS$_4$ or AdS$_4$ metric. Greek indexes $\mu, \nu$ refer to four-dimensional coordinates, while italic indexes $m, n$ refer to internal coordinates.\\

In order not to break the maximal symmetry of the four-dimensional space, the VEVs of all the fermionic fields have to vanish. Furthermore, in order to have unbroken supersymmetry in four dimensions, it is also required that the VEVs of the supersymmetric variations of the fermionic fields vanish. In Type II theories there are two gravitinos $\psi_M^A$ ($A = 1,2$) and two dilatinos $\lambda^A$ ($A = 1,2$), where uppercase italic indexes $M, N$ refer to ten-dimensional coordinates. Given these phenomenological requirements, two possibilities can be explored. In Sec.~\ref{ssec:CYCompactifications} we review the simplest one of them, leaving the analysis of the most generic possibility to the subsequent sections.

\subsection{Calabi-Yau Compactifications}
\label{ssec:CYCompactifications}

In the simplest case, only the metric has a non-trivial background along the internal space $\chi$~\cite{Candelas:1985en}. Demanding vanishing VEVs for the supersymmetric variations of the gravitinos and the dilatino leads to
\begin{align}
\label{eq:SUSYconditionFluxlessCase}
\langle \delta \psi_M^A \rangle = \nabla_M \epsilon^A\,, \qquad \langle \delta \lambda^A \rangle = \slashed{\partial} \phi \, \epsilon^A \,,
\end{align}
where the slash denotes a contraction with a ten-dimensional gamma matrix: $\slashed{\partial} = \partial_M \Gamma^M$\footnote{$\Gamma$ denotes the ten-dimensional chirality matrix. For the expressions involving spinors in higher dimensions we refer to Appendix B of~\cite{Polchinski:1998rr}.}, while $\epsilon^A$ are the parameters of the supersymmetric transformation.\\

It is a well known result that Type II string compactifications with a non-trivial background only for the metric along the internal manifold lead to an unwarped Minkovski four-dimensional space, with a Calabi-Yau manifold as internal space \cite{Candelas:1985en}. Indeed, the first condition in eq. \eqref{eq:SUSYconditionFluxlessCase} translates into the requirement of the existence of two covariantly constant spinors on the ten-dimensional space-time
\begin{align}
\label{eq:CovariantlyConstantSpinor}	
\nabla_M \epsilon^A = 0 \,,
\end{align}
whose four-dimensional component reads
\begin{align}
\label{eq:UnwarpedCYFluxlessCase}
k + \nabla_m A \nabla^m A = 0\,.
\end{align}
The solution of eq. \eqref{eq:UnwarpedCYFluxlessCase} on a compact space is $A = 0$, which implies a vanishing four-dimensional curvature $k = 0$. On the other hand, in order to study the constraints arising from the internal components of eq. \eqref{eq:CovariantlyConstantSpinor}, it is necessary to decompose $\epsilon^A$ as
\begin{align}
\label{eq:SpinorDecompositionBareIIA}
\epsilon^1_{\text{IIA}} = \xi^1_+ \otimes \eta_+ \oplus \xi^1_- \otimes \eta_- \,, \nonumber \\
\epsilon^2_{\text{IIA}} = \xi^2_+ \otimes \eta_- \oplus \xi^2_- \otimes \eta_+\,,
\end{align}
in the Type IIA case, such that $\Gamma \epsilon^1_{\text{IIA}} = \epsilon^1_{\text{IIA}}$ and $\Gamma \epsilon^2_{\text{IIA}} =- \epsilon^2_{\text{IIA}}$, where $\xi^{1,2}_- = \left(\xi^{1,2}_+\right)^*$ and $\eta_- = \left(\eta_+\right)^*$. $\eta$ is a six-dimensional Weyl spinor, while $\xi$ is a four-dimensional Weyl spinor. We denote both the chiralities under $\Gamma_5$ and under $\gamma$ by the subscripts $\pm$: $\Gamma_5 \xi^A_\pm = \pm \xi_\pm$ and $\gamma \eta_\pm = \pm \eta_\pm$\footnote{$\Gamma_5$ and $\gamma$ are the chirality matrices respectively in four (space-time) and six (internal) dimensions.}. In the Type IIB case instead
\begin{align}
\label{eq:SpinorDecompositionBareIIB}
\epsilon^A_{\text{IIB}} = \xi^A_+ \otimes \eta_+ \oplus \xi^A_- \otimes \eta_-\,, \qquad A= 1,2\,,
\end{align}
so that $\Gamma \epsilon^A_{\text{IIB}} = \epsilon^A_{\text{IIB}}$.\\

As a consequence the internal component of eq. \eqref{eq:CovariantlyConstantSpinor} can be written as
\begin{align}
\label{eq:CYCondition}
\nabla_m \eta_\pm = 0 \,,
\end{align}
which implies that the internal manifold $\chi$ has to admit a covariantly constant spinor $\eta$. This constraint can be rephrased into the condition that $\chi$ must have $SU(3)$ holonomy, or equivalently, that it has to be Ricci-flat: $\mathcal{R}_{i \overline{j}} = 0$. Such manifolds are called \textit{Calabi-Yau} (CY) spaces \cite{Candelas:1985en}.\\

CY manifolds\footnote{In this thesis we always deal with internal manifolds of complex dimension $3$.} are both complex and \Kahler\footnote{A CY manifold is also symplectic.}, so that we can always choose a complex basis $z^i$ ($i = 1,2,3$) in which the metric has only mixed components. The \Kahler form is defined as
\begin{align}
\label{eq:KahlerForm}
J = i g_{i \overline{j}} dz^i \wedge d\overline{z}^{\overline{j}} \,,
\end{align}
and it is closed: $dJ = 0$. Moreover, a CY manifold possesses a unique closed $(3,0)$-form, which we denote by $\Omega$. In general a CY space can be defined as a \Kahler manifold which admits both a closed, globally defined and non-vanishing $(3,0)$-form $\Omega$ and a closed, globally defined and non-vanishing $(1,1)$-form $J$. These can be written also in terms of the covariantly constant spinor $\eta$ which satisfies eq. \eqref{eq:CYCondition}:
\begin{align}
\label{eq:JandOmegaEta}
J_{ij} = \mp 2i \eta^\dagger_\pm \Gamma_{ij} \eta_\pm\,, \qquad \Omega_{ijk} = -2i \eta^\dagger_- \Gamma_{ijk} \eta_+\,.
\end{align}
The entire cohomology structure can be summarized in the Hodge diamond for CY manifolds, which takes the form
\begin{small}
\begin{align}
\begin{array}{ccccccc}
&  &  & h^{(0,0)} &  &  &  \\
&  & h^{(1,0)} &  & h^{(0,1)} &  &  \\
& h^{(2,0)} &  & h^{(1,1)} &  & h^{(0,2)} &  \\
h^{(3,0)} &  & h^{(2,1)} &  & h^{(1,2)} &  & h^{(0,3)} \\
& h^{(3,1)} &  & h^{(2,2)} &  & h^{(1,3)} &  \\
&  & h^{(3,2)} &  & h^{(2,3)} &  &  \\
&  &  & h^{(3,3)} &  &  &
\end{array}
=
\begin{array}{ccccccc}
&  &  & 1 &  &  &  \\
&  & 0 &  & 0 &  &  \\
& 0 &  & h^{(1,1)} &  & 0 &  \\
1 &  & h^{(1,2)} &  & h^{(1,2)} &  & 1 \\
& 0 &  & h^{(1,1)} &  & 0 &  \\
&  & 0 &  & 0 &  &  \\
&  &  & 1 &  &  &
\end{array}
\label{eq:HodgeDiamond}
\end{align}
\end{small}
The first observation is that on a CY manifold the only undetermined $h^{(i,j)}$ are $h^{(1,1)}$ and $h^{(1,2)}$. This feature considerably simplifies the compactification procedure. Since there are not harmonic one-forms and harmonic five-forms, then $J \wedge \Omega = 0$. Furthermore, since $h^{(3,0)} = 1$ there is a unique $\Omega$, and then $\Omega \wedge \overline{\Omega}$ and $J \wedge J \wedge J$ have to be proportional:
\begin{align}
\label{eq:ConventionJOmega}
J \wedge J \wedge J = \frac{3 i}{4} \Omega \wedge \overline{\Omega}\,.
\end{align}

Unfortunately CY compactifications feature the presence of many massless scalar fields in the low-energy spectrum, which are called moduli and span a manifold called \textit{moduli space}. They are related to the possible deformations of the size and of the shape of the sub-manifolds of the internal space which do not cost any energy, and manifest themselves in the EFT as flat directions of the scalar potential. In order to infer the structure of the moduli space it is necessary to look for deformations $h_{ij}$ of the background metric $\langle g_{i \overline{j}} \rangle$ which preserve the Ricci-flat condition
\begin{align}
\label{eq:DeformationRicciFlat}
\mathcal{R}(\langle g_{i \overline{j}} \rangle + h_{ij}) = 0 \,.
\end{align}
Since we are not interested in deformations which can be re-absorbed through a general coordinate transformation, it is necessary to fix the gauge, for instance by imposing $\nabla^i h_{ij} = 0$. Such deformations can be divided into two categories:	
\begin{enumerate}
\item Deformations with mixed indexes $h_{i \overline{j}}$: eq. \eqref{eq:DeformationRicciFlat} implies that $\Delta h_{i \overline{j}} = 0$, namely $h_{i \overline{j}}$ is a harmonic $(1,1)$-form. The number of linearly independent deformations with mixed indexes then is counted by the Hodge number $h^{(1,1)}$, and they correspond to different choices of the \Kahler class. It is possible to expand the deformation
\begin{align}
\label{eq:KahlerModuli}
h_{i \overline{j}} = i \sum_{I = 1}^{h^{(1,1)}} t^I(x) \left(\omega_I\right)_{i \overline{j}} \,,
\end{align}
where $\omega_I$ is a basis of harmonic $(1,1)$-forms. The fields $t^I(x)$ are real scalar fields, called \textit{\Kahler moduli} and they depend only on the four-dimensional coordinates. In order to ensure that the resulting metric is positive definite the following conditions have to be imposed
\begin{align}
\label{eq:KahlerConeCondition}
\int_{\gamma \in \chi} J > 0 \,, \quad \int_{\Sigma \in \chi} J \wedge J > 0 \,, \quad \int_{\chi} J \wedge J \wedge J > 0 \,,
\end{align}
where $\gamma$ denotes any curve in $\chi$ and $\Sigma$ denotes any surface in $\chi$. The conditions in eq. \eqref{eq:KahlerConeCondition} define a cone in the linear space spanned by $t^I$, which is called \textit{\Kahler cone} or \textit{\Kahler moduli space}.
\item Deformations with non-mixed indexes $h_{ij}$: eq. \eqref{eq:DeformationRicciFlat} implies that $\Delta h_{ij} = 0$, which means that $h_{ij}$ is a harmonic $(2,0)$-form. Since CY manifolds feature $h^{(2,0)} = 0$, it is useful to relate this kind of deformations to $(1,2)$-forms as follows
\begin{align}
\label{eq:ComplexStructureDeformations}
h_{ij} = \sum_{A = 1}^{h^{(1,2)}} \frac{i}{||\Omega||^2} \overline{U}^A(x) \left(\overline{\chi}_A\right)_{i \overline{i} \overline{j}} {\Omega^{\overline{i} \overline{j}}}_j \,,
\end{align}
where $\chi_A$ is a basis of $H^{(1,2)}$, $\Omega$ is the holomorphic $(3,0)$-form and $||\Omega||^2 = \frac{1}{3!} \Omega_{ijk} \overline{\Omega}^{ijk}$. The four-dimensional fields $U^A$ are called \textit{complex structure moduli}, since they parametrize deformations of the complex structure. This can be understood as follows: as soon as we deform the metric with $h_{ij}$, in order for the new metric to be \Kahler, there must be a basis in which it can be written with mixed indexes components only. Since holomorphic transformations do not affect the index structure, the only transformation that can remove non-mixed components is a transformation of the complex structure.\\
\end{enumerate}

The presence of such massless fields is strongly disliked from a phenomenological point of view, for a couple of reasons. The first one is that they would mediate long distance interactions, which are not observed in nature. The second one is that since all the parameters of the four-dimensional EFT depend on the VEVs of the moduli, without fixing them it would be impossible to get predictions from string theory.\\

Finally, a few words about four-dimensional supersymmetry in Type II CY compactifications. Starting from a ten dimensional spinor, it is possible to decompose its $\mathbf{16}$ representation as\\
\begin{center}
\begin{tabular}{ccccc}
$SO(10)$ & $\longrightarrow$ & $SO(6) \times SO(1,3)$ & $\longrightarrow$ & $SU(3) \times SO(1,3)$ \\
$\mathbf{16}$ &  & $(\mathbf{4},\mathbf{2}) + (\mathbf{\overline{4}},\mathbf{2'})$ & \empty  & $(\mathbf{3},\mathbf{2}) + (\mathbf{\overline{3}},\mathbf{2'}) + (\mathbf{1},\mathbf{2}) + (\mathbf{1},\mathbf{2'}) \,$,
\end{tabular}
\end{center}
\vspace{0.4cm}
which implies that the four-dimensional theory preserves a $\mathcal{N} = 2$ supersymmetry, given that there are two ten-dimensional gravitinos both in Type IIA and in Type IIB. $\mathcal{N} = 2$ supersymmetry in four dimensions is strongly unwelcome, since it does not lead to a chiral theory and then it precludes the possibility of getting a realistic four-dimensional low-energy theory. Consequently, additional ingredients are needed in the compactification, in order to break supersymmetry further to $\mathcal{N} = 1$.

\subsection{Basic Definitions About Fluxes}
\label{ssec:BasicFluxes}

An alternative which has been extensively studied in the last decade is the possibility to allow for a non-trivial background also for some of the ten-dimensional $p$-forms fields of the massless string spectrum, in addition to the metric. Such compactifications are called \textit{flux compactifications} (see ~\cite{Grana:2005jc, Douglas:2006es, Blumenhagen:2006ci} and references therein). It turns out that fluxes back-react on the geometry of the compact space, so that Calabi-Yau manifolds are no longer solutions of the supersymmetry conditions, and it is necessary to re-analyze the constraints coming from requiring vanishing VEVs for the supersymmetric variations of dilatinos and gravitinos. The solutions to these requirements, supplemented by the request of breaking half of the supersymmetry from $\mathcal{N} = 2$ to $\mathcal{N} = 1$ in the four-dimensional EFT, are typically less constrained manifolds, ranging from complex manifolds to conformal Calabi-Yau's. We discuss such solutions in Sec.~\ref{ssec:SUSYConditions}. In addition, the non-trivial background of the form-fields also constrains the shape of the sub-manifolds of the compact space, leading to the stabilization of some of the moduli and giving them a large mass, as we will review in Sec.~\ref{ssec:ModuliStabilization}. Flux compactifications have been extensively studied also in the context of generalized complex geometry~\cite{Hitchin:2004ut, Gualtieri:2007ng, Grana:2004bg, Grana:2005sn, Martucci:2005ht, Martucci:2006ij, Koerber:2007xk, Martucci:2009sf}. In the present section we review some basic notions about fluxes in string theory.\\

In order to preserve the four-dimensional maximal symmetry, fluxes can either be only present along the internal directions of the ten-dimensional space-time, or they have to fill the four-dimensional space-time (which is only possible for four-forms at least). As a consequence the NSNS flux $H_3$ can only be internal, while $F_4$ and $F_5$ can fill the four-dimensional space for Type IIA and Type IIB respectively. In the following we assume that these requirements are satisfied.\\

Bianchi identities for fluxes in the democratic formulation introduced in Sec.~\ref{sssec:EffectiveActionsTypeII} take the form
\begin{align}
\label{eq:BianchiIdentities}
d H_3 = 0\,, \qquad d F_p - H_3 \wedge F_{p-2} = 0\,, \qquad \text{(almost everywhere)} \,,
\end{align}
where the note ``almost everywhere'' is to include possible sources. If these are present the right hand side of eq.~\eqref{eq:BianchiIdentities} is modified by delta functions with support on the source world-volume (i.e. D-branes and/or O-planes). In presence of sources $p$-form fields are not well-defined, so that the integral of the corresponding field strength over a cycle can be different from zero, i.e. there is a \textit{non-vanishing flux}. Non-vanishing fluxes can also arise in the absence of sources, as soon as the cycle supporting them is a non-contractible one. In general, non-vanishing fluxes have to obey a Dirac quantization condition, which reads
\begin{align}
\label{eq:DiracQuantization}
\frac{1}{(2 \pi)^2 \alpha'} \int_{\Sigma_4} H_3 \in \mathbb{Z}\,, \qquad \frac{1}{\left(2 \pi \sqrt{\alpha'}\right)^{p-1}} \int_{\Sigma_p} \hat{F}_p \in \mathbb{Z}\,,
\end{align}
where $\Sigma_p$ denotes a non-contractible $p$-cycle. As a consequence of Hodge and Poincar\'e duality there are as many two-cycles as four-cycles, while three-cycles come in dual pairs, as summarized in Tab.~\ref{tab:BasisII} where we also provide a basis for the corresponding (co-)homology groups. $(\alpha_{\hat{A}}, \beta^{\hat{B}})$ form a real, symplectic basis on $H^{(3)}$.\\

\begin{table}[h!]
\begin{center}
\begin{tabular}{|c|c|c|c|c|}
\hline
\begin{tabular}[x]{@{}c@{}} Cohomology\\group\end{tabular} & Dimension & \begin{tabular}[x]{@{}c@{}} Basis of\\harmonic forms\end{tabular} & \begin{tabular}[x]{@{}c@{}} Basis of\\non-trivial cycles\end{tabular} & Indexes\\
\hline
$H^{(1,1)}$ & $h^{(1,1)}$ & $\omega_I$ & $d_I$ & $I, J, K, ...$ \\
\hline
$H^{(2,2)}$ & $h^{(1,1)}$ & $\tilde{\omega}^I$ & $D^I$ & $I, J, K, ...$ \\
\hline
$H^{(2,1)}$ & $h^{(2,1)}$ & $\chi_A$ & $A^A$ & $A, B, C, ...$ \\
\hline
$H^{(3)}$ & $2 h^{(2,1)} + 2$ & $(\alpha_{\hat{A}}, \beta^{\hat{B}})$ & $(A_{\hat{A}}, B^{\hat{B}})$ & $\hat{A}, \hat{B}, \hat{C}, ...$ \\
\hline
\end{tabular}
\caption[]{\label{tab:BasisII} Basis of the cohomology group of a CY manifold and their Poincar\'e dual non-trivial cycles.}
\end{center}
\end{table}

It is possible to define electric and magnetic fluxes for each field strength as follows
\begin{align}
\label{eq:ElectricMagneticFluxes}
&\frac{1}{(2 \pi)^2 \alpha'} \int_{A_{\hat{A}}} H_3 = m^{\hat{A}}\,, \quad \frac{1}{(2 \pi)^2 \alpha'} \int_{B^{\hat{A}}} H_3 = e_{\hat{A}} \,, \qquad {\hat{A}} = 1, \dots, \frac{h^3}{2} \,,& \nonumber \\
&\frac{1}{(2 \pi)^2 \alpha'} \int_{A_{\hat{A}}} \hat{F}_3 = m^{\hat{A}}_{\rm RR}\,, \quad \frac{1}{(2 \pi)^2 \alpha'} \int_{B^{\hat{A}}} \hat{F}_3 = e_{\text{RR}\, {\hat{A}}} \,,& \\
&\frac{1}{2 \pi \sqrt{\alpha'}} \int_{d_I} \hat{F}_2 = m^I_{\rm RR}\,, \quad \frac{1}{\left(2 \pi \sqrt{\alpha'}\right)^3} \int_{D^I} \hat{F}_4 = e_{\text	{RR}\, I} \,, \quad I = 1, \dots, h^2 \,.& \nonumber
\end{align}
On a CY manifold the relevant cohomology groups decompose as
\begin{align}
H^{(3)} = H^{(3,0)} \oplus H^{(2,1)} \oplus H^{(1,2)} \oplus H^{(0,3)}\,, \qquad H^{(2)} = H^{(1,1)}\,,
\end{align}
since there are not $(2,0)$ or $(0,2)$-forms. Note that we did not define the integrals of $\hat{F}_1$ and $\hat{F}_5$ because there are not non-trivial one- and five-cycles in CY manifolds. Poincar\'e dualities are encoded in the following relations
\begin{align}
\label{eq:PoincareDuals}
\int_{A_{\hat{B}}} \alpha_{\hat{A}} = \int_{\chi} \alpha_{\hat{A}} \wedge \beta^{\hat{B}} = - \int_{B^{\hat{A}}} \beta_{\hat{B}} = \delta^{\hat{B}}_{\hat{A}}\,, \nonumber \\
\int_{d_J} \omega_I = \int_{\chi} \omega_I \wedge \tilde{\omega}^J = - \int_{D^I} \tilde{\omega}^J = \delta^J_I\,.
\end{align}

\subsection{Supersymmetry Conditions}
\label{ssec:SUSYConditions}

In the presence of non-vanishing fluxes the supersymmetry conditions in eq. \eqref{eq:SUSYconditionFluxlessCase} get much more complicated:
\begin{align}
\label{eq:SUSYConditionsWithFluxes}
\delta& \psi_M^A = \nabla_M \epsilon^A + \frac{1}{4} \slashed{H}_M \mathcal{P} \epsilon^A + \frac{1}{16} e^{\phi} \sum_n \slashed{F}_n \Gamma_M  \mathcal{P}_n \epsilon^A \,, \nonumber \\
\delta \lambda^A &= \left(\slashed{\partial} \phi + \frac{1}{2} \slashed{H} \mathcal{P}\right) \epsilon^A + \frac{1}{8} e^{\phi} \sum_n \left(-1\right)^n \left(5 - n\right) \slashed{F}_n \mathcal{P}_n \epsilon^A \,,
\end{align}
where the $2 \times 2$ matrices $\mathcal{P}$ and $\mathcal{P}_n$ are different in Type IIA and Type IIB. In Type IIA $\mathcal{P} = \Gamma$ and $\mathcal{P}_n = \Gamma^{n/2} \sigma^1$, while for Type IIB $\mathcal{P} = - \sigma^3$ and $\mathcal{P}_n = \sigma^1$ if $\frac{n + 1}{2}$ is even, and $\mathcal{P}_n = i\sigma^2$ if $\frac{n + 1}{2}$ is odd. $\sigma^i$ are the Pauli matrices. The sum runs over $n = 0, 2, 4, 6, 8, 10$ for Type IIA, and over $n = 3, 5, 7 , 9$ for Type IIB.\\

Interestingly, supersymmetry conditions in eq. \eqref{eq:SUSYConditionsWithFluxes} split into two requirements. The first one is a topological requirement, while the second one is a constraint on the differential structure of the compact space. Let us analyze them more in detail.
\begin{itemize}
\item[a)] \textit{Topological condition} - The topological condition consists in requiring the existence of a globally well defined non-vanishing spinor. It can be easily understood in terms of the \textit{structure group} of the internal manifold~\cite{Gauntlett:2002fz, Bovy:2005qq, Behrndt:2003uq, Kaste:2003zd, Martelli:2003ki, MacConamhna:2004fb, Gillard:2004xq, Dall'Agata:2004dk}. The structure group of a manifold is the group of transformations required to patch the orthonormal frame bundle in such a way that some structures are preserved on it. For instance, on a Riemannian six-manifold, the structure that has to be preserved is the metric, and this requirement automatically reduces the structure group from the general coordinate transformations group $GL(6)$ to $SO(6)$. As soon as new structures are introduced on the manifold, further restrictions on the structure group take place. In the case of string compactifications, as long as we require some preserved supersymmetry in four dimensions, or even in the case in which all the supersymmetries are spontaneously broken by fluxes in the four-dimensional EFT, supercurrents have to be globally well defined on the compactification space. This means that the internal manifold must admit as many globally defined spinors as the number of supercurrents is. A globally well defined non-vanishing spinor exists only if the structure group of the compact space is reduced to $SU(3)$. This can be easily understood, since the spinor representation in six dimensions is in the $\mathbf{4}$ of $SO(6) \simeq SU(4)$. Under $SU(3)$ this can be decomposed as
\begin{align}
\label{eq:SpinorDecompositionSU3}
\bf{4} \rightarrow \bf{3} + \bf{1} \,,
\end{align}
so that if the structure group is reduced to $SU(3)$, then the orthonormal frame patchings admit a singlet, namely a globally well defined non-vanishing spinor. It is possible to decompose also vectors, two-forms and three-forms under $SU(3)$, and we get respectively
\begin{align}
\text{Vector}: &\quad &&\bf{6} \, \rightarrow \, \bf{3} + \bf{\overline{3}} \,, \nonumber \\
\text{Two-form}: &\quad &&\bf{15} \, \rightarrow \, \bf{8} + \bf{3} + \bf{\overline{3}} + \bf{1}\,, \nonumber \\
\text{Three-form}: &\quad &&\bf{20} \, \rightarrow \, \bf{6} + \bf{\overline{6}} + \bf{3} + \bf{\overline{3}} + \bf{1} + \bf{1}\,,
\end{align}
from which we realize that there are also a globally well defined non-vanishing real two-form and a globally well defined non-vanishing complex three-form. They correspond to the \Kahler two-form $J$ and to the holomorphic three-form $\Omega$ which are present also in the case of CY manifolds, as we pointed out in Sec.~\ref{ssec:CYCompactifications}. The topological condition leads to the same results as in the fluxless case. A manifold whose structure group is reduced to $SU(3)$ is a $SU(3)$-\textit{structure manifold}~\cite{Kaste:2003dh, Frey:2004rn, Behrndt:2005bv}.
\item[b)] \textit{Differential condition} - The differential condition consists in imposing eq. \eqref{eq:SUSYConditionsWithFluxes}. As we have reviewed in Sec.~\ref{ssec:CYCompactifications}, in the fluxless case they boil down to the requirement of reduced $SU(3)$ holonomy for the internal manifold, which is equivalent to require the existence of a covariantly constant spinor such that $\nabla \eta_\pm = 0$ on the compact space. In general on a $SU(3)$-structure manifold, it is possible to find a connection $\Gamma'$ such that the covariant derivative $\nabla'(\Gamma')$ (possibly with a non-vanishing torsion)~\cite{LopesCardoso:2002vpf, Gauntlett:2003cy, Grana:2004sv}
\begin{itemize}
\item is compatible with the metric: $\nabla' g = 0$,
\item has reduced $SU(3)$ holonomy: $\nabla' \eta_\pm = 0$,
\end{itemize}
where $\eta$ is the six-dimensional spinor in eq.s \eqref{eq:SpinorDecompositionBareIIA} and \eqref{eq:SpinorDecompositionBareIIB}. In other words a CY space is a particular example of a $SU(3)$-structure manifold, in which the globally well defined spinor is also covariantly constant with respect to the Levi-Civita connection.
\end{itemize}

The torsion tensor associated with the new connection $\Gamma'$ can be decomposed as	
\begin{align}
\label{eq:TorsionDecomposition}
{T_{mn}}^p \in \Lambda^1 \otimes \left(su(3) \oplus su(3)^{\perp}\right) \,,
\end{align}
where $\Lambda^1$ is the space of one-forms and we used the decomposition $so(6) = su(3) \oplus su(3)^{\perp}$. Since its relevant action is on the $SU(3)$-invariant component of $\eta_\pm$, the $su(3)$ piece can be ignored. The corresponding decomposition is called \textit{intrinsic torsion} ${T^0_{mn}}^p$ and takes the form
\begin{eqnarray}
\label{eq:IntrinsicTorsion}
T^0_{mn}\,^p \in \Lambda^1 \otimes su(3)^\perp &=& ({\bf 3} \oplus {\bf \bar
3}) \otimes  ({\bf 1} \oplus {\bf 3} \oplus {\bf \bar 3}) \nn \\
&=& ({\bf 1} \oplus {\bf 
1}) \oplus  ({\bf 8} \oplus {\bf 8}) \oplus ({\bf 6} \oplus {\bf \bar
6}) \oplus  2 \, ({\bf 3} \oplus {\bf \bar 3} ) \\
&& \quad \mathcal{W}_1 \quad  \quad\quad   \mathcal{W}_2 \quad  \quad\quad \mathcal{W}_3 \quad
\quad \,\, \mathcal{W}_4, \mathcal{W}_5 \nn
\end{eqnarray}

$\mathcal{W}_1, \dots, \mathcal{W}_5$ are the five torsion classes appearing in the covariant derivative of the globally well defined spinor, which characterize the differential properties of the $SU(3)$-structure manifold~\cite{Dall'Agata:2004dk, LopesCardoso:2002vpf, Gauntlett:2003cy, Grana:2004sv, Dall'Agata:2003ir}. As inferable from eq. \eqref{eq:IntrinsicTorsion}, $\mathcal{W}_1$ is a complex scalar, $\mathcal{W}_2$ is a complex two-form, $\mathcal{W}_3$ is a real primitive $(1,2) + (2,1)$-form, while $\mathcal{W}_4$ and $\mathcal{W}_5$ are real vectors. Depending on which component of the intrinsic torsion is non-zero, the internal space deviates from being a CY manifold. This can be clearly seen by writing down the exterior derivatives of the \Kahler two-form $J$ and of the holomorphic three-form $\Omega$ in terms of the torsion classes:
\begin{align}
\label{eq:dJdOmegaTorsion}
dJ = \frac{3}{2} \text{Im} \left(\overline{\mathcal{W}}_1 \Omega\right) + \mathcal{W}_4 \wedge J + \mathcal{W}_3 \,, \\
d\Omega = \mathcal{W}_1 J \wedge J + \mathcal{W}_2 \wedge J + \overline{\mathcal{W}_5} \wedge \Omega \,.
\end{align}
In the fluxless case we know that the internal space is a CY manifold and that both $J$ and $\Omega$ are closed forms, so that all the torsion classes vanish. A $SU(3)$-structure manifold ranges from being a CY manifold when $\mathcal{W}_1 = \dots = \mathcal{W}_5 = 0$ to being simply a complex manifold when only $\mathcal{W}_1 = \mathcal{W}_2 = 0$. In Tab.~\ref{tab:TorsionClasses} we report more details.

\begin{table}
\begin{center}
\begin{tabular}{|c|c|}
\hline
{\bf Manifold} & {\bf Vanishing torsion class}\\
\hline
Complex & $\mathcal{W}_1 = \mathcal{W}_2 = 0$ \\ 
\hline
Symplectic & $\mathcal{W}_1 = \mathcal{W}_3 = \mathcal{W}_4 = 0$ \\ 
\hline
\Kahler & $ \mathcal{W}_1 = \mathcal{W}_2 = \mathcal{W}_3 = \mathcal{W}_4 = 0$ \\
\hline
Calabi-Yau & $ \mathcal{W}_1 = \mathcal{W}_2 = \mathcal{W}_3 = \mathcal{W}_4 = \mathcal{W}_5 = 0$ \\
\hline
``Conformal'' Calabi-Yau & $ \mathcal{W}_1 = \mathcal{W}_2 = \mathcal{W}_3 = 3 \mathcal{W}_4 - 2 \mathcal{W}_5 = 0$ \\
\hline
\end{tabular}
\caption{ \label{tab:TorsionClasses}
\text{Vanishing torsion classes in special SU(3)-structure manifolds.}}
\end{center}
\end{table}

\subsubsection*{$\mathcal{N} = 1$ Supersymmetry}
\label{sssec:N1SUSY}

The main message from the previous section is that the presence of fluxes back-reacts on the geometry of the compact space, making some components of the torsion different from zero, so that the manifold is no longer a CY space. Furthermore, non-vanishing fluxes can break supersymmetry. In order to study the four-dimensional $\mathcal{N} = 1$ Minkovski vacua which are compatible with flux compactifications, it is necessary to impose a relation between the four-dimensional spinor $\xi^{A}$ ($A = 1, 2$) in eq.s \eqref{eq:SpinorDecompositionBareIIA} and \eqref{eq:SpinorDecompositionBareIIB}. The only possible choice which allows to preserve the four-dimensional maximal symmetry is to take the four-dimensional spinors proportional to each other \cite{Grana:2005jc}
\begin{align}
\label{eq:SpinorProportionality}
a(y) \xi_+^1 = b(y) \xi_+^2 = \xi_+ \,,
\end{align}
where the proportionality complex functions $a$ and $b$ depend on the internal coordinates $y$. Then it is possible to decompose the ten-dimensional spinors as
\begin{align}
\label{eq:SpinorDecompositionIIAN1}
\epsilon^1_{\text{IIA}} = \xi_+ \otimes \left(a \eta_+\right) \oplus \xi_- \otimes \left(\overline{a}\eta_-\right) \,, \nonumber \\
\epsilon^2_{\text{IIA}} = \xi_+ \otimes \left(\overline{b}\eta_-\right) \oplus \xi_- \otimes \left(b \eta_+\right)\,,
\end{align}
and
\begin{align}
\label{eq:SpinorDecompositionIIBN1}
\epsilon^1_{\text{IIB}} = \xi_+ \otimes \left(a \eta_+\right) \oplus \xi_- \otimes \left(\overline{a}\eta_-\right) \,, \nonumber \\
\epsilon^2_{\text{IIB}} = \xi_+ \otimes \left(b \eta_+\right) \oplus \xi_- \otimes \left(\overline{b} \eta_-\right)\,.
\end{align}
Using eq.s \eqref{eq:SpinorDecompositionIIAN1} and \eqref{eq:SpinorDecompositionIIBN1} to impose the supersymmetry conditions in eq. \eqref{eq:SUSYConditionsWithFluxes} provides a set of relations between the torsion, the fluxes, the cosmological constant and the warp-factor, which describe how the four-dimensional $\mathcal{N} = 1$ supersymmetric vacuum sits in the underlying $\mathcal{N} = 2$ EFT. One degree of freedom of the complex functions $a$ and $b$ can be fixed by the normalization: $|a|^2 + |b|^2 = e^A$, where $e^{2 A}$ is the warp-factor. A second degree of freedom is redundant and can be fixed by a gauge choice~\cite{Grana:2005jc, Grana:2005sn}, then all $\mathcal{N} = 1$ flux vacua are parametrized by a couple of angles $\alpha$ and $\beta$ as follows
\begin{align}
\label{eq:VacuaParameterization}
a = e^{A/2} \, \cos \alpha \, e^{i \frac{\beta}{2}}\,, \quad b = e^{A/2} \, \sin \alpha \, e^{-i \frac{\beta}{2}} \,.
\end{align}
In Tab.s~\ref{tab:IIAN1vacua} and~\ref{tab:IIBN1vacua} we report the results for Type IIA and Type IIB respectively. We can discuss them separately:

\begin{table}
\begin{center}
\begin{tabular}{|c|c|c|c|}
\hline 
{\bf IIA}& $a=0$ or $b=0$ (A)& \multicolumn{2}{c|}{$a=b\, e^{i\beta}
$  (BC)} \\\hline
{\bf 1}&\multicolumn{3}{c|}{$\mathcal{W}_1=H_3^{(1)}=0\,$.}\\ \cline{2-4}
&\minicent{3.1}{\vspace{.2cm}$F_0^{(1)} = \mp F_2^{(1)} = \quad = \,F_4^{(1)} = \mp F_6^{(1)}\,$.\vspace{.2cm}} &
\multicolumn{2}{c|}{$F_{2n}^{(1)}=0\,$.}\\\hline
{\bf 8}&  &generic $\beta$ & $\beta=0$\\\cline{3-4}
&\minicent3{\vspace{.2cm} $\mathcal{W}_2= F_2^{(8)} = \quad = F_4^{(8)}=0\,$.}&\minicent3{\vspace{.2cm}$\mathcal{W}_2^+=e^{\phi} F_2^{(8)}\,$; \\ $\mathcal{W}_2^-=0\,.$\vspace{.2cm}}& 
\minicent5{\vspace{.2cm}$\mathcal{W}_2^+=e^{\phi} F_2^{(8)}+e^{\phi}F_4^{(8)}\,$; \\ $\mathcal{W}_2^-=0\,$.}
\\\hline
{\bf 6}&$\mathcal{W}_3=\mp *_6 H_3^{(6)}\,$. & \multicolumn{2}{c|}{$\mathcal{W}_3= H_3^{(6)}=0\,$.}\\\hline
{\bf 3}&\minicent{3.2}{\vspace{.2cm} $\overline{\mathcal{W}}_5=2\mathcal{W}_4= \qquad \qquad=\mp 2iH_3^{(\overline 3)}=\overline{\del}\phi\,$;\\
$\overline{\del} A=\overline{\del} a=0\,$.\vspace{.2cm}}&\multicolumn{2}{c|}{
\minicent6{\vspace{.2cm}$F_2^{(\overline{3})}=2i \overline{\mathcal{W}}_5=-2i \overline{\partial} A =
\frac{2}{3} i\overline{\partial} \phi\,$; $\mathcal{W}_4=0\,$.\vspace{.2cm}}}\\\hline
\end{tabular}
\caption{\label{tab:IIAN1vacua} Possible $N=1$ vacua in IIA. We defined $\mathcal{W}_2^+ = \text{Re}(\mathcal{W}_2)$ and $\mathcal{W}_2^- = \text{Im}(\mathcal{W}_2)$.}
\end{center}
\end{table}

\begin{itemize}
\item \textbf{Type IIA -} In Type IIA, as can be observed in Tab.~\ref{tab:IIAN1vacua}, there are two classes of solutions to the supersymmetry conditions in eq. \eqref{eq:SUSYConditionsWithFluxes}:
\begin{itemize}
\item[(A)] It corresponds to a solution with the NSNS flux and the singlet components of RR fluxes. It features $\mathcal{W}_1 = \mathcal{W}_2 = 0$, but $\mathcal{W}_3 \neq 0$ and $\overline{\mathcal{W}}_5 = 2 \mathcal{W}_4 = \overline{\partial} \phi$, meaning that the compact space is complex but it is not even \Kahler or symplectic.
\item[(BC)] It is a solution with only RR fluxes, and $\mathcal{W}_1 = \mathcal{W}_3 = \mathcal{W}_4 = 0$, such that the compact space is symplectic. It corresponds to the dimensional reduction of an M-theory solution on a seven-dimensional manifold with $G_2$ holonomy \cite{Kaste:2003dh}. In case of a eleven-dimensional torsion-free $G_2$ structure the internal space becomes a \Kahler manifold, corresponding to the second column of Tab. \ref{tab:IIAN1vacua} with $F_2^{(8)} = 0$.
\end{itemize}
We can conclude that the introduction of fluxes in Type IIA causes a substantial back-reaction on the geometry of the compact space, which deviates significantly from being a CY manifold.
\item \textbf{Type IIB -} In Type IIB there are three\footnote{Actually the classes are four. There is an additional class called (ABC). We refer to~\cite{Grana:2004sv} for more details.} different classes of solutions to the supersymmetry conditions in eq. \eqref{eq:SUSYConditionsWithFluxes}, as shown in Tab.~\ref{tab:IIBN1vacua}:
\begin{itemize}
\item[(A)] This class of solutions is very similar to class (A) of Type IIA. In particular the vanishing torsion classes are again $\mathcal{W}_1 = \mathcal{W}_2 = 0$, so that the internal manifold is complex. In Type IIB all components of RR fluxes are zero.
\item[(B)] This class of solutions is the most interesting one. One of its main properties is that it allows for the RR five-form and for both RR and NSNS $3$-forms. The latter are related by a Hodge duality. It is convenient to define the \textit{three-form flux}
\begin{align}
\label{eq:G3Definition}
G_3 = F_3 - i e^{-\phi} H_3 = \hat{F}_3 - \tau H_3 \,,
\end{align}
where $\tau = C_0 + i e^{-\phi}$ is the \textit{axio-dilaton}. The constraint $e^{\phi}F_3^{(6)}=\mp * H_3^{(6)}$ translates into the fact that $G_3$ is \textit{Imaginary Self Dual} (ISD), namely
\begin{align}
\label{eq:ISD}
*_6 G_3 = i G_3 \,.
\end{align}
Finally, since in the class (B) there are not non-vanishing fluxes in the singlet representation, then $G_3^{(0,3)} = 0$.
\begin{itemize}
\item[(B$_{\rm NF}$)] This sub-class features $\mathcal{W}_1 = \mathcal{W}_2 = \mathcal{W}_3 = 0$, and $3 \mathcal{W}_4 = 2 \mathcal{W}_5 = 6 \overline{\partial} A$, so that the manifold is a \textit{conformal Calabi-Yau} space, accordingly to Tab.~\ref{tab:TorsionClasses}. Non-vanishing torsion classes source a slight deviation from a CY space, since the metric can be written as
\begin{align}
\label{eq:WarpedMetric}
ds^2 = e^{2 A} \eta_{\mu \nu} dx^\mu dx^\nu + e^{-2 A} \tilde{g}_{mn} dx^m dx^n\,,
\end{align}
where $\tilde{g}_{mn}$ is a CY metric. The \textit{conformal factor} $e^{-2 A}$ is the inverse of the warp-factor and as we will discuss below, in the limit of large volume it can be safely neglected. As a consequence, all the mathematical tools developed for CY compactifications can be used in this case to compute the four-dimensional effective field theory.
\item[(B$_{\rm F}$)] In this case $\mathcal{W}_4 = \mathcal{W}_5$, so the internal manifold is not a conformal CY, even if it is still complex. Furthermore it features a non-constant holomorphic dilaton $\tau$. This class gives rise to F-theory solutions.
\end{itemize}
\item[(C)] This class is S-dual to class (A), hence it has only RR fluxes and the same vanishing torsion classes: $\mathcal{W}_1 = \mathcal{W}_2 = 0$. 
\end{itemize}
\end{itemize}

\begin{table}
\begin{center}
\begin{tabular}{|c|c|c|c|c|}\hline
{\bf IIB}& $a=0$ or $b=0$  (A) & \multicolumn{2}{c|}{$a=\pm ib$  (B)}& 
$a=\pm b $ (C)\\\hline
{\bf 1}&\multicolumn{4}{c|}{$\mathcal{W}_1=F^{(1)}_3=H_3^{(1)}=0$.} \\ \cline{1-5}
{\bf 8}&\multicolumn{4}{c|}{$\mathcal{W}_2=0$.}\\ \cline{1-5}
{\bf 6}&\minicent{3}{\vspace{.2cm}$F_3^{(6)}=0\,$;\\$\mathcal{W}_3=\pm*H_3^{(6)}\,$.\vspace{.2cm}}&
\multicolumn{2}{c|}{\minicent{3.2}{$\mathcal{W}_3=0\,$;\\ 
$e^{\phi}F_3^{(6)}=\mp * H_3^{(6)}\,$.}}&
\minicent{3}{$H_3^{(6)}=0\,$;\\ $\mathcal{W}_3= \pm e^{\phi} * F_3^{(6)}\,$.}\\\hline
{\bf 3}& {\minicent{3.5}{\vspace{0.3cm}$\overline{\mathcal{W}}_5=2\mathcal{W}_4= \qquad=\mp 2iH_3^{(\overline 3)}= 2 \overline\del\phi\,$;\\\vspace{0.3cm} $\overline\del A=\overline\del a=0\,$.}}
 & 
\multicolumn{2}{c|}{\minicent{4}{\vspace{.2cm}$ e^\phi F_5^{(\overline{3})}= 
\frac{2}{3} i \overline{\mathcal{W}}_5 = \qquad = i \mathcal{W}_4=-2i\overline{\partial} A = \quad =-4i\overline\partial \log a\,$;\\
$\overline{\partial} \phi =0\,$. \vspace{.2cm}}} &
{\minicent{3.4}{\vspace{0.5cm}$\pm e^\phi F_3^{(\overline 3)}=2i \overline{\mathcal{W}}_5=$\\
$=- 2i\overline\partial A = \qquad \quad =- 4i\overline\partial \log a=$\\
$=-i \overline{\partial} \phi\,$.}}\\\cline{3-4}
&& 
F & \minicent{4}{\vspace{.2cm}$e^\phi F_1^{(\overline 3)}=2 e^\phi F_5^{(\overline 3)}=$ \\ 
$= i \overline{\mathcal{W}}_5 = i \mathcal{W}_4=i\overline\partial \phi\,$.\vspace{.2cm}}
&\\\hline
\end{tabular}
\caption{\label{tab:IIBN1vacua} \text{Possible $N=1$ vacua in IIB.}}
\end{center}
\end{table}

We can conclude that as soon as fluxes are introduced into the compactification, they back-react on the geometry so that typically the internal manifold is far from being a CY space. As we argued, the most interesting class of solutions is the Type IIB (B$_{\rm NF}$) class, to which we refer for simplicity as \textit{warped compactifications}~\cite{Giddings:2001yu, Curio:2000dw, Gubser:2000vg, Grana:2000jj, DeWolfe:2002nn, Frey:2003tf}, since it leads to warped (or conformal) CY manifolds. In this case, the conformal factor is the inverse of the warp-factor, and in the large radius limit it can be written as
\begin{align}
\label{eq:ConformalFactorLimit}
e^{2 A} \sim 1 + \mathcal{O}\left(\frac{g_s N \alpha'^{2}}{R^4}\right) \,,
\end{align}
where $N$ measures the units of three-form flux $G_3$, while $R$ is the radius of the compact space (assuming it is isotropic). Hence, in the large radius limit, corrections to the CY metric can be safely neglected, and the KK reduction can be performed using the mathematical tools we illustrated in Sec.~\ref{ssec:CYCompactifications}. In particular the moduli space can be approximated with that of a CY manifold. Even if this is a very good approximation in the large volume regime, a proper computation of the moduli space should include the warping~\cite{Giddings:2005ff, deAlwis:2003sn, Shiu:2008ry, Douglas:2008jx, Frey:2008xw, Chen:2009zi, Underwood:2010pm, Frey:2013bha, Martucci:2014ska}. In the present thesis we focus on warped compactifications, and we perform computations in the large volume regime, in order to trust the supergravity approximation, so that we can also safely neglect warping effects on the moduli space of the conformal CY manifold.\\

Any additional object introduced in the compactification can further back-react on the geometry, possibly destroying the warped compactification solution. For instance, let us recall what supersymmetries are preserved by sources for fluxes, such as D-branes and O-planes \cite{Polchinski:1998rr}. In general D-branes and O-planes preserve supersymmetries whose transformation parameters are related in the following way
\begin{align}
\label{eq:DbranesSUSY}
\epsilon^1 = \Gamma^\perp \epsilon^2\,,
\end{align}
where $\Gamma^\perp$ denotes the product of ten-dimensional gamma matrices in the directions orthogonal to the source itself. It turns out that in general the product of six or two gamma matrices in euclidean space has eigenvalue $\pm i$, so that D3/D7-branes and O$3$/O$7$-branes preserve supersymmetries of the type $a = \pm ib$. Such supersymmetries correspond exactly to those preserved in warped compactifications, as reported in Tab.~\ref{tab:IIBN1vacua} (the minus sign is the right one for anti D3-branes). On the other hand the products of $4$ gamma matrices have eigenvalue $\pm 1$, so that D$5$/D$9$-branes and O$5$/O$9$-planes preserve supersymmetries of the class Type IIB (C). For this reason, in this thesis we consider warped compactifications with D3/D7-branes and O$3$/O$7$-branes.

\subsection{Equations of Motion}
\label{ssec:EqOfMotion}

After having determined the class of internal manifolds which admit $\mathcal{N} = 1$ supersymmetry, we have to check that in such backgrounds the equations of motion of string theory can be satisfied. We have already observed that in the fluxless case, the integrated equations of motion in eq. \eqref{eq:UnwarpedCYFluxlessCase} let us rule out warped CY compactifications. In this section we review some generic constraints arising from the integrated equations of motion in presence of fluxes. As we will observe, the first consequence is the necessity to introduce sources for fluxes in the compactification.\\

The relevant constraints for the ten-dimensional bosonic massless fields of Type IIB strings are:
\begin{itemize}
 \item Einstein's equation,
 \item Equation of motion for the dilaton,
 \item Bianchi identities for fluxes,
 \item Equations of motion for fluxes,
 \end{itemize}

As a first step, let us focus on the constraints coming from RR fields. Due to the self-duality constraint in eq. \eqref{eq:RRSelfDuality} it turns out that the equations of motion for the RR fields, which take the form
\begin{align}
 \label{eq:RREquationsOfMotion}
 d\left(*_{10} F_p\right) + H_3 \wedge *_{10} F_{p + 2} = 0\,,
 \end{align}
are contained into the Bianchi identities
\begin{align}
\label{eq:RRBianchiIdentities}
dF_p - H_3 \wedge F_{p-2} = 0\,.
\end{align}
Every time we have to impose differential conditions such as in eq. \eqref{eq:RRBianchiIdentities} on a compact manifold we face with constraints coming from their integrated version, as a consequence of the Gauss' law. The simplest example in quantum field theory arises in presence of a scalar field $\phi$ with a source $J$ on a compact space $\chi$. The equation of motion is of the form $\Box \phi = - J$, hence it constrains the integral of the source to vanish $\int_\chi J = 0$. The same happens when we consider the constraints in eq. \eqref{eq:RRBianchiIdentities}. As reported in Tab.~\ref{tab:IIBN1vacua}, due to supersymmetry $F_3 \sim *_{6} H_3$, and then from eq. \eqref{eq:RRBianchiIdentities} we can immediately infer that the compactification is inconsistent, since the integral of $dF_5$ on the compact space $\chi$ vanishes, while the integral of $H_3 \wedge F_3$ is positive definite. This kind of inconsistency can be avoided in string theory due to the presence of BPS sources, like O3-planes, which contribute to the right hand side of eq. \eqref{eq:RRBianchiIdentities} with the right sign. Including the contributions coming both from D3-branes and from O3-planes, which are sources for the D3-brane density charge, the Bianchi identity reads
\begin{align}
\label{eq:TadpoleF3}
dF_5 - H_3 \wedge F_3 = \left(2 \pi \sqrt{\alpha'}\right)^4 \rho^{\text{loc}}_3\,,
\end{align}
where $\rho^{\text{loc}}_3$ is the dimensionless D3 charge density of the localized sources. In general it can be written as
\begin{align}
\label{eq:D3ChargeDensity}
\rho^{\rm loc}_3 = \mu_3 \sum_a \pi^a_6 + \mu_3 Q_3 \sum_a \pi^{a, \rm Op}_6 \,,
\end{align}
where $\pi^a_6$ and $\pi^{a, \rm Op}$ are six-forms Poincar\'e dual to the supports of D3-branes and O3-planes respectively. For D3-branes the coefficient $\mu_3$ is given by
\begin{align}
\label{eq:DpTension}
\mu_3 = \frac{1}{(2 \pi)^3 \left(\alpha^{'}\right)^2}\,,
\end{align}
while for a O3-plane $Q_3$ is negative \cite{Giddings:2001yu}. We are finally led to the schematic \textit{tadpole cancellation condition}:
\begin{align}
\label{eq:TadpoleCancellation}
N_{\text{D3}} - \frac{1}{2} N_{\text{O3}} + \frac{1}{\left(2 \pi\right)^4 \alpha^{'2}} \int_{\chi} H_3 \wedge F_3 = 0\,,
\end{align}
which is a constraint on the number of sources present in the compactifications. In eq. \eqref{eq:TadpoleCancellation} we are neglecting possible contributions to the D3-charge density arising from D7-branes. We will provide more details on tadpole cancellation in a compactification in Sec.~\ref{ssec:ExplicitGlobalModel}.\\

It is also possible to get a constraint on the geometry of the compact space from the integrated equations of motion in a way which does not make use of the supersymmetry conditions of Sec.~\ref{ssec:SUSYConditions}, but with the only assumption of warped metric as in eq. \eqref{eq:WarpedMetric}\footnote{Here we assume that the metric has the warped form of \eqref{eq:WarpedMetric}, but $\tilde{g}_{mn}$ is not necessarily CY.}~\cite{Giddings:2001yu}. In fact, Poincar\'e invariance and self-duality of $F_5$ constrain the form of $F_5$ itself:
\begin{align}
\label{eq:F5Constrained}
F_5 = (1 + *_{10})\left(d \alpha(y) \wedge dx^0 \wedge dx^1 \wedge dx^2 \wedge dx^3\right)\,,
\end{align}
where $\alpha$ is a generic function which depends only on the internal coordinates, collectively denoted by $y$. Subtracting from eq. \eqref{eq:F5Constrained} the trace of the non-compact components of the Einstein's equation we get (in the Einstein frame)\footnote{A $\widetilde{}$ denotes quantities computed using the metric $\widetilde{g}$ in eq. \eqref{eq:WarpedMetric}.}
\begin{align}
\label{eq:EOMConstraint}
\tilde{\nabla}^2 \left(e^{4A} - \alpha\right) =&\, \frac{e^{2A + \phi}}{6} |*_6 G_3 - i G_3|^2 + e^{-6A} |\partial\left(e^{4A} - \alpha\right)|^2 + \nonumber\\
&+ (4 \pi^2 \alpha')^4 e^{2A} \left[\frac{1}{2} \left(T^i_i - T^\mu_\mu\right)^{\rm loc} - T_3 \rho_3^{\rm loc}\right] \,,
\end{align}
where $T_{ij}$ is the stress tensor of the localized sources defined in the usual way as
\begin{align}
\label{eq:StressTensorSources}
T_{ij} = -\frac{2}{\sqrt{-g}}\frac{\delta S_{\rm loc}}{\delta g^{ij}} \,.
\end{align}
Since the left-hand-side of eq. \eqref{eq:EOMConstraint} has to vanish upon integration on a compact space, we can observe that a warped compactification to a four-dimensional Minkovski space is possible only if:
\begin{enumerate}
 \item the warp-factor and the five-form potential are related: $e^{4A} = \alpha$,
 \item $G_3$ is ISD: $*_6 G_3 = i G_3$,
 \item the BPS condition $\frac{1}{2} \left(T^i_i - T^\mu_\mu\right)^{\rm loc} = T_3 \rho_3^{\rm loc}$ holds, where $T_3 = \left(2 \pi \sqrt{\alpha'}\right)^{-4}$ is the D3-brane tension.
\end{enumerate}
Condition 3. holds for BPS objects like D3-branes and O3-planes. Also D7-branes satisfy this condition, since as we have previously observed, they preserve the same $\mathcal{N} = 1$ supersymmetry of D3-branes. Condition 1. implies that the five-form $F_5$ is constrained to be of the same form as in warped compactifications of Tab.~\ref{tab:IIBN1vacua}, while condition 2. ensures that the three-form $G_3$ has to be ISD, again as in warped compactifications. We infer that the constraints from the equations of motion in case of a warped metric as in eq. \eqref{eq:WarpedMetric} lead almost to the same background geometry that we got from supersymmetry conditions, assuming that all the localized sources present in the compactification are BPS objects. The slight difference is that supersymmetry constraints allow neither for a non-vanishing singlet component $G_3^{(0,3)}$ of the three-form flux nor for a non-vanishing non-primitive $(1,2)$ component of $G_3$, while equations of motion do. The non-primitive $(1,2)$ component of $G_3$ is always absent on a CY manifold, because there are not non-trivial five-forms. On the other hand the $G^{(0,3)}_3$ component can be used to break supersymmetry in a controllable way, as we explain below.\\

Finally, let us come back to the Einstein's equation and to the equation of motion for the dilaton. In the Einstein frame they read
\begin{align}
\label{eq:DilatonEinsteinEOM}
\tilde{R}_{ij} =&\, \frac{\left(4 \pi^2 \alpha'\right)^4}{4} e^{2 \phi} \partial_{[i} \tau \partial_{j]} + (2 \pi)^7 \left(T_{ij}^{\rm D7} - \frac{1}{8} \tilde{g}_{ij} T^{\rm D7}\right) \,, \nonumber \\
&\tilde{\nabla}^2 \tau = i e^\phi \left(\tilde{\nabla} \tau\right)^2 - 4 (2 \pi)^7 e^{-2 \phi} \frac{1}{\sqrt{-g}} \frac{\delta S_{\rm D7}}{\delta \tau} \,,
\end{align}
where $T^{\rm D7}_{ij}$ is the stress tensor of D7-branes ($T^{\rm D7}$ is its trace) and $S_{\rm D7}$ is the D7-branes action. In the absence of the $G^{(0,3)}_3$ component of the three-form flux, and also in the absence of D7-branes in the compactification, the background solution is exactly the same as in warped compactifications of Tab.~\ref{tab:IIBN1vacua}. $G^{(0,3)}_3$ can be used to break supersymmetry from $\mathcal{N} = 1$ to $\mathcal{N} = 0$. However, from eq. \eqref{eq:DilatonEinsteinEOM} we immediately realize that, as soon as we introduce D7-branes into the compactification, the internal space is no longer a warped CY, since it is not Ricci-flat $\tilde{\mathcal{R}}_{ij} \neq 0$. In order for the constraints arising from the integrated equations of motion to be satisfied, the dilaton field has to acquire a dependence on the internal coordinates. This means that we are bound to consider F-theory solutions, lying in the class ($B_F$) of supersymmetric solutions of Tab.~\ref{tab:IIBN1vacua}. In the following we will work in a halfway case, since we will compute the spectrum of the EFT assuming that the internal space is a conformal CY, but at the same time we will introduce a limited number of D7-branes (which are usually needed for the consistency of the compactification, as we will see for example in Sec.~\ref{ssec:ExplicitGlobalModel}), so that we can consider the right hand sides of eq.s \eqref{eq:DilatonEinsteinEOM} as small perturbations of the conformal CY background. At the same time, in order to get SUSY-breaking in a controllable way, also the $G^{(0,3)}_3$ component of the three-form flux has to be a small perturbation of the warped compactification.

\subsection{Moduli Space and Kaluza-Klein Reduction}
\label{ssec:ModuliSpace}

In order to get the four-dimensional spectrum of the compactification, it is necessary to perform a Kaluza-Klein (KK) reduction of the ten-dimensional theory on a compact internal manifold $\chi$. This procedure typically gives rise to a tower of states, called \textit{Kaluza-Klein (KK) states} whose masses scale as\footnote{In this rough estimate we assume that the compactification manifold is isotropic.} $m_n \sim n M_s/R$, where $R$ is the radius of the compact space. $R$ can be rewritten in terms of the volume $\V$ of the compact space as $R = \ell_s \V^{1/6}$. As we have have observed in Sec.~\ref{sec:StringTheory} $M_s \sim M_{\rm P} \V^{-1/2}$, so that the KK states masses are given by
\begin{align}
\label{eq:KKTower}
M_n \sim \frac{n M_{\rm P}}{\V^{2/3}}\,.
\end{align}
Since massive KK states have not been observed in nature, we are interested in the EFT which is valid below the scale at which the massive KK states are excited 
\begin{align}
\label{eq:KKScale}
E \ll M_{KK} \simeq \frac{M_{\rm P}}{\V^{2/3}}\,,
\end{align}
namely we consider only massless KK fields, which correspond to harmonic forms of the compact internal space~\cite{Candelas:1990pi}. Eq. \eqref{eq:KKScale} is called \textit{KK approximation}. In order to trust the supergravity approximation we require $\V \gg 1$, then there is a hierarchy between the string scale and the KK scale
\begin{align}
\label{eq:HierarchyMsMKK}
M_{KK} \ll M_s \ll M_{\rm P}\,.
\end{align}
The computation of the four-dimensional spectrum requires the knowledge of the moduli space of the compact manifold, and this is exactly the reason why it is necessary to restrict to the class of warped compactifications in Tab.~\ref{tab:IIBN1vacua}. In the other cases the moduli space is poorly known. As we stressed in the last section, even in the conformal CY case the moduli space is not exactly the same as in CY manifolds, but assuming that the warping effects are small corrections we can safely use the mathematical tools of Sec.~\ref{ssec:CYCompactifications} to compute the spectrum.\\

The first step is to expand the ten-dimensional form-fields in the basis provided in Tab.~\ref{tab:BasisII}, and we get~\cite{grimm:2004uq}
\begin{align}
\label{eq:ExpansionII}
B_2(x,y) &\,= B_2(x) + b^I(x) \omega_I\,, \quad C_0(x,y) = C_0(x) \,, \quad C_2(x,y) = C_2(x) + c^I(x) \omega_I \,, \nonumber \\
&C_4(x,y) = D^I_2(x) \wedge \omega_I + V^{\hat{A}}(x) \wedge \alpha_{\hat{A}} - Y_{\hat{A}}(x) \wedge \beta^{\hat{A}} + \psi_I(x) \tilde{\omega}^I\,,
\end{align}
where the fields on the left hand sides of the equations represent the ten-dimensional fields, which depend both on the four-dimensional coordinates $x$ and on the internal coordinates $y$. On the other hand, the coefficients of the expansions which depend only on the four-dimensional coordinates $x$, are four-dimensional fields. Along with the fields $t^I(x)$ and $U^A(x)$ in the expansions of eq.s \eqref{eq:KahlerModuli} and \eqref{eq:ComplexStructureDeformations}, the fields $B_2(x)$, $b^I(x)$, $C_0(x)$, $C_2(x)$, $c^I(x)$, $D^I_2(x)$, $V^{\hat{A}}(x)$, $Y_{\hat{A}}(x)$, $\psi_I(x)$ and the graviton $g_{\mu\nu}(x)$ constitute the four-dimensional spectrum of the KK reduction\footnote{Due to the self-duality of the five-form $F_5$ it is necessary to eliminate half of the degrees of freedom in the expansion of $C_4(x,y)$. We choose to eliminate $D^I_2(x)$ and $Y_{\hat{A}}(x)$. We further discuss this point in Sec.~\ref{ssec:EFTfromIIB}.}. They can be organized in $\mathcal{N} = 2$ supermultiplets as follows
\begin{itemize}
\item 1 gravity multiplet: $(g_{\mu\nu}, V^0)$,
\item $h^{(2,1)}$ vector multiplets: $(V^A, U^A)$,
\item $h^{(1,1)}$ hypermultiplets: $(t^I, b^I, c^I, \psi_I)$,
\item 1 double-tensor multiplet: $(B_2, C_2, \phi, C_0)$.
\end{itemize}
The four-dimensional action can be obtained simply by inserting the expansion of the ten-dimensional field in the ten-dimensional supergravity action and then integrating over the internal compact space. The result is
\begin{footnotesize}
\begin{align}
\label{eq:N2effectiveaction}
S^{(4D)}_{\rm IIB} &= \int d^4 x\, \left[-\frac{1}{2} R *_4 1 + \frac{1}{4} \text{Re} \mathcal{M}_{\hat{A} \hat{B}} F^{\hat{A}} \wedge F^{\hat{B}} + \frac{1}{4} \text{Im} \mathcal{M}_{\hat{A} \hat{B}} F^{\hat{A}} \wedge *_4 F^{\hat{B}} -\right. \nonumber \\
&- G_{AB} dU^A \wedge *_4 d \overline{U}^B - G_{IJ} dt^I \wedge *_4 dt^J - \frac{1}{4} d \text{ln} \mathcal{K} \wedge *_4 d \text{ln} \mathcal{K} - \frac{1}{4} d\phi \wedge *_4 d\phi - \nonumber \\
&- \frac{1}{4} e^{2 \phi} d C_0 \wedge *_4 d C_0 - e^{-\phi} G_{IJ} db^I \wedge *_4 db^J - e^{\phi} G_{IJ} \left(dc^I - C_0 db^I\right) \wedge *_4 \left(dc^J - C_0 db^J\right) - \nonumber \\
&- \frac{9 G^{IL}}{4 \mathcal{K}} \left(d\psi_I - \frac{1}{2} \mathcal{K}_{IJK} \left(c^J db^K - b^J dc^K\right)\right) \wedge *_4 \left(d\psi_L - \frac{1}{2} \mathcal{K}_{LMN} \left(c^M db^N - b^M dc^N\right)\right) - \nonumber \\
&- \frac{\mathcal{K}^2}{144} e^{-\phi} dB_2 \wedge *_4 dB_2 - \frac{\mathcal{K}^2}{144} e^{\phi} \left(dC_2 - C_0 dB_2\right) \wedge *_4 \left(dC_2 - C_0 dB_2\right) + \nonumber \\
&\left. + \frac{1}{2} \left(db^I \wedge C_2 + c^I dB_2\right) \wedge \left(d\psi_I - \mathcal{K}_{IJK} c^J db^K\right) + \frac{1}{4} \mathcal{K}_{IJK} c^I c^J dB_2 \wedge db^K \right]\,,
\end{align}
\end{footnotesize}
where $F^{\hat{A}} = d V^{\hat{A}}$ and $\mathcal{M}_{\hat{A} \hat{B}}$ can be written as
\begin{align}
\label{eq:MKL}
\mathcal{M}_{\hat{A} \hat{B}} = \overline{\mathcal{F}}_{\hat{A} \hat{B}} + 2i \frac{\left(\text{Im} \mathcal{F}\right)_{\hat{A} \hat{C}} X^{\hat{C}} \left(\text{Im} \mathcal{F}\right)_{\hat{B} \hat{D}} X^{\hat{D}}}{X^{\hat{A}} \left(\text{Im} \mathcal{F}\right)_{\hat{A} \hat{B}} X^{\hat{B}}} \,,
\end{align}
and $X^{\hat{A}}$, $\mathcal{F}_{\hat{A}}$ are the periods of the holomorphic three-form $\Omega(U)$. $\mathcal{F}_{\hat{A} \hat{B}}$ is the period matrix defined as
\begin{align}
\label{eq:Periods}
X^{\hat{A}} = \int_{\chi} \Omega \wedge \beta^{\hat{A}} \,, \quad \mathcal{F}_{\hat{A}} = \int_{\chi} \Omega \wedge \alpha_{\hat{A}} \,, \quad \mathcal{F}_{\hat{A} \hat{B}} = \frac{\partial \mathcal{F}_{\hat{A}}}{\partial X^{\hat{B}}} \,,
\end{align}
so that in general the holomorphic three-form can be expanded as
\begin{align}
\Omega(U) = X^{\hat{A}}(U) \alpha_{\hat{A}} - \mathcal{F}_{\hat{A}}(U) \beta^{\hat{A}}\,.
\end{align}
Interestingly, $\mathcal{F}_{\hat{A}}$ can be seen as the derivative of a holomorphic prepotential $\mathcal{F}_{\hat{A}} = \frac{\partial \mathcal{F}}{\partial X^{\hat{A}}}$. As a consequence the metric $G_{AB}$ on the space of complex structure deformations
\begin{align}
\label{eq:MetricComplexStructure}
G_{AB} = \frac{\partial}{\partial U^A} \frac{\partial}{\partial \overline{U}^B} K_{\rm cs}\,,
\end{align}
is completely determined by the holomorphic prepotential $\mathcal{F}$ and the space of complex structure deformations is a special \Kahler manifold. In fact $K_{\rm cs}$ is the \Kahler po-\\	tential of the space of complex deformations, and
\begin{align}
\label{eq:KahlerPotentialComplexDeformations}
K_{\rm cs} = - \text{ln} \left[-i \int_{\chi} \Omega \wedge \overline{\Omega}\right] = - \text{ln} \left[i \left(\overline{X}^{\hat{A}} \mathcal{F}_{\hat{A}} - X^{\hat{A}} \overline{\mathcal{F}}_{\hat{A}}\right)\right] \,,
\end{align}

On the other hand, $G_{IJ}$ is the metric on the space of \Kahler deformations, and it takes the form
\begin{align}
\label{eq:MetricKahlerModuli}
G_{IJ} = \frac{3}{2 \mathcal{K}} \int_{\chi} \omega_I \wedge *_6 \omega_J = -\frac{3}{2} \left(\frac{\mathcal{K}_{IJ}}{\mathcal{K}} - \frac{3}{2} \frac{\mathcal{K}_I \mathcal{K}_J}{\mathcal{K}^2}\right)\,,
\end{align}
where we have defined
\begin{align}
\label{eq:IntersectionNumbers}
&\mathcal{K}_{IJK} = \int_{\chi} \omega_I \wedge \omega_J \wedge \omega_K \,, \qquad \mathcal{K}_{IJ} = \int_{\chi} \omega_I \wedge \omega_J \wedge J \,,& \nonumber \\
&\mathcal{K}_I = \int_{\chi} \omega_I \wedge J \wedge J \,, \qquad \mathcal{K} = \int_{\chi} J \wedge J \wedge J = \mathcal{K}_{IJK} v^I v^J v^K \,.&
\end{align}
Let us notice that with these conventions the volume $\V$ of the compact space (measured in units of string length $\ell_s = 2 \pi \sqrt{\alpha'}$) is given by
\begin{align}
\label{eq:VolumeFormal}
\V = \frac{\mathcal{K}}{6} \,,
\end{align}
As expected, the bare compactification of the ten-dimensional supergravity on a CY space gives us a $\mathcal{N} = 2$ supersymmetric theory in four dimensions. In the next section we will review how to reduce it to $\mathcal{N} = 1$ theory.

\subsection{$\mathcal{N} = 1$ EFT from Type IIB Orientifolds}
\label{ssec:EFTfromIIB}

In Sec.~\ref{ssec:SUSYConditions} we have observed that a broad set of flux configurations is compatible with a $\mathcal{N} = 1$ four-dimensional supersymmetric EFT, and as we already stressed, we restricted to the class warped compactifications in Tab.~\ref{tab:IIBN1vacua} because in the limit of large volume for the compact space, it is known how to safely perform the KK reduction. Furthermore, in Sec.~\ref{ssec:EqOfMotion} we have noticed that, in order to satisfy the constraints coming from the integrated equations of motion, the compactification must include negative tension sources for the D3-brane charge density, namely O3-planes. Very interestingly, these two observations nicely fit together. In fact, as we mentioned at the end of Sec.~\ref{ssec:SUSYConditions} the introduction of O3/O7-planes is compatible with warped compactifications, so that on the one hand they leave us with a $\mathcal{N} = 1$ low-energy theory, and on the other hand it is possible to keep trusting the EFT. From a practical point of view, the $\mathcal{N} = 1$ spectrum is obtained by truncating the $\mathcal{N} = 2$ spectrum of the previous section, projecting out the state which are not invariant under an involution introduced by the presence of the O-planes. In the present section we briefly review this procedure following~\cite{Grana:2003ek}.\\

As a first step we require that $\chi$ is symmetric under an action $\sigma$ such that:
\begin{enumerate}
\item it leaves invariant the four-dimensional non-compact space-time,
\item it is involutive $\sigma^2 = 1$,
\item it is isometric: $\sigma g = g$,
\item it is holomorphic: $\sigma {J_m}^n = {J_m}^n$,
\item it acts on the holomorphic three-form as $\sigma^* \Omega = - \Omega$,
\end{enumerate}
where ${J_m}^n = J_{m \overline{p}} g^{\overline{p} n}$ is the complex structure, and $\sigma^*$ is the pull-back of $\sigma$. $\sigma$ can be interpreted as a reflection with respect to a fixed plane, which is the O-plane. Since it leaves the four-dimensional space-time invariant, O-planes have to span the non-compact directions. Let us notice that the requirement of holomorphicity in principle allows all even-dimensional O-planes, but in order to lie in the class of warped compactifications, we additionally require condition 5., which leaves out O5/O9-planes. In fact, in complex coordinates $z^k$, we can write $\Omega \propto dz^1 \wedge dz^2 \wedge dz^3$. Because of condition 5., we have two possibilities: either one single complex coordinates gets reversed under $\sigma$, or all the complex internal coordinates get reversed under $\sigma$. The first possibility implies that the O-plane spans 8 dimensions (namely it is a O7-plane), while the second possibility means that the O-plane spans 4 dimension (namely it is a O3-plane). Hence $\sigma$ is compatible with supersymmetry in presence of O3/O7 planes.\\

The spectrum of the $\mathcal{N} = 2$ theory can be consistently truncated by the action of the operator
\begin{align}
\label{eq:OrientifoldInvolution}
\mathcal{O}_{\rm O3/O7} = \left(-1\right)^{F_L} \Omega_p \sigma\,,
\end{align}
where $\Omega_p$ is the usual world-sheet parity, while $F_L$ is the space-time fermion number in the left moving sector.  We summarize the parity of the Type IIB ten-dimensional bosonic fields under the action of $\Omega_p$ and $\left(-1\right)^{F_L}$ in Tab.~\ref{tab:ParityOrientifold}.
\begin{table}[h!]
\begin{center}
\begin{tabular}{|c|c|c|c|c|c|c|}
\hline
 & $\phi$ & $g$ & $B_2$ & $C_0$ & $C_2$ & $C_4$\\
\hline
$\left(-1\right)^{F_L}$ & $+$ & $+$ & $+$ & $-$ & $-$ & $-$ \\
\hline
$\Omega_p$ & $+$ & $+$ & $-$ & $-$ & $+$ & $-$ \\
\hline
\end{tabular}
\caption{\label{tab:ParityOrientifold} 
Parity of the Type IIB ten-dimensional bosonic fields under the action of $\Omega_p$ and $\left(-1\right)^{F_L}$.}
\end{center}
\end{table}

Since $\sigma$ is an involution, then the cohomology groups of the CY manifold splits into two eigenspaces under the action of $\sigma^*$, corresponding to the eigenvalues $\pm 1$
\begin{align}
\label{eq:SplitCohomologyGroups}
H^{(p,q)} = H^{(p,q)}_+ \oplus H^{(p,q)}_- \,.
\end{align}
In Tab.~\ref{tab:SplitCohomologyGroups} we report the dimensions of the cohomology groups, and also the splitting of the relative basis, which plays a crucial role in the truncation of the $\mathcal{N} = 2$ spectrum.\\
\begin{small}
\begin{table}
\begin{center}
\begin{tabular}{|c|c||c|c||c|c||c|c|}
\hline
\multicolumn{2}{|c||}{cohomology groups} & \multicolumn{2}{c||}{dimensions} & \multicolumn{2}{c||}{basis} & \multicolumn{2}{c|}{indexes}\\
\hline
$H^{(1,1)}_+$ & $H^{(1,1)}_-$ & $h^{(1,1)}_+$ & $h^{(1,1)}_-$ & $\omega_i$ & $\omega_\iota$ & $i, j, k$ & $\iota, \kappa, \lambda$ \\
\hline
$H^{(2,2)}_+$ & $H^{(2,2)}_-$ & $h^{(2,2)}_+$ & $h^{(2,2)}_-$ & $\tilde{\omega}^i$ & $\tilde{\omega}^\iota$ & $i, j, k$ & $\iota, \kappa, \lambda$ \\
\hline
$H^{(2,1)}_+$ & $H^{(2,1)}_-$ & $h^{(2,1)}_+$ & $h^{(2,1)}_-$ & $\chi_\alpha$ & $\chi_a$ & $\alpha, \beta, \gamma$ & $a, b, c$ \\
\hline
$H^{(3)}_+$ & $H^{(3)}_-$ & $2 h^{(2,1)}_+$ & $2 h^{(2,1)}_- + 2$ & $(\alpha_\alpha, \beta^{\beta})$ & $(\alpha_{\hat{a}}, \beta^{\hat{b}})$ & $\alpha, \beta, \gamma$ & $\hat{a}, \hat{b}, \hat{c}$ \\
\hline
\end{tabular} 
\caption{\label{tab:SplitCohomologyGroups} Splitting of the cohomology groups under the action of $\sigma^*$, with relative dimensions and basis.}
\end{center}
\end{table}
\end{small}

All the split indexes run from 1 to the dimension of the relative eigenspace of the cohomology group. Again, $(\alpha_\alpha, \beta^{\beta})$ and $(\alpha_{\hat{\alpha}}, \beta^{\hat{\beta}})$ form symplectic basis for the corresponding cohomology groups, namely the only non-vanishing intersections are
\begin{align}
\label{eq:OrientifoldedSymplecticBasis}
\int_\chi \alpha_\alpha \wedge \beta^\beta = \delta_\alpha^\beta \,, \quad \int_\chi \alpha_{\hat{a}} \wedge \beta^{\hat{b}} = \delta_{\hat{a}}^{\hat{b}}\,.
\end{align}

It is easy now to infer the orientifolded spectrum of the four-dimensional EFT.
\begin{itemize}
\item $\phi(x,y)$ and $C_0(x,y)$ are invariant under $\mathcal{O}_{\rm O3/O7}$ and then the four-dimensional fields $\phi(x)$ and $C_0(x)$ remain in the spectrum.
\item Since $B_2(x,y)$ are odd under the combined action of $\left(-1\right)^{F_L} \Omega_p$, only the odd components under the action of $\sigma^*$ remain in the spectrum. Their expansions look like
\begin{align}
\label{eq:OrientifoldedB2C2}
B_2(x,y) = b^i(x) \omega_i\,, \qquad C_2(x,y) = c^i(x) \omega_i \,.
\end{align}
\item $C_4(x,y)$ is even under the combined action of $\left(-1\right)^{F_L} \Omega_p$, then only the even components under $\sigma^*$ remain in the spectrum. Its expansion looks like
\begin{align}
\label{eq:OrientifoldedC4}
C_4(x,y) = D_2^i(x) \wedge \omega_i + V^\alpha(x) \wedge \alpha_\alpha + Y_\alpha(x) \wedge \beta^\alpha + \psi_i(x) \tilde{\omega}^i \,.
\end{align}
\item Since the action of $\sigma^*$ is holomorphic, then only the even components under $\sigma^*$ of the \Kahler form $J$ remain in the spectrum
\begin{align}
\label{eq:OrientifoldedJ}
J = t^i(x) \omega_i \,.
\end{align}
\item Given that the action of $\sigma$ leaves the metric invariant, while $\Omega$ is odd under the action of $\sigma^*$, then only the odd components of the complex structure moduli remain in the spectrum, and the expansion of the metric perturbation in eq. \eqref{eq:ComplexStructureDeformations} takes the form
\begin{align}
\label{eq:OrientifoldedComplexStructure}
h_{ij} = \sum_{a = 1}^{h_-^{(2,1)}} \frac{i}{||\Omega||^2} \overline{U}^a(x) \left(\overline{\chi}_a\right)_{i \overline{i} \overline{j}} {\Omega^{\overline{i} \overline{j}}}_j \,.
\end{align}
\end{itemize}

The surviving fields constitute a $\mathcal{N} = 1$ EFT, and they assemble into the multiplets reported in Tab.~\ref{tab:OrientifoldedMultiplets}\\
\begin{table}
\begin{center}
\begin{tabular}{|c|c|c|}
\hline 
Multiplets & Number & Fields \\
\hline
gravity multiplet & $1$ & $g_{\mu\nu}$ \\
\hline
vector multiplets & $h^{(2,1)}_+$ & $V^\alpha$ \\
\hline
& $h^{(2,1)}_-$ & $U^a$ \\
\cline{2-3} chiral &
 $h^{(1,1)}_+$ & $(t^i, \psi_i)$ \\
 \cline{2-3} multiplets &
 $h^{(1,1)}_-$ & $(b^\iota, c^\iota)$ \\
\cline{2-3}
 & $1$ & $(\phi, C_0)$ \\
\hline
\end{tabular} 
\caption{\label{tab:OrientifoldedMultiplets} Orientifolded spectrum arranged in $\mathcal{N} = 1$ multiplets.}
\end{center}
\end{table}

A few observations are in order at this point. First, notice that the four-dimensional forms $B_2(x)$ and $C_2(x)$ are projected out of the $\mathcal{N} = 1$ spectrum. Furthermore, the non-vanishing of the scalar fields $c^\iota$ and $b^\iota$ are related to the presence of O7-planes. In fact, since O-planes are fixed loci under the orientifold involution, in presence of only O3-planes all the tangent vectors are odd under the action of $\sigma$. As a consequence the expansions in eq. \eqref{eq:OrientifoldedB2C2} are not admitted. On the contrary, if also O7-planes are present in the compactification, then it is possible to get harmonic forms with the correct transformation behavior so that the expansions in eq. \eqref{eq:OrientifoldedB2C2} are admitted, and the scalar fields $c^\iota$ and $b^\iota$ remain in the spectrum.\\

The four-dimensional effective action can be computed by performing again a KK reduction of the orientifolded spectrum. Since it is just a long but straightforward operation, we mention only a couple of key points:
\begin{itemize}
\item Once we take into account the orientifold projection, the metric on the complex structure deformations space becomes
\begin{align}
\label{eq:OrientifoldedComplexStructureMetric}
G_{ab} = \frac{\partial}{\partial U^a} \frac{\partial}{\partial \overline{U}^b} K_{\rm cs} \,, \quad K_{\rm cs} = - \text{ln} \left[i\left(\overline{X}^{\hat{a}} \mathcal{F}_{\hat{a}} - X^{\hat{a}} \overline{\mathcal{F}}_{\hat{a}}\right)\right] \,,
\end{align}
\item Due to the orientifold the following intersection numbers vanish
\begin{align}
\label{eq:VanishingIntersectionNumbersOrientifold}
\mathcal{K}_{ij \lambda} = \mathcal{K}_{\iota \kappa \lambda} = \mathcal{K}_{i \kappa} = \mathcal{K}_{\iota} = 0\,,
\end{align}
and then the metric on the \Kahler deformations space takes the form
\begin{align}
\label{eq:OrientifoldedKahlerMetric}
G_{ij} = -\frac{3}{2} \left(\frac{\mathcal{K}_{ij}}{\mathcal{K}} - \frac{3}{2} \frac{\mathcal{K}_i \mathcal{K}_j}{\mathcal{K}^2}\right) \,, \quad G_{\iota \kappa} = -\frac{3}{2} \frac{\mathcal{K}_{\iota \kappa}}{\mathcal{K}} \,, \quad G_{i \kappa} = G_{\iota j} = 0 \,,
\end{align}
where
\begin{align}
\label{eq:NonVanishingOrientifoldedIntersectionNumbers}
\mathcal{K}_{ij} = \mathcal{K}_{ijk} t^k \,, \quad \mathcal{K}_{\iota \kappa} = \mathcal{K}_{\iota \kappa l} t^l\,, \quad \mathcal{K}_{i} = \mathcal{K}_{ijk} t^j t^k \,, \quad \mathcal{K} = \mathcal{K}_{ijk} t^i t^j t^k \,.
\end{align}
\item The self-duality condition of $F_5$ can be imposed by adding to the four-dimensional action the following total derivative
\begin{align}
\label{eq:TotalDerivative}
\delta S^{(4)} = \frac{1}{4} dV^\alpha \wedge dY_\alpha + \frac{1}{4} dD_2^i \wedge d\psi_i\,.
\end{align}
the equations of motion for $D_2^i$ and $Y_\alpha$ (or equivalently for $\psi_i$ and $V^\alpha$) coincide with the self-duality condition and then it is possible to eliminate $D_2^i$ and $Y_\alpha$ (or $\psi_i$ and $V^\alpha$) by inserting their equations of motion into the action. The choice of eliminating the former or the latter corresponds to the choice of expressing the four-dimensional action in terms of linear or chiral multiplets respectively. Since we want to express the four-dimensional action in terms of chiral multiplets, we choose to eliminate $D^i_2$ and $Y_\alpha$.
\end{itemize}
We report the four-dimensional action expressed in terms of chiral multiplets:
\begin{small}
\begin{align}
\label{eq:FourDimensionalOrientifoldedAction}
S^{(4)} &= \int d^4x\, \left[-\frac{1}{2} R *_4 1 - G_{ab} dU^a \wedge *_4 d\overline{U}^b - G_{ij} dt^i \wedge *_4 dt^j - \frac{1}{4} d \text{ln} \mathcal{K} \wedge *_4 d \text{ln} \mathcal{K} - \right.& \nonumber \\
&- \frac{1}{4} d \phi \wedge *_4 d \phi - \frac{1}{4} e^{2 \phi} d C_0 \wedge *_4 dC_0 - e^{-\phi} G_{\iota \kappa} db^\iota \wedge *_4 d b^\kappa -& \nonumber \\
&- e^{\phi} G_{\iota \kappa} (dc^\iota - C_0 db^\iota) \wedge *_4 (dc^\kappa - C_0 db^\kappa) -& \nonumber \\
&- \frac{9 G^{ij}}{4 \mathcal{K}^2} \left(d \psi_i - \frac{1}{2} \mathcal{K}_{i \iota \kappa} (c^\iota db^\kappa - b^\iota dc^\kappa )\right) \wedge *_4 \left(d \psi_j - \frac{1}{2} \mathcal{K}_{j \iota \kappa} (c^\iota db^\kappa - b^\iota dc^\kappa)\right) +& \nonumber \\
&\left.+ \frac{1}{4} \text{Im} \mathcal{M}_{\alpha \beta} F^\alpha \wedge *_4 F^\beta + \frac{1}{4} \mathcal{M}_{\alpha \beta} F^\alpha \wedge F^\beta - V *_4 1 \right] \,,&
\end{align}
\end{small}
where $F^\alpha = dV^\alpha$ and $\mathcal{M}_{\alpha \beta}$ is the same as in eq. \eqref{eq:MKL} but evaluated in $U^\alpha = \overline{U}^\beta = 0$ to take into account the orientifold projection. The scalar potential turns out to be semi-definite positive, and it takes the form
\begin{align}
\label{eq:ScalarPotentialOrientifolded}
V = \frac{18i e^\phi}{\mathcal{K}^2 \int_\chi \Omega \wedge \overline{\Omega}} \left(\int_\chi \Omega \wedge \overline{G}_3\int_\chi \overline{\Omega} \wedge G_3 + G^{ab} \int_\chi \chi_a \wedge G_3 \int_\chi \overline{\chi}_b \wedge \overline{G}_3\right) \,.
\end{align}

The next step is to bring the four-dimensional action in eq. \eqref{eq:FourDimensionalOrientifoldedAction} in the standard $\mathcal{N} = 1$, which boils down the the choice of the proper coordinates on the moduli space such that it can be written as
\begin{align}
\label{eq:StandardAction}
S^{(4)} =& - \int d^4x \, \left[\frac{1}{2} R *_4 1 + K_{I \overline{J}} DM^I \wedge *_4 D\overline{M}^{\overline{J}} +\right. \nonumber \\
&\left. + \frac{1}{2} \text{Re} f_{\alpha \beta} F^{\kappa} \wedge *_4 F^{\lambda} + \frac{1}{2} \text{Im} f_{\alpha \beta} F^{\kappa} \wedge F^{\lambda} + V *_4 1 \right] \,,
\end{align}
where
\begin{align}
\label{eq:ScalarPotential}
V = e^K \left(K^{I \overline{J}} D_I W D_{\overline{J}} \overline{W} - 3 |W|^2\right) + \frac{1}{2} \left(\text{Re} f\right)^{-1 \, \alpha \beta} D_\alpha D_{\beta} \,.
\end{align}
In eq.s \eqref{eq:StandardAction} and \eqref{eq:ScalarPotential} $M^I$ denotes collectively all complex scalars of the theory and $K_{I \overline{J}}$ is a \Kahler metric which satisfies $K_{I \overline{J}} = \partial_I \partial_{\overline{J}} K(M, \overline{M})$, where $K$ is the \Kahler potential of the theory. The \Kahler covariant derivative takes the form $D_I W = \partial_I W + (\partial_I K) W$.\\

It is necessary to find a complex structure on the moduli space such that the metric of the action in eq. \eqref{eq:FourDimensionalOrientifoldedAction} is manifestly \Kahler. It turns out that the coordinates $U^a$ on the space of the complex structure deformations are already proper \Kahler coordinates and $G_{ab}$ is the corresponding \Kahler metric. On the contrary, the remaining fields need a non-trivial redefinition~\cite{Grana:2003ek, Becker:2002nn}:
\begin{align}
\label{eq:FieldRedefinition}
\tau = C_0 + i e^{-\phi} \,, \qquad G^\iota = c^\iota - \tau b^\iota \,, \nonumber \\
T_i = \frac{1}{2} \mathcal{K}_i(t^i) - \zeta_i(\tau, \overline{\tau}, G, \overline{G}) + i \psi_i\,,
\end{align}
where $\mathcal{K}_i$ is defined in eq. \eqref{eq:NonVanishingOrientifoldedIntersectionNumbers}, while
\begin{align}
\label{eq:Zeta}
\zeta_i = - \frac{i}{2 (\tau - \overline{\tau})} \mathcal{K}_{i \kappa \lambda} G^\kappa (G - \overline{G})^{\lambda}\,.
\end{align}

$\psi_i$ are axionic fields. Since they correspond to the integral of $C_4$ over the four-cycles of the compact manifold, the $\mathcal{N} = 1$ EFT inherits a perturbative \textit{shift-symmetry} under the transformations
\begin{align}
\label{eq:ShiftSymmetry}
\psi_i \rightarrow \psi_i + \tilde{\psi}_i \,, \qquad \tilde{\psi}_i = \text{const.}\,,
\end{align}
from the gauge symmetry in eq. \eqref{eq:GaugeTransformations} of the ten-dimensional theory. Also the field $C_0$ is an axion. It inherits the same shift-symmetry as the fields $\psi_i$ as a consequence of the gauge symmetry of the ten-dimensional $C_0(x,y)$ field. As a consequence $\psi_i$ and $C_0$ do not enter the expression for the tree-level \Kahler potential $K_0$, which in terms of the coordinates in eq. \eqref{eq:FieldRedefinition} can be written as
\begin{align}
\label{eq:KahlerPotential}
K_0 = - 2 \text{ln} \V - \text{ln} \left(-i \left(\tau - \overline{\tau}\right)\right) - \text{ln} \left(-i \int_\chi \Omega \wedge \overline{\Omega} \right)\,,
\end{align}
where in order to emphasize the physical meaning of $\mathcal{K}$, which is essentially the volume of the compact space, we redefined
\begin{align}
\label{eq:VolumeStandardForm}
\V \equiv \frac{\mathcal{K}\left(t^i(T_i + \overline{T}_i, \tau - \overline{\tau}, G^\kappa - \overline{G}^\kappa)\right)}{6} \,,
\end{align}
where $\V$ is the volume of the CY space measured in the Einstein frame and in string units $\ell_s = 2 \pi \sqrt{\alpha'}$. Even if it is not always possible to invert eq. \eqref{eq:FieldRedefinition} in order to get $t^i(\tau, T, G)$, in eq. \eqref{eq:KahlerPotential} we should regard $\V$ as a function of $(\tau - \overline{\tau}, T_i + \overline{T}_i, G^{\kappa} - \overline{G}^\kappa)$, where these combinations of the fields $\tau, T_i, G^{\kappa}$ are due to the shift-symmetry of axions of eq. \eqref{eq:ShiftSymmetry}.\\

In the simplest case with $h_+^{(1,1)} = 1$, so that one single \Kahler field $T$ parametrizes the volume of the compact space, eq. \eqref{eq:FieldRedefinition} can be solved explicitly for $t$ and the final result is
\begin{align}
\label{eq:KahlerPotentialSingleField}
-2 \text{ln} \mathcal{K} = -3 \text{ln} \left[T + \overline{T} - \frac{i}{2(\tau - \overline{\tau})} \mathcal{K}_{1 \kappa \lambda} (G - \overline{G})^\kappa (G - \overline{G})^\lambda\right]\,.
\end{align}

An interesting point is that the moduli space takes the diagonal form
\begin{align}
\label{eq:ModuliSpace}
\mathcal{M} = \mathcal{M}_{\text{cs}}^{h_-^{(1,2)}} \times \mathcal{M}_{\text{k}}^{h_-^{(1,1)} + 1}\,,
\end{align}
where each component is a \Kahler manifold. In~\cite{Martucci:2014ska} the author computed the \Kahler potential without neglecting warping effects and he showed that they deform the \Kahler component of the moduli space. In the present thesis we do not consider this possibility. As we will observe in Sec.~\ref{ssec:VisibleSector}, the introduction of the visible sector supported on D3-branes at singularities spoils the product structure of the moduli space at sub-leading order in the volume expansion. In the simplified case with $G^{\iota} = 0$ the tree-level moduli space undergoes a further split, since the first term depends only on the \Kahler moduli, while the second term depends only on the axio-dilaton. Such splitting is however broken as soon as quantum corrections to the tree-level \Kahler potential in eq. \eqref{eq:KahlerPotential} are considered.\\

The tree-level \Kahler potential $K_0$ in eq. \eqref{eq:KahlerPotential} obeys the \textit{no-scale structure} condition, which takes the form
\begin{align}
\label{eq:NoScaleStructure}
K_0^{I \overline{J}} \frac{\partial K_0}{\partial M^I} \frac{\partial K_0}{\partial \overline{M}_{\overline{J}}} = 3\,,
\end{align}
where the sum runs over the moduli $(T_i, G^{\kappa})$. Eq. \eqref{eq:NoScaleStructure} implies that the scalar potential in eq. \eqref{eq:ScalarPotential} is positive semi-definite.\\

The gauge kinetic functions can be written as
\begin{align}
\label{eq:GaugeKineticFunctions}
f_{\alpha \beta} = -\frac{i}{2} \left.\overline{\mathcal{M}}_{\alpha \beta}\right|_{U^\alpha = \overline{U}^\alpha = 0} \,,
\end{align}
where $\mathcal{M}_{\alpha \beta}$ is reported in eq. \eqref{eq:MKL}. It turns out that $f_{\alpha \beta}$ are holomorphic in $U^a$, as it should be. Finally the scalar potential in eq. \eqref{eq:ScalarPotentialOrientifolded} can be inferred from a tree-level superpotential of the form
\begin{align}
\label{eq:TreeLevelSuperpotential}
W_0 (\tau, U^a) = \int_\chi \Omega \wedge G_3 \,,
\end{align}
which is called Gukov-Vafa-Witten (GVW) superpotential~\cite{Gukov:1999ya}. As we will see in detail in Sec.~\ref{sec:ModelBuilding}, the generation of this superpotential, which can be traced back to the presence of fluxes, gives mass to some of the moduli of the compactification. Notice that \Kahler moduli do not appear in $W_0$, as a consequence of its holomorphicity and of the shift-symmetry in eq. \eqref{eq:ShiftSymmetry} of axion fields.

\section{Model Building}
\label{sec:ModelBuilding}

The aim of the present section is to show how to build a semi-realistic model of particle physics starting from the EFT built in Sec.~\ref{ssec:EFTfromIIB}. There are two main issues in doing model building following the bottom-up approach~\cite{Aldazabal:2000sa}:
\begin{itemize}
 \item Global issues: as we already mentioned, moduli are massless scalar fields which are unwelcome from a phenomenological point of view, since on the one hand they would mediate a fifth force which is not observed in nature, and on the other hand all the parameters of the action in eq. \eqref{eq:StandardAction} depend on their VEVs, so that without fixing them it is not possible to get predictions from string theory. The procedure used to fix the moduli is called \textit{moduli stabilization} and it is the main subject of Sec.~\ref{ssec:ModuliStabilization}.
 \item Local issues: as we argued, in the bottom-up approach to string phenomenology, the visible sector is localized on stacks of D-branes wrapping some cycles in the compact space. It is anyway necessary to look for a D-branes configuration which reproduces the desired extension of the SM. This is the main subject of Sec.~\ref{ssec:VisibleSector}.
\end{itemize}

Given that the visible sector is localized in a small region of the compact space, a nice feature of these models is that they allow to get some degree of decoupling between global and local issues, depending on the details of the model. However in the end, in order to embed the local D-branes configuration within a global construction, it is necessary to perform all the consistency checks which are required to ensure that the compactification is well-defined. We will report an explicit example in Sec.~\ref{ssec:ExplicitGlobalModel}.

\subsection{Moduli Stabilization}
\label{ssec:ModuliStabilization}

The present section is organized as follows: in the first part we show how to stabilize many moduli by simply using the presence of fluxes in the compactification. \Kahler moduli remain unfixed, and in order to stabilize them it is necessary to introduce quantum corrections to the tree-level scalar potential as we show in the second part of the section. Finally, we analyze the moduli stabilization procedure which is extensively used in the present thesis: the \textit{Large Volume Scenario}.

\subsubsection{Tree-level Moduli Stabilization by Fluxes}
\label{sssec:TreeLevelModuliStabilization}

The first step consists in the analysis of the effects of fluxes on the spectrum of the EFT~\cite{Giddings:2001yu, Kachru:2003aw, Lust:2005bd, Antoniadis:2005nu, Gomis:2005wc, GarciadelMoral:2005js, Camara:2004jj}.\\

As we have seen in Sec.~\ref{ssec:EFTfromIIB} the presence of fluxes generates a tree-level superpotential $W_0$ which depends on complex structure moduli $U^a$ and on the dilaton $S$\footnote{Henceforth we use a slightly different notation for the axio-dilaton field: $S = -i \tau = e^{- \phi} - i C_0$.}, as in eq. \eqref{eq:TreeLevelSuperpotential}. The dependence on the complex structure moduli is encoded in $\Omega$, while the dependence on the axio-dilaton field is explicit in the definition of $G_3$. $W_0$ generates a scalar potential for the complex structure moduli and the dilaton. We analyze the scalar potential in an inverse volume expansion, since in order to trust the supergravity approximation $\V \gg 1$. The tree-level expression in eq. \eqref{eq:ScalarPotentialOrientifolded} can be rewritten in a more compact form as
\begin{align}
\label{eq:ScalarPotentialNoScale}
V_{\mathcal{O}\left(\V^{-2}\right)} = e^{K} \left(\left|D_S W_0\right|^2 + \left|D_{U} W_0\right|^2\right)\,,
\end{align}
where the first term in the bracket corresponds to the first term in the bracket of eq. \eqref{eq:ScalarPotentialOrientifolded}, while the second term corresponds to the second term in eq. \eqref{eq:ScalarPotentialOrientifolded}. The sum over the \Kahler moduli does not appear due to the no-scale structure of eq. \eqref{eq:NoScaleStructure}, which makes the \Kahler moduli-dependent terms in the bracket vanish. The only \Kahler moduli dependence lies into the pre-factor $e^K \sim \V^{-2}$, which induces a run-away in the \Kahler directions, as it can be easily understood in the simplest example of a single \Kahler modulus $T$, which implies $\V = \left(T + \overline{T}\right)^{3/2}$. This pre-factor also fixes the order of $V_{\mathcal{O}\left(\V^{-2}\right)}$ in the volume expansion. The only allowed global minimum is then at $V_{\mathcal{O}\left(\V^{-2}\right)} = 0$, namely at
\begin{align}
\label{eq:GlobalMinimum}
D_S W_0 = \int_\chi \Omega \wedge \overline{G}_3 = 0\,, \qquad D_{U^a} W_0 = i \int_{\chi} \chi_a \wedge G_3 = 0\,.
\end{align}
These two conditions correspond to having a ISD $G_3$
\begin{align}
\label{eq:SupersymmetricMinimum}
\left\{\begin{matrix} D_S W_0 = 0 \\ D_{U^a} W_0 = 0 \end{matrix}\right. \quad \Rightarrow \quad \left\{\begin{matrix} G_{3}^{(0,3)} = 0 \\ G_{3}^{(1,2)} = 0 \end{matrix}\right. \quad \Rightarrow \quad *_6 G_3 = i G_3\,.
\end{align}
Given that F-terms govern the breaking of supersymmetry, and they are defined as
\begin{align}
\label{eq:FTermsDefinition}
F^i = e^{K/2} K^{i \overline{j}} D_{\overline{j}} \overline{W} \,,
\end{align}
the minimum in eq. \eqref{eq:GlobalMinimum} is clearly supersymmetric if the additional requirement is satisfied:
\begin{align}
\label{eq:Vanishing03Component}
D_{T_i} W_0 = D_{G^\kappa} W_0 \propto W_0 = 0 \,,
\end{align}
which translates into $G_3^{(0,3)} = 0$, since $W_0 = \int_\chi G_3 \wedge \Omega_3$. As expected, the ISD condition supplemented with $G_3^{(0,3)} = 0$ corresponds exactly to the class of warped compactifications in Tab.~\ref{tab:IIBN1vacua}.\\

The masses of complex structure moduli and of the axio-dilaton field are determined by the flux energy density, which is
\begin{align}
\label{eq:FluxesMasses}
m_{U^a} \sim m_S \simeq \int_{\Sigma_3} F_3 \simeq \int_{\Sigma_3} H_3 \sim \frac{\alpha'}{R^3} = \frac{M_{\rm P}}{\V} \,,
\end{align}
where $\Sigma_3$ denotes a generic a three-cycle inside $\chi$.\\

Eq. \eqref{eq:SupersymmetricMinimum} implies the existence of a landscape of vacua, since it gives rise to $2 h^{(1,2)} + 2$ real equations, which have hundreds or thousands of solutions for $(S,U)$\footnote{In the sensible range of parameters, namely for $s \gg 1$, since it fixes the string coupling constant to small values, as required in the perturbative approach.}, once the parameter space given by $(m^{\hat{a}}, e_{\hat{a}}, m_{\rm RR}^{\hat{a}}, e_{\text{RR}\, \hat{a}}) \in \mathbb{Z}^4$ is scanned.\\

We can conclude that the presence of fluxes allows for the stabilization of complex structure moduli $U^a$ and the axio-dilaton $S$ at supersymmetric global minima, for which eq.s \eqref{eq:SupersymmetricMinimum}, \eqref{eq:Vanishing03Component} hold. This is an intuitive result, since for a given set of fluxes $(m^{\hat{a}}, e_{\hat{a}}, m_{\rm RR}^{\hat{a}}, e_{\text{RR}\, \hat{a}}) \in \mathbb{Z}^4$ there exist only a limited number of complex structures and values of the axio-dilaton field which constrain $G_3$ to have only $(2,1)$ non-vanishing component. On the other hand, it is also intuitive that the presence of fluxes does not stabilize \Kahler moduli, since the ISD condition in eq. \eqref{eq:ISD} is invariant under a rescaling of the internal metric contained into $*_6$, and then a rescaling of the size of the internal cycles is allowed.

\subsubsection{Corrections beyond Tree Level}
\label{sssec:CorrectionsToTreeLevel}

The presence of fluxes allows to stabilize complex structure moduli and the axio-dilaton, but it leaves the \Kahler directions flat as a consequence of the no-scale structure of eq. \eqref{eq:NoScaleStructure}. Nevertheless, there exist various corrections which break the no-scale structure and then produce a non-vanishing potential for the \Kahler moduli. In particular, due to the non-renormalization theorem which protects the superpotential from perturbative corrections~\cite{dine:1986vd, Burgess:2005jx, Becker:2006ks}, we can write schematically
\begin{align}
\label{eq:SchematicRenormalization}
W = W_0 + W_{\rm np}\,, \qquad K = K_0 + K_{\rm p} + K_{\rm np} \,,
\end{align}
where the subscript np stands for ``non-perturbative'', while the subscript p stands for ``perturbative''. $W_0$ and $K_0$ are the tree-level expressions, respectively given in eq. \eqref{eq:TreeLevelSuperpotential} and eq. \eqref{eq:KahlerPotential}.\\

As we have mentioned in Sec.~\ref{sec:StringTheory}, string theory naturally contains two different dimensionless expansion parameters: the string coupling constant $g_s$ and the string-over-internal-size $\frac{\alpha'}{R^2}$. The first one governs the strength of string interactions, while the second one governs the effects due to the one-dimensional nature of strings. In the perturbative approach which we are using, both of them have to be small. As a consequence two different perturbative expansions are allowed in the EFT descending from string compactifications. Non-perturbative corrections instead are due to either D3-brane instantons or gaugino condensation on stacks of D7-branes. At the moment they are well understood only from a four-dimensional point of view.\\

The corrections we are going to introduce break both four-dimensional supersymmetry and the no-scale structure, in order to generate a non-vanishing scalar potential for the \Kahler moduli. From a ten-dimensional perspective this amounts to allow for non-vanishing components $G_{3}^{(3,0)}$, $G_3^{(1,2)}$, $G_3^{(0,3)}$ of the three-form flux. As we will observe in Sec.~\ref{sssec:LVS}, such corrections produce a sub-leading effect in the effective field theory, since the scalar potential for \Kahler moduli is generated at order $\mathcal{O}\left(\V^{-3}\right)$, while the supersymmetric stabilization due to fluxes takes place at order $\mathcal{O}\left(\V^{-2}\right)$. Consequently, in the large volume regime $\V \gg 1$ these corrections can be considered as small perturbations around the supersymmetric background of warped compactifications, so that the EFT is still trustable.

\paragraph{$\mathbf{\alpha'}$-corrections - } The tree-level ten-dimensional action of Type IIB supergravity in eq. \eqref{eq:EffectiveAction} receives corrections from higher derivatives operators. Schematically they can be written as
\begin{align}
\label{eq:CorrectedEffectiveAction}
S_{\rm IIB} = &\, S_{0,\rm tree} + \left(\alpha'\right)^3 S_{0, (3)} + \dots + \left(\alpha'\right)^n S_{0, (n)} + S_{\text{CS}, \rm tree} + \nonumber \\
&+ S_{\rm loc, tree} + \left(\alpha'\right)^2 S_{\text{loc}, (2)} + \dots + \left(\alpha'\right)^n S_{\text{loc}, (n)} \,,
\end{align}
where
\begin{align}
\label{eq:EffectiveActionRRNSNS}
S_{0, \rm tree} = \frac{1}{2 \kappa_{10}^2} \int d^{10}x\,\sqrt{-g} \left[\La_{\rm NSNS} + \La_{\rm RR}\right]\,,
\end{align}
as defined in eq. \eqref{eq:EffectiveAction}, while $S_{\text{CS}, \rm tree} = \frac{1}{2 \kappa_{10}^2} \int d^{10}x\,\sqrt{-g} \La_{\rm CS}$. Each subscript within brackets $(i)$ denotes the $i$-th order correction in the $\alpha'$-expansion. $\alpha'$-corrections to the local action give a non-vanishing potential for D7-branes, but not for D3-branes. In particular it turns out that the $\left(\alpha'\right)^2 S_{\text{loc}, (2)}$ correction produces an effective D3-branes charge for D7-branes~\cite{Giddings:2001yu}. We focus on the leading $\alpha'$-correction $\left(\alpha'\right)^3 S_{0, (3)}$. Supersymmetry and invariance under worldsheet parity constrain $S_{0, (3)}$ to take the schematic form~\cite{Conlon:2005ki}
\begin{align}
\label{eq:LeadingAlphaAction}
\left(\alpha'\right)^3 S_{0, (3)} \sim&\, \frac{1}{2 \kappa_{10}^2} \int d^{10}x\,\sqrt{-g} \left[\mathcal{R}^4 + \mathcal{R}^3 \left(G_3 G_3 + G_3 \overline{G}_3 + \overline{G}_3 \overline{G}_3 + F_5^2 + \left(\nabla \tau\right)^2\right) + \right. \nonumber \\
&+ \mathcal{R}^2 \left(G_3^4 + G_3^2 \overline{G}_3^2 + \dots + \left(\nabla G_3\right)^2 + \left(\nabla F_5\right)^2 + \dots\right) + \nonumber \\
&\left. + \mathcal{R} \left(G_3^6 + \dots + G_3^2 \left(\nabla G_3\right)^2 + \dots\right) + G_3^8 + \dots \right] \,,
\end{align}
where the precise tensorial structure is fully known only for a few contributions~\cite{Kehagias:1997cq, Policastro:2006vt, Liu:2013dna}. For example, the tensorial structure of the $\mathcal{R}^4$ term is completely understood~\cite{Becker:2002nn, Antoniadis:1997eg}. It can be written as the combination
\begin{align}
\label{eq:AlphaPrimeTensorStructure}
J_0 = t_8 t_8 \mathcal{R}^4 - \frac{1}{8} \epsilon_{10} \epsilon_{10} \mathcal{R}^4 \,,
\end{align}
where the tensor $t_8$ is defined in~\cite{Green:1987sp, Tseytlin:1995bi, Minasian:2015bxa}\footnote{For the explicit definition of $t_8$ in terms of the metric see for example Appendix A of~\cite{Minasian:2015bxa}}, while $\epsilon_{10}$ is the totally antisymmetric tensor in 10 dimensions. The correction in eq. \eqref{eq:AlphaPrimeTensorStructure} gives rise to a correction of the \Kahler potential of the form~\cite{Becker:2002nn}
\be
\label{eq:AlphaPrimeKahlerPotential}
K \supset - 2 \log \left(\V + \frac{\xi \left(S + \overline{S}\right)^{3/2}}{2}\right) \,,
\ee
where
\begin{align}
\label{eq:XiHat}
\xi = - \frac{\left(\alpha'\right)^3 \zeta(3) \chi(\chi)}{2^{5/2} (2 \pi)^3} \,,
\end{align}
and $\zeta(3) \simeq 1.202$ is the Riemann zeta function computed in $3$, while $\chi(\chi)$ is the Euler characteristic of the CY manifold $\chi$ considered, which can be written in terms of the Hodge numbers as
\begin{align}
\label{eq:EulerCY}
\chi(\chi) = 2 \left(h^{(1,1)} - h^{(1,2)}\right) \,.
\end{align}
In~\cite{Ciupke:2015msa} further $\alpha'$-corrections at the same order $\left(\alpha'\right)^3$ have been computed. We report on them in Sec.~\ref{sssec:HigherDerivativeAlphaCorrections}. Furthermore, in~\cite{Minasian:2015bxa} the authors showed that the orientifold planes present in the compactification can affect $(\alpha')^3$-correction in eq. \eqref{eq:AlphaPrimeKahlerPotential} by shifting the Euler characteristic in eq. \eqref{eq:XiHat} in the following way
\begin{align}
\chi(\chi) \quad \longrightarrow \qquad \chi(\chi) + 2 \int_{\chi} D_{\rm O7}^3 \,,
\end{align}
where $D_{\rm O7}$ is the class Poincar\'e dual to the divisor wrapped by the O7-plane. This correction does not introduce qualitative changes in the moduli stabilization procedure, and consequently we ignore it.

\paragraph{String Loop Corrections - } The effective ten-dimensional action of Type IIB strings receives additional corrections from string loop effects, both in the bulk part and in the local one. Such corrections, which are poorly understood, are governed by the string coupling constant $g_s$. In the bulk part of the ten-dimensional action they first appear at order $\left(\alpha'\right)^3$, so that they can be though about as a further $g_s$-expansion of each term in the action $S_{\rm IIB}$ in eq. \eqref{eq:CorrectedEffectiveAction}, starting from $\left(\alpha'\right)^3 S_{0, (3)}$. String loop corrections have been computed for Type IIB orientifolds on tori with D5/D9 and D3/D7-branes in~\cite{Berg:2004ek, Berg:2005ja}. Starting from those results, it is possible to guess the volume and dilaton dependence of string loop corrections in the case of a generic CY~\cite{vonGersdorff:2005bf, Cicoli:2007xp, Cicoli:2008va, Berg:2007wt}. In general there are two contributions to the \Kahler potential:
\bi
\item The first contribution comes from the exchange between D3-branes (or O3-planes) and D7-branes (or O7-planes) of closed strings which carry KK momentum. For this reason they are labeled with a index KK:
\begin{align}
\label{eq:KKStringLoopCorrections}
\delta K^{\rm KK}_{(g_s)} = - \frac{1}{128 \pi^4} \sum_{i=1}^{h^{(1,1)}} \frac{\mathcal{C}_i^{\rm KK}(U, \overline{U}) \left(a_{il} t^l\right)}{\text{Re}(S) \V} \,,
\end{align}
where $\mathcal{C}_i^{\rm KK}$ are complicate unknown functions of the complex structure moduli which can be computed explicitly only with a full stringy calculation. $a_{il} t^l$ is a linear combination of the two-cycles volumes $t^i$.
\item The second contribution from string loops comes from the exchange of winding strings between intersecting stacks of D7-branes (or between intersecting D7-branes and O7-planes). It takes the form
\begin{align}
\label{eq:WindingStringLoopCorrections}
\delta K^{\rm W}_{(g_s)} = - \frac{1}{128 \pi^4} \sum_{i=1}^{h^{(1,1)}} \frac{\mathcal{C}_i^{\rm W}(U, \overline{U})}{\left(a_{il} t^l\right) \V} \,,
\end{align}
where again $\mathcal{C}_i^{\rm W}$ requires a full stringy computation to be calculated. $a_{il} t^l$ is a linear combination of the two-cycles volumes $t^i$. 
\ei

\paragraph{Non-perturbative effects - } As we already mentioned, non-perturbative effects are well understood only from a four-dimensional point of view. In case non-perturbative effects take place, the superpotential can be written as~\cite{Blumenhagen:2006ci, Bergshoeff:2005yp, Lust:2005cu, Lust:2005dy, Lust:2006zg, Blumenhagen:2009qh}
\begin{align}
\label{eq:NonPerturbativeEffects}
W = W_0 + \sum_{j = 1} A_j (\phi, S,U) e^{-a_j T_j} \,,
\end{align}
where the sum takes into account all the contributions from both D3-brane instantons and from gaugino condensation on stacks of D7-branes. $\phi$ collectively denotes the (possibly present) open strings degrees of freedom associated with the stack of D7-branes on which gaugino condensation takes place. In general $A(\phi, S, U)$ is an unknown function of $\phi, U, S$. More in detail:
\begin{itemize}
 \item[a)] \textbf{D3 branes instantons - } They can be obtained by wrapping Euclidean D3-branes on four-cycles of the compact space $\chi$. In this case it turns out that $a_j = 2 \pi$. Such superpotential is generated only if precisely two fermionic zero modes are present on the world volume of the Euclidean D3-branes. In F-theory and M-theory such condition can be rephrased in terms of the arithmetic genus $\chi$ of the divisor $\Sigma$ wrapped by the Euclidean M5-branes \cite{Witten:1996bn}
\begin{align}
\chi(\Sigma) = h^{(0,0)} - h^{(0,1)} + h^{(0,2)} - h^{(0,3)} = 1\,.
\end{align}
A similar condition can be stated in Type IIB in some specific cases. It is important to notice that the zero-modes counting can be substantially modified in presence of three-form fluxes and orientifolds \cite{Bergshoeff:2005yp, Lust:2006zg, Kallosh:2005yu, Kallosh:2005gs}. For example in \cite{Lust:2005cu} it has been shown that the component $G^{(1,2)}_3$ of the three-form flux can lift some zero modes.
\item[b)] \textbf{Gaugino condensation on stacks of D7-branes -} We consider a stack of space-time filling D7-branes wrapped around a four-cycle $D$ in the internal space. The relevant physics is given by the open string spectrum on the D7-branes world volume. There are several possibilities. We mention two of them \cite{Lust:2005dy, Intriligator:1995au}:
 \begin{itemize}
 \item \textit{Pure $\mathcal{N} = 1$ Yang-Mills theory with gauge group G.}
 In this case gaugino condensation generates a non-perturbative contribution to the superpotential of the form
 \begin{align}
 \label{eq:PureYMSuperpotential}
 W_{\rm np} \sim \Lambda^3 = e^{- \frac{8 \pi^2}{b g^2}} = e^{- 2 \pi \text{Vol}(D)}\,,
 \end{align}
 where we used that $\text{Vol}(D) = \text{Re}(T) = \frac{4 \pi}{g^2}$ and $b$ is the $\beta$-function of $g$:
 \begin{align}
 \label{eq:BetaFunction}
 \frac{1}{g^2(\mu)} = \frac{1}{g_0^2} - \frac{b}{16 \pi^2} \log \left(\frac{\Lambda_{\rm UV}}{\mu^2}\right)\,,
 \end{align}
 where $g_0 = \sqrt{\frac{4 \pi}{\text{Re}(T)}}$ is the bare gauge coupling fixed at $\Lambda_{\rm UV}$, while $\mu$ is a generic energy scale.
 \item \textit{$\mathcal{N} = 1$ Supersymmetric QCD (SQCD) with gauge group $SU(N_c)$ and $N_F$ matter fields $Q, \tilde{Q}$ in the representations $N_F \left(N_c \oplus \overline{N}_c\right)$.}
 In this case a superpotential of the form
 \begin{align}
 \label{eq:SuperpotentialYMMatter}
 W_{\rm np} = \left(N_c - N_F\right) \left(\frac{\Lambda^b}{\text{det} Q \tilde{Q}}\right)^{\frac{1}{N_c - N_F}}\,,
 \end{align}
is generated if $N_c > N_F$, where $b = 3 N_c - N_F$ is the $\beta$-function of SQCD. $\Lambda$ is the energy scale at which the VEV of $Q \tilde{Q}$ breaks the gauge group $SU(N_c)$ to $SU(N_c - N_F)$. If $N_c \leq N_F$ no non-perturbative superpotential is generated.
 \end{itemize}
\end{itemize}

As a final comment, let us remark that it is not possible to generate a non-perturbative effect on a geometric cycle which also supports a stack of D7-branes giving rise to the visible sector~\cite{Blumenhagen:2007sm}. This is a consequence of the chirality of the MSSM (or some extensions thereof) which makes the \Kahler modulus $T$ associated to the four-cycle wrapped by D7-branes charged under an anomalous $U(1)$. As a consequence of gauge invariance of the superpotential, the non-perturbative contribution can only have the following schematic form~\cite{Conlon:2008wa}
\begin{align}
\label{eq:NPSuperpotentialMatter}
W_{\rm np} \sim \left(\prod_i \phi^i_{\rm hidden}\right) \left(\prod_j \phi^j_{\rm visible}\right) e^{-a T}\,.
\end{align}
However, since the visible fields are required to have vanishing VEV, such a non-perturbative contribution to the superpotential can not arise. This argument pushes towards the possibility of building the visible sector on top of a singularity of the compact space, where this problem can be avoided, as we will see in Sec.~\ref{ssec:ExplicitGlobalModel}.

\subsubsection{Large Volume Scenario}
\label{sssec:LVS}

The \textit{Large Volume Scenario} (LVS) provides a quite generic way to stabilize \Kahler\\moduli by using an interplay of non-perturbative and $\alpha'$-corrections, which break the no-scale structure of eq. \eqref{eq:NoScaleStructure}~\cite{Balasubramanian:2005zx}. In this picture both the dilaton $S$ and the complex structure moduli $U^a$ are stabilized at tree-level, as already explained in Sec.~\ref{sssec:TreeLevelModuliStabilization}. In the present section we take for simplicity $h^{(1,1)}_- = 0$. In the simplest version of the LVS the volume of the compact space takes the so-called ``swiss cheese'' form:
\begin{align}
\label{eq:VolumeLVS}
\V = \alpha_b \tau_b^{3/2} - \sum_{i=2}^{h^{(1,1)}_+} \alpha_i \tau_i^{3/2}\,,
\end{align}
where $\tau_b = \text{Re}(T_b)$ and $\tau_i = \text{Re}(T_i)$ govern respectively the size of the so called ``big'' four-cycle and ``small'' four-cycles, such that $\tau_b \gg \tau_i$. Since the big cycle is much larger than the small ones, the volume is dominated by its value. $\alpha_b$, $\alpha_i$ are constants determined by the intersection numbers of the compact space. In the following we will fix $\alpha_b = \alpha_i = 1$, unless differently specified. Non-perturbative effects due to E$3$-instantons or gaugino condensation on stacks of D7-branes take place on the small cycles, so that the superpotential takes the form
\begin{align}
\label{eq:LVSSuperpotential}
W = W_0(U,S) + \sum_{i = 2}^{h^{(1,1)}_+} A_i(U,S,\phi) e^{-a_i T_i}\,,
\end{align}
while the \Kahler potential
\begin{align}
\label{eq:LVsKahlerPotential}
K = - 2 \log\left(\V + \frac{\hat{\xi}}{2}\right) - \log\left(S + \overline{S}\right) + \log\left(i \int_\chi \Omega \wedge \overline{\Omega}\right) \,,
\end{align}
where we have included the leading $\alpha'$-corrections introduced in Sec.~\ref{sssec:CorrectionsToTreeLevel}. Notice that in order to use LVS it is necessary that $\xi$ in eq. \eqref{eq:XiHat} is negative, namely the CY manifold needs to have more complex structure moduli than \Kahler moduli~\cite{Balasubramanian:2005zx}: $h^{(1,2)}_- > h^{(1,1)}_+$.\\

For the sake of simplicity we consider the case in which $h^{(1,1)}_+ = 2$, and we denote the \Kahler modulus associated to the small cycle $T_2 \equiv T_s$. The F-terms scalar potential at leading order in the volume expansion has the same form as in eq. \eqref{eq:ScalarPotentialNoScale}, despite the presence of $\alpha'$-corrections. Such a scalar potential fixes the dilaton and the complex structure moduli at the supersymmetric minimum in eq. \eqref{eq:GlobalMinimum} which we now rewrite as
\begin{align}
\label{susyminim}
\left.D_S W_0\right|_{\xi  = 0} = 0\ , \qquad D_{U^a} W_0 \big{|}_{\xi = 0} = 0\, ,
\end{align}
where the subscript $\xi = 0$ means that $\alpha'$-corrections can be ignored at this level of approximation, since they are sub-leading. The K\"ahler moduli are stabilized using $\alpha'$-corrections to $K$ \eqref{generalk} and non-perturbative corrections to $W_0$ which give rise to $\mc{O}(\mathcal{V}^{-3})$ contributions to the scalar potential of the form
\begin{align}
\label{eq:LVSScalarPotential}
V_{\mathcal{O}\left(\V^{-3}\right)} = \frac{1}{2 s} \left[\frac{8}{3}(a_s A_s)^2 \sqrt{\tau_s} \frac{e^{- 2 a_s \tau_s}}{\mathcal{V}}
- 4 a_s A_s W_0 	\tau_s \frac{e^{-a_s \tau_s}}{\mathcal{V}^2 } + \frac{3\hat\xi W_0^2}{4\mathcal{V}^3}\right]\, ,
\end{align}
where as a generic feature of LVS, the axion field in the exponents of the non-perturbative effect is stabilized in such a way that the scalar potential in eq. \eqref{eq:LVSScalarPotential} admits a minimum\footnote{The first term in the scalar potential of eq. \eqref{eq:LVSScalarPotential} has to be positive, while the second one has to be negative.}. The last term in the bracket is due to $\alpha'$-corrections. The scalar potential in eq. \eqref{eq:LVSScalarPotential} admits an AdS global minimum which breaks SUSY. Minimization with respect to $\tau_s$ yields
\begin{align}
e^{- a_s \tau_s} = \frac{3 \sqrt{\tau_s} W_0}{4 a_s A_s \mathcal{V}} \frac{\left(1-4\epsilon_s\right)}{\left(1-\epsilon_s\right)}\qquad\text{with}\qquad \epsilon_s \equiv\frac{1}{4 a_s \tau_s}\sim\mc{O}\left(\frac{1}{\ln\mathcal{V}}\right)\ll 1\,.
\label{eq:TaubMinimum}
\end{align}
On the other hand, minimization with respect to $\tau_b$ gives
\begin{align}
\tau_s^{3/2} \simeq \frac{\hat\xi}{2}\,.
\label{eq:TausMinimum}
\end{align}
From eq. \eqref{eq:TaubMinimum} we infer that the minimum of the LVS scalar potential lies at exponentially large volume $\V \gg 1$. Furthermore, since $\V \sim e^{a_s \tau_s}$ and $\tau_s \sim \hat{\xi}$, we understand that the existence of a hierarchy is determined by the smallness of the string coupling constant $g_s \ll 1$ in the perturbative regime. A nice feature of LVS is that $W_0$ does not require to be fine-tuned in order to get a minimum, as it happens in the KKLT setup \cite{Kachru:2003aw}, where $W_0 \sim e^{- \tau}$, with $\tau$ denoting here the size of the four cycle which governs the volume of the compact space. Unfortunately the minimum in eq.s \eqref{eq:TaubMinimum} and \eqref{eq:TausMinimum} is not de Sitter, so that for the compactification to reproduce the real world some additional ingredients need to be added. Further ingredients are also required in order to embed the visible sector into the compactification, since this can not be supported on a four-cycle whose size is stabilized through non-perturbative effects, as the small cycles in the LVS setup.\\

Let us briefly summarize the main properties of LVS relevant for soft SUSY-breaking:
\begin{itemize}
\item{\it Hierarchy of scales:} LVS leads to a hierarchy of scales for masses and soft-terms~\cite{Conlon:2005ki}.
In terms of the volume $\V$, the string scale is (see Appendix~A of~\cite{Conlon:2005ki} for the derivation of the exact pre-factors)
\begin{align}
M_s = \frac{g_s^{1/4} M_{\rm P}}{\sqrt{4\pi\mathcal{V}}}\,,
\label{Ms}
\end{align}
the Kaluza-Klein scale is
\begin{align}
M_\KK\simeq \frac{M_{\rm P}}{\sqrt{4\pi}\mathcal{V}^{2/3}}\,,
\label{MKK}
\end{align}
and the gravitino mass is\footnote{We set the VEV of the K\"ahler potential for complex structure moduli such that $e^{{K_{\rm cs}}/2}=1$.}
\begin{align}
m_{3/2} = e^{K/2} |W|= \left(\frac{g_s^{1/2}}{2\sqrt{2\pi}}\right) \frac{W_0 M_{\rm P}}{\mathcal{V}} +\ldots\, ,
\label{m32}
\end{align}
where the dots indicate suppressed corrections in the inverse volume expansion. Most of the moduli receive a mass of order $m_{3/2}$ except for the volume mode whose mass is $m_\mathcal{V} \simeq m_{3/2}/\sqrt{\mathcal{V}}$. Hence there is a natural hierarchy of scales $M_s\gg M_\KK\gg m_{3/2}\gg m_\mathcal{V}$ for the flux superpotential $W_0$ taking generic values between $1-100$.

\item {\it Bottom-up model building:} The D-branes configuration of the visible sector is localized in a particular corner of the bulk geometry, allowing for a realization of the bottom-up approach to string model building~\cite{Aldazabal:2000sa}. The structure of soft-terms does not depend on the gauge theory realized in the visible sector but only on the type of D-branes configuration (e.g. branes at singularities, D7-branes in the geometric regime) as in the modular approach to string model building. The realization of the visible sector on a cycle different from the one supporting non-perturbative effects allows to achieve compatibility of chirality and moduli stabilization~\cite{Blumenhagen:2007sm,Cicoli:2011qg}.

\item {\it SUSY-breaking:} Assuming a D-branes configuration that leads to the MSSM, the effective field theory allows to analyze the structure of soft-masses. In particular, the pattern of soft masses depends on the location and type of the MSSM D-branes construction in the CY orientifold compactification. If the MSSM is located at a divisor geometrically separated from the main sources of SUSY-breaking in the bulk, e.g.~on a shrinking divisor, there can be a hierarchical suppression of the soft masses below the gravitino mass and the lightest modulus~\cite{Blumenhagen:2009gk}. If the dominant source of SUSY-breaking is in the proximity of the visible sector brane configuration (as it happens if the F-term of the modulus of the cycle wrapped by the SM brane breaks SUSY), the soft masses are of order the gravitino mass with only mild suppressions~\cite{Conlon:2005ki, Conlon:2010ji,Choi:2010gm,Shin:2011uk}.
\end{itemize}

Generically moduli masses tend to be of order the gravitino mass, as we have seen in eq. \eqref{eq:FluxesMasses} and \eqref{m32}. In view of the Cosmological Moduli Problem (CMP) which sets a lower bound on moduli masses of order $100$ TeV~\cite{Coughlan:1983ci,Banks:1993en,deCarlos:1993ja}, it is often desirable to have soft masses well below the gravitino/moduli masses although achieving this requires a special mechanism at play. We will refer to models which have hierarchically suppressed soft masses (not just by loop factors) as sequestered models.\footnote{A similar suppression appears also in the context of realizations of the KKLT scenario~\cite{Choi:2005ge}.} Depending on the location of SM particles and the value of the CY volume, we distinguish three interesting LVS scenarios for SUSY-breaking:
\begin{enumerate}
\item{\bf Unsequestered GUT scale string models:}
Motivated by unification, if one takes the string scale to be close to the GUT scale $10^{14}-10^{16}$ GeV, where the range in the volume captures the uncertainty about high-scale threshold corrections, then the volume is of order $\mathcal{V}\simeq 10^{3}-10^{7}$ for $g_s\simeq 0.1$. This implies a large gravitino mass, $m_{3/2}\simeq 10^{10}-10^{14}$ GeV, i.e.~unobservable sparticles, unless the flux superpotential is tuned to extremely small values (tuning of up to $W_0\sim 10^{-10}$) to get TeV soft-terms.\footnote{TeV scale soft masses in this scenario would lead to light moduli which suffer from the CMP since $m_\mathcal{V} \simeq 10$ GeV.} So the generic situation without tuning $W_0$ is that soft-terms are at an intermediate scale, roughly in the range $10^{10}-10^{14}$ GeV. This scenario is safe from the CMP. The string landscape can in principle address the hierarchy problem.

\item{\bf Unsequestered intermediate scale strings:}
Requiring TeV scale soft-terms in an unsequestered setting leads to a volume of order $\mathcal{V}\simeq 10^{14}$ for $W_0\sim 10$, implying an intermediate string scale, $M_s\simeq 5\cdot 10^{10}$ GeV. This scenario addresses the hierarchy problem, although unification has to work differently from the MSSM (see~\cite{Aldazabal:2000sk, Dolan:2011qu,Cicoli:2013mpa} for concrete string examples with intermediate scale unification). Its spectrum of soft-terms at the electroweak scale has been studied in~\cite{Conlon:2007xv}. It suffers from the CMP since the volume modulus mass is slightly below 1 MeV.

\item{\bf Sequestered high scale string models:}
There can be special situations in which the soft-terms are hierarchically smaller than the gravitino mass, referred to as sequestered scenarios~\cite{Blumenhagen:2009gk}. In LVS this happens in configurations in which the Standard Model degrees of freedom are localized in the extra-dimensions, such as in models where the visible sector arises from open strings on D3-branes at a singularity. In particular, in this setup the F-term of the Standard Model cycle vanishes and the dominant F-terms are associated with other moduli (the volume modulus, the dilaton and other K\"ahler and complex structure moduli).
However the dominant F-terms couple very weakly to the visible sector because of their bulk separation, and this produces a hierarchy between the soft-terms and the gravitino mass. Typically gaugino masses are of order $M_{1/2}\simeq m_{3/2}/\mathcal{V}$, whereas scalar masses can be as suppressed as the gaugino masses or hierarchically larger by a power $\mathcal{V}^{1/2}$ (leading to a Split-SUSY scenario in this last case). This makes these models very attractive for phenomenology since they feature TeV scale soft-terms and no CMP for $\mathcal{V}\simeq 10^7$ and $W_0\simeq 50$ which give
$M_{1/2}\simeq 1$~TeV, $m_{3/2}\simeq 10^{10}$ GeV and $m_\mathcal{V} \simeq 5\cdot 10^6$ GeV. The unification scale in these models is set by the winding scale $M_W = 2 \pi\sqrt{\pi g_s} M_{\rm P} / \mathcal{V}^{1/3}$~\cite{Conlon:2009xf,Conlon:2009kt} which turns out to be of the same order of the standard GUT scale. The appearance of this hierarchical suppression of soft masses is subject to the structure of the effective supergravity. Changes to the EFT at loop or non-perturbative level (see for instance~\cite{Conlon:2010ji,Choi:2010gm, Berg:2010ha, Conlon:2011jq,Berg:2012aq}) can lead to desequestering. In Sec.~\ref{sec:DesequesteringSources} we comment more explicitly on possible sources for desequestering while in the rest of the chapter we assume that these desequestering effects are absent.
\end{enumerate}

In this thesis we focus our attention on sequestered models, hence we need to introduce D3-branes at singularities.

\subsection{Visible Sector}
\label{ssec:VisibleSector}

As we already discussed, the visible sector in Type IIB can be obtained through a D-branes configuration~\cite{Maharana:2012tu}. In order to get a sequestered scenario as described in the last section, we focus on space-time filling D3-branes configurations. The spectrum of a D3-brane is summarized in Tab.~\ref{tab:D3Spectrum}. 
\begin{table}[h!]
\begin{center}
\begin{tabular}{cc}
\hline
Sector & $4$D field \\
\hline
NS & 1 Gauge Boson: $A_\mu$ \\
NS & 6 Real Scalars: $\phi^i$ \\
\hline
RR & 4 Majorana Fermions: $\lambda_\alpha$ \\
\hline
\end{tabular}
\end{center}
\caption{Spectrum of a D3-brane worldvolume theory.}
\label{tab:D3Spectrum}
\end{table}
The fields in Tab.~\ref{tab:D3Spectrum} compose a $U(1)$ multiplet of $\mathcal{N} = 4$ four-dimensional supersymmetry. As a consequence, the worldvolume theory of a D3-brane is not chiral, and the same happens for the straightforward generalization to a stack of $N$ D3-branes, whose spectrum can be arranged in $U(N)$\footnote{Assuming that the D3-branes are not placed on top of an orientifold plane. In the last case also the gauge groups $SO(N)$ and $Sp(N)$ could be obtained.} multiplets of $\mathcal{N} = 4$ supersymmetry. The only known way to allow for chirality in the case of space-time filling D3-branes, is by placing the stack on top of a singularity inside the compact space~\cite{Aldazabal:2000sa, Douglas:1996sw, Douglas:1997de, Kachru:1998ys, Lawrence:1998ja, Hanany:1998it}. D3-branes at singularities constitute the quintessential realization of the bottom-up approach, since the visible degrees of freedom arise from a single point in the compact space. In this case the largest possible separation between gauge degrees of freedom and bulk ones takes place, allowing to tackle moduli stabilization and D-branes model building almost independently. The separation is more accentuated if D7 flavour branes are absent in the compactification, since they would couple directly visible fields to the bulk. For this reason we avoid the presence of D7 flavour branes, and we assume that the MSSM comes from a D3-branes construction.\\

As an example of the power of the bottom-up approach we report a very simple argument to deduce the volume scaling of the \Kahler matter metric of visible scalar fields~\cite{Conlon:2008wa}. As it is well known, physical Yukawa couplings $\hat{Y}_{\alpha \beta \gamma}$ take the form
\begin{align}
\label{eq:HolomorphicYukawaCoupling}
\hat{Y}_{\alpha \beta \gamma} = e^{K/2} \frac{Y_{\alpha \beta \gamma}}{\sqrt{\tilde{K}_\alpha \tilde{K}_\beta \tilde{K}_\gamma}} \,,
\end{align}
in terms of the holomorphic Yukawa couplings $Y_{\alpha \beta \gamma}$ and of the \Kahler matter metric (assuming it is flavour diagonal) $\tilde{K}_\alpha$. Since physical Yukawa couplings should be generated by gauge interactions taking place on top of the D3-branes which correspond to a single point in the compact space, then it is reasonable to assume that they do not depend on the volume of the compactification space. This translates into the following volume scaling for the \Kahler matter metric
\begin{align}
\label{eq:KahlerMatterMetricScaling}
\tilde{K}_\alpha \approx \frac{1}{\V^{2/3}} \,.
\end{align}
As we will observe in the next section, this is exactly the leading order scaling obtained by performing the Kaluza-Klein reduction.\\

As we consider a specific kind of singularities, called \textit{del Pezzo singularities}, from a geometrical point of view the most intuitive description of D3-branes at singularities can be given in terms of the del Pezzo divisor hosting them, in the non-singular (or ``blow-up'') limit~\cite{Malyshev:2007zz}. A del Pezzo surface $dP_n$ ($0 \leq n \leq 8$) is a four-cycle inside the CY manifold which contains several non-trivial two-cycles $H_i$. In general, for a $dP_n$ surface, the non-vanishing Hodge numbers are given by
\begin{align}
\label{eq:HodgeDelPezzo}
h^{(0,0)}_{\rm dP} = 1 \,, \qquad h^{(1,1)}_{\rm dP} = n + 1 \,, \qquad h^{(2,2)}_{\rm dP} = 1 \,,
\end{align}
so that the Euler characteristic of a $dP_n$ is $\chi(dP_n) = n + 3$. In the non-singular limit $dP_n$ divisors can be wrapped by D7-branes, $H_i$ can be wrapped by D5-branes, while $0$-cycles inside $dP_n$ can support D3-branes. In general, not every D-branes configuration is a stable one: only BPS states are minimum-energy states and then they are stable. A stack of D3-branes forms a stable configuration if placed at a smooth point of a CY manifold, but D-branes get recombined into \textit{fractional branes} if placed on top of a singularity. Fractional branes configurations can be described as a stable bound states of D3/D5/D7-branes respectively wrapping the $0/2/4$-cycles of the del Pezzo divisor hosting the D-branes configuration~\cite{Malyshev:2007zz}.

\subsubsection{Dimensional Reduction}

In order to study the EFT arising from a stack of $N$ D3-branes, it is necessary to perform a Kaluza-Klein reduction~\cite{Grana:2003ek}, as we already reviewed for the bulk theory in Sec.~\ref{ssec:ModuliSpace}. The action governing the dynamics of a stack of $N$ D3-branes contains two contributions. The first one is the non-abelian Dirac-Born-Infeld (DBI) action, which in the string frame reads
\begin{align}
\label{eq:DBIAction}
S_{\rm DBI} = - \mu_3 \int_{\mathcal{W}} d^{4}\xi \,\, \text{Tr} e^{-\phi} \sqrt{- \text{det} \left[\varphi^* \mathbf{T}_{\mu \nu} + 2 \pi \alpha' F_{\mu \nu}\right]	 \text{det} Q^n_m}\,,
\end{align}
where
\begin{align}
\label{eq:E}
\mathbf{T}_{\mu \nu} = E_{\mu \nu} + E_{\mu n} \left(Q^{-1} -\delta\right)^{nm} E_{m \nu}\,, \qquad E = g + B \,,
\end{align}
$\mu_3$ has already been introduced in eq. \eqref{eq:DpTension}, $\mathcal{W}$ denotes the D3-brane space-time filling worldvolume, $\xi$ collectively denotes the coordinates on $\mathcal{W}$ and $\varphi: \mathcal{W} \hookrightarrow M$ is the embedding map of the D3-branes into the ten-dimensional space-time $M$. $F_{\mu \nu}$ denotes the field strength of the $U(N)$ gauge theory described by the D3-branes configuration. Moreover we define
\begin{align}
\label{eq:Qnm}
Q^n_m = \delta^n_m + 2 i \pi \alpha' \left[\phi^n, \phi^k\right] E_{km} \,, \quad n, m = 1, \dots, 6 \,,
\end{align}
where the six scalar fields $\phi^i$ parametrize the position of the D3-brane inside the compact space, and transform in the adjoint representation of $U(N)$. The DBI action provides the coupling between the open strings degrees of freedom and the NSNS fields of the bulk theory.\\

The second contribution to the D3-branes action is given by the Chern-Simon action, which reads
\begin{align}
\label{eq:D3CS}
S_{\rm CS} = \mu_3 \int_{\mathcal{W}} \text{Tr} \left(\varphi^* \left(e^{2 i \pi \alpha' \, i_\phi i_\phi} \sum_{q \,\text{even}} C_q e^B\right) e^{2 \pi \alpha' F}\right) \,,
\end{align}
where $i_\phi$ denotes the interior multiplication of a form with the field $\phi^n$, which can be written as
\begin{align}
\label{eq:InteriorProduct}
i_{\phi} C_q = \frac{1}{q!} \sum_{k = 1}^q \, (-1)^{k+1} \phi^n \left(C_q\right)_{\nu_1 \dots \nu_{k-1} n \nu_{k+1} \dots \nu_q} dx^{\nu_1} \wedge \dots \wedge \widehat{dx^k} \wedge \dots \wedge dx^{\nu_q}\,,
\end{align}
where the differential with the $\widehat{\quad}$ is omitted. The CS action provides the coupling between the opens strings degrees of freedom and the RR fields of the bulk theory.\\

The KK reduction is a quite long procedure and it contains many subtleties. Here we just want to emphasize the main points \cite{Grana:2003ek}:
\begin{itemize}
\item As a first step it is necessary to compute the pull-back of the ten-dimensional fields $g$ and $B$ which appear in the actions in eq.s \eqref{eq:DBIAction} and \eqref{eq:D3CS}. However, in order to capture all the degrees of freedom arising from a D3-branes configuration, it is necessary to perturb the background map $\varphi$. Schematically
\begin{align}
\label{eq:PerturbedEmbeddingMap}
\varphi \quad \rightarrow \quad \varphi + \delta \varphi \,,
\end{align}
so that $\delta \varphi$ takes into account the fluctuations of the D3-brane in the directions orthogonal to its world-volume. Such a procedure is called \textit{normal coordinate expansion}.
\item It is necessary to expand the square root of the determinant of the field $Q^n_m$, taking into account the non-abelian nature of the fields $\phi^i$, which leads to
\begin{align}
\sqrt{\text{det} Q^i_j} = 1 + \frac{i}{4 \pi \alpha'} \left[\phi^m, \phi^n\right] \phi^k \partial_k B_{nm} + \left(\pi \alpha'\right)^2 g_{mn} g_{pq} \left[\phi^p, \phi^m\right] \left[\phi^n, \phi^q\right] + \dots\,, \nonumber
\end{align}
where the $\dots$ denote fields which vanish after taking the trace in the lagrangian.
\item In the KK reduction of the Chern-Simon action, the four-form $C_4$ is expanded around its background value, which is determined by the condition $\alpha = e^{4 A}$. We allow for small perturbations around this background value, which means taking into account a small, non-ISD component for $G_3$. As a consequence the solution breaks supersymmetry and deviates from the class of warped compactifications in Tab.~\ref{tab:IIBN1vacua}. We assume that this is a very small departure from the supersymmetric background, so that we can treat it perturbatively, without destroying the solution.
\item Once the reduction has been performed, it is necessary to recast all the expressions in terms of chiral superfields, and then it is necessary to find a complex structure which is compatible with $\mathcal{N} = 1$ supersymmetry. It turns out that the complex structure $J$ of the compact space is the correct choice. A key point in this step is that a mixing between complex structure moduli and open strings degrees of freedom takes place, so that \Kahler moduli need to be redefined differently from eq. \eqref{eq:FieldRedefinition}, as follows~\cite{Grana:2003ek}
\begin{align}
\label{eq:D3FieldRedefinition}
T_i = \frac{1}{2} \mathcal{K}_i - \zeta_i + 4i \pi^2 \left(\alpha'\right)^2 \mu_3 \left(\omega_i\right)_{m \overline{n}} \text{Tr} \left[\phi^m \left(\overline{\phi}^{\overline{n}} - \frac{i}{2} \overline{U}^a \left(\chi_a\right)^{\overline{n}}_p \phi^p\right)\right] \,,
\end{align}
where the trace is performed on the gauge indexes. In the case of a single \Kahler modulus $T$, the moduli space takes the same form as in eq. \eqref{eq:KahlerPotential}, but with a different $\mathcal{K} \equiv \mathcal{K}(\tau,T_i,G^\kappa,U^a,\phi^i)$:
\begin{align}
\label{eq:KahlerPotentialD3}
K=-2 \text{log}&\,  \mathcal{K} = -3 \text{log} \left[T + \overline{T} + \zeta_1 + 12i \pi^2 \left(\alpha'\right)^2 \mu_3 \left(\omega_1\right)_{m \overline{n}} \text{Tr} \left(\phi^m \overline{\phi}^{\overline{n}}\right) +\right. \nonumber \\
&+\left. 3 \pi^2 \left(\alpha'\right)^2 \mu_3 \left(\left(\omega_1\right)_{m \overline{n}} \overline{U}^a \left(\overline{\chi}_a\right)^{\overline{n}}_p \text{Tr} \left(\phi^m \phi^p + \text{h.c.}\right)\right)\right]\,.&
\end{align}
\end{itemize}
From the \Kahler potential in eq. \eqref{eq:KahlerPotentialD3} it is possible to infer the same scaling behavior of eq. \eqref{eq:KahlerMatterMetricScaling} for the \Kahler matter metric:
\begin{align}
\label{eq:KahlerMatterMetricScalingKK}
\tilde{K}_{m \overline{n}} \approx \frac{\left(\omega_1\right)_{m \overline{n}}}{T + \overline{T}} \,.
\end{align}

Notice that in eq. \eqref{eq:KahlerPotentialD3} the complex structure moduli mix with the \Kahler moduli in $\mathcal{K}$ so that the moduli space is no longer of the diagonal form as in eq. \eqref{eq:ModuliSpace}. If warping effects are taken into account as in~\cite{Martucci:2014ska}, then an additional correction with the same volume scaling as the \Kahler matter metric in eq. \eqref{eq:KahlerMatterMetricScalingKK} can appear. In the present thesis we neglect this possibility.

\subsubsection{D3-branes at Singularities}
\label{sssec:D3atSingularities}

Every time a manifold features a discrete symmetry, it is possible to build a new manifold by identifying points under the corresponding transformation. The possible fixed points under the action of the transformation are singularities of the new compactification manifold, called \textit{orbifold singularities}. Orbifold singularities are the simplest singularities which can be encountered, and in particular $\mathbb{C}^3/\mathbb{Z}_3$ which is equivalent to a $dP_0$ divisor whose size is shrunk to zero, is an orbifold singularity. In such a singularity fractional branes have to form a representation of $\mathbb{Z}_3$, namely the $\mathbb{Z}_3$ action has to act by interchanging fractional branes placed at three image points. A D3-brane at a singularity splits into 3 fractional branes, each of which carrying $1/3$ of the original D3-brane mass.\\

The interesting feature of singularities, from a phenomenological point of view, is that by placing a stack of D3-branes on top of a singular point in the compact space it is possible to get a chiral spectrum, starting from a non-chiral $\mathcal{N} = 4$ supersymmetric theory. There is also an additional reason to consider such models. As we discussed in Sec.~\ref{sssec:CorrectionsToTreeLevel}, it is not possible to conciliate the stabilization of the size of an internal four-cycle of the manifold with chirality, which is an essential feature of our world~\cite{Blumenhagen:2007sm}. As a consequence it is not possible to both build the visible sector on top of a stack of branes wrapping a four-cycle in the geometric regime (as the ``small'' cycle in the simplest LVS setup), and stabilize it through non-perturbative effects like E$3$-instantons or gaugino condensation on D7-branes. A possible way-out is represented by D3-branes at singularities constructions \cite{Conlon:2008wa}. In fact, let us assume that the visible gauge sector is hosted by a stack of D7-branes wrapping a four-cycle (whose associated \Kahler modulus is $T_{\rm vis}$) inside the compact space. The chiral nature of the observable world implies the existence of anomalous $U(1)$'s, under which the modulus $T_{\rm vis}$ is charged. We assume for the simplicity of the argument that there is only a single anomalous $U(1)$, so that $T_{\rm vis}$ transforms as
\begin{align}
\label{eq:ChargedVisibleModulus}
\delta_\lambda T_{\rm vis} = T_{\rm vis} + i Q_{T_{\rm vis}} \lambda\,,
\end{align}
where $Q_{T_{\rm vis}}$ is the charge of the modulus $T_{\rm vis}$ under the anomalous $U(1)$, and $\lambda$ is the parameter of the transformation. It generates a D-terms scalar potential which takes the schematic form
\begin{align}
\label{eq:DTermPotential}
V_D \sim \left(\sum_i c_i \left|\Phi^i_{\rm vis}\right|^2 - \xi\right)^2\,,
\end{align}
where $\Phi^i_{\rm vis}$ represent the visible scalar fields, $c_i$ denotes generic coefficients and $\xi$ is the $T_{\rm vis}$-dependent Fayet-Iliopoulos term. This part of the scalar potential fixes $T_{\rm vis}$ such that the combination $\sum_i c_i \left|\Phi^i_{\rm vis}\right|^2 = \xi$, leaving as many flat directions as the number of visible fields charged under the anomalous $U(1)$. This degeneracy is then lifted by sub-leading supersymmetry breaking effects, which contribute to the scalar potential with terms like $m^2 \left|\Phi^i_{\rm vis}\right|^2$. In order for the visible fields not to roll down to charge or color breaking minima, the mass squared $m^2$ has to be positive. In such a way the visible scalar fields are fixed at $\Phi^i_{\rm vis} = 0$, which implies also $\xi = T_{\rm vis} = 0$. As a consequence the size of the four-cycle which supports the visible sector is fixed at zero size: the singularity is obtained in a dynamical way. We will study the details of such a model in Chap.~\ref{chap:Soft-Terms}.\\

Interestingly, from the previous argument it is also possible to infer the tree-level contribution of $T_{\rm vis}$ to the \Kahler potential \cite{Conlon:2008wa}. Indeed, the D-terms scalar potential in eq. \eqref{eq:DTermPotential} arises from an anomalous $U(1)$ which get massive due to the Green-Schwarz mechanism. It is possible to compute that the mass of such a field is given by the string scale $m_{U(1)}^2 \simeq M_s^2$, and since
\begin{align}
\label{eq:MassU1}
m_{U(1)}^2 \simeq M_{\rm P}^2 \frac{\partial}{\partial \overline{T}_{\rm vis}}\frac{\partial}{\partial T_{\rm vis}} K(T_{\rm vis} + \overline{T}_{\rm vis}) \,,
\end{align}
then the leading contribution to the \Kahler potential is given by
\begin{align}
K \supset \lambda_{\rm vis} \frac{\left(T_{\rm vis} + \overline{T}_{\rm vis} + q_{\rm vis} V_{\rm vis}\right)^2}{\V} \,,
\end{align}
where $q_{\rm vis}$ is the charge of the modulus $T_{\rm vis}$ under the anomalous $U(1)$, while $V_{\rm vis}$ is the corresponding vector multiplet. $\lambda_{\rm vis}$ is a $\mathcal{O}\left(1\right)$ unknown constant, which can not be computed in the EFT approach.\\

The crucial point is that models of D3-branes at singularities can be adjusted to support a realistic visible sector. We report here the gauge theories which is possible to get in presence of a stack of space-time filling D3-branes, placed on top of a generic $\mathbb{R}^6/\mathbb{Z}_n$ singularity \cite{Aldazabal:2000sa}. In terms of $\mathcal{N} = 1$ supersymmetry, the fields arising from a stack of $N$ D3-branes can be arranged in $U(N)$ vector multiplets, and three adjoint chiral multiplets $\Phi^r$, whose components are the $6$ real scalar fields $\phi^i$ of Tab. \ref{tab:D3Spectrum}, transforming in the $\mathbf{6}$ of $SU(4) \simeq SO(6)$ plus $4$ adjoint fermions transforming in the $\mathbf{4}$ of $SU(4)$. In order to get a non-abelian gauge theory when a stack of $N$ D3-branes is considered, the states reported in Tab.~\ref{tab:D3Spectrum} for the single D3-brane case have to be supplemented by a Chan-Paton matrix $\lambda \equiv \lambda_{ij}$, where the indexes $i, j = 1, \dots, N$ refer to the branes on which open strings have their endpoints. Each D3-brane placed on top of a $\mathbb{R}^6/\mathbb{Z}_n$ singularity split into $n$ fractional branes, placed at $n$ image points under the action of $\mathbb{Z}_n$. The resulting spectrum is given by states which are invariant under the $\mathbb{Z}_n$ action.\\

The action of $\mathbb{Z}_n$ on fermions and scalars is given respectively by the matrices
\begin{align}
\label{eq:ZnAction}
\mathbf{R}_{\rm ferm}& = \text{diag} \left(e^{2 \pi i a_1/n}, e^{2 \pi i a_2/n}, e^{2 \pi i a_3/n}, e^{2 \pi i a_4/n}\right) \,, \nonumber\\
&\mathbf{R}_{\rm scal} = \text{diag} \left(e^{2 \pi i b_1/n}, e^{2 \pi i b_2/n}, e^{2 \pi i b_3/n}\right) \,,
\end{align}
where
\begin{align}
\label{eq:ZnActionConstraints}
&a_1 + a_2 + a_3 + a_4 = 0 \,\, \text{mod}\, n\,, \nonumber\\
b_1 = a_2 +& a_3 \,, \quad b_2 = a_1 + a_3 \,, \quad b_3 = a_1 + a_2 \,,
\end{align}
Furthermore, the action of $\mathbb{Z}_n$ needs to be embedded on the Chan-Paton indexes. The embedding is provided by the matrix
\begin{align}
\label{eq:ZnActionChanPaton}
\gamma_{\theta,\,3} = \text{diag} \left(\mathbf{1}_{n_0}, e^{2 \pi i/n} \mathbf{1}_{n_1}, \dots, e^{2 \pi (n - 1) i/n} \mathbf{1}_{n_{n-1}}\right)\,,
\end{align}
where $\sum_i n_i = n N$ and the matrices $\mathbf{1}_{n_i}$ are the $n_i \times n_i$ unit matrices. The spectrum of the gauge theory living on D3-branes at a $\mathbb{Z}_n$ singularity is obtained by keeping the states of the $\mathcal{N} = 4$ theory which are invariant under the $\mathbb{Z}_n$ action. This is defined on the states as follows
\begin{itemize}
\item Gauge bosons:
\begin{align}
\label{eq:ZnProjectionBosons}
\lambda = \gamma_{\theta,\,3} \lambda \gamma_{\theta,\,3}^{-1} \,,
\end{align}
\item Fermions:
\begin{align}
\label{eq:ZnProjectionFermions}
\lambda = e^{2 \pi i a_\alpha/n} \gamma_{\theta,\,3} \lambda \gamma_{\theta,\,3}^{-1} \,, \quad \alpha = 1, \dots, 4\,,
\end{align}
\item Complex scalars:
\begin{align}
\label{eq:ZnProjectionScalars}
\lambda = e^{-2 \pi i b_r/n} \gamma_{\theta,\,3} \lambda \gamma_{\theta,\,3}^{-1} \,, \quad r = 1,2,3 \,.
\end{align}
\end{itemize}
The spectrum is finally given by
\begin{align}
\label{eq:33Spectrum}
\text{Vectors}& &&\prod_{i=0}^{n-1} U(n_i) \,, \nn\\
\text{Complex scalars}& &&\sum_{r=1}^3 \prod_{i=0}^{n-1} \left(n_i, \overline{n}_{i-b_r}\right) \,,\\
\text{Fermions}& &&\sum_{\alpha=1}^4 \prod_{i=0}^{n-1} \left(n_i, \overline{n}_{i+a_{\alpha}}\right) \nn\,.
\end{align}
In general this spectrum is not supersymmetric, unless $b_1 + b_2 + b_3 = 0$, in which case $a_4 = 0$ and the fermions corresponding to $\alpha = 4$ turn out to transform as gauginos (in the adjoint representation of $U(n_i)$), while the fermions corresponding to $\alpha = 1,2,3$ transform in the same bi-fundamental representations as complex scalars. In this case complex scalars and fermions fill out a set of chiral multiplets, while gauge bosons and gauginos compose vector multiplets of the resulting $\mathcal{N} = 1$ supersymmetric theory. The condition $b_1 + b_2 + b_3 = 0$ corresponds to the requirement that the $\mathbb{Z}_n$ action is contained in $SU(3)$, as expected given that it gives rise to a supersymmetric theory.\\

We are interested in the case $n = 3$, with $N = 3$ D3-branes at singularities each of which splits into 3 fractional branes placed at image points under the $\mathbb{Z}_3$ action, so that $n_i = 3$ for each $i = 0, 1, 2$. Generically the spectrum reported in eq. \eqref{eq:33Spectrum} is not free from anomalies, and it is necessary to include D7 flavour branes in the compactification in order to erase them. However, in the specific case with $n = 3$ and $n_i = 3$ for $i = 0, 1, 2$ anomalies are automatically canceled~\cite{Aldazabal:2000sa}. The gauge group in this case is that of a \textit{trinification model}:
\begin{align}
\label{eq:TrinificationGaugeGroup}
G = SU(3)_{\rm col} \times SU(3)_{\rm L} \times SU(3)_{\rm R} \,,
\end{align}
and the spectrum arise from the following $27$ states
\begin{align}
\label{eq:TrinificationStates}
3 \, \left[(\mathbf{3},\mathbf{\overline{3}},\mathbf{1}), (\mathbf{1}, \mathbf{3},\mathbf{\overline{3}}), (\mathbf{\overline{3}}, \mathbf{1}, \mathbf{3})\right] \,.
\end{align}
Finally, since there is only a trivial overall $U(1)$ which decouples, the hypercharge has to arise from the $SU(3)_{\rm L} \times SU(3)_{\rm R}$ sector. The main problem of this model is that the structure of Yukawa couplings always generates a wrong hierarchy for fermion masses of the form $(M, M, 0)$, with two heavy and one massless generations~\cite{Conlon:2008wa, Maharana:2012tu}. 

\subsection{An Explicit Global Model}
\label{ssec:ExplicitGlobalModel}

In the present section we exhibit an explicit global embedding of the D3-branes at singularities setup, in which all the consistency conditions have been checked~\cite{Cicoli:2012vw}. In particular it is required that:
\begin{itemize}
\item moduli stabilization is performed in the LVS setup,
\item the visible sector is supported by D3-branes at singularities,
\item the compactification is self-consistent (tadpole cancellation, anomaly cancellation, etc.) and the EFT is trustable.
\end{itemize}
We consider Type IIB warped compactifications with O3/O7-planes. Since we require the gauge theory of the visible sector to be $SU(N)$, the singularity has to lie away from the O7-plane. As a consequence it is necessary to look for a CY manifold which contains two singularities, exchanged by the orientifold action whose fixed locus is an O7-plane. As we have discussed, LVS moduli stabilization requires at least a couple of additional four-cycles, so that the minimal setup contains at least four four-cycles. Finally, as explained in Sec.~\ref{ssec:VisibleSector}, it is required not to have D7 flavour branes, in order to maximize the effect of decoupling between gauge and bulk degrees of freedom. Gauge anomaly cancellation is automatically achieved in models of D3-branes at $\mathbb{C}^3/\mathbb{Z}_3$ singularities without D7 flavour branes.\\

In~\cite{Cicoli:2012vw} the authors considered a scan among CY manifolds which are hypersurfaces in four dimensional toric ambient spaces~\cite{Kreuzer:2006ax, Skarke:1998yk, Denef:2008wq, CYData, Kreuzer:2002uu, Braun:2011ik, Braun:2012vh}, looking for one which satisfies the above requirements. The Hodge numbers of such a CY are $h^{(1,1)}_+ = 3$ and $h^{(1,1)}_- = 1$. From a technical point of view the authors looked for a CY containing two $dP_0$ divisors exchanged by the orientifold involution. They eventually checked that the point of the moduli space where these del Pezzo divisors shrink to zero size lies inside the \Kahler cone.\\

The CY manifold $\chi$ considered contains several divisors which we call $D_4$, $D_5$, $D_6$, $D_7$, $D_8$, of which $D_4$, $D_7$, $D_8$ are three non-intersecting $dP_0$ divisors. A basis for the four-cycles can be chosen as
\begin{align}
\label{eq:BasisH11}
\Gamma_b = D_6 + D_7 = D_4 + D_5 \,, \quad \Gamma_{q_1} = D_4 \,, \quad \Gamma_{q_2} = D_7 \,, \quad \Gamma_s = D_8 \,,
\end{align}
where the subscripts $b$ and $s$ refer respectively to the ``big'' and ``small'' cycles of LVS, while $q_i$ refer to the cycles supporting the visible sector.\\

The expression for the volume of the CY takes the form
\begin{align}
\label{eq:VolumeExplicitCY}
\V = \frac{1}{9} \sqrt{\frac{2}{3}} \left(\tau_b^{3/2} - \sqrt{3} \left(\tau_{q_1}^{3/2} + \tau_{q_2}^{3/2} + \tau_s^{3/2}\right)\right)\,,
\end{align}
where
\begin{align}
\label{eq:CyclesExplicit}
\tau_{q_1} = \text{Vol}(\Gamma_{q_1}) \,, \quad \tau_{q_2} = \text{Vol}(\Gamma_{q_2}) \,, \quad \tau_s = \text{Vol}(\Gamma_s) \,, \quad \tau_b = \text{Vol}(\Gamma_b)\,.
\end{align}

We summarize the D-branes/O-planes configuration of the model:
\begin{itemize}
\item There are two non-intersecting O7-planes: one in the homology class $\Gamma_b$ and one in the homology class $\Gamma_s$.
\item In order to cancel the D7-charge generated by O7-planes, four D7-branes are placed on top of each O7-plane. This generates two hidden sectors $SO(8) \times SO(8)$, which can give rise to gaugino condensation.
\item The visible sector is realized by placing three D3-branes on top of the two singularities exchanged by the orientifold involution. In this way the gauge theory obtained is a trinification model $SU(3)^3$, which can be further broken to the MSSM gauge group~\cite{Cicoli:2012vw}. Gauge anomalies are automatically canceled.
\end{itemize}

As we have seen in Sec.~\ref{ssec:EqOfMotion} it is necessary to cancel the D3-branes charge. This is achieved taking into account the D3-branes charge generated by the presence of D7-branes. Schematically, the condition that is satisfied by the D-branes configuration takes the following form
\begin{align}
\label{eq:D3ChargeCancellationExplicit}
Q^{\rm quiver}_{D3} + Q^{\rm (b)}_{D3} + Q^{\rm (s)}_{D3} + Q^{\rm fluxes} = 0\,,
\end{align}
where $Q^{\rm quiver}_{D3}$ takes into account the contributions from the fractional D3-branes placed at the singularities, while $Q^{\rm (b)/(s)}_{D3}$ correspond to the contributions from D7-branes and O$7$-planes in the homology class $\Gamma_b$ and $\Gamma_s$ respectively.\\

After the involution and in the singular limit the relevant \Kahler moduli are
\begin{align}
\label{eq:KahlerModuliPaper}
T_b = \tau_b + i \psi_b \,, \qquad T_s = \tau_s + i \psi_s \,, \qquad T_{\rm SM} = \tau_{\rm SM} + i \psi_{\rm SM} \,, \qquad G = b + i c \,.
\end{align}
Both $\tau_{\rm SM}$ and $b$ tend to zero in the singular limit, so that they do not enter the expression of the volume in the \Kahler potential
\begin{align}
\label{eq:KahlerPotentialExplicit}
K = - 2 \text{log} \left(\V + \frac{\hat{\xi}}{2}\right) - \text{log} \left(S + \overline{S}\right) + \text{log} \left(i \int_\chi \Omega \wedge \overline{\Omega}\right) + \nonumber \\
+ \frac{\left(T_{\rm SM} + \overline{T}_{\rm SM} + q_{\rm SM} V_{\rm SM}\right)^2}{\V} + \frac{\left(G + \overline{G} + q_G V_G\right)^2}{\V} + \tilde{K_\alpha} C^\alpha \overline{C}^{\overline{\alpha}}\,,
\end{align}
In the definition of $T_{\rm SM}$ and $G$ we have considered the invariant cycle $D_{\rm SM} = D_4 + D_7$ such that
\begin{align}
\label{eq:InvariantCycle}
\tau_{\rm SM} = \frac{1}{2} \int_{D_{\rm SM}} J \wedge J \,, \quad \psi_{\rm SM} = \int_{D_{\rm SM}} C_4 \,,
\end{align}
and the anti-invariant cycle $D_- = D_4 - D_7$ so that
\begin{align}
\label{eq:AntiInvariantCycle}
b = \int_{\hat{D}_-} B_2 \,, \quad c = \int_{\hat{D}_-} C_2 \,.
\end{align}
Finally $q_{\rm SM}$, $q_G$ are the charges of the moduli $T_{\rm SM}$ and $G$ under the anomalous $U(1)$'s with vector multiplets $V_{\rm SM}$ and $V_G$ as we already discussed in the last section.\\

Finally, an important point is to check whether it is possible to get a non-perturbative effect at least on top of the small cycle, in order to perform \Kahler moduli stabilization in the LVS setup. The presence of chiral states on the stack of D7-branes in the homology classes $\Gamma_s$ and $\Gamma_b$, can destroy the possibility of having gaugino condensation. The simplest way to avoid this issue is by setting to zero on both stacks of D7-branes the gauge flux
\begin{align}
\label{eq:GaugeFlux}
\mathcal{F} = F - B\,,
\end{align}
where $F$ denotes the gauge flux of the D7-brane gauge theory. In general this is not a straightforward step, since once the condition $\mathcal{F} = 0$ is imposed on one stack of D7-branes, it could be not possible to impose the same condition on the second stack. However, in the properly chosen configuration of~\cite{Cicoli:2012vw} this is not the case, because the divisors belonging to the homology classes $\Gamma_s$ and $\Gamma_b$ have no intersections. There are two possibilities:
\ben
\item If the field $B$ is chosen such that $\mathcal{F}_s = 0$ and $\mathcal{F}_b \neq 0$, then no non-perturbative effect for $T_b$ is generated. In this case it is possible to get a dS vacuum, since the non-vanishing flux on $\Gamma_b$ induces a non-zero $U(1)$ charge for $T_b$. As a consequence a D-terms in the scalar potential is generated, whose interplay with the F-terms part of the scalar potential leads to an uplifting contribution of the usual AdS vacuum of LVS. This kind of solution is that explicitly built in~\cite{Cicoli:2012vw}. Further details on the de Sitter mechanism are provided in Sec.~\ref{sssec:MatterUplift}.
\item If the field $B$ is chosen in such a way that $\mathcal{F} = 0$ on both $\Gamma_s$ and $\Gamma_b$, depending on the details of the model a non-perturbative effect for $T_b$ can be generated, but in general it is not possible to get a dS vacuum without the introduction of new elements in the compactification, such as anti-branes at the tip of warped throats~\cite{Kachru:2003aw} or new non-perturbative effects taking place on an additional singularity placed on top of the O7-plane~\cite{Cicoli:2012fh}. In Sec.~\ref{sssec:DilatonUplift} we take into account the latter possibility. A global construction for such a de Sitter sector is still missing.
\een
In both cases the superpotential contains a non-perturbative term for the ``small'' cycle
\begin{align}
\label{eq:SuperpotentialExplicit}
W \supset A e^{- a_s T_s}\,,
\end{align}
which is needed in the LVS setup for \Kahler moduli stabilization. Finally in \cite{Cicoli:2012vw} the authors checked that in this setup it is possible to choose gauge fluxes and $B$-field in order to cancel the Freed-Witten anomaly in a way consistent with the generation of the non-perturbative superpotential terms.\\

In conclusion the configuration of D-branes/O-planes considered in \cite{Cicoli:2012vw} gives rise to a consistent global compactification, in which the visible sector is realized on top of D3-branes at singularities. The consistency of the compactification is ensured by tadpole cancellation, cancellation of Freed-Witten anomalies and cancellation of gauge anomalies. As we will observe in chapter \ref{chap:Soft-Terms} in this setup it is possible to stabilize all the closed moduli in the LVS setup, since a term as in eq. \eqref{eq:SuperpotentialExplicit} is always present in the superpotential. A realistic de Sitter vacuum can be achieved thanks to the presence of hidden matter on the D7-branes wrapping the ``big'' cycle of the swiss-cheese CY. It is also possible to get a de Sitter vacuum in a second way, by exploiting a perturbative effect taking place on top of a four-cycle invariant under the orientifold involution, but this setup requires a different CY manifold and a complete realization of the global embedding is still missing.

\part{Supersymmetry Breaking in Sequestered Models}
\label{part:SUSYBreaking}

\chapter{Sequestered de Sitter String Models: Soft-Terms}
\label{chap:Soft-Terms}

The simplest models of low-energy supersymmetry as a solution to the hierarchy problem are in tension with the latest LHC results (see e.g.~\cite{Craig:2013cxa} and references therein) which are moving the bounds for sparticle masses beyond the TeV scale. We are then either in a situation where we accept two to three orders of magnitude of tuning as still `natural', or we are at a very particular corner in the MSSM parameter space with less fine-tuning (e.g.~natural SUSY~\cite{Kitano:2005ew,Papucci:2011wy,Brust:2011tb}, compressed spectra~\cite{Lebedev:2005ge,LeCompte:2011cn}, RPV models~\cite{Allanach:2012vj,Evans:2012bf}), or we need alternatives to the conventional MSSM. Given this, there are various avenues to explore for addressing the electroweak hierarchy problem:
\begin{enumerate}
\item The simplest MSSM models (e.g.~CMSSM) need to be modified at low energies to account for particular corners in the MSSM parameter space with reduced fine-tuning. Or, one step further, extensions of the MSSM including extra matter and/or interactions at the TeV scale may relax the tuning of the MSSM (see e.g.~\cite{Ross:2012nr}).
\item The MSSM is the correct description for beyond the Standard Model physics but the hierarchy problem is addressed by different amounts of fine-tuning through the multiverse just like the cosmological constant problem~\cite{Bousso:2000xa}, where we can distinguish the following classes:
\begin{enumerate}
\item The simplest models of low-energy SUSY are realized with some two to three orders magnitude of fine-tuning.
\item One just keeps the appealing features of low-energy SUSY of realizing the correct DM density and gauge coupling unification whereas the hierarchy problem is no longer addressed. This proposal is commonly referred to as Split-SUSY~\cite{ArkaniHamed:2004fb} where gauginos are at the TeV scale while the scalar superpartners are hierarchically heavier.
\item DM and gauge coupling unification are achieved by other mechanisms and the SUSY particles are at a scale far above the electroweak scale such as an intermediate scale.
\end{enumerate}
\item One can consider alternative solutions to the hierarchy problem such as composite models or extra-dimensional models.
\end{enumerate}

Each of these scenarios has its own virtues and demerits. The first one aims at avoiding fine-tuning in the parameter space of the MSSM, but without a principle on why to favor a particular extension in a UV theory, it is in some sense a tuning in theory space which is as appealing as fine-tuning in parameter space, the others simply accept some sort of tuning.\footnote{Particular interesting corners of parameter space for soft-terms can be obtained by invoking principles such as precision gauge coupling unification~\cite{Krippendorf:2013dqa} or by identifying pattern in underlying UV theories (e.g.~realization of natural SUSY and compressed spectra in the Heterotic mini-landscape~\cite{Krippendorf:2012ir,Badziak:2012yg}).} Given this state of affairs, we are left with the unpleasant situation that at present the best argument in favor of low-energy SUSY is that other alternatives, like large extra-dimensions or composite models, are looking even worse.\\

This is a golden opportunity for string theoretical scenarios to play a role. Being the only explicit scenarios that provide a UV completion of the Standard Model, they should be able to address the problems of the scenarios mentioned above,
provide guidance towards their explicit realization and maybe even suggest other alternative avenues.\\

Consistent string theories are typically supersymmetric. Unfortunately, low-energy SUSY or the MSSM are not a prediction of string theory and its potential discovery or lack of will not directly test string theory. Moreover for a high string scale of order $10^{16}$~GeV (as hinted by standard MSSM gauge coupling unification), obtaining at the same time low-energy SUSY can be a challenge for model building. Another important feature is the string landscape which can potentially have an impact on the hierarchy problem. These are very important issues which can impact LHC and future collider observations. They need to be addressed systematically and within a complete string framework, as we do in one specific example in the present chapter.\\

Fortunately progress in the understanding of SUSY-breaking in string compactifications is maturing right on time to play a role. Several scenarios in which most of the string moduli have been stabilized with SUSY-breaking and computable soft-terms have emerged~\cite{Choi:2005ge,Nilles:1997cm,Conlon:2006us,Conlon:2006wz, Lowen:2008fm, deAlwis:2009fn, Cicoli:2013rwa,Acharya:2008zi}. Some of them are also consistent with cosmological constraints such as the CMP and the realization of de Sitter (dS) vacua. In particular, the LARGE Volume Scenario (LVS)~\cite{Balasubramanian:2005zx}, on which we focus, allows for several of the above SUSY-breaking scenarios in which soft-terms can be explicitly computed. Moreover, LVS is an ideal framework to build globally consistent MSSM-like chiral models for explicit CY compactifications with all closed string moduli stabilized~\cite{Cicoli:2011qg,Cicoli:2013mpa,Cicoli:2012vw,Cicoli:2013cha}. It is also possible to obtain dS vacua from supersymmetric effective actions~\cite{Cicoli:2012vw,Cicoli:2012fh} and the string landscape allows for a controllable fine-tuning of the cosmological constant and potentially the electroweak hierarchy problem.\\

In this chapter our focus shall be on sequestered models, in order to explore the possibility of getting low-energy SUSY from string compactifications. In~\cite{Blumenhagen:2009gk} it was realized that soft-terms can potentially be sensitive to the mechanism responsible for achieving a dS minimum. Lack of a controlled understanding of the way to get dS vacua made it difficult to present a complete analysis of the SUSY phenomenology. Recently there has been progress in obtaining dS vacua from supersymmetric effective actions~\cite{Cicoli:2013mpa, Cicoli:2012vw, Cicoli:2012fh, Cicoli:2013cha,Krippendorf:2009zza}. In this chapter we work out this dependence on the uplifting mechanism in sequestered models. As previously, we assume an MSSM spectrum from the local D-branes configuration for simplicity.\\

We explicitly compute all soft-terms for sequestered scenarios identifying different cases depending on the mechanism to obtain dS vacua and the moduli-dependence of the K\"ahler metric for matter fields. Broadly, we find two classes of models: scenarios in which all soft-terms are of order $m_{3/2}/\mathcal{V}$ and scenarios where gaugino masses and A-terms are of this order but scalar masses are of order $m_{3/2}/\mathcal{V}^{1/2}$. In both cases the numerical coefficients of the soft-terms are determined by background fluxes and therefore can be tuned by scanning through the landscape. This provides an explicit mechanism for the (small) tuning that might be necessary to confront LHC data. In the first class of models the spectrum is similar to standard MSSM spectra with soft-terms of the same order but with the potential of extra non-universal flux-dependent contribution. The second one gives a universal Mini-Split scenario with negligible non-universalities.\\

The rest of this chapter is organized as follows. Sec.~\ref{sec:SeqLVS} contains the detailed setup that leads to sequestered LVS models and a presentation of two mechanisms to obtain dS vacua. We then compute the leading order expressions of the associated F-terms and soft-masses for these scenarios in Sec.~\ref{sec:ftermsandsoftmasses}. Finally in Sec.~\ref{sec:DesequesteringSources} we comment on possible sources of desequestering before concluding in Sec.~\ref{sec:conclusions}. This chapter is based on~\cite{Aparicio:2014wxa}.

\section{Sequestered LVS Scenarios}
\label{sec:SeqLVS}

After having introduced many ingredients needed to build a phenomenologically viable compactification, let us summarize a setup in Type IIB CY flux compactifications with O3/O7-planes that leads to dS moduli stabilization
\`a la LVS and a visible sector sequestered from SUSY-breaking:
\begin{itemize}
\item The simplest LVS vacua can be obtained for a CY with negative Euler number and at least one blow-up
of a point-like singularity~\cite{Cicoli:2008va}, as explained in Sec.~\ref{sssec:LVS}. For these manifolds the volume~$\mathcal{V}$ is of swiss-cheese type as in eq. \eqref{eq:VolumeLVS}. In the explicit case of Sec.~\ref{ssec:ExplicitGlobalModel} the intersection numbers are given in eq. \eqref{eq:VolumeExplicitCY}, but the results of this chapter are independent of the particular values of $\alpha_i$.\footnote{It is possible to implement LVS in CYs which have a more general volume form~\cite{Cicoli:2008va} but this does not alter the structure of soft-masses and so we do not consider these cases.}
\item The visible sector can be realized with appropriate D-branes configurations on blow-up moduli. Concrete D-branes realizations with D3/D7 branes at del Pezzo singularities can lead to interesting gauge/matter extensions of the MSSM. As qualitatively discussed in Sec.~\ref{sssec:D3atSingularities} the size of the associated four-cycle can shrink to zero value due to D-terms stabilization. Because of this shrinking, the F-term of the corresponding blow-up K\"ahler modulus is vanishing at leading order giving rise to a sequestered scenario. We report all the details of this mechanism in Sec.~\ref{sssec:Shrinking}  In order to get the maximal degree of sequestering, in this thesis we consider D3-branes at singularities constructions without D7 flavour branes.
\item In order to realize a dS vacuum one introduces further ingredients in the compactification. Here we concentrate on two options:
(i) Hidden sector matter fields on the large cycle which acquire non-zero F-terms because of D-terms fixing~\cite{Cicoli:2012vw}; (ii) E(-1) instantons at a second singularity whose blow-up mode develops non-vanishing F-term due to new dilaton-dependent non-perturbative effects~\cite{Cicoli:2012fh}. These mechanisms will be discussed respectively in Sec.~\ref{sssec:MatterUplift} and in Sec.~\ref{sssec:DilatonUplift}.
\end{itemize}
The setup with de Sitter scenario arising from matter fields on the large cycle has been realized in concrete CY orientifold compactifications with D3(/D7) branes at singularities~\cite{Cicoli:2013mpa,Cicoli:2012vw,Cicoli:2013cha} that satisfy all global consistency conditions (e.g.~tadpole cancellation), as reported in Sec.~\ref{ssec:ExplicitGlobalModel}.\\ 

The minimal setup that allows this realization includes at least four K\"ahler moduli: a `big' four-cycle $T_b$ controlling the size of the CY volume, a `small' blow up mode $T_s$ supporting non-perturbative effects, the visible sector cycle $T_\SM$ and its orientifold image $G$. These last two moduli are associated to two del Pezzo divisors which collapse to zero size due to D-terms fixing\footnote{The positivity of soft scalar masses for visible sector fields fixes all remaining flat-directions after D-terms stabilization~\cite{Cicoli:2012vw}.} and are exchanged by the orientifold involution. This setup leads to $h^{(1,1)}_+ = 3$ and $h^{(1,1)}_- = 1$ with the K\"ahler moduli already reported in eq. \eqref{eq:KahlerModuliPaper}.

\subsection*{${\cal N} = 1$ Supergravity Effective Field Theory}

In this section we briefly review the low-energy effective action relevant for our construction in the language of 4D ${\cal N}=1$ supergravity, that we have already studied in Sec. \ref{chap:Compactifications&ModelBuilding}. We write the expressions in such a way that a couple of alternative scenarios for de Sitter sectors are included, the first one of which has already a fully consistent global embedding \cite{Cicoli:2012vw} as explained in Sec. \ref{ssec:ExplicitGlobalModel}. Including de Sitter and matter contributions, the superpotential takes the following form
\begin{align}
W= W_0(U,S) + A_s(U, S)\, e^{- a_s T_s} +  W_\dS+W_{\rm matter}\,.
\label{eq:FullSuperpotential}
\end{align}
As explained in Sec.~\ref{sssec:CorrectionsToTreeLevel}, the pre-factor $A_s(U,S)$ depends on both complex structure moduli $U$ and the dilaton $S$, and it is an $\mathcal{O}(1)$ function.\footnote{The dependence on $S$ and $U$-moduli is structurally different, i.e.~the dependence on the dilaton is generated when including the back-reaction of sources and warping on the geometry~\cite{Baumann:2006th}.} The term $W_\dS$ involves the contribution from the mechanism used to obtain a dS vacuum (see Sec.~\ref{ssec:deSitterScenarios}) while $W_{\rm matter}$ is the visible sector superpotential
\begin{align}
W_{\rm matter} = \mu(M) H_u H_d + \frac 16 Y_{\alpha\beta\gamma}(M) C^\alpha C^\beta C^\gamma + \cdots\,,
\label{Wmatter}
\end{align}
where we denoted the moduli as $M$ and the MSSM superfields as $C^\alpha$. Moreover, the dots refer to higher dimensional operators. We also separated the two higgs doublets $H_u$ and $H_d$ from the rest of matter fields in the moduli-dependent $\mu$-term. Because of the holomorphicity of $W$ and the perturbative shift-symmetry of the axionic components of the K\"ahler moduli, the Yukawa couplings and the $\mu$-term can depend only on $S$ and $U$ at the perturbative level with the $T$-moduli appearing only non-perturbatively. We discuss this dependence in more detail in Sec.~\ref{sec:ftermsandsoftmasses} and ~\ref{sec:DesequesteringSources}.\\

For the reasons explained in Sec.~\ref{sssec:D3atSingularities}~\cite{Conlon:2008wa,Blumenhagen:2009gk}, we assume the following form of the K\"ahler potential which describes the regime for the visible sector near the singularity
\begin{align}
\label{generalk}
K = - 2 \ln\left(\mathcal{V} + \frac{\hat\xi}{2}\right) - \ln(2s) + \lambda_\SM \frac{\tau_\SM^2}{\mathcal{V}} + \lambda_b \frac{b^2}{\mathcal{V}} + K_\dS + K_{\rm cs}(U) + K_{\rm matter}\, ,
\end{align}
where $\hat\xi\equiv \xi s^{3/2}$, the $\lambda$'s are $\mc{O}(1)$ coefficients, $K_{\rm cs}(U)$ is the tree-level K\"ahler potential for complex structure moduli and $K_\dS$ encodes the dependence on the sector responsible for obtaining a dS vacuum (see Sec.~\ref{ssec:deSitterScenarios}). The matter K\"ahler potential $K_{\rm matter}$ is taken to be
\begin{align}
K_{\rm matter} = \tilde{K}_{\alpha}(M, \overline{M}) \overline{C}^{\overline{\alpha}} C^{\alpha} + [Z(M, \overline{M}) H_u H_d + \text{h.c.}]\,.
\label{mm}
\end{align}
We assume at this stage that the matter metric is flavour diagonal beyond the leading order structure which was highlighted in~\cite{Conlon:2007dw}.\footnote{Sub-leading flavour off-diagonal entries which can in principle appear~\cite{Camara:2013fta} are taken to be absent. This is motivated by the appearance of additional anomalous $U(1)$ symmetries in D-branes models, in particular also in the context of del Pezzo singularities~\cite{Dolan:2011qu}.} The only exception is that we allow for the higgs bilinear to appear in $K_{\rm matter}$ which we parametrize with the function $Z$. Note that $\tilde{K}_\alpha$ is the matter metric for the visible sector which we will parametrize as~\cite{Blumenhagen:2009gk}
\begin{align}
\tilde{K}_{\alpha} = \frac{f_\alpha(U,S)}{\mathcal{V}^{2/3}} \left(1 - c_s \frac{\hat\xi}{\mathcal{V}}
+ \tilde{K}_\dS + c_\SM \tau_\SM^p + c_b b^p\right), \qquad p > 0\,,
\label{mattermetric}
\end{align}
where we have used $\tilde{K}_\dS$ to parametrize the dependence on the dS mechanism (details will be given in Sec.~\ref{soft}). The $c$'s are taken as constants for simplicity while $p$ is taken to be positive in order to have a well-behaved metric in the singular limit $b,\tau_\SM \to 0$. As they can in principle depend on $U$ and $S$, we comment in due course on the influence on the soft-terms of such a dependence. The appearance of the higgs bilinear and its potential parametrization are discussed in Sec.~\ref{sssec:SoftTerms} when we analyze the $\mu$-term in this scenario. In general the functions $f_\alpha(U,S)$ could be non-universal. Such non-universality can have interesting phenomenological implications (e.g.~mass hierarchies among families of sfermion masses needed for a realization of natural SUSY). As we are interested in soft-terms arising for D-branes at singularities, we take the gauge kinetic function to be
\begin{align}
f_a=\delta_a S+ \kappa_a \,T_\SM\, ,
\label{eq:gaugekineticf}
\end{align}
where $\delta_a$ are universal constants for $\mathbb{Z}_n$ singularities but can be non-universal for more general singularities.\\

\subsection{Moduli Stabilization}

As outlined earlier in this section, we stabilize the moduli following the LVS procedure. The complex structure moduli and the dilaton are fixed at tree-level by background fluxes while the K\"ahler moduli are fixed using higher order corrections to the effective action~\cite{Cicoli:2013cha}. In this section we first explain the D-terms stabilization in detail, and then we perform the F-terms stabilization taking into account both the effects coming from the de Sitter sector, and the small shift of the supersymmetric minimum $D_S W_0 = D_{U^a} W_0 = 0$ due to non-perturbative and $\alpha'$-corrections which induces a non-vanishing IASD component of the three-form flux $G_3$.

\subsubsection{D-terms Stabilization}
\label{sssec:Shrinking}

The K\"ahler moduli where the visible sector D-branes configuration is located are stabilized using D-terms which are the leading order contribution to the potential. Remaining flat directions are stabilized using sub-leading F-terms contributions. To set the notation, let us review D-terms stabilization~\cite{Cicoli:2012vw}. The moduli $T_\SM$ and $G$ are charged under two anomalous $U(1)$ symmetries with charges $q_1$ and $q_2$. The corresponding D-terms potential reads
\begin{align}
V_D = \frac{1}{2{\rm Re}(f_1)} \left(\sum_\alpha q_{1 \alpha} \frac{\partial K}{\partial C^\alpha} C^\alpha - \xi_1\right)^2
+ \frac{1}{2{\rm Re}(f_2)} \left(\sum_\alpha q_{2 \alpha} \frac{\partial K}{\partial C^\alpha} C^\alpha - \xi_2\right)^2\,,
\label{dtermpot}
\end{align}
where $f_1$ and $f_2$ are the gauge kinetic functions of the two $U(1)$s.
The Fayet-Iliopoulos~(FI) terms are given by (see appendix of~\cite{Cicoli:2011yh} for the exact numerical factors)
\begin{align}
\xi_1 &= - \frac{q_1}{4\pi} \frac{\partial K}{\partial T_\SM} = - \frac{q_1  \lambda_\SM}{4\pi} \frac{\tau_\SM}{\mathcal{V}}\,, \label{xiSM} \\
&\xi_2 = - \frac{q_2}{4\pi} \frac{\partial K}{\partial G} = - \frac{q_2 \lambda_b}{4\pi} \frac{b}{\mathcal{V}}\,.
\end{align}
The vanishing D-terms condition fixes therefore $a_\SM$ and $b$ in terms of visible sector matter fields.
The remaining flat directions are fixed by sub-leading F-terms contributions which give vanishing VEVs to the $C^\alpha$ if they develop non-tachyonic soft masses from SUSY-breaking~\cite{Cicoli:2012vw}.\footnote{If the soft scalar masses of some $C^\alpha$ are tachyonic, they develop non-zero VEVs (which could be phenomenologically allowed for some Standard Model singlets) that, in turn, induce non-zero FI-terms~\cite{Cicoli:2013cha}. However $\tau_\SM$ and $b$ would still be fixed in the singular regime since their VEVs would be volume-suppressed~\cite{Cicoli:2013cha}.} Hence the D-terms potential \eqref{dtermpot} vanishes in the vacuum since it is fixed to a supersymmetric minimum at $\xi_1 = \xi_2 = 0$. This corresponds to the singular limit $\tau_\SM = b = 0$. In turn, the axions $a_\SM$ and $c$ are eaten up by the two $U(1)$ gauge bosons in the process of anomaly cancellation.

\subsubsection{F-terms Stabilization}

F-terms stabilization proceeds as explained in Sec.~\ref{sssec:LVS}. The de Sitter sector generates a correction to the minimum in eq. \eqref{eq:TausMinimum}
\begin{align}
\tau_s^{3/2} = \frac{\hat\xi}{2} \left[1 + f_\dS(\epsilon_s) \right]\,,
\label{eq:TausMinimumCorrected}
\end{align}
where $f_\dS$ is a sub-dominant function of $\epsilon_s = \frac{1}{4 a_s \tau_s}$ which depends on the particular mechanism used to obtain a dS vacuum, as reported below. The relation \eqref{eq:TausMinimumCorrected} implies that at the minimum (neglecting $f_\dS$)
\begin{align}
\hat\xi\simeq \frac{1}{4\left(a_s \epsilon_s\right)^{3/2}}\sim\mc{O}\left[\left(\ln\mathcal{V}\right)^{3/2}\right] \gg 1\,.
\label{tausms2}
\end{align}
Given that the potential (\ref{eq:LVSScalarPotential}) depends on $S$ and $U$ (via $A_s(U,S)$ and $s$-dependent $\alpha'$ effects), the minimum (\ref{susyminim}) is slightly shifted from its supersymmetric locus. This shift is fundamental for the soft-term computation in sequestered scenarios since non-vanishing F-terms of $U$ and $S$ at sub-leading order can actually provide the main contribution to soft-terms~\cite{Blumenhagen:2009gk}.

\subsubsection{Shift in the Minimum}
\label{shiftsection}

Let us try to estimate the shift of $S$ and $U$ from their supersymmetric minimum (\ref{susyminim}) because of $\alpha'$ and non-perturbative effects. The K\"ahler covariant derivative of the total superpotential evaluated at the minimum (\ref{eq:TaubMinimum}) and (\ref{eq:TausMinimumCorrected}) reads (we neglect $\mc{O}(\epsilon_s)$ effects)
\begin{align}
D_S W \simeq \left.D_S W_0\right|_{\xi = 0} - \frac{3\hat\xi W_0}{4s\mathcal{V}}
\left[1 + \epsilon_s s\partial_s \ln A_s(U,S)\right]\,.
\end{align}
Since we do not know the functional dependence of $A_s(U,S)$ and since the $W_\dS$ term in eq. \eqref{eq:FullSuperpotential} can also potentially shift the dilaton minimum, it is not possible for us to compute this shift explicitly. We will parametrize it by using the parameter $\omega_S(U,S)$ defined as
\begin{align}
\label{dilmin}
D_S W = - \frac{3 \omega_S(U,S)}{4} \frac{\hat\xi W_0}{s\mathcal{V}}\,.
\end{align}
The dependence of $A_s(U, S)$ on the complex structure moduli is also responsible for shifting the $U$-moduli
from their supersymmetric minimum. After imposing the minimization conditions, the total $D_{U^a} W$ looks like (denoting $u\equiv {\rm Re}(U)$)
\begin{align}
D_{U^a} W \simeq \left.D_{U^a} W_0\right|_{\xi = 0} - \frac{3\hat\xi W_0}{4\mathcal{V}}
\epsilon_s \partial_u\left[K_{\rm cs}(U)+  \ln A_s(U,S)\right]\,,
\end{align}
and so we can parametrize this shift by $\omega_{U^a}(U,S)$ as
\begin{align}
\label{csmin}
D_{U^a} W = - \frac{3 \omega_{U^a} (U,S)}{4} \frac{\hat\xi W_0}{s\mathcal{V}} \quad\Rightarrow\quad D_{U^a} W =\frac{\omega_{U^a} (U,S)}{\omega_S(U,S)} D_S W\sim \mc{O}(\mathcal{V}^{-1})\,,
\end{align}
where both $\omega_S$ and $\omega_{U^a}$ are expected to be $\mc{O}(1)$ functions of $S$ and $U$. Note that both functions $\omega_{S/U^a}$ depend also on the dS mechanism. These corrections give rise to $\mathcal{O}\left(\V^{-4}\right)$ contributions to the scalar potential, and then as already mentioned the EFT is self-consistent. They generate a non-vanishing IASD and $G_3^{(0,3)}$ components of the three-form flux $G_3$, which can be seen as small perturbations around the supersymmetric background.

\subsection{Scenarios for de Sitter Vacua}
\label{ssec:deSitterScenarios}

In this section we review two mechanisms which can lead to dS vacua in LVS.

\subsubsection{Case 1: dS Vacua from Hidden Matter Fields}
\label{sssec:MatterUplift}

\begin{figure}[t]
\begin{center}
\includegraphics[width=0.4\textwidth, angle=270]{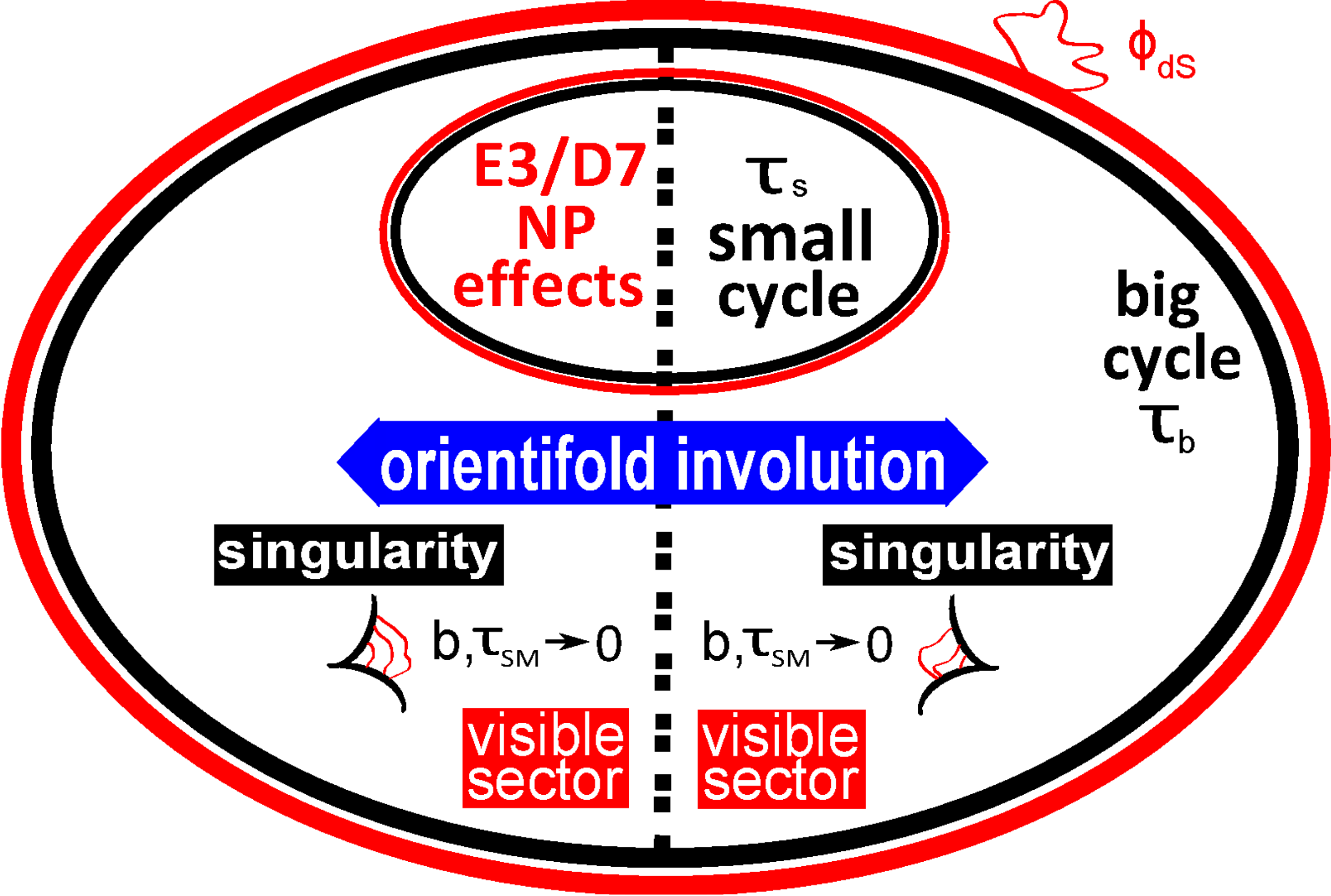}
\caption{Pictorial sketch of our CY setup for dS vacua from hidden matter fields.} \label{Fig1}
\end{center}
\end{figure}

In the LVS setting, dS vacua can arise if some hidden matter fields acquire non-vanishing F-terms which provide a positive definite contribution to the scalar potential~\cite{Cicoli:2012vw}. The models constructed in~\cite{Cicoli:2012vw} provide globally consistent explicit examples of string models with a semi-realistic visible sector, moduli stabilization and a positive cosmological constant (see Fig.~\ref{Fig1} for a pictorial sketch of this setup).\\

Generically, the choice of $B_2$ which cancels the Freed-Witten anomaly on the small cycle $T_s$, leads to non-vanishing gauge fluxes on the big cycle $T_b$. As a consequence, $T_b$ acquires a non-zero $U(1)$-charge $q_b$
generating a moduli-dependent FI-term. The D-terms potential becomes (focusing for simplicity on a single matter field $\phi_\dS$ with K\"ahler metric $K_\dS= s^{-1}|\phi_\dS|^2$~\cite{Conlon:2006tj, Aparicio:2008wh} and $U(1)$-charge $q_\phi$)
\begin{align}
\label{dmatter}
V_D = \frac{1}{2{\rm Re}(f_b)} \left(\frac{q_\phi}{s} |\phi_\dS|^2 - \xi_b\right)^2\,,
\end{align}
where $f_b = T_b$ (neglecting $S$-dependent flux corrections) and the FI-term is given by
\begin{align}
\xi_b = - \frac{q_b}{4\pi} \frac{\partial K}{\partial T_b} = \frac{3 q_b}{8\pi}\frac{1}{\mathcal{V}^{2/3}}\,,
\end{align}
Therefore the total scalar potential takes the form
\begin{align}
V_{\rm tot} = V_D + V_F =  \frac{1}{2\mathcal{V}^{2/3}} \left(\frac{q_\phi}{s}  |\phi_\dS|^2 - \frac{3 q_b}{8\pi\mathcal{V}^{2/3}}\right)^2
+ \frac{1}{s} m_{3/2}^2 |\phi_\dS|^2 + V_{\mc{O}(\mathcal{V}^{-3})}\,,
\label{Vtot}
\end{align}
where $m_{3/2}$ is the gravitino mass as in eq.~(\ref{m32}) and $V_{\mc{O}(\mathcal{V}^{-3})}$ is given in (\ref{eq:LVSScalarPotential}). If the two $U(1)$-charges $q_\phi$ and $q_b$ have the same sign, $\phi_\dS$ develops a non-vanishing VEV
\begin{align}
\label{minphi}
\frac{q_\phi}{s} |\phi_\dS|^2 = \xi_b - \frac{m_{3/2}^2 \mathcal{V}^{2/3}}{q_\phi}\,.
\end{align}
Substituting this VEV in (\ref{Vtot}) we obtain
\begin{align}
V_{\rm tot} = V_{D,0}+\frac{3 q_b}{16\pi q_\phi} \frac{W_0^2}{s  \mathcal{V}^{8/3}}+V_{\mc{O}(\mathcal{V}^{-3})}\,,
\label{VTOT}
\end{align}
where the new positive contribution can lead to an LVS dS vacuum while the D-terms potential gives rise only to a sub-leading effect of order
\begin{align}
V_{D,0} = \frac{m_{3/2}^4 \mathcal{V}^{2/3}}{2 q_\phi^2}\sim \mc{O}\left(\mathcal{V}^{-10/3}\right)\,.
\label{VD0ds1}
\end{align}
Following~\cite{Cicoli:2012vw}, we can minimize the total scalar potential (\ref{VTOT}) with respect to $\tau_s$ and $\mathcal{V}$, finding the following value of the vacuum energy (neglecting the sub-leading effect of $V_{D,0}$)
\begin{align}
\langle V_{\rm tot} \rangle \simeq \frac{3 W_0^2}{8s a_s^{3/2}\langle\mathcal{V}\rangle^3}\left[\delta\, \mathcal{V}^{1/3}
-\sqrt{\ln\left(\frac{\langle\mathcal{V}\rangle}{W_0}\right)}\right]\,,
\label{VTOTfin}
\end{align}
where
\begin{align}
\delta = \frac{1}{18\pi}\frac{q_b\,a_s^{3/2}}{q_\phi} \simeq 0.02 \left(\frac{q_b\,a_s^{3/2}}{q_\phi}\right)\,.
\end{align}
A cancellation of the vacuum energy at $\mc{O}(\mathcal{V}^{-3})$ requires therefore to tune $W_0$ so that
(a sub-leading tuning is needed to cancel also $V_{D,0}$)
\begin{align}
\left[\ln\left(\frac{\langle\mathcal{V}\rangle}{W_0}\right)\right]^{3/2}= \delta^3\, \langle\mathcal{V}\rangle \sim 5\cdot 10^{-6}\, \langle\mathcal{V}\rangle\quad\Leftrightarrow\quad
|\phi_\dS|^2 = \frac{27 s}{4 a_s^{3/2}\mathcal{V}}\sqrt{\ln\left(\frac{\langle\mathcal{V}\rangle}{W_0}\right)}\sim \frac{1}{\mathcal{V}\sqrt{\epsilon_s}}\,.
\label{CCconddS1}
\end{align}
For natural $\mc{O}(1)$ values of all underlying parameters, this relation gives a minimum for $\mathcal{V}$ at order $10^6 - 10^7$ (see~\cite{Cicoli:2012vw}). In~\cite{Cicoli:2015ylx} the authors showed that this de Sitter scenario can be generically obtained in presence of D7-branes with non-zero gauge flux, which lead to a T-brane configuration~\cite{Cecotti:2010bp}. Expanding the D7-branes action around the T-brane background gives a positive definite term in the scalar potential, which can be used to get a de Sitter vacuum.

\subsubsection{Case 2: dS Vacua from Non-perturbative Effects at Singularities}
\label{sssec:DilatonUplift}

\begin{figure}[t]
\begin{center}
\includegraphics[width=0.45\textwidth]{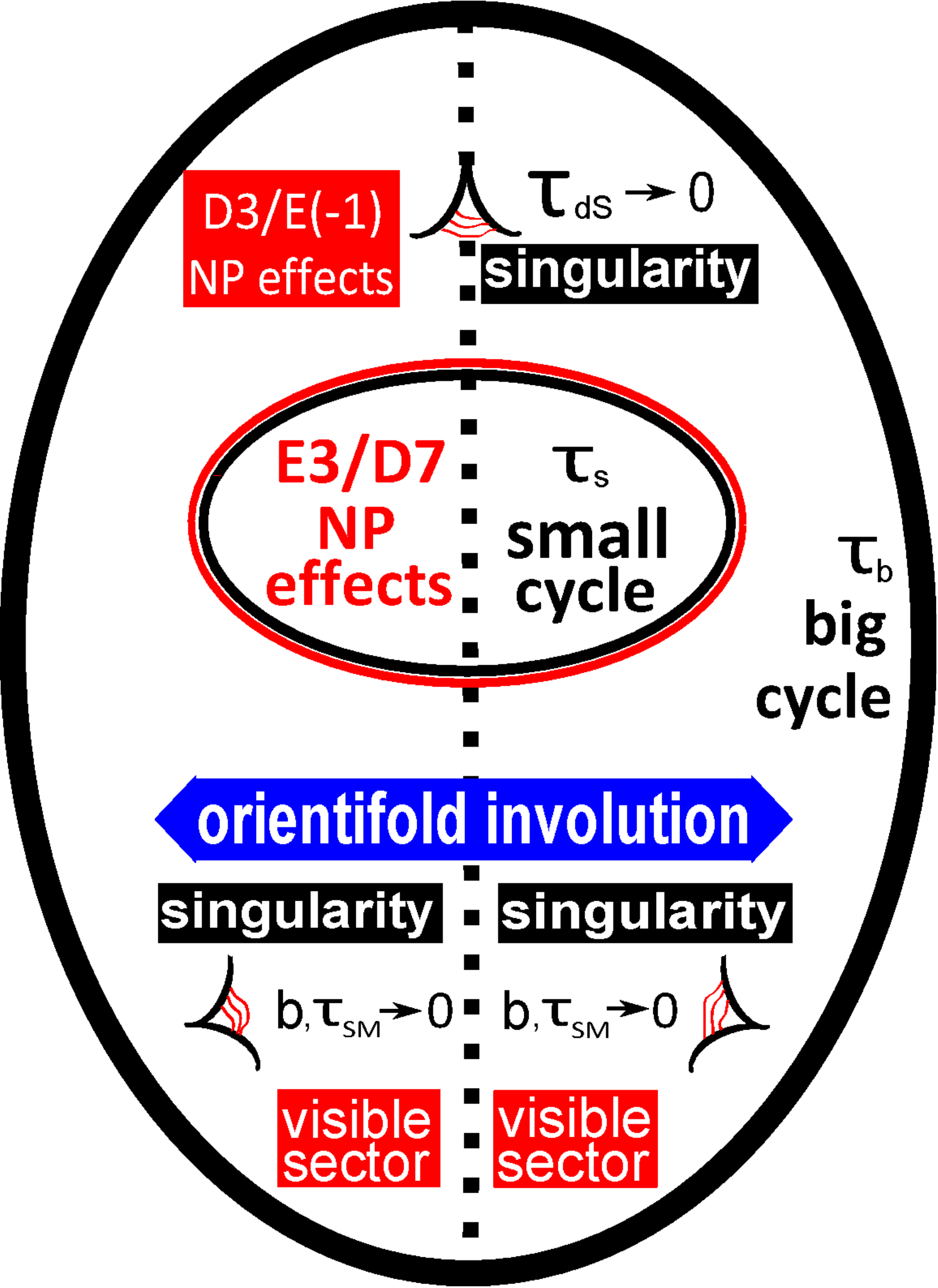}
\caption{Pictorial view of our CY setup for dS vacua from non-perturbative effects at singularities.} \label{Fig2}
\end{center}
\end{figure}

Reference~\cite{Cicoli:2012fh} provided a novel method for obtaining LVS dS vacua (see Fig.~\ref{Fig2} for a pictorial sketch of this setup).
The additional contribution to the scalar potential needed to achieve a positive cosmological constant arises
from non-perturbative effects at singularities (like gaugino condensation on space-time filling D3-branes or E(-1) instantons). These effects generate a new contribution to the superpotential in eq. \eqref{eq:FullSuperpotential} of the form
\begin{align}
W_\dS = A_\dS(U, S) \,e^{-a_\dS (S + \kappa_\dS T_\dS)}\,.
\label{Wds2}
\end{align}
Because of the presence of an additional K\"ahler modulus, the K\"ahler potential (\ref{generalk}) has to be supplemented with
\begin{align}
\label{dk2}
K_\dS =  \lambda_\dS\frac{\tau_\dS^2}{\mathcal{V}}\,,
\end{align}
with $\tau_\dS={\rm Re}(T_\dS)$. This blow-up mode can be fixed in the singular regime by minimizing the hidden sector D-terms potential (focusing for simplicity on canonically normalized hidden fields $\phi_{{\rm h},i}$ with charges $q_{{\rm h},i}$ under an anomalous $U(1)$)
\begin{align}
\label{DtermPot1}
V_D = \frac{1}{2{\rm Re}(f_{\rm h})} \left(\sum_i q_{{\rm h},i} |\phi_{{\rm h},i}|^2 - \xi_{\rm h}\right)^2\,,
\end{align}
where $f_{\rm h} = S$ (neglecting $T_\dS$-dependent corrections) and the FI-term is given by
($q_\dS$ is the $U(1)$-charge of $T_\dS$ and from now on we set $\lambda_\dS=1$ for simplicity)
\begin{align}
\xi_{\rm h} = - \frac{q_\dS}{4\pi} \frac{\partial K}{\partial T_\dS} = - \frac{q_\dS}{4\pi} \frac{\tau_\dS}{\mathcal{V}}\,.
\end{align}
In fact, the total scalar potential takes the leading order form (after fixing the axionic phase of $T_\dS$)~\cite{Cicoli:2012fh}
\begin{align}
V_{\rm tot} = \frac{1}{2s} \left(\sum_i q_{{\rm h},i} |\phi_{{\rm h},i}|^2 + \frac{q_\dS}{4\pi} \frac{\tau_\dS}{\mathcal{V}}\right)^2
+ \frac{\left(\kappa_\dS a_\dS A_\dS\right)^2}{s}\frac{e^{-2 a_\dS \left(s+\kappa_\dS\tau_\dS\right)}}{\mathcal{V}} + V_{\mc{O}(\mathcal{V}^{-3})}\,,
\label{Vtot1}
\end{align}
where the second term comes from the new superpotential (\ref{Wds2}) and $V_{\mc{O}(\mathcal{V}^{-3})}$ is given in~(\ref{eq:LVSScalarPotential}). Minimization with respect to $\tau_\dS$ gives
\begin{align}
\frac{q_\dS}{4\pi} \frac{\tau_\dS}{\mathcal{V}}=- \sum_i q_{{\rm h},i} |\phi_{{\rm h},i}|^2
+ \frac{a_\dS \kappa_\dS}{q_\dS}\left(\kappa_\dS a_\dS A_\dS\right)^2 e^{-2 a_\dS s}\,.
\label{tdsVEV}
\end{align}
Assuming that model-dependent contributions from F-terms of hidden matter fields fix some $\phi_{{\rm h}, i}$ at non-zero VEVs such that $\langle\sum_i q_{{\rm hid},i} |\phi_{{\rm hid},i}|^2\rangle = 0$ but $A_\dS\neq 0$,\footnote{In order to make $W_\dS$ gauge invariant, $A_\dS$ has to depend on the $\phi_{{\rm h},i}$ which can develop non-zero VEVs for appropriate hidden field F-terms contributions, giving $A_\dS\neq 0$ with $\tau_\dS$ in the singular regime~\cite{Cicoli:2012fh}.} and substituting the VEV (\ref{tdsVEV}) in (\ref{Vtot1}) we obtain at leading order
\begin{align}
V_{\rm tot} = V_{D,0}+ \frac{(\kappa_\dS a_\dS A_\dS)^2}{s} \frac{e^{-2 a_\dS s}}{\mathcal{V}} + V_{\mc{O}(\mathcal{V}^{-3})}\,.
\end{align}
Given that the dilaton is fixed by a ratio of flux quanta, the extra positive definite contribution can easily be tuned to obtain a dS minimum. Following~\cite{Cicoli:2012fh}, a cancellation of the vacuum energy at $\mc{O}(\mathcal{V}^{-3})$ requires to tune three-form fluxes such that
\begin{align}
\left(\frac{\kappa_\dS a_\dS A_\dS}{W_0}\right)^2 \,e^{-2 a_\dS s}
= \frac{9}{32} \frac{\epsilon_s \hat\xi}{\mathcal{V}^2}\,.
\label{CCconddS2}
\end{align}
On the other hand, the D-terms potential gives rise only to a sub-leading effect of order
\begin{align}
V_{D,0} =\frac{1}{2s} \left(\frac{a_\dS \kappa_\dS}{q_\dS}\right)^2\left(\kappa_\dS a_\dS A_\dS\right)^4 e^{-4 a_\dS s}\sim \mc{O}\left(\mathcal{V}^{-4}\right)\,.
\label{VD0ds2}
\end{align}

\section{F-terms and Soft-terms}
\label{sec:ftermsandsoftmasses}

In this section we list the leading order contributions to the F-terms relevant for the computation of all soft-terms, and we also report the sub-leading corrections to the F-terms in a subsection. After defining our parametrization for the K\"ahler matter metric in local and ultra-local scenarios, we then calculate the soft-terms.

\subsection{Summary of F-terms}
\label{fterms}

The general supergravity expression for an F-term is~\cite{Kaplunovsky:1993rd, Brignole:1993dj}
\begin{align}
F^I = e^{K/2} K^{I \overline{J}} D_{\overline{J}} \overline{W}\,.
\end{align}
The exact expressions for the F-terms are rather complicated. Considerable simplifications occur if we perform an expansion in $\mathcal{V}^{-1}$ and $\epsilon_s$. We also factor out the gravitino mass which is given by the following expression
\begin{align}
m_{3/2}=e^{K/2} |W|= \frac{g_s^{1/2} M_{\rm P}}{2\sqrt{2\pi}}\,\frac{W_0}{\mathcal{V}} \left[ 1-\frac{\hat{\xi}}{2\mathcal{V}}
\left(1+3 y_\dS \epsilon_s+\mc{O}(\epsilon_s^2)\right)+\mc{O}\left(\frac{1}{\mathcal{V}^2}\right)\right]\,,
\label{expandm32}
\end{align}
where $y_\dS =1$ for the dS case 1 of Sec.~\ref{sssec:MatterUplift} while $y_\dS = 1+\frac{\sqrt{2} a_s^{3/4}}{\kappa_\dS a_\dS}$ for the dS case 2 of Sec.~\ref{sssec:DilatonUplift}. The leading order F-terms for $T_b$ and $T_s$ turn out to be
\begin{align}
\frac{F^{T_b}}{\tau_b} \simeq - 2 m_{3/2} \left(1+\frac{x_\dS}{a_s^{3/2} \mathcal{V}\sqrt{\epsilon_s}}\right)\,,
\qquad
\frac{F^{T_s}}{\tau_s} \simeq - 6   m_{3/2} \epsilon_s\,,
\label{FTb}
\end{align}
where $x_\dS = - 45/16$ for the dS case 1 of Sec.~\ref{sssec:MatterUplift} while $x_\dS \sim \mc{O}(1/\mathcal{V})$ for the dS case 2 of Sec.~\ref{sssec:DilatonUplift}, as reported below. Because of the shift from their supersymmetric minimum, also $S$ and $U$ develop non-vanishing F-terms whose leading order
expressions are
\begin{align}
\label{ds}
\frac{F^S}{s} \simeq \frac{3 \omega'_S(U,S) }{8 a_s^{3/2}}  \frac{m_{3/2}}{\mathcal{V}\epsilon_s^{3/2}}\,, \qquad
F^{U^a} \simeq - \frac{K^{U^a \overline{U}^{\overline{b}}}}{2 s^2}\frac{\omega_{\overline{U}^{\overline{b}}}(U,S)}{\omega'_S(U,S)}F^S\equiv \beta^{U^a}(U,S)F^S\,,
\end{align}
where $\omega'_S(U,S) \equiv 3 - 2 \omega_S(U,S)$ with $\omega_S$ as defined in (\ref{dilmin}) and $\beta^{U^a}$ are unknown $\mc{O}(1)$ functions of $U$ and $S$. Additional non-zero F-terms are associated to fields responsible for achieving a dS solution. For the dS case 1 of Sec.~\ref{sssec:MatterUplift} there is an F-term associated to $\phi_\dS$
\begin{align}
\frac{F^{\phi_\dS}}{\phi_\dS} \simeq m_{3/2}\,,
\end{align}
with $\phi_\dS$ given in (\ref{CCconddS1}) (up to an irrelevant phase). On the other hand, in the dS case 2 of Sec.~\ref{sssec:DilatonUplift} the blow-up mode $T_\dS$ has a non-vanishing F-term (using the condition (\ref{CCconddS2}))
\begin{align}
F^{T_\dS} \simeq \frac{3}{4\sqrt{2}a_s^{3/4}}\frac{m_{3/2}}{\epsilon_s^{1/4}} \,.
\end{align}
Finally, the F-terms associated to the MSSM cycles $T_\SM$ and $G$ vanish:
\begin{align}
F^G = F^{T_\SM} = 0\, .
\end{align}
This result is crucial for sequestering since the dominant F-terms are then associated with moduli which couple weakly to the visible sector via Planck-suppressed interactions.\footnote{$T_\SM$ and $G$ can develop non-zero F-terms only in the presence of tachyonic scalar masses~\cite{Cicoli:2013cha}. However, also in this case, their contribution to soft-terms turns out to be negligible.}

\subsubsection{Sub-leading Corrections to F-terms}
\label{sssec:SubleadingCorrectionsToFTerms}

In the present section we first describe the shift of the LVS minimum after including an extra term responsible to achieve a dS vacuum, and then provide sub-leading corrections to F-terms. As described in Sec.~\ref{ssec:deSitterScenarios}, the mechanism which realizes a dS vacuum gives rise effectively to an extra term of the form
\begin{align}
V_\dS=\frac{r}{\mathcal{V}^m}\qquad\text{with}\quad r>0\quad\text{and}\quad m<3\,.
\end{align}
We are interested in minimizing the combined system
\begin{align}
V=V_{\mc{O}(\mathcal{V}^{-3})}+V_\dS\,,
\end{align}
with the additional constraint of vanishing vacuum energy. This constraint relates the coefficient $r$ with the tunable flux parameters in the LVS potential such as $W_0$ or $g_s$. A concrete dS scenario, such as the ones in Sec.~\ref{ssec:deSitterScenarios}, typically fixes $r$ by construction with only moderate tuning. However the real tuning can be achieved by simply tuning the flux superpotential and the string coupling in agreement with the flux landscape.\\

The expressions for the moduli VEVs are largely independent on the way to get dS vacua. In fact, the relation (\ref{eq:TaubMinimum}) is generic whereas the expression (\ref{eq:TausMinimumCorrected}) for the VEV of $\tau_s$ depends on the way to get a dS vacuum only at sub-leading order. The exact minimum for $\tau_s$ is given by
\begin{align}
\tau_s^{3/2}=\frac{\hat{\xi}}{2}\, \frac{(1-\epsilon_s)^2}{(1-4\epsilon_s)}\, \frac{1}{1+\frac{2m}{m-3} \epsilon_s}
=\frac{\hat{\xi}}{2}\left[1+f_\dS(\epsilon_s)\right]\,,
\label{eq:shiftds}
\end{align}
and so the function $f_\dS$ is $f_\dS= 18 \epsilon_s+297 \epsilon_s^2$ in the case of dS vacua from hidden matter fields ($m=8/3$),
while $f_\dS=3 \epsilon_s+12 \epsilon_s^2$ for the case of non-perturbative effects at singularities ($m=1$).
Note that as a consequence of the shift in $\tau_s$, also the overall volume in (\ref{eq:TaubMinimum}) is shifted and, as the shift is in the exponential, this shift can be parametrically large.\\

Equipped with the minimum, we can evaluate the F-terms. To simplify the notation we factor out an overall factor of the gravitino mass $m_{3/2}$ which is given by (\ref{expandm32}). The F-terms turn out to have the following expressions:
\begin{align}
\frac{F^{T_b}}{\tau_b}&=-2 m_{3/2}\left[1+\frac{9 \hat{\xi} \epsilon_s}{4 \mathcal{V}} \frac{m-1}{m-3+2m\epsilon_s} +\mc{O}\left(\frac{1}{\mathcal{V}^2}\right)\right]\,, \\
\frac{F^{T_s}}{\tau_s}&= -2 m_{3/2}\left[\frac{3\epsilon_s}{(1-\epsilon_s)}-\frac{\hat{\xi}}{2\mathcal{V}}
\left(1-\frac{9\epsilon_s}{2}\frac{m-1}{m-3}+\mc{O}\left(\epsilon_s^2\right)\right)\right]\,, \\
\frac{F^S}{s}&= \frac{3 \omega'_S }{8 a_s^{3/2}}  \frac{m_{3/2}}{\mathcal{V}\epsilon_s^{3/2}}\left[1+\mc{O}\left(\epsilon_s\right)\right]\,, \\
F^U&= - \frac{K^{U^a \overline{U}_{\overline{b}}}}{2 s^2}\frac{\omega_{\overline{U}_{\overline{b}}}}{\omega'_S}F^S\equiv \beta^{U^a}\,F^S\,, \\
F^{\phi_\dS} &= \phi_\dS m_{3/2}\left[1+ \mc{O}\left(\frac{1}{\mathcal{V}}\right)\right] \,, \\
F^{T_{\dS}}&= \frac{3}{4\sqrt{2}a_s^{3/4}}\frac{m_{3/2}}{\epsilon_s^{1/4}}\left[1+\mc{O}\left(\epsilon_s\right)\right]\,.
\end{align}

\subsection{Local and Ultra-local Scenarios}
\label{soft}

Our analysis of soft-terms will distinguish between two classes of models: local and ultra-local. This classification is motivated by locality arguments already discussed in Sec.~\ref{sssec:D3atSingularities}: the \Kahler matter metric should scale as $\tilde{K}_\alpha \sim  \V^{-2/3}$ at leading order. In the present section we expand the constraint on the \Kahler matter metric coming from the structure of physical Yukawa couplings in eq. \eqref{eq:HolomorphicYukawaCoupling} at sub-leading order, and we get (for $\tau_\SM = b= C^\alpha= 0$)
\begin{align}
\label{ul}
\tilde{K}_\alpha = h_\alpha(U,S)\,e^{K/3} \simeq  \frac{h_\alpha(U,S)\,e^{K_{\rm cs}/3}}{(2s)^{1/3}\mathcal{V}^{2/3}} \left(1- \frac{\hat\xi}{3\mathcal{V}}  + \frac 13 K_\dS\right)\,,
\end{align}
where $h_\alpha(U,S)$ is an unknown function of $U$ and $S$ and in the approximation we focus on the first sub-leading order corrections, e.g.~neglecting higher order corrections of $\mc{O}\left(1/\mathcal{V}^{8/3}\right)$. Note that this result has the same volume scaling of our formula for the matter metric (\ref{mattermetric}) which for $\tau_\SM = b= 0$ reduces to
\begin{align}
\tilde{K}_{\alpha} = \frac{f_\alpha(U,S)}{\mathcal{V}^{2/3}} \left(1 - c_s \frac{\hat\xi}{\mathcal{V}}
+ \tilde{K}_\dS\right)\equiv f_\alpha(U,S)\tilde{K}\,.
\label{newKtilde}
\end{align}
As found in~\cite{Blumenhagen:2009gk}, our soft-terms computation is sensitive to the form of $\tilde{K}_\alpha-$
beyond leading order in a $\mathcal{V}^{-1}$ expansion. There is no reason to expect that (\ref{ul}) still holds beyond leading order since we cannot use locality to fix the form of $\tilde{K}_\alpha$ (although there is some evidence from perturbative string computations~\cite{Conlon:2011jq}). It was noted in~\cite{Blumenhagen:2009gk} that the relation (\ref{ul}) has interesting implications for the soft-terms. Guided by this, we organize our analysis of models into two classes of phenomenological models:
\begin{itemize}
\item {\it Local}: We call a scenario `local' if (\ref{ul}) holds only to leading order in $\mathcal{V}^{-1}$;
\item {\it Ultra-local}: We call a scenario `ultra-local' if (\ref{ul}) holds exactly.
\end{itemize}
If we parametrize $\tilde{K}_\dS$ as $\tilde{K}_\dS  =  c_\dS K_\dS$,
comparing (\ref{ul}) with (\ref{newKtilde}), we find that in the ultra-local case
\begin{align}
f_{\alpha}(U,S) = \frac{h_\alpha(U,S)\,e^{K_{\rm cs}/3}}{(2s)^{1/3}}\qquad \text{and}\qquad c_s=c_\dS=\frac 13\,.
\label{fdef}
\end{align}
Subleading deviations from the approximation in~(\ref{ul}) can be accounted for by small changes in $c_s$ and $c_\dS$ at the appropriate sub-leading order.

\subsection{Soft-terms}
\label{sssec:SoftTerms}

We now proceed to compute all soft-terms distinguishing between ultra-local and local scenarios. Throughout this section we work to leading order in $\mathcal{V}^{-1}$ and $\epsilon_s$.

\bi
\item \textbf{Gaugino masses}\\

The general expression for gaugino masses in gravity mediation is
\begin{align}
\label{gauginomass}
M_a = \frac{1}{2 {\rm Re}\left(f_a\right)} F^I \partial_I f_a\, ,
\end{align}
where $f_a=\delta_a S+ \kappa_a \,T_\SM$ is the gauge kinetic function as in~\eqref{eq:gaugekineticf}. As $F^{T_\SM}=0$, we obtain universal gaugino masses, $M_1=M_2=M_3=M_{1/2}$, which are generated by the dilaton F-term. Potential non-universalities can arise through anomaly mediated contributions which turn out to be sub-leading (see Sec.~\ref{sec:DesequesteringSources} for more details). The leading order expression for the gaugino masses is
\begin{align}
M_{1/2} = \frac{F^S}{2 s} \simeq \frac{3 \omega'_S(U,S)}{16 a_s^{3/2}}  \frac{m_{3/2}}{\mathcal{V}\epsilon_s^{3/2}}
\sim \mc{O}\left(m_{3/2}\frac{\left(\ln\mathcal{V}\right)^{3/2}}{\mathcal{V}}\right)\ll m_{3/2}\,.
\label{ggmm}
\end{align}
Note that this leading order result depends on the shift of the dilaton minimum induced by $\alpha'$ and non-perturbative effects (see Sec.~\ref{shiftsection}). We neglect possible phases of gaugino masses. We will return to this question in the context of the low-energy analysis of soft-terms~\cite{Aparicio:2015sda}.

\item \textbf{Scalar masses}\\

Scalar masses in gravity mediation receive both F- and D-terms contributions. Let us study them separately, presenting their leading order expressions.\\

\ben
\item \textbf{F-terms contributions}\\
Assuming a diagonal K\"ahler matter metric as in~(\ref{mm}), the general expression for the F-terms contributions to scalar masses in gravity mediation is~\cite{Brignole:1993dj}
\begin{align}
\label{genscalF}
\left.m_\alpha^2\right|_F = m_{3/2}^2 - F^I \overline{F}^{\overline{J}} \partial_I \partial_{\overline{J}} \ln \tilde{K}_\alpha\,.
\end{align}

\emph{Local limit}: In the local limit we obtain universal scalar masses, $m_\alpha^2=m_0^2$ $\forall \alpha$, where
\begin{align}
\left.m_0^2\right|_F &\simeq m_{3/2}^2 - \left(\frac{F^{T_b}}{2}\right)^2 \partial^2_{\tau_b} \ln\tilde{K}
\simeq \nn \\
&\simeq \frac{15}{2}\left(c_s-\frac{1}{3}\right)\, \frac{m_{3/2}^2 \tau_s^{3/2}}{\vo}\sim m_{3/2} M_{1/2}\,,
\label{localscalar}
\end{align}
The dominant contribution to this expression comes from the F-term of $T_b$. More precisely, the leading term of $F^{T_b}$ in (\ref{FTb}) together with the leading term of $\tilde{K}$ in (\ref{newKtilde}) give a contribution which cancels against $m_{3/2}^2$ in (\ref{localscalar}) because of the underlying no-scale structure. The first non-vanishing term in (\ref{localscalar}) originates from the leading term of $F^{T_b}$
together with the first sub-leading correction to $\tilde{K}$. On the other hand, the sub-leading correction to $F^{T_b}$ in (\ref{FTb}) yields a contribution suppressed by $\epsilon_s$, and so turns out to be negligible.\\

Scalar masses are universal since they get generated by the F-term of $T_b$. Non-universal effects can arise from $F^S$ and $F^{U^a}$ but they are volume suppressed since they would give contributions of order $m_{3/2}^2/\mathcal{V}^2$. If $c_s>1/3,$ the scalar masses are non-tachyonic.

\medskip
\emph{Ultra-local limit}: An interesting feature of (\ref{localscalar}) is that it vanishes if one takes the ultra-local limit $c_s= 1/3$.\footnote{We neglect potential higher order corrections at this stage.} In fact, there is a general argument~\cite{Blumenhagen:2009gk} which guarantees the vanishing of $m_0^2$ at $\mc{O}(\mathcal{V}^{-3})$. Using (\ref{ul}) (the defining property of ultra-local models) in the general expression for the F-terms contributions to scalar masses (\ref{genscalF}) we find
\begin{align}
\left.m_\alpha^2\right|_F = - \frac 13 V_{F,0} - F^I \overline{F}^{\overline{J}} \partial_I \partial_{\overline{J}} \ln h_\alpha(U,S)\,,
\label{ulsm2}
\end{align}
where we used the fact that $V_F= K_{I\overline{J}}F^I \overline{F}^{\overline{J}}- 3 m_{3/2}^2$. Recalling that $V_0 = V_{F,0} + V_{D,0}$ and setting the cosmological constant to zero (in the dS constructions of Sec.~\ref{ssec:deSitterScenarios} we showed how to cancel $V_0$ at $\mc{O}(\mathcal{V}^{-3})$ but this can in principle be done at any order in the $\mathcal{V}^{-1}$ expansion), $V_{F,0}$ can be traded for $V_{D,0}$,
and we so shall include it in our analysis of D-terms contributions to scalar masses.\\

On the other hand, if the functions $h_\alpha(S,U)$ are not constants, there is a non-vanishing contribution
from the F-terms of the dilaton and the complex structure moduli at $\mc{O}(\mathcal{V}^{-2})$. Using (\ref{ds}), the $S$ and $U$-dependent contribution to scalar masses turns out to be
\begin{align}
\left.m_\alpha^2\right|_F =  - M_{1/2}^2 s^2 \left(\partial_s^2+\beta^{U^a}\partial_{u^a, s}
+ \beta^{U^a}\beta^{\overline{U}^b}\partial_{u^a,u^b}\right) \ln h_\alpha\sim \mc{O}\left(M_{1/2}^2\right)\,,
\label{ulsm3}
\end{align}
where $M_{1/2}$ is the gaugino mass in (\ref{ggmm}). Note that this contribution is generically non-universal
and might also give rise to tachyonic scalars depending on the explicit functional dependence of the functions $h_\alpha(U,S)$.\\

\item \textbf{D-terms contributions}\\
Assuming a diagonal K\"ahler matter metric as in~\eqref{mm}, the general expression for the D-terms contributions to scalar masses in gravity mediation is~\cite{Dudas:2005vv}
\begin{align}
\left.m_\alpha^2\right|_D = \tilde{K}_\alpha^{-1} \sum_i g_i^2 D_i \partial^2_{\alpha \overline{\alpha}} D_i- V_{D,0}\,.
\label{genscalD}
\end{align}
Given that this result depends on the value of the D-terms potential at the minimum, this contribution depends on the way to achieve a dS vacuum. As explained in Sec.~\ref{sssec:Shrinking}, the VEV of the D-terms potential associated to visible sector $U(1)$s is vanishing in the absence of tachyonic scalars.\footnote{Even in the presence of tachyonic scalars, the contribution to scalar masses from visible sector D-terms turns out to be a negligible effect since visible matter fields, $\tau_\SM$ and $b$ would still be stabilized at zero at leading order.}

\medskip
\emph{dS case 1}: In the dS case 1 of Sec.~\ref{sssec:MatterUplift} the relevant D-term is the one associated with the anomalous $U(1)$ living on the big cycle. As can be seen from (\ref{VD0ds1}), $V_{D,0}$ scales as $\mathcal{V}^{-10/3}$ which is sub-dominant with respect to the first term in (\ref{genscalD}) that gives
\begin{align}
\left.m_0^2\right|_D &= \frac{q_b}{2 f_\alpha(U,S)} D_{\dS_1} \partial_{\tau_b} \tilde{K}_{\alpha}= \frac{m_{3/2}^2}{3s}|\phi_\dS|^2 = \nn \\
&= \frac{6\epsilon_s}{\omega'_S}m_{3/2}M_{1/2}\sim  \mc{O}\left(m_{3/2}^2\frac{\sqrt{\ln\mathcal{V}}}{\mathcal{V}}\right)\,,
\label{sc1}
\end{align}
once we impose the condition (\ref{CCconddS1}) to have a vanishing cosmological constant at $\mc{O}(\mathcal{V}^{-3})$. In the local limit, this result is suppressed with respect to the F-terms contribution (\ref{localscalar}) by a factor of $\epsilon_s$. On the other hand, in the ultra-local limit, this D-terms contribution dominates over the F-terms one given in (\ref{ulsm3}) which scales as $m_{3/2}^2\epsilon^2$.
Hence it leads to universal and non-tachyonic scalar masses.

\medskip
\emph{dS case 2}: In the dS case 2 of Sec.~\ref{sssec:DilatonUplift} the relevant D-term is the one associated with the anomalous $U(1)$ which belongs to the hidden sector responsible for achieving a dS vacuum. In this case both terms in (\ref{genscalD}) have the same scaling since
\begin{align}
\left.m_0^2\right|_D &= \frac{q_\dS \mathcal{V}^{2/3}}{2 s f_\alpha(U,S)} D_{\dS_2} \partial_{\tau_\dS} \tilde{K}_{\alpha}-V_{D,0} = \nn \\
&= \frac{c_\dS}{s} D_{\dS_2} q_\dS\frac{\tau_\dS}{\mathcal{V}} -V_{D,0} = \left(2c_\dS-1\right)V_{D,0}\,.
\label{Vds22}
\end{align}
As can be seen from (\ref{VD0ds2}), $V_{D,0}$ scales as $\mathcal{V}^{-4}$. Hence in the local limit the D-terms contribution is sub-leading with respect to the F-terms one given in (\ref{localscalar}) which scales as $\mathcal{V}^{-3}$. In the ultra-local limit the F-terms contribution to scalar masses is given by (\ref{ulsm2}).
Adding $-V_{F,0}/3= V_{D,0}/3$ to (\ref{Vds22}) we find that the total D-terms contribution to scalar masses vanishes in the ultra-local limit once we impose $c_\dS=1/3$ as in (\ref{fdef}) since
\begin{align}
\left.m_0^2\right|_D =  2\left(c_\dS-\frac 13\right)V_{D,0}=0 \quad\text{for}\quad c_\dS=\frac 13\,.
\label{Vds2}
\end{align}
Hence scalar masses get generated by F-terms also in the ultra-local limit. Their expression is given in (\ref{ulsm3}) and scales as $\mathcal{V}^{-4}$.\\

\noindent \textbf{Summary of the results for scalar masses}

Let us summarize our results for soft scalar masses. The expression for $m_0^2$ in the local limit does not depend on the way to obtain a dS vacuum since in each case it is given by the F-terms contribution (\ref{localscalar}) that scales as $\mathcal{V}^{-3}$. Scalar masses are non-tachyonic if $c_s>1/3$ and universal. On the other hand, the result for the ultra-local limit depends on the dS mechanism. In the dS case 1 of Sec.~\ref{sssec:MatterUplift}, scalar masses get generated by the D-terms contribution (\ref{sc1}) which has again an overall $\mathcal{V}^{-3}$ scaling but with an $\epsilon_s$ suppression with respect to the local case. Scalar masses turn out to be non-tachyonic and universal. On the contrary, in the dS case 2 of Sec.~\ref{sssec:DilatonUplift}, the main contribution to scalar masses comes from F-terms and it is given by (\ref{ulsm3}) which scales as $\mathcal{V}^{-4}$. This result could potentially lead to tachyonic and non-universal scalar masses depending on the exact functional dependence of the functions $h_\alpha(U,S)$.
\een

\item \textbf{A-terms}\\
For the current discussion, we assume that the Yukawa couplings $Y_{\alpha \beta \gamma}$ receive no non-perturbative contributions from the K\"ahler moduli and are hence only functions of the complex structure moduli and the dilaton $Y_{\alpha \beta \gamma}=Y_{\alpha \beta \gamma}(U,S)$. The trilinear A-terms in gravity mediation receive both F- and D-terms contributions. The D-terms contributions turn out to be zero for vanishing VEVs of visible sector matter fields~\cite{Dudas:2005vv}. On the other hand, the general formula for the F-terms contribution is~\cite{Brignole:1993dj}
\begin{align}
\label{tril}
A_{\alpha\beta\gamma} &= F^I \partial_I \left[K + \ln \left(\frac{Y_{\alpha \beta \gamma}(U,S)}{\tilde{K}_\alpha \tilde{K}_\beta \tilde{K}_\gamma}\right)\right] = \nn \\
&= F^I \partial_I \left[K -3\ln\tilde{K}+ \ln \left(\frac{Y_{\alpha \beta \gamma}(U,S)}{f_\alpha f_\beta f_\gamma}\right)\right],
\end{align}
where the holomorphic Yukawas $Y_{\alpha \beta \gamma}(U,S)$ do not depend on the K\"ahler moduli because of their axionic shift-symmetry. Let us present the expression for $A_{\alpha\beta\gamma}$ at leading order in $\mathcal{V}^{-1}$ and $\epsilon_s$.

\medskip
\emph{Local limit}: In the local limit we find
\begin{align}
\label{tril1}
A_{\alpha\beta\gamma}= &- \left[1 -s\beta^{U^a}\partial_{u^a} K_{\rm cs} - \frac{6}{\omega'_S}\left(c_s-\frac 13\right) - \right. \nn \\
&\left.- s \partial_{s,u} \ln \left(\frac{Y_{\alpha \beta \gamma}}{f_\alpha f_\beta f_\gamma} \right) \right] M_{1/2}\sim\mc{O}\left(M_{1/2}\right)\,,
\end{align}
with $\partial_{s,u}\equiv \partial_s+\beta^{U^a}\partial_{u^a}$.
Note that there is a cancellation at $\mc{O}(\mathcal{V}^{-1})$ between $K$ and $3\ln\tilde{K}$ in (\ref{tril}).
The dominant contributions to (\ref{tril1}) come from the F-terms of $T_b$, $S$ and $U$.

\medskip
\emph{Ultra-local limit}: In the ultra-local limit defined by (\ref{ul}), the contribution to $A_{\alpha\beta\gamma}$ from $F^{T_b}$ vanishes, as can be seen at leading order in (\ref{tril1}) by setting $c_s=1/3$ and $f_{\alpha} = h_\alpha\,e^{K_{\rm cs}/3}(2s)^{-1/3}$. In this limit, the general expression (\ref{tril}) simplifies to
\begin{align}
\label{trilul}
A_{\alpha\beta\gamma} = s \partial_{s,u}
\ln \left(\frac{Y_{\alpha \beta \gamma}(U,S)}{h_\alpha h_\beta h_\gamma}\right) M_{1/2}\sim\mc{O}\left(M_{1/2}\right)\,.
\end{align}

\item \textbf{$\hat{\mu}$ and $B\hat{\mu}$ terms}

Let us discuss different effects that can contribute to the superpotential and K\"ahler potential higgs bi-linear terms. Whether they are present or not is model-dependent and a concrete realization or combination of various mechanisms might not be possible. The following list should be understood as a list of possible effects that can lead to a non-vanishing $\mu$-term. The canonically normalized $\hat{\mu}$ and $B\hat{\mu}$-terms receive contributions from both K\"ahler potential and superpotential effects.
Let us discuss these two different effects separately.\\

\ben
\item \textbf{K\"ahler potential contributions}

Non-zero $\hat{\mu}$ and $B \hat{\mu}$ get generated from a non-vanishing pre-factor $Z$ in the matter K\"ahler potential (\ref{mm})~\cite{Kim:1983dt,Giudice:1988yz}. Their general expression in gravity mediation is~\cite{Brignole:1993dj,Dudas:2005vv}
\begin{align}
&\hat{\mu} = \left(m_{3/2} Z - \overline{F}^{\overline{I}} \partial_{\overline{I}} Z\right)
\left(\tilde{K}_{H_u} \tilde{K}_{H_d}\right)^{-1/2} \nn \\
&B\hat{\mu} = \left.B\hat{\mu}\right|_F + \left.B\hat{\mu}\right|_D\,,
\label{muK}
\end{align}
where
\begin{align}
&\left.B\hat{\mu}\right|_F = \left\{2 m_{3/2}^2 Z - m_{3/2} \overline{F}^{\overline{I}} \partial_{\overline{I}} Z + m_{3/2} F^I \left[\partial_I Z - Z \partial_I \ln\left(\tilde{K}_{H_u} \tilde{K}_{H_d}\right)\right]\right. \nonumber \\
&\left. - F^I \overline{F}^{\overline{J}} \left[\partial_I \partial_{\overline{J}} Z - \partial_I Z \partial_{\overline{J}} \ln\left(\tilde{K}_{H_u} \tilde{K}_{H_d}\right)\right] \right\}\left(\tilde{K}_{H_u} \tilde{K}_{H_d}\right)^{-1/2} \,, \label{BmuKF} \\
&\left.B\hat{\mu}\right|_D = \left(\tilde{K}_{H_u} \tilde{K}_{H_d}\right)^{-1/2}\left(\sum_i g_i^2 D_i \partial_{H_u} \partial_{H_d} D_i - V_{D,0} Z\right)\,.
\label{BmuKD}
\end{align}
Motivated by the fact that we are at the singular regime, we take $Z$ of the same form as the matter metric (\ref{mattermetric}) with $f_\alpha(U,S)$ replaced by a different unknown function of $S$ and $U$ which we denote $\gamma(U,S)$. We stress that $Z=\gamma(U,S)\tilde{K}$ is just the simplest ansatz for $Z$ given our present knowledge but its form could in general be different from $\tilde{K}_\alpha$.\footnote{However in models with a shift-symmetric higgs sector $f_{H_u}=f_{H_d}=\gamma$~\cite{LopesCardoso:1994is,Antoniadis:1994hg,Brignole:1995fb,Brignole:1996xb,Hebecker:2012qp}.}\\

Let us compute the leading expressions (in an expansion in $\mathcal{V}^{-1}$ and $\epsilon_s$) for both $\hat\mu$ and $B\hat\mu$ in the local and ultra-local limit.

\medskip
\emph{Local limit}: In the local limit we find
\begin{align}
\hat\mu  =  \frac{\gamma}{\sqrt{f_{H_u } f_{H_d}}} \left[\frac{6}{3 \omega'_S} \left(c_s-\frac 13\right)
- s \partial_{s,u} \ln\gamma \right] M_{1/2}\sim\mc{O}\left(M_{1/2}\right)\,,
\label{mu11}
\end{align}
where again there is a cancellation at $\mc{O}(\mathcal{V}^{-1})$ between the term proportional to $m_{3/2}$ in (\ref{muK}) and the leading order contribution from $\overline{F}^{\overline{T}_b}\partial_{\overline{T}_b}Z$. The dominant contributions to (\ref{mu11}) come from the F-terms of $T_b$, $S$ and $U$. On the other hand, the $B\hat\mu$-term behaves as the soft scalar masses since both F- and D-terms contributions can be rewritten as
\begin{align}
\left.B\hat{\mu}\right|_{F,D} = \frac{\gamma}{\sqrt{f_{H_u} f_{H_d}}}\, \left.m_0^2\right|_{F,D}\,.
\label{bmu11}
\end{align}
Recalling our results for $m_0^2$, we realize that in the local limit the leading contribution to $B\hat\mu$ comes from F-terms and scales as $\left.m_0^2\right|_F$ in (\ref{localscalar}). Hence the final result for $B\hat\mu$ is
\begin{align}
B\hat{\mu} = \frac{\gamma}{\sqrt{f_{H_u} f_{H_d}}}\, \frac{5\left(c_s-\frac 13\right)}{\omega'_S} m_{3/2}M_{1/2}\sim \mc{O}\left(m_{3/2}^2\frac{\left(\ln\mathcal{V}\right)^{3/2}}{\mathcal{V}}\right)\,.
\label{bmu11local}
\end{align}

\medskip
\emph{Ultra-local limit}: Similarly to the ultra-local limit for $\tilde{K}_\alpha$ defined by (\ref{ul}), we can define also an ultra-local limit for $Z=\gamma(U,S)\tilde{K}$ as $Z\equiv z(U,S)\,e^{K/3}$ which implies
\begin{align}
\label{fdef2}
\gamma(U,S) = \frac{z(U,S)\,e^{K_{\rm cs}/3}}{(2s)^{1/3}}\qquad \text{and}\qquad c_s=c_\dS=\frac 13\,.
\end{align}
In this limit the F-term of $T_b$ does not contribute to $\hat\mu$ whose expression simplifies to
\begin{align}
\label{mu2}
\hat\mu  =  - \frac{z\,s\partial_{s,u} \ln\gamma}{\sqrt{h_{H_u } h_{H_d}}}\,M_{1/2}\sim\mc{O}\left(M_{1/2}\right)\, .
\end{align}
In this ultra-local case the expression (\ref{BmuKF}) for $\left.B\hat\mu\right|_F$ gives
\begin{align}
\left.B\hat\mu\right|_F = \frac{z}{\sqrt{h_{H_u } h_{H_d}}}\left[\sigma(U,S) \,M_{1/2}^2-\frac 13 V_{F,0}\right]\,,
\label{Bmu2}
\end{align}
where $\sigma(U,S)$ is a complicated $\mc{O}(1)$ function of $S$ and $U$ which looks like
\begin{align}
\sigma(U,S) =&\, \frac 19 \left(1-s\beta^{U^a}\partial_{u^a}K_{\rm cs} \right)
\left[1 - 3 s \partial_{s,u} \ln \left(h_{H_u} h_{H_d}\right)\right] + \nonumber \\
&+ s \partial_{s,u} \ln \left(h_{H_u} h_{H_d}\right) s \partial_{s,u}\ln z- \nn \\
&- s^2\left[\partial_s \ln z \,\partial_{s,u} \ln \partial_s z +\beta^{U^a} \partial_{u^a} \ln z \,\partial_{s,u} \ln \partial_{u^a} z\right]\,. \nonumber
\end{align}
Recalling that $V_0 = V_{F,0} + V_{D,0}=0$, $V_{F,0}$ can be traded for $V_{D,0}$, and so we shall include it in our analysis of D-terms contributions to $B\hat\mu$.
\begin{enumerate}
\item In the dS case 1 of Sec.~\ref{sssec:MatterUplift}, the D-terms generated $B\hat\mu$ is
\begin{align}
\left.B\hat\mu\right|_D &= \frac{z}{\sqrt{h_{H_u} h_{H_d}}}\,\left.m_0^2\right|_D = \nn \\
&= \frac{z}{\sqrt{h_{H_u} h_{H_d}}}\,\frac{6\epsilon_s}{\omega'_S}m_{3/2}M_{1/2}\sim  \mc{O}\left(m_{3/2}^2\frac{\sqrt{\ln\mathcal{V}}}{\mathcal{V}}\right)\,,
\label{bmu2dS1}
\end{align}
where we used the result in (\ref{sc1}). This term dominates over the F-terms contribution given in (\ref{Bmu2}).

\item In the dS case 2 of Sec.~\ref{sssec:DilatonUplift}, the D-terms generated $B\hat\mu$ is vanishing since
\begin{align}
\left.B\hat\mu\right|_D &= \frac{z}{\sqrt{h_{H_u} h_{H_d}}}\,\left.m_0^2\right|_D = \nn \\
&= \frac{z}{\sqrt{h_{H_u} h_{H_d}}}\,2\left(c_\dS-\frac 13\right)V_{D,0}=0 \quad\text{for}\quad c_\dS=\frac 13\,,
\label{bmu2dS2}
\end{align}
where we used the result in (\ref{Vds2}). Thus in this case $B\hat\mu$ is generated purely by F-terms and it is given by (\ref{Bmu2}) without the term proportional to $V_{F,0}$ that we included in the D-terms contribution. Hence the final result for $B\hat\mu$ is
\begin{align}
B\hat\mu = \frac{z}{\sqrt{h_{H_u} h_{H_d}}}\,\sigma(U,S) \,M_{1/2}^2\sim\mc{O}\left(M_{1/2}^2\right)\,.
\label{Bmu22}
\end{align}
\end{enumerate}

\item \textbf{Superpotential contributions}

Let us discuss the contributions to $\hat\mu$ and $B\hat\mu$ from $\mu\neq 0$ in $W_{\rm matter}$ given by (\ref{Wmatter}). Their general expression in gravity mediation reads~\cite{Brignole:1993dj}
\begin{align}
\label{muW}
&\hat{\mu} = \frac{\mu\,e^{K/2}}{\left(\tilde{K}_{H_u} \tilde{K}_{H_d}\right)^{1/2}}\,,& \\
\label{BmuW}
&B\hat{\mu} = \frac{\mu\,e^{K/2}}{\left(\tilde{K}_{H_u} \tilde{K}_{H_d}\right)^{1/2}} \left[F^I \left(K_I + \partial_I \ln\mu - \partial_I \ln\left(\tilde{K}_{H_u} \tilde{K}_{H_d}\right)\right) - m_{3/2}\right]\,.&
\end{align}

\medskip
\emph{Non-perturbative effects}: Non-perturbative effects can generate in the low-energy action an effective $\mu$-term of the form (up to pre-factors)
\begin{align}
W\supset e^{-a T} H_u H_d\quad\Rightarrow\quad \mu_{\rm eff} = e^{-a T}\,,
\label{Wnpmu}
\end{align}
if the cycle $\tau={\rm Re}(T)$ is in the geometric regime~\cite{Ibanez:2007tu} or
\begin{align}
W\supset e^{-b \left(S+\kappa T\right)} H_u H_d\quad\Rightarrow\quad \mu_{\rm eff} = e^{-b\left(S+\kappa T\right)}\,,
\label{Wnpmu2}
\end{align}
if the cycle $\tau={\rm Re}(T)$ is in the singular regime, i.e. $\tau\to 0$~\cite{Berenstein:2005xa}. Note that there are two distinct classes of non-perturbative contributions leading to the above EFT coupling: if the higgs bi-linear is forbidden by anomalous $U(1)$ symmetries, charged instanton contributions for instance via ED3 can realize such a coupling~\cite{Ibanez:2007tu,Berenstein:2005xa}. Alternatively, if the higgs-bilinear is forbidden by an approximate global symmetry of the local model, this global symmetry is broken by compactification effects. For the latter case, ref.~\cite{Berg:2012aq} studied the topological conditions under which non-perturbative effects of the form (\ref{Wnpmu}) and (\ref{Wnpmu2}) contribute to the effective action. If $T$ is a bulk cycle, the coupling (\ref{Wnpmu}) is always generated but in our case it would be negligible since this effect would be proportional to $e^{-\mathcal{V}^{2/3}}$. On the other hand, if $T$ is the blow-up of a local singularity, the couplings (\ref{Wnpmu}) and (\ref{Wnpmu2}) get generated only if this divisor shares a homologous two-cycle with the blow-up mode $T_\SM$ of the MSSM singularity. This condition is not satisfied if either $T$ or $T_\SM$ is a very simple divisor like a dP$_0$ which has been used in the explicit global models of~\cite{Cicoli:2012vw} and~\cite{Cicoli:2013cha}.\\

If in both cases the appropriate conditions are satisfied, both (\ref{Wnpmu}) and (\ref{Wnpmu2}) would lead to a non-vanishing contribution which can be parametrized as follows
\begin{align}
\hat{\mu} \simeq c_{\mu,W}(U,S)\, \frac{M_{\rm P}}{\mathcal{V}^{n+\frac 13}} \qquad\text{and}\qquad B \hat{\mu} \simeq c_{B,W}(U,S) \,\frac{M_{\rm P}^2}{\mathcal{V}^{n+ \frac 43}}\,,
\label{mumag}
\end{align}
where in (\ref{Wnpmu}) we have set $T=T_s$ and $a=n a_s$ with $n>0$, while in (\ref{Wnpmu2}) we have parametrized $b=n a_s$ recalling that $s\simeq \tau_s$ from (\ref{eq:TausMinimumCorrected}). $c_{\mu,W}$ and $c_{B,W}$ are constants which absorb the dependence on the pre-factor of the instanton contribution, the complex structure moduli and the dilaton. Note that for different values of $n$ non-perturbative effects could generate $\hat{\mu}$ and $B \hat{\mu}$ in the complete range interesting for MSSM phenomenology regardless of the size of the other soft-terms. However these effects can be in competition with K\"ahler potential contributions for $n \geq 5/3$.\\

\emph{Background fluxes}: Primitive $(1,2)$ IASD fluxes
can generate $\hat\mu$ for D3-branes at singularities~\cite{Grana:2003ek,Grana:2002nq,Camara:2003ku}. However, given that their contribution is proportional to the F-terms of the complex structure moduli, this effect has already been included in the contributions from the K\"ahler potential. In other words, direct computations of soft-terms by reducing the D3-brane action in a fluxed background show that $\mu=0$~\cite{Grana:2003ek}.\\

\newpage
\item \textbf{Anomalous $U(1)$ symmetries}

A term proportional to $H_u H_d$ in $K$ or $W$ could be forbidden in the presence of an anomalous $U(1)$ symmetry. In this case, the only way to generate a higgs bilinear would be to multiply this term by an operator involving a $U(1)$-charged field which makes the whole contribution gauge invariant. As already discussed above, the only closed string moduli that can lead to such an effect are K\"ahler moduli.\\

Alternatively, the $U(1)$-charged field could be an open string mode $\Phi$ appearing in $K$ and $W$
in a gauge invariant combination of the form ($\Lambda$ denotes the appropriate moduli-dependent cut-off)
\begin{align}
K\supset \left(\frac{\Phi}{\Lambda}\right)^m H_u H_d\,,\quad \quad W\supset\frac{\Phi^m}{\Lambda^{m-1}} H_u H_d\,.
\label{KWphi}
\end{align}
Thus the field $\Phi$ has to be an MSSM singlet since a higgs bilinear gets generated only when $\Phi$ develops a non-zero VEV breaking the $U(1)$ symmetry. However, as can be seen from eqs. (\ref{dtermpot}) and (\ref{xiSM}), D-terms stabilization fixes the VEV of $\Phi$ in terms of $\tau_\SM$: $|\Phi|^2 \propto \tau_\SM/\mathcal{V}$, and so the couplings in (\ref{KWphi}) would give rise to effective $\mu$ and $Z$-terms which depend only on closed string moduli
\begin{align}
Z_{\rm eff}\propto \frac{1}{\Lambda^m}\left(\frac{\tau_\SM}{\mathcal{V}}\right)^{m/2}\,,
\quad \quad\mu_{\rm eff}\propto \frac{1}{\Lambda^{m-1}}\left(\frac{\tau_\SM}{\mathcal{V}}\right)^{m/2}\,.
\label{Eff}
\end{align}
Once the cut-off $\Lambda$ is explicitly written in terms of $T$-moduli, one could plug (\ref{Eff}) into the standard supergravity formulas to work out the final contribution to $\hat\mu$ and $B\hat\mu$. The result will depend on the VEV and the F-term of $T_\SM$. As discussed in~\cite{Cicoli:2013cha}, $\Phi$ needs to receive tachyonic contribution from soft-terms in order for $T_\SM$ to develop a non-zero VEV. If this condition is satisfied, $\tau_\SM \sim \mathcal{V}^{-1}$ implying $F^{T_\SM}\sim \mathcal{V}^{-2}$ for the local case
and $\tau_\SM \sim \mathcal{V}^{-3}$ implying $F^{T_\SM}\sim \mathcal{V}^{-4}$ for the ultra-local case.
This effect corresponds to switching on an FI-term, and so to breaking the anomalous $U(1)$ by moving slightly away from the singularity. However in both cases the VEV of $\tau_\SM$ is smaller than unity, and so we are still consistently in the singular regime.\\

Given that all these results are clearly model-dependent and require physics beyond the MSSM, at this stage we do not pursue these options in more detail and leave them for future work. Let us just mention that the only case where the effective $\mu$-term in (\ref{Eff}) does not depend on $\Lambda$ is for $m=1$. In this situation $\hat\mu$ would scale as $\mathcal{V}^{-4/3}$ in the local case and as $\mathcal{V}^{-7/3}$ in the ultra-local case. If instead $\Phi$ does not receive tachyonic contributions from soft-terms, another option would be to consider models where $\Phi$ develops a non-zero VEV because of radiative effects.
\een
\ei

\subsection{Summary of Soft-terms}

Let us summarize our results for the soft-terms in the two cases to obtain dS vacua (see also Table~\ref{tab:PhenoScenarios}). Given that in each case the gaugino masses turn out to have the same value, we will use $M_{1/2}$ as a useful parameter which can be rewritten as
\begin{align}
M_{1/2} = c_{1/2} m_{3/2}\,\frac{m_{3/2}}{M_{\rm P}}\, \left[ \ln\left(\frac{M_{\rm P}}{m_{3/2}}\right)\right]^{3/2}\, ,
\end{align}
where $c_{1/2}$ is a flux-dependent tunable coefficient. We will state our results for the model-independent case where $\hat\mu$ and $B\hat\mu$ are generated from moduli induced K\"ahler potential contributions. If these contributions are absent (for example if these terms are forbidden by anomalous $U(1)$ symmetries), then $\hat\mu$ and $B\hat\mu$ can take different values because of model-dependent contributions from either $K$ or $W$ as previously discussed. Let us discuss the local and ultra-local limits separately.

\medskip
\emph{Local limit}: In the local limit, the soft-terms turn out to be the same in both dS mechanisms (all the $c$'s are flux-dependent parameters)
\begin{align}
\label{eq:SummarySoftLocal}
&m_0^2 = c_0\,m_{3/2} M_{1/2}\,,\quad &A_{\alpha\beta\gamma} = (c_A)_{\alpha\beta\gamma}\,M_{1/2}\,, \nn \\
&\hat\mu= c_{\mu,K}\,Z\,M_{1/2}\,,\quad &B\hat\mu = c_{B,K}\,Z\,m_0^2\,.
\end{align}

\medskip
\emph{Ultra-local limit}: In the ultra-local limit, the soft-terms take different forms in the two dS cases
(again all the $c$'s are flux-dependent coefficients which are distinct in different scenarios)
\begin{enumerate}
\item dS vacua from hidden matter fields
\begin{align}
&m_0^2 = c_{0}\,\frac{m_{3/2}M_{1/2}}{\ln\left(M_{\rm P}/m_{3/2}\right)}\,,\quad &A_{\alpha\beta\gamma} = (c_A)_{\alpha\beta\gamma}\,M_{1/2}\,, \nn \\
&\hat\mu= c_{\mu,K}\,Z\,M_{1/2}\,,\quad &B\hat\mu = c_{B,K}\,Z\,m_0^2\,;
\end{align}
\item dS vacua from non-perturbative effects at singularities
\begin{align}
&m_\alpha = (c_{0})_\alpha\,M_{1/2}\,,\quad &A_{\alpha\beta\gamma} = (c_A)_{\alpha\beta\gamma}\,M_{1/2}\,, \nn \\
&\hat\mu= c_{\mu,K}\,Z\,M_{1/2}\,,\quad &B\hat\mu = c_{B,K}\,Z\,M_{1/2}^2\,.
\end{align}
\end{enumerate}
Clearly, the local limit and the dS case 1 for the ultra-local limit correspond to typical (mini-)Split-SUSY scenarios whereas the dS case 2 for the ultra-local limit reproduces a standard MSSM picture with universal gaugino masses and soft masses all of the same order. If the flux-dependent coefficients for the scalar masses are universal $(c_0)_\alpha=c_0,$ a standard CMSSM scenario emerges. Non-universalities in the flux-dependent coefficients can lead to interesting soft-terms patterns such as in NUHM or natural SUSY scenarios. We discuss the dark matter phenomenology related to these scenarios in Chap.~\ref{chap:NonThermalDM}~\cite{Aparicio:2015sda}.\\

\begin{table}
\begin{center}
{\tabulinesep=1.4mm
   \begin{tabu}{|c|c|c|c|}
\hline
Soft term &  Local Models & Ultra Local dS$_1$ & Ultra Local dS$_2$ \\ \hline \hline
$M_{1/2}$ & \multicolumn{3}{c|} { $c_{1/2}\,m_{3/2}\,\frac{m_{3/2}}{M_{\rm P}}\, \left[ \ln\left(\frac{M_{\rm P}}{m_{3/2}}\right)\right]^{3/2}$}\\ \hline
$m_{\alpha}^2$ & $c_0\, m_{3/2} M_{1/2}$ &  $c_0\, \frac{m_{3/2}M_{1/2}}{\ln (M_{\rm P}/m_{3/2})}$ & $(c_0)_\alpha\, M_{1/2}^2$ \\ \hline
$A_{\alpha\beta\gamma}$ & \multicolumn{3}{c|}{$(c_A)_{\alpha\beta\gamma}\, M_{1/2} $}
\\ \hline
$\hat\mu$ & \multicolumn{3}{c|}{$\begin{array}{c}c_{\mu,K}\,Z\,M_{1/2}\, \qquad \text{ (contribution from $K$)} \\
c_{\mu,W} \frac{M_{\rm P}}{\V^{n + \frac 13}}\, \qquad\,\,\,\,\,\,\,\, \text{(contribution from $W$)}
\end{array}$}
\\ \hline
$B\hat\mu$ & \multicolumn{3}{c|}{$\begin{array}{c}c_{B,K}\,Z\,m_{0}^2\, \qquad  \text{ (contribution from $K$)} \\
c_{B,W} \frac{M^2_{\rm P}}{\V^{n + \frac 43}} \qquad\,\,\,\, \text{(contribution from $W$)}
\end{array}$}
\\
\hline
\end{tabu}}
\caption{Summary of soft-terms for different sequestered scenarios for the two dS mechanisms discussed in the text: hidden sector matter (dS$_1$) and non-perturbative effects at singularities~(dS$_2$). All soft-terms are hierarchically smaller than $m_{3/2}$. Gaugino masses, A-terms and the $\hat{\mu}$-term take the same value in each case whereas scalar masses and hence the $B\hat{\mu}$-term are model-dependent. The coefficients $c$ are flux-dependent and can generically take different values in each scenario presented here while $n$ is a positive model-dependent parameter. They can be tuned to compare with data. Local and ultra-local 1 cases give a Split-SUSY spectrum while ultra local 2 implies a standard MSSM spectrum with soft-masses of the same order and possible small non-universalities due to the flux-dependent parameters $c$.}
\label{tab:PhenoScenarios}
\end{center}
\end{table}

For illustrative purposes, we just mention here two simple benchmark models for the dS case 2 in the ultra-local limit. Setting all the $\beta$'s to zero, we find

\medskip
\emph{Benchmark model 1: $h_\alpha=z=1$}
\begin{align}
m_\alpha \simeq 0\,\,\,\forall\alpha\,,\quad A_{\alpha\beta\gamma} = (c_A)_{\alpha\beta\gamma}\,M_{1/2}\,,
\quad\hat\mu= \frac{M_{1/2}}{3} \,,\quad B\hat\mu = \hat\mu^2\,,
\end{align}
where $(c_A)_{\alpha\beta\gamma}=s\partial_s \ln Y_{\alpha\beta\gamma}$.
This reproduces a typical gaugino mediation scenario~\cite{Schmaltz:2000gy,Yanagida:2013ah}.

\medskip
\emph{Benchmark model 2: $f_\alpha=\gamma=1$}
\begin{align}
m_\alpha = m_0 = \frac{M_{1/2}}{\sqrt{3}}\,\,\,\forall\alpha\,,\quad A = -\sqrt{3}\,m_0\,,
\quad\hat\mu \simeq \frac{m_0}{\ln\left(M_{\rm P}/m_{3/2}\right)}\,,\quad B\hat\mu = m_0^2\,,
\end{align}
if the holomorphic Yukawas do not depend on $S$. This leads to a typical natural SUSY scenario for example if we allow $m_{H_u}$ to be slightly larger than $m_0$ together with a light third generation~\cite{Baer:2012uy}. This can be done by considering the more general case with non-zero $\beta$'s and allowing for a $U$-dependence in $f_\alpha$. The $\ln\left(M_{\rm P}/m_{3/2}\right)$ suppression of $\hat\mu$ with respect to $m_0$ comes from sub-leading contributions to $\hat\mu$ from $F^{T_s}$.

\section{Possible Sources of Desequestering}
\label{sec:DesequesteringSources}

There is a general belief that in any supergravity theory once SUSY is broken all sparticles should get a mass at least of the order of the scale determined by the split in the gravity multiplet. In particular, all soft masses are expected to be of order the gravitino mass. Furthermore, if for some reason some of the sparticle masses are found to be smaller than $m_{3/2}$ at tree level, since SUSY no longer protects these masses against quantum corrections, they should be lifted to a loop factor times $m_{3/2}$. So soft masses are expected to be at most one order of magnitude lighter than the gravitino mass but not much smaller.\footnote{This separation between $m_{3/2}$ and soft masses occurs for example in the case of mirage (mixed moduli and anomaly) mediation~\cite{LoaizaBrito:2005fa,Choi:2005hd}.} Effects which tend to push the soft masses to the scale of the gravitino mass are referred to as sources of \textit{desequestering}. In this section we will argue that our models can be stable against desequestering effects.

\subsection{Loop Corrections}

For sequestered string scenarios, it is natural to expect that loop corrections bring soft masses to a magnitude of order a loop factor times $m_{3/2}$. However there can be exceptions since the couplings can be Planck suppressed. A detailed calculation of loop corrections to the mass of bulk scalars like the volume modulus (its tree level mass $m_\mathcal{V} \sim m_{3/2}/\mathcal{V}^{1/2}$ is hierarchically smaller than $m_{3/2}$) was presented in~\cite{Burgess:2010sy}.\\

The size of loop corrections can be estimated by realizing that, if SUSY is broken, loop corrections to the mass should be given by the heaviest particles circulating in the loop (or the cut-off scale) which is the Kaluza-Klein scale $M_\KK\sim M_{\rm P}/\mathcal{V}^{2/3}$. In the absence of SUSY there is a need of a SUSY-breaking insertion (a spurion field representing the relevant F-term) in the loop and the correction to the mass is at most
\begin{align}
\delta m=\alpha_\lp \frac{M_\KK m_{3/2}}{M_{\rm P}}\sim \alpha_\lp\frac{W_0}{\mathcal{V}^{5/3}}\ll \alpha_\lp m_{3/2}\,,
\end{align}
with $\alpha_\lp\sim g^2/(16\pi^2)$ a loop factor. Note that the ratio $\delta m/m\sim \alpha_\lp \mathcal{V}^{-1/6}$ is very small and therefore the volume modulus mass is stable against loop corrections.\\

For matter fields located at the Standard Model brane, loop corrections should be even further suppressed.
The effective field theory on the brane is supersymmetric and feels the effects of SUSY-breaking in the bulk only via Planck suppressed couplings. Therefore masses as small as $M_{\rm soft}\sim W_0/\mathcal{V}^2$ are still stable under standard loop corrections (since volume suppressed brane-bulk couplings imply $\delta M_{\rm soft}\ll \delta m$).\\

Over the years explicit calculations have been performed estimating loop corrections to soft masses in no-scale
and general gravity mediated models. See for example~\cite{Ellis:1986nr,Antoniadis:1997ic} in which loop corrections to scalar and gaugino masses were estimated in supergravity and M-theory frameworks with results of order $\delta m \sim \alpha_\lp m_{3/2}^2/M_{\rm P}\sim \alpha_\lp M_{\rm P}/\mathcal{V}^2$. More recently, explicit calculations for gravitino loop contributions to gaugino masses was performed in~\cite{Lee:2013aia}. The diagrams are quadratically divergent and proportional to the gravitino mass:
\begin{align}
\delta M_{1/2} = \frac{m_{3/2}}{16\pi^2}\left(\frac{\Lambda^2}{M_{\rm P}^2}+\dots\right)\,,
\end{align}
where $\Lambda$ is the cut-off scale and the dots represent sub-leading logarithmically divergent terms. In string theory we expect that $\Lambda\leq M_s\sim M_{\rm P}/\mathcal{V}^{1/2}$ which then corrects the gaugino masses to order $\delta M_{1/2}\leq \alpha_\lp M_{\rm P}/\mathcal{V}^2$ which is smaller than the sequestered gaugino masses $M_{1/2}\sim M_{\rm P}/\mathcal{V}^2$.\\

This behavior of sequestered models motivated the work of Randall and Sundrum to introduce anomaly mediation.
However, as we will illustrate below, the approximate no-scale structure of LVS makes anomaly mediated corrections to soft-terms sub-leading (they vanish identically for no-scale models) in generic points of parameter space.

\subsection{Anomaly Mediated Contributions}
\label{anosec}

In this section we examine anomaly mediated contributions to soft-terms and compute their strength in the dS constructions discussed in Sec.~\ref{ssec:deSitterScenarios}. The anomaly mediated gaugino masses~\cite{Bagger:1999rd} are given by\footnote{Note that there is a certain discussion on the validity of this formula~\cite{deAlwis:2008aq}. For the purpose of the present chapter we assume that the standard derivation from field theory or string theory~\cite{Conlon:2010qy} is valid.}
\begin{align}
M_{1/2}^{\rm anom}=\frac{g^2}{16 \pi^2}\left[\left(T_R-3T_G\right)m_{3/2}+\left(T_G-T_R\right)F^I\partial_I K
+\frac{2T_R}{d_R}F^I\partial_I\ln\det \tilde{K}_{\alpha\beta}\right],
\label{eq:anomalygauginomasses}
\end{align}
where $T_{G,R}$ are the Dynkin indexes of the adjoint representation and the matter representation $R$ of dimension $d_R$ (summation over all matter representations is understood). Assuming that the K\"ahler metric for matter fields can be written as $\tilde{K}_{\alpha\beta}=\delta_{\alpha\beta} f_\alpha\tilde{K}$,
the expression (\ref{eq:anomalygauginomasses}) reduces to
\begin{align}
M_{1/2}^{\rm anom}=&\frac{g^2}{16 \pi^2}\left[\left(T_R-3T_G\right)m_{3/2}+\left(T_G-T_R\right)F^I\partial_I K
+ \right. \nn \\
&\left.+\frac{2T_R}{\tilde{K}} F^I\partial_I\tilde{K}+\frac{2T_R}{d_R}\sum_{\alpha=1}^{d_R} F^I\partial_I\ln f_\alpha\right].
\label{eq:anomalygauginomasses2}
\end{align}
In the local case, we find that the leading order anomaly mediated contribution can be written in terms
of the modulus dominated gaugino mass $M_{1/2}$ given in (\ref{ggmm})
\begin{align}
M_{1/2}^{\rm anom} =& \frac{g^2}{16\pi^2} \left[\left(T_R - T_G\right)\left(1-s\beta^{U^a}\partial_{u^a} K_{\rm cs}\right) - \right. \nn \\
&\left. -\frac{4 T_R}{\omega_S'} \left(c_s-\frac 13 \right)+\frac{2sT_R}{d_R}\sum_{\alpha=1}^{d_R} \partial_{s,u}\ln f_\alpha\right] M_{1/2},
\label{anorel}
\end{align}
with $\omega_S'$ as defined below (\ref{ds}). For the ultra-local case we find instead
\begin{align}
M_{1/2}^{\rm anom} = \frac{g^2}{16\pi^2} \left[\left(\frac{T_R}{3} - T_G\right)\left(1-s \beta^{U^a}\partial_{u^a} K_{\rm cs}\right) +\frac{2sT_R}{d_R}\sum_{\alpha=1}^{d_R} \partial_{s,u}\ln h_\alpha\right] M_{1/2}\, .
\label{anorel2}
\end{align}
Therefore in both cases the anomaly mediated contribution is loop suppressed with respect to the moduli mediated one. This result is the consequence of the approximate no-scale structure of LVS models.\\

A more careful analysis is needed for a very particular point in the underlying parameter space:
$\omega_S' \to 0$, i.e. in the very tuned situation where the F-term of the dilaton is vanishing at leading order because of a special compensation between the contribution to $F^S$ from $D_S W$ and $D_{T_b} W$. In this case the leading contribution to gaugino masses given in (\ref{ggmm}) is zero and the first non-vanishing moduli mediated contribution can be estimated to scale as $M_{1/2}^{\rm new}\sim m_{3/2} \sqrt{\ln\mathcal{V}}/\mathcal{V} $. On the other hand, the anomaly mediated contribution scales as $M_{1/2}^{\rm anom}\sim c \,M_{1/2}^{\rm new}$ where $c = c' \left(\frac{g^2}{16\pi^2}\right) \ln\mathcal{V}$ and $c'$ denotes a numerical factor arising from evaluating~\eqref{anorel2} exactly. For $g\simeq 0.1$ and $\mathcal{V} \simeq 5\cdot 10^6$ (the value needed to get $M_{1/2}^{\rm new}$ approximately around the TeV-scale), $c$ is roughly of order $c'\times10^{-3}.$ Depending on the exact value of $c',$ which is beyond the scope of this analysis, we can achieve competing contributions from moduli mediation and anomaly mediation.

\subsection{Moduli Redefinitions}

Desequestering can also potentially occur due to moduli redefinitions which might be necessary order by order in perturbation theory. This desequestering effect can for example arise due to a shift of the local cycle $\tau_\SM\rightarrow \tau_\SM+\alpha\ln\mathcal{V}$ which has the effect of making the soft-terms of the same order as the gravitino mass~\cite{Conlon:2010ji,Choi:2010gm}.\\

Such moduli redefinitions depend on the structure of the D-branes configuration. In particular, it has been argued that redefinitions are absent for configurations involving only D3-branes at orbifold singularities but are present for D3-branes at orientifold singularities and in cases with both D3- and D7-branes (see~\cite{Conlon:2010ji}).\\

We emphasize that desequestering occurs only if the moduli redefinition leads to a change in the functional form of the K\"ahler potential. Arguments suggesting a change in the functional form were presented in~\cite{Conlon:2010ji} but an explicit computation of such a change is still not available in the literature. Some recent explicit computations of the K\"ahler potential~\cite{Grimm:2013bha, Junghans:2014zla} have shown that perturbative corrections can be such that, along with a field redefinition, there is also an additional term generated in the K\"ahler potential. In these cases, however, the two effects conspire to leave the functional form of the K\"ahler potential invariant. More detailed studies of perturbative corrections to the K\"ahler potential are crucial to get a comprehensive understanding  of the relationship between moduli redefinitions and desequestering.

\subsection{Superpotential Desequestering}

Apart from potentially destroying the hierarchy between soft masses and $m_{3/2}$, various sub-leading effects can have important phenomenological consequences. Interesting constraints arise from non-perturbative terms in the superpotential involving visible sector fields~\cite{Berg:2010ha}. Superpotential terms of the type
\begin{align}
\label{sud}
\hat{W} = \left( \hat{\mu} H_u H_d + \hat {\lambda}^u_{ij} Q^i u^j H_u
+ \hat {\lambda}^d_{ij}  Q^i d^j H_d + \hat {\lambda}^u_{ij} L^i e^j H_d \right) e^{-a_s T_s}\,,
\end{align}
would lead to flavour violation and CP-violation via A-terms with a strength sensitive to the hierarchy between soft masses and $m_{3/2}$. For $M_{\rm soft}^2 \sim m_{3/2}^2/\mathcal{V}^n$ the strength of CP and flavour violation induced by A-terms would be
\begin{align}
\label{flav}
\delta \sim  \mathcal{V}^n   10^{-16} \left( {v \over 100 \phantom{.} {\rm GeV}} \right)\,,
\end{align}
with $v$ equal to the higgs VEV. CP violation and FCNC bounds then require $\mathcal{V} < 10^5$. This gives a slight tension with our results but there can be several ways around this issue. The estimate~(\ref{flav}) is based on effective field theory arguments; it assumes generic order one coefficients for the superpotential terms in (\ref{sud}). A string computation of  the coefficients was done in~\cite{Berg:2012aq}. This indicates that the coefficients are suppressed unless the Standard Model cycle and the cycle on which the instanton is supported share a homologous two-cycle. The presence of flavour symmetries~\cite{Conlon:2008wa, Dolan:2011qu, Krippendorf:2010hj, Maharana:2011wx} in the visible sector can also alleviate this tension.

\section{Conclusions}
\label{sec:conclusions}

In this chapter we have analyzed soft-terms for LVS sequestered models with dS moduli stabilization. These models are particularly attractive for phenomenological reasons: the string scale is around the GUT scale, soft masses are at the TeV scale and the lightest modulus is much heavier than the bound imposed by the CMP. The volume of the compactification is of order $\mathcal{V}\sim 10^7$ in string units and the visible sector can be localized on D3-branes at a singularity.\\

The pattern of soft-terms for these models has been previously studied in~\cite{Blumenhagen:2009gk}. In this chapter we have deepened the analysis of~\cite{Blumenhagen:2009gk} by studying the effect on soft-terms of the sector responsible to realize a dS vacuum, and by classifying in a systematic way any possible correction to the leading no-scale structure of soft-terms. In particular, given that soft-terms depend on the moduli-dependence of the K\"ahler metric for matter fields $\tilde{K}_\alpha$, we defined two possible limits for $\tilde{K}_\alpha$: \emph{Local scenarios} where $\tilde{K}_\alpha$ is such that the visible sector Yukawa couplings $Y_{\alpha\beta\gamma}$ do not depend on $\mathcal{V}$ only at leading order in an inverse volume expansion, and \emph{Ultra-local scenarios} where $Y_{\alpha\beta\gamma}$ are exactly independent on $\mathcal{V}$ at all orders.\footnote{Evidence in favor of ultralocality has been obtained from explicit string computations in toroidal orbifolds~\cite{Conlon:2011jq}. The case of realistic CY compactifications remains to be explored.} Moreover, due to the present lack of explicit string computations of $\tilde{K}_\alpha$, we parametrized its dependence on the dilaton and complex structure moduli as an unknown function $f_\alpha(U,S)$.\\

The computation of soft-terms has produced a wide range of phenomenological possibilities depending on the exact moduli-dependence of the matter K\"ahler metric and the way to achieve a dS vacuum. We considered two dS realizations based on supersymmetric effective actions: \emph{dS case 1} where hidden sector matter fields living on a bulk cycle develop non-vanishing F-terms because of D-terms fixing, and \emph{dS case 2} where the blow-up mode of a singularity different from the visible sector one develops non-zero F-terms due to non-perturbative effects. Broadly speaking, we found two classes of models:
\begin{enumerate}
\item \emph{Split-SUSY}: Local models and ultra-local models in the dS case 1 yield gaugino masses and A-terms which are suppressed with respect to scalar masses: $M_{1/2} \sim m_{3/2}\epsilon\ll m_0 \sim m_{3/2} \sqrt{\epsilon}\ll m_{3/2}$ for $\epsilon \sim m_{3/2}/M_{\rm P}\ll 1$. For volumes of order of $10^7$ in string units these models provide a version of the Split-SUSY scenario with a `largish' splitting between gauginos and scalars (according to current experimental bounds). Non-universalities are present but suppressed by inverse powers of the volume.
\item \emph{Standard MSSM}: For ultra-local models in the dS case 2, all soft-terms are of the same order: $M_{1/2} \sim m_0 \sim m_{3/2}\epsilon\ll m_{3/2}$. Therefore these models include the CMSSM parameter space and its possible generalizations since each soft-term comes along with a tunable flux-dependent coefficient. Moreover, depending on the exact functional dependence of the K\"ahler metric for matter fields, these models can also feature non-universalities which are constrained by the experimental bounds on flavour changing neutral currents.
\end{enumerate}

Let us stress again that the exact numerical coefficients of the soft-terms are functions of the dilaton and complex structure moduli which are fixed in terms of flux quanta. Hence soft-terms vary as one scans through the string landscape. This crucial property of our scenarios gives supersymmetric models the freedom to perform any tuning which is needed for phenomenological reasons. In particular, it is low-energy SUSY that addresses the hierarchy problem by stabilizing the higgs mass at the weak scale, while scanning through the landscape provides small variations in the size of soft-terms as necessary to reproduce all the detailed features of experimental data. This tuning at low energies can be viewed as a choice of parameters in the high scale theory. There is a large freedom of choice in the high scale theory which is however not arbitrary since this freedom is provided by the theory itself (by having a computable landscape of vacua).\\

Note that in the ultra-local case the two ways to achieve a dS vacuum give rise to a different pattern of soft-terms. This can intuitively be understood as follows: the depth of the LVS AdS vacuum is of order $m_{3/2}^2\epsilon$, and so any extra term which yields a dS solution has to be of this order of magnitude. In turn, if the field $\phi$ responsible for dS uplifting is not decoupled from the visible sector, scalar masses of order $m_{3/2} \sqrt{\epsilon}$ are expected to arise because of this new contribution to the scalar potential. This is actually what happens in the dS case 1 since $\phi$ lives on a bulk cycle, and so it is not decoupled from the visible sector. On the other hand, in the dS case 2 $\phi$ lives on a singularity which is geometrically separated from the one supporting the visible sector. This gives rise to an effective decoupling between $\phi$ and the visible sector, resulting in suppressed scalar masses.\\

We would also like to emphasize that our analysis for the dS case 1, together with~\cite{Cicoli:2012vw}
(which provided visible sector models embedded in moduli stabilized compact CYs), provides a very comprehensive study of SUSY-breaking in string theory.\\

The soft-terms which are more complicated to estimate are the $\hat{\mu}$ and $B\hat{\mu}$-terms since they receive contributions from both the K\"ahler potential and the superpotential. Moreover, these contributions could generically be forbidden in models with branes at singularities because of the presence of anomalous $U(1)$ symmetries. In this case, effective $\hat{\mu}$ and $B\hat{\mu}$-terms could still be generated due to non-perturbative corrections (e.g. D-branes instantons) or matter fields which develop non-vanishing VEVs. However in this last case, besides the need to go beyond the MSSM by including additional matter fields, any prediction for $\hat{\mu}$ and $B\hat{\mu}$-terms would necessarily be model-dependent.

\part{Non-Standard Cosmology}
\label{part:Cosmology}

\chapter{Non-Thermal Dark Matter}
\label{chap:NonThermalDM}

\section{Motivation for Non-thermal Dark Matter}
\label{sec:motivation}

One of the main particle physics candidates for DM is a stable neutralino $\chi$ which emerges as the lightest supersymmetric particle (LSP) in several scenarios beyond the Standard Model. The DM relic abundance is generically \emph{assumed} to be produced thermally by the following process: the LSP is in a thermal bath in the early universe, subsequently drops out of thermal equilibrium and freezes-out at temperatures of order
$T_f \simeq m_{\rm DM}/20$ when DM annihilation becomes inefficient.\\

However, we have no direct observational evidence of the history of the universe before Big Bang Nucleosynthesis (BBN) for temperatures above $T_{\rm BBN} \simeq 3$ MeV. There is therefore no reason to assume a very simple cosmological history characterized by just a single period of radiation dominance from the end of inflation until BBN. In fact, the presence of a period of matter domination between the end of inflation and BBN could completely change the final prediction for the DM relic density if the reheating temperature at the end of this period of matter dominance is below $T_f$~\cite{Barrow:1982ei, Kamionkowski:1990ni, Moroi:1999zb, Fujii:2001xp, Kitano:2008tk, Dutta:2009uf}.\\

This non-thermal picture emerges generically in UV theories like string theory due to the ubiquitous presence of gravitationally coupled scalars~\cite{Coughlan:1983ci, Banks:1993en, deCarlos:1993wie, Acharya:2008bk, Acharya:2009zt, Acharya:2010af, Allahverdi:2013noa}. During inflation these fields, called moduli, get a displacement from their minimum that is in general of order $M_{\rm P}$~\cite{Dine:1995kz}. After the end of inflation, when the Hubble constant reaches their mass, $H\sim m_{\rm mod}$, they start oscillating around their minimum and store energy. Redshifting as matter, they quickly dominate the energy density of the universe which gets reheated when the moduli decay. Being only gravitationally coupled, the moduli tend to decay very late when $H\sim \Gamma \sim m_{\rm mod}^3/M_{\rm P}^2$. The corresponding reheating temperature
\begin{align}
T_{\rm rh} \sim \sqrt{\Gamma M_{\rm P}}\sim 0.1 \,m_{\rm mod} \sqrt{\frac{m_{\rm mod}}{M_{\rm P}}}\,,
\end{align}
has to be larger than $T_{\rm BBN}$ in order to preserve the successful BBN predictions.\footnote{$T_{\rm rh}$ has also to be lower than the temperature above which the internal space decompactifies~\cite{Buchmuller:2004tz, Anguelova:2009ht}.} This requirement sets a lower bound on the moduli masses of order $m_{\rm mod}\gtrsim 1.3 \times 10^5$ GeV~\cite{Coughlan:1983ci, Banks:1993en, deCarlos:1993wie}.\\

Generically in string compactifications SUSY-breaking effects develop a mass for the moduli and generate by gravity or anomaly mediation soft-terms of order $M_{\rm soft}$. Due to their common origin, the mass of the lightest modulus $m_{\rm mod}$ is therefore related to the scale of the soft-terms as $M_{\rm soft} = \kappa m_{\rm mod}$. Given the cosmological constraint $m_{\rm mod}\gtrsim 1.3 \times 10^5$ GeV, only models with $\kappa\ll 1$ can allow for low-energy SUSY to solve the hierarchy problem. Values of $\kappa\sim\mc{O}(10^{-2})$ can come from loop suppression factors~\cite{Choi:2005ge, Lowen:2008fm, Acharya:2008zi,Dudas:2012wi} while much smaller values $\kappa\sim\mc{O}(10^{-3}-10^{-4})$ can arise due to sequestering effects~\cite{Aparicio:2014wxa, Blumenhagen:2009gk}. For $M_{\rm soft}\sim \mc{O}(1)$ TeV, the corresponding reheating temperature becomes
\begin{align}
T_R \sim  \frac{M_{\rm soft}}{\kappa^{3/2}} \sqrt{\frac{M_{\rm soft}}{M_{\rm P}}} \sim \,\kappa^{-3/2}\,\mc{O}(10^{-2})\,\,{\rm MeV}\,,
\end{align}
which for $10^{-2}\lesssim\kappa\lesssim 10^{-4}$ is between $\mc{O}(10)$ MeV and $\mc{O}(10)$ GeV. This is below the freeze-out temperature for LSP masses between $\mc{O}(100)$ GeV and $\mc{O}(1)$ TeV which is
$T_f \sim \mc{O}(10-100)$ GeV. Therefore any DM relic density previously produced via the standard thermal mechanism gets erased by the late-time decay of the lightest modulus. In this new scenario, the LSP gets produced non-thermally from the modulus decay.\\

From a bottom-up perspective, non-thermal cosmological histories can also enlarge the available parameter space of different DM models consistent with direct and indirect detection experiments, due to the presence of the additional parameter $T_R$. This is appealing as it is very hard to reproduce a correct thermal relic density in the CMSSM/mSUGRA (see for instance~\cite{Baer:2012uya}) since a bino-like LSP tends to overproduce DM
(apart from some fine-tuned cases like stau co-annihilation and A-funnel or in the case of precision gauge coupling unification~\cite{Krippendorf:2013dqa}) while for a higgsino- or wino-like LSP the relic density is in general under-abundant (except for cases like well tempered bino/higgsino or bino/wino DM~\cite{ArkaniHamed:2006mb}).\\

The purpose of this chapter is to study the production of Non-Thermal DM in the CMMS, where the free parameters are: the standard parameters of the CMSSM/mSUGRA~\cite{Arnowitt:1992aq, Barger:1993gh, Kane:1993td, Baer:1996kv}, i.e.~the universal scalar mass $m$, universal gaugino mass $M$ and trilinear coupling $A$ defined at the GUT scale, $\tan\beta$ and the sign of $\mu$, with in addition the reheating temperature $T_R$ from the decay of the lightest modulus. We shall follow the RG running of these parameters from the GUT to the electroweak (EW) scale and require a higgs mass $m_h \simeq 125$ GeV, a correct radiative EW symmetry breaking (REWSB) and no DM non-thermal overproduction. We shall then focus on the points satisfying all these requirements and we will impose on them several phenomenological constraints coming from LEP~\cite{LEP1, LEP2, LEP3, LEP4}, LHC~\cite{Aad:2014wea,Aad:2012hm, Chatrchyan:2012lia, Aad:2014vma, Khachatryan:2014qwa}, Planck~\cite{Ade:2015xua}, Fermi (pass 8 limit)~\cite{Ackermann:2015tah}, XENON100~\cite{Orrigo:2015cha}, CDMS~\cite{Agnese:2013rvf}, IceCube~\cite{Aartsen:2012kia} and LUX~\cite{Akerib:2013tjd}. Moreover we shall focus only on cases where the LSP has a non-negligible higgsino component since bino-like DM requires a very low reheating temperature which is strongly disfavored by dark radiation bounds in the context of many string models~\cite{Allahverdi:2014ppa}. Interestingly we shall find that the constraints from Fermi and LUX are very severe and do not rule out the entire non-thermal CMSSM parameter space only for reheating temperatures $T_R\gtrsim \mc{O}(1)$ GeV. The best case scenario is realized for $T_R = 2$ GeV where a higgsino-like LSP with a mass around $300$ GeV can saturate the observed DM relic abundance. For larger reheating temperatures the LSP bino component has to increase, resulting in strong direct detection bounds which allow only for cases with DM underproduction. Values of $T_R$ above $1$ GeV require values of $\kappa\sim\mc{O}(10^{-3}-10^{-4})$ which can be realized only in models where the CMSSM is sequestered from the sources of SUSY-breaking~\cite{Aparicio:2014wxa, Blumenhagen:2009gk}. In these models the contribution to gaugino masses from anomaly mediation turns out to be negligibly small and hence is distinct from other scenarios~\cite{Choi:2005ge, Lowen:2008fm, Acharya:2008zi, Dudas:2012wi} where contributions from anomaly mediation are significant and where a wino LSP can be realized. Apart from DM, these models are very promising since they can be embedded in globally consistent CY compactifications~\cite{Cicoli:2011qg, Cicoli:2012vw, Cicoli:2013cha}, allow for TeV-scale SUSY and successful inflationary models~\cite{Conlon:2005jm, Burgess:2013sla, Cicoli:2011zz}, do not feature any CMP,\footnote{Ref.~\cite{Dutta:2014tya} provides a significant bound on moduli masses and the number of e-foldings during inflation which can be a challenge for many models.} are compatible with gauge coupling unification and do not suffer from any moduli-induced gravitino problem~\cite{Endo:2006zj, Nakamura:2006uc}. Non-standard thermal history using CMSSM was considered  in ref.~\cite{Roszkowski:2014lga} where the entropy due to the decay of the inflaton reduces the thermal relic abundance. In this thesis, however, we consider that the DM particles from the moduli decay constitute the DM abundance and the thermal  DM abundance contribution to the net DM content is negligible since $T_r$ is not required to be very close to $T_f$.\\

In Sec.~\ref{ssec:CMSSMSoftTerms} we discuss CMSSM soft-terms, in Sec.~\ref{ssec:NTDMCMSSMResults} we analyze the non-thermal CMSSM, in Sec.~\ref{ssec:DiscussionResults} we discuss our results.

\section{CMSSM Soft-Terms}
\label{ssec:CMSSMSoftTerms}

In a UV completion of the MSSM like string theory, SUSY is spontaneously broken by some dynamical mechanism which generates particular relations between the soft-terms via gravity, anomaly or gauge mediation. In the case when the soft-terms are universal at the GUT scale, they are given by the scalar mass $m$, the gaugino mass $M$, the trilinear coupling $A$ and the bilinear higgs mixing $B$. We can generically parametrize these soft-terms and the $\mu$ parameter at the GUT scale as:
\begin{align}
m = a\ |M|\,,\quad A = b\ M\,,\quad B = c \ M\,, \quad \mu = d\ M\,,
\label{CMSSM}
\end{align}
where, in a stringy embedding, the coefficients $a$, $b$, $c$ and $d$ are functions of the underlying parameters while the gaugino mass $M$ sets the overall scale of the soft-terms in terms of the gravitino mass $m_{3/2}$. In order to perform a phenomenological analysis of this scenario one has to follow the renormalization group (RG) evolution of these soft-terms from the GUT to the EW scale and impose the following constraints: a correct REWSB, a higgs mass of order $m_h \simeq 125$ GeV, no DM overproduction and no contradiction with flavour observables and with any experimental result in either particle physics or cosmology.\\

A viable REWSB can be obtained if at the EW scale the following two relations are satisfied:
\begin{align}
\mu^2 = \frac{m_{H_d}^2-m_{H_u}^2 
\tan^2\beta}{\tan^2\beta-1}-\frac{m_Z^2}{2}\,,
\label{resb1}
\end{align}
where:
\begin{align}
\sin(2\beta)=\frac{2 |B\mu|}{m_{H_d}^2+m_{H_u}^2 + 2\mu^2}\,.
\label{resb2}
\end{align}
Given that the requirement of a correct REWSB fixes only the magnitude of $\mu$ leaving its sign as free, the parameters (\ref{CMSSM}) are typically traded for the standard CMSSM/mSUGRA parameters:
\begin{align}
m = a\ |M|\,,\quad A = b\ M\,,\quad \tan\beta\equiv \frac{\langle H_u^0\rangle}{\langle H_d^0\rangle}\,, \quad {\rm sign}(\mu)\,,
\label{CMSSMparameters}
\end{align}
where one runs $m$, $M$ and $A$ (or $a$, $b$ and $M$ in our case) from the GUT to the EW scale with a particular choice of $\tan\beta$ and sign$(\mu)$. eq.~\eqref{resb1} and \eqref{resb2} then give the value of $B$ and $\mu$
at the EW scale. This is the way in which typical spectrum generators operate.\footnote{Here we use \texttt{SPheno} v3.3.3~\cite{Porod:2003um, Porod:2011nf}.} The boundary values of $B$ and $\mu$ at the GUT scale which give a correct REWSB can be obtained by running back $B$ and $\mu$ from the EW to the GUT scale. In this way we obtain the values of the coefficients $c$ and $d$. In a viable UV model, these values of $c$ and $d$ have to be compatible with the values allowed by the stringy dynamics responsible for SUSY-breaking and the generation of soft-terms. The expressions in eq.~\eqref{resb1} and \eqref{resb2} are tree level relations. We use loop corrections to these relations in our calculations.

\section{Non-thermal CMSSM}
\label{ssec:NTDMCMSSMResults}

As motivated in Sec.~\ref{sec:motivation}, we shall consider scenarios where the LSP is produced non-thermally like in the case of string compactifications where the reheating temperature $T_R$ from the decay of the lightest modulus is generically below the thermal freeze-out temperature~\cite{Acharya:2008bk, Acharya:2009zt, Acharya:2010af, Allahverdi:2013noa}. This reheating temperature represents an additional parameter which has to be supplemented to the standard free parameters of the CMSSM ($a$, $b$, $M$, $\tan\beta$ and the sign of $\mu$). We call this new scenario the `non-thermal CMSSM' which is characterized by the following free parameters: $T_R$, $a$, $b$, the gaugino mass $M$, $\tan \beta$ and the sign of $\mu$.

\subsection{Non-thermal Dark Matter Relic Density}

The abundance of DM particles $\chi$ produced non-thermally by the decay of the lightest modulus is given by~\cite{Moroi:1999zb}:
\begin{align}
\label{dmdens}
\left({n_{\rm DM} \over s}\right)^{\rm NT} = {\rm min} \left[\left({n_{\rm DM} \over s}\right)_{\rm obs} {\langle \sigma_{\rm ann} v \rangle_f^{\rm Th} \over \langle \sigma_{\rm ann} v \rangle_f} \sqrt{\frac{g_*(T_f)}{g_*(T_R)}}\left({T_f \over T_R}\right), \, Y_{\rm mod}~ {\rm Br}_{\rm DM} \right] \,,
\end{align}
where $g_*$ is the number of relativistic degrees of freedom, $\langle \sigma_{\rm ann} v \rangle_f^{\rm Th}\simeq 2 \times 10^{-26} {\rm cm}^3\,{\rm s}^{-1}$ is the annihilation rate at the time of freeze-out needed in the thermal case to reproduce the observed DM abundance:
\begin{align}
\left({n_{\rm DM} \over s}\right)_{\rm obs} = \left(\Omega_{\rm DM} h^2\right)_{\rm obs} \left(\frac{\rho_{\rm crit}}{m_{\rm DM} s h^2}\right) \simeq 0.12 \left(\frac{\rho_{\rm crit}}{m_{\rm DM} s h^2}\right),
\label{obs}
\end{align}
whereas the yield of particle abundance from modulus decay is:
\begin{align}
Y_{\rm mod} \equiv {3 T_R \over 4 m_{\rm mod}} \sim \sqrt{\frac{m_{\rm mod}}{M_{\rm P}}}\,.
\label{yield}
\end{align}
${\rm Br}_{\rm DM}$ denotes the branching ratio for modulus decays into $R$-parity odd particles which subsequently decay to DM.\\

The expression in eq. \eqref{dmdens} leads to two scenarios for non-thermal DM:
\ben
\item `Annihilation scenario': in this case the DM abundance is given by the first term on the right-hand side of eq.~\eqref{dmdens} and the DM particles undergo some annihilation after their initial production by modulus decay.
In order to avoid DM overabundance one needs
\begin{align}
\langle \sigma_{\rm ann} v \rangle_f \geq \langle \sigma_{\rm ann} v \rangle_f^{\rm Th}\,\sqrt{\frac{g_*(T_f)}{g_*(T_R)}} \left(\frac{T_f}{T_R}\right)\,.
\end{align}
Given that $T_R < T_f$ and $g_*(T_R) < g_*(T_f)$, this scenario requires $\langle \sigma_{\rm ann} v \rangle_f > \langle \sigma_{\rm ann} v \rangle_f^{\rm Th}$ as in the case of thermal underproduction. This condition is satisfied by a higgsino- or wino-like LSP but not by a pure bino-like LSP which would generically lead to non-thermal overproduction (apart from the aforementioned cases). However, given that we shall focus on models with just gravity mediated SUSY-breaking (contributions from anomaly mediation are sub-leading) and universal gaugino masses at the GUT scale as in~\cite{Aparicio:2014wxa, Blumenhagen:2009gk}, the LSP can never be wino-like due to the RG running of the gauginos. Bino is the lightest gaugino.  Moreover a wino-like LSP has a significantly larger annihilation cross section than a higgsino-like LSP resulting in a strong conflict with Fermi bounds for sub-TeV wino-like DM~\cite{Cohen:2013ama}. In this context, the `Annihilation scenario' requires a higgsino-like DM. Let us finally point out that the non-thermal DM relic density can be written in terms of the thermal one as:
\begin{align}
\Omega^{\rm NT}_{\rm DM} h^2 = \sqrt{\frac{g_*(T_f)}{g_*(T_R)}}\left({T_f \over T_R}\right) \Omega^{\rm T}_{\rm DM} h^2\,,
\label{NTdm}
\end{align}
where NT stands for \textit{Non-Thermal}, while T means \textit{Thermal}. For $5$ GeV $< T_f < 80$ GeV (corresponding to $100$ GeV $< m_{\rm DM} < 1.6$ TeV), the top, the higgs, the Z and the W$^\pm$ are not relativistic, giving $g_*(T_f)=86.25$ and:
\begin{align}
\Omega^{\rm NT}_{\rm DM} h^2 = 0.142 \,\sqrt{\frac{10.75}{g_*(T_R)}}\left({m_{\rm DM} \over T_R}\right) \Omega^{\rm T}_{\rm DM} h^2\,.
\label{nonthermaldm}
\end{align}

\item `Branching scenario': in this case the DM abundance is given by the second term on the right-hand side of eq.~\eqref{dmdens} and the DM particles are produced directly from the modulus decay since their residual annihilation is inefficient. In this case both large and small cross sections can satisfy the DM content since the annihilation cross section does not have any impact on the DM content. This scenario is very effective to understand the DM and baryon abundance coincidence problem~\cite{Allahverdi:2010im}. Given that in general we have ${\rm Br}_{\rm DM}\gtrsim 10^{-3}$~\footnote{In this case, the dominant decay mode of moduli is gauge Bosons, ${\phi \rightarrow gg}$. However the gauginos appear in three body final state, i.e., ${\phi\rightarrow g\ \tilde g\tilde g}$  with ${\rm Br}_{\rm DM}\gtrsim 10^{-3}$~\cite{Allahverdi:2010im}.}, in order not to overproduce DM for LSP masses of order hundreds of GeV, one needs $Y_{\rm mod}\lesssim 10^{-9}$. This condition requires a very low reheating temperature:
\begin{align}
T_R\lesssim 10^{-9} m_{\rm mod} = 10^{-9}\kappa^{-1} M_{\rm soft}\,.
\end{align}
For $M_{\rm soft}\sim \mc{O}(1)$ TeV and $\kappa\sim\mc{O}(10^{-2}-10^{-4})$ we find $T_R\lesssim \mc{O}(10)$ MeV. In order to obtain such a low reheating temperature one has in general to consider models where the modulus coupling to visible sector fields is loop suppressed~\cite{Allahverdi:2013noa}. However in this case it is very challenging to avoid a large modulus branching ratio into hidden sector light fields like stringy axions~\cite{Cicoli:2015bpq, Cicoli:2012aq, Higaki:2012ar} and so typically dark radiation is overproduced~\cite{Allahverdi:2014ppa}. Therefore the `Branching scenario' does not seem very promising from the phenomenological point of view.
\een

\subsection{Collider and CMB Constraints}

\begin{figure}
\centering
\subfloat[]
{\includegraphics[width=.45\textwidth]{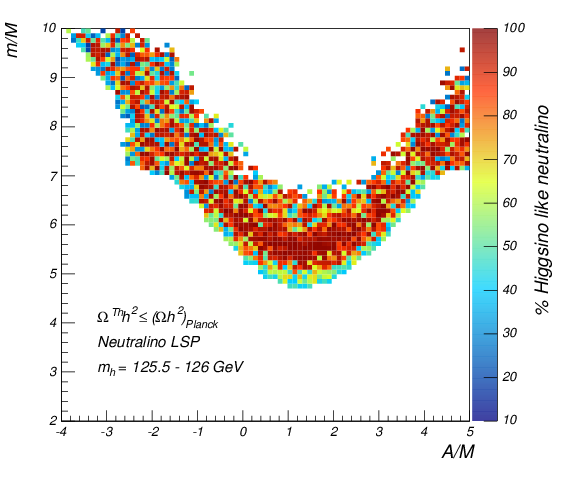}} \quad
\subfloat[]
{\includegraphics[width=.45\textwidth]{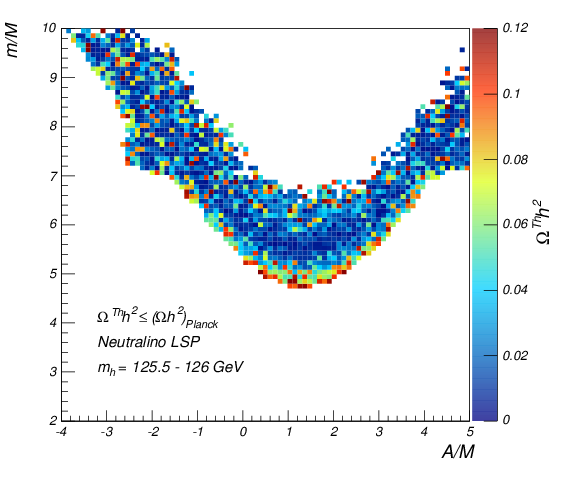}} 
\caption{Correlation between $a=m/M$ and $b=A/M$ for different LSP compositions (left) and DM thermal relic densities (right) in the region where the LSP is at least 10\% higgsino.}
\label{Fig3}
\end{figure}

\begin{figure}
\centering
\subfloat[]
{\includegraphics[width=.45\textwidth]{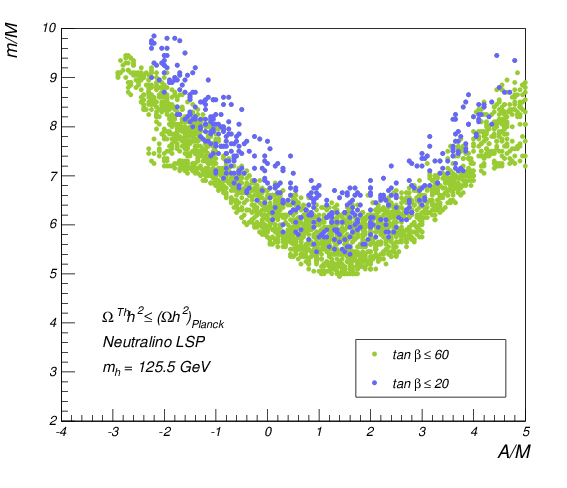}} \quad
\subfloat[]
{\includegraphics[width=.45\textwidth]{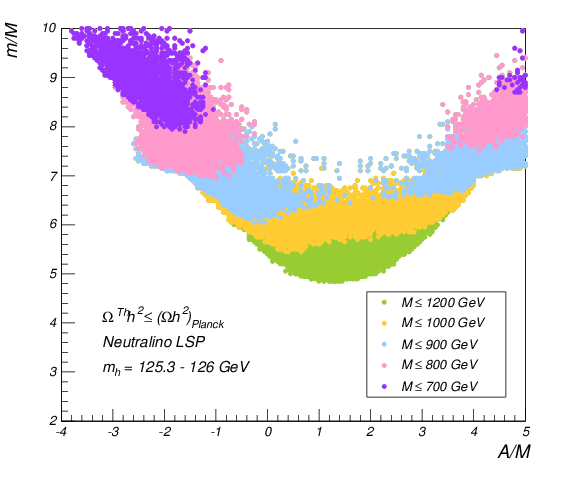}} 
\caption{Correlation between $a=m/M$ and $b=A/M$ for different values of $\tan\beta$ (left) and gaugino mass (right) in the region where the LSP is at least 10\% higgsino.}
\label{Fig4}
\end{figure}

\begin{figure}
\centering
\subfloat[]
{\includegraphics[width=.45\textwidth]{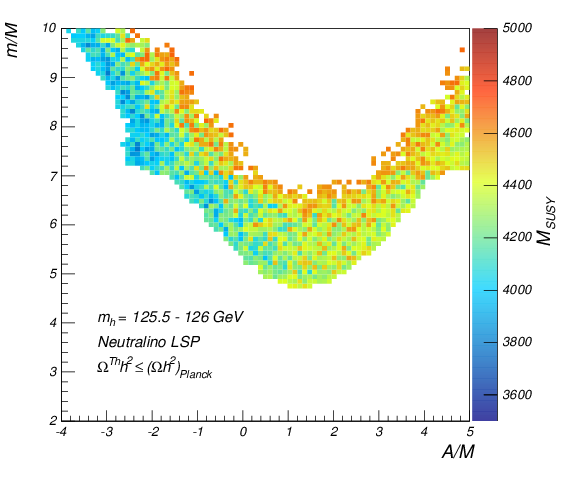}} \quad
\subfloat[]
{\includegraphics[width=.45\textwidth]{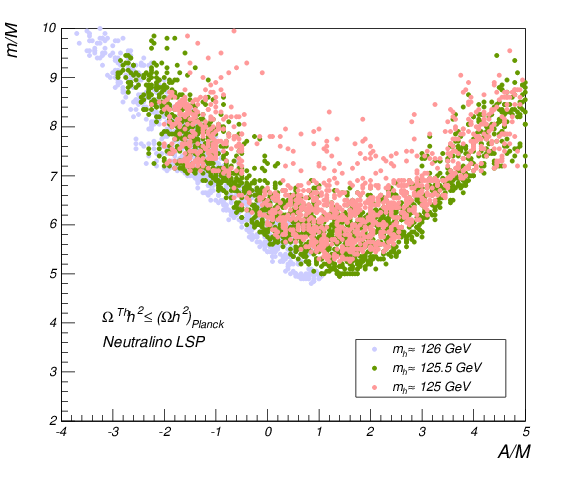}} 
\caption{Correlation between $a=m/M$ and $b=A/M$ for different values of the averaged stop mass $M_{\rm SUSY}$ (left) and the higgs mass (right) in the region where the LSP is at least 10\% higgsino.}
\label{Fig5}
\end{figure}

Due to the considerations mentioned above, if the LSP is bino-like we generically get DM overproduction also in the non-thermal case. The CMSSM boundary condition requires bino to be the lightest among the gauginos. We shall therefore look for particular regions in the non-thermal CMSSM parameter space where the LSP has a non-negligible higgsino component.  We have developed a Monte Carlo programme to find the regions of this parameter space where the LSP is higgsino-like, the higgs mass is around $125.5$ GeV,\footnote{Both ATLAS and CMS give values of the higgs mass between $125$ and $126$ GeV. In what follows,  we will consider ranges of values in this region, allowing to some extent for the uncertainty in the spectrum generators as well. Allowing for a larger uncertainty in the higgs mass does not alter the following results qualitatively.} REWSB takes place correctly and the following phenomenological constraints are satisfied: 
\bi
\item LEP~\cite{LEP1, LEP2, LEP3, LEP4} and LHC~\cite{Aad:2014wea, Aad:2012hm, Chatrchyan:2012lia, Aad:2014vma, Khachatryan:2014qwa} constraints on neutralino and chargino direct production: $m_{\rm DM} \gtrsim 100$ GeV;
\item LHC~\cite{Aad:2014wea, Aad:2012hm, Chatrchyan:2012lia, Aad:2014vma, Khachatryan:2014qwa} bounds on gluino and squark masses: $m_{\tilde{g}} \gtrsim 1300$ GeV;
\item LHC constraints from flavour physics: BR$(B_s \rightarrow \mu^+ \mu^-)$~\cite{CMS:2014xfa} and the constraint on BR$(b\rightarrow s\gamma)$~\cite{Agashe:2014kda};
\item Planck data on DM relic density~\cite{Ade:2015xua}.
\ei
To avoid complications with the applicability of the standard \texttt{Spheno} version, we restrict our scan on the following parameter ranges: $\tan\beta=1$ to $55,$ $a=0\,{\rm to}\,10,$ $b=-5\,{\rm to}\,5,$ and the universal gaugino mass at the high scale $M=0.3-3$~TeV. The results are shown in Figs.~\ref{Fig3}-\ref{Fig5} for positive $\mu$ (the LSP relic density has been calculated using \verb"micrOMEGAS" v3~\cite{Belanger:2013oya}).\\

The plots in Fig.~\ref{Fig3} show the points surviving the above constraints in the $A/M$-$m/M$ plane (at the GUT scale). The points fit into a V-shaped band illustrating a slight hierarchy between scalar and gaugino masses ($m\gtrsim 5 M$) and values of $A$ almost symmetric around $A\simeq M$. The regions shown in the plot are mostly for $T_R$ rather smaller than $T_f$ which keeps mostly the focus point regions in the allowed parameter space. The coannihilation and A-funnel regions can also contribute to the allowed parameter space but they are very fine tuned. The V-shape of our plots is caused by the focus point region which can be obtained by setting $\mu\sim m_Z$ in the EWSB condition with loop corrections. In fact, the dependence of $\mu^2$ on $A$, $M$ and $m$ arises through $m_{H_u}^2$ which depends on the UV soft-terms in the following way: $M^2(f(Q) + g(Q) A/M+ h(Q) (A/M)^2 + e(Q) (m/M)^2)$, where $f$, $g$, $h$ and $e$ depend on dimensionless gauge and Yukawa couplings ($e$ also includes the tadpole correction from the stop loop)  and $Q$ is the SUSY-breaking scale~\cite{Feng:1999zg}. A leading order cancellation in this expression, as needed to achieve a small $\mu$-term in eq. \eqref{resb1}, gives a V-shaped band in the $A/M$-$m/M$ plane. We also apply the higgs mass constraint in this parameter space which depends on the square of $X_t\equiv A_t-\mu \cot\beta$~\cite{Carena:1995bx} and $X_t^2$ preserves the V-shape due to its dependence on $(A/M)^2$.

To illustrate the allowed parameter range and to illustrate different aspects of the surviving points, we show the following plots (for positive sign of $\mu$):
\begin{enumerate}
\item In Fig.~\ref{Fig3}, the color codes illustrate the percentage of higgsino-like neutralino on the left plot and the neutralino contribution to the thermal DM relic density on the right. Note that in most of the points the neutralinos contribute only a small percentage of the total DM relic density and other DM candidates, such as axions, have to be present. The thermal DM relic density is close to the observed Planck value only in a small region corresponding to an LSP that is approximately 50\% bino and 50\% higgsino. However, as we will see in the next section, after imposing indirect detection constraints from Fermi, the only region which survives is the one where thermal DM is under-abundant (by about 10\% of the observed relic density). On the other hand, as we shall see in the next section, non-thermal higgsino-like DM can lead to larger relic densities which can saturate the Planck value for reheating temperatures around $2$ - $3$ GeV.

\item The color codes in Fig.~\ref{Fig4} illustrate the dependence on $\tan\beta$ on the left plot and different values of gaugino masses on the right plot with well-defined domains for different ranges of gaugino masses inside the V-shaped band. Note that $\tan\beta$ tends to have larger values as expected from the fit of $m_h$ and the parameter ranges in this scan. Smaller values of $A/M$ and $m/M$ are preferred for larger gaugino mass due to RG flow of masses to fit the experimental value of $m_h$.

\item For Fig.~\ref{Fig5} the colors illustrate on the left the different values of the typical scale of SUSY particles $M_{\rm SUSY}$, defined here as the averaged stop mass $M_{\rm SUSY}^2=m_{\tilde{t}_1} m_{\tilde{t}_2}$. Notice that $M_{\rm SUSY}$ is around $4$ - $5$ TeV. In principle, we could explore values larger than $5$ TeV however it would bring us beyond the level of applicability of the spectrum generator \verb"SPheno" we have been using which assume similar values for all soft-terms. An analysis for a Split-SUSY-like case with larger differences between sfermions and gaugino masses would be required in that case but this goes beyond the scope of the analysis of this chapter. The colors on the right plot illustrate the dependence on the higgs mass for which we have taken $m_h=125, 125.5, 126$ GeV respectively. Note that for $m_h=126$ GeV there are allowed points only on the left of the V-shaped band because of the above mentioned cut-off on $M_{\rm SUSY}$. Generally speaking we see that by allowing a larger 
range for the higgs mass, we widen the V-shaped region. 
\end{enumerate}

\subsection{Direct and Indirect Detection Constraints}

In the figures above we have set $\mu>0$ but their pattern does not change for $\mu<0$. The next step is to impose the following phenomenological constraints for the separate case of positive and negative $\mu$ since the DM direct detection cross section depends on sign$(\mu)$:
\ben
\item Fermi bounds on DM indirect detection~\cite{Ackermann:2015tah};
\item IceCube~\cite{Aartsen:2012kia} and XENON100~\cite{Orrigo:2015cha} bounds for spin-dependent DM direct detection;
\item LUX~\cite{Akerib:2013tjd}, CDMS~\cite{Agnese:2013rvf} bounds on spin-independent DM direct detection.
\een

\begin{itemize}
\item \textbf{Results for positive} $\mathbf{\mu}$.\\
If we impose the above constraints, indirect detection bounds turn out to be very severe. Using the new Fermi bounds (pass 8 limit) coming from data collected until 2014 (the pass 7 limit includes only data until 2012), we do not find any allowed point for $T_R\lesssim 2$ GeV.

\begin{figure}
\centering
\subfloat[]
{\includegraphics[width=.45\textwidth]{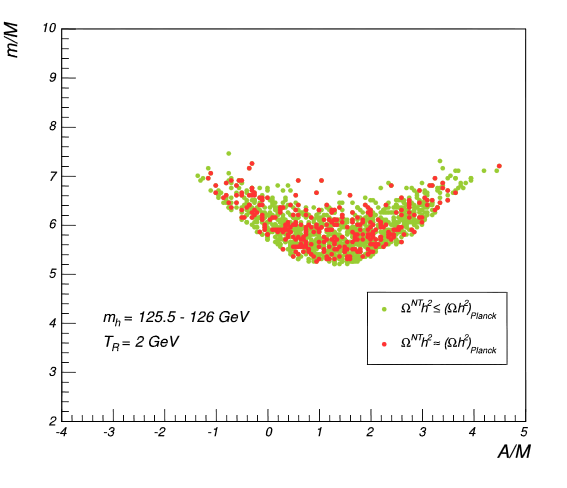}} \quad
\subfloat[]
{\includegraphics[width=.45\textwidth]{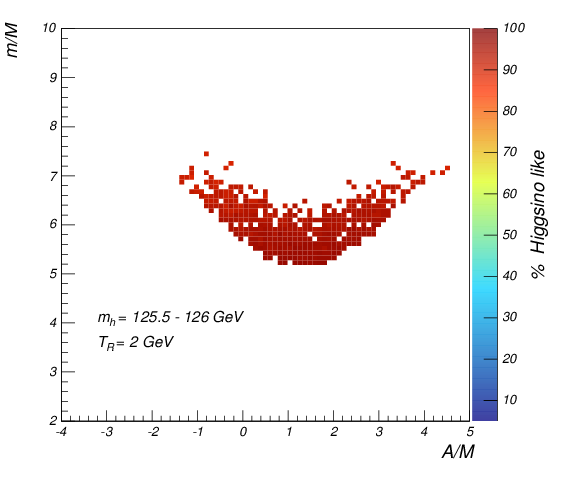}} \\
\subfloat[]
{\includegraphics[width=.45\textwidth]{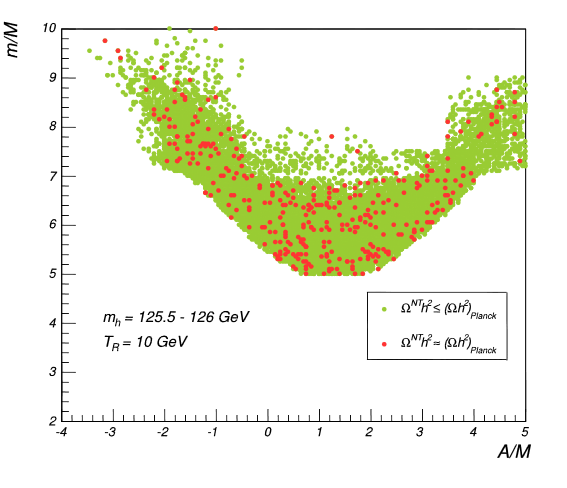}} \quad
\subfloat[]
{\includegraphics[width=.45\textwidth]{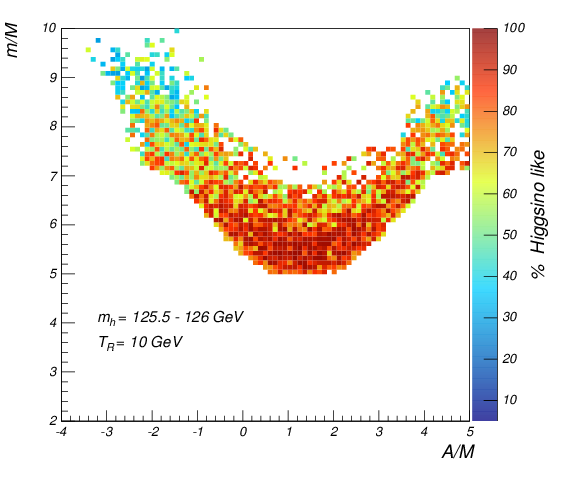}}
\caption{Case with $\mu>0$: $m/M$ vs $A/M$ after imposing LEP, LHC, Planck and Fermi constraints (left)
and corresponding LSP composition (right) for $m_h = 125.5$ - $126$ GeV and $T_R = 2, 10$ GeV.}
\label{Fig6}
\end{figure}

In Fig.~\ref{Fig6} we show the results for different reheating temperatures. The red points show the parameter space where we saturate the DM content measured from Planck~\cite{Ade:2015xua}. We find more allowed points for $T_R=10$ GeV compared to $T_R=2$ GeV since the ratio of $T_f/T_R$ becomes smaller for large $T_R$, and so a smaller annihilation cross section is needed, resulting in a better chance to satisfy the bounds from Fermi. We only see the large higgsino dominated regions for smaller $T_R$ since in this case a larger annihilation cross section is needed to saturate the DM content. For larger $T_R$, an LSP with a smaller higgsino component becomes allowed (higher bino component) and this region appears for smaller values of gaugino mass. We finally mention that the Planck constraints on indirect detection through DM annihilation during the recombination epoch are less stringent than those of Fermi for our scenarios.

Fig.~\ref{Fig7} shows the spin-independent and spin-dependent WIMP-nucleon cross section after imposing LEP, LHC, Planck and Fermi constraints for $T_R = 10$ GeV. IceCube bounds rule out the orange and red regions of the right-hand side plot of Fig.~\ref{Fig7}.

\begin{figure}
\centering
\subfloat[]
{\includegraphics[width=.45\textwidth]{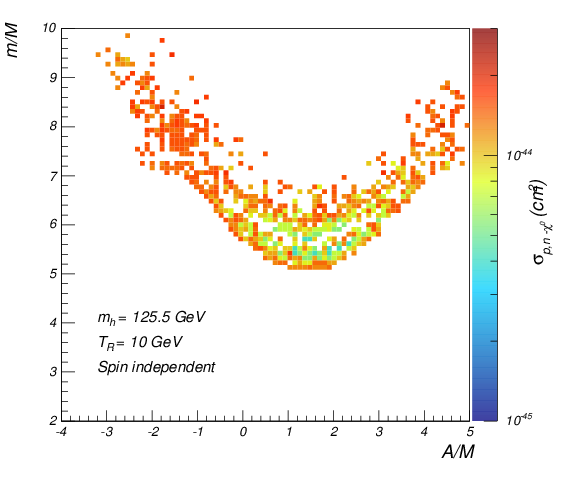}} \quad
\subfloat[]
{\includegraphics[width=.45\textwidth]{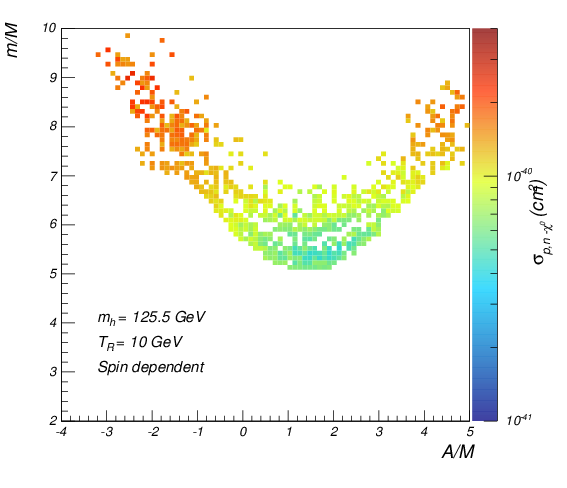}} 
\caption{Case with $\mu>0$: correlation between $a=m/M$ and $b=A/M$ after imposing LEP, LHC, Planck and Fermi (pass 8 limit) constraints with the corresponding spin-independent (left) and spin-dependent (right) WIMP-nucleon cross section for $m_h = 125.5$ GeV and $T_R = 10$ GeV.}
\label{Fig7}
\end{figure}

Fig.~\ref{Fig9} shows the inclusion of LUX bounds on the spin-independent direct detection constraints which rule out most of the points in Fig.~\ref{Fig7} apart from a region corresponding to LSP masses around $300$ GeV which is at the border of detectability.

\begin{figure}
\centering
\subfloat[]
{\includegraphics[width=.45\textwidth]{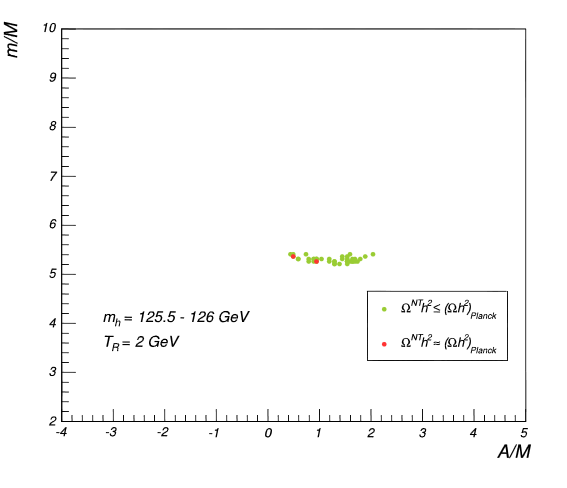}} \quad
\subfloat[]
{\includegraphics[width=.45\textwidth]{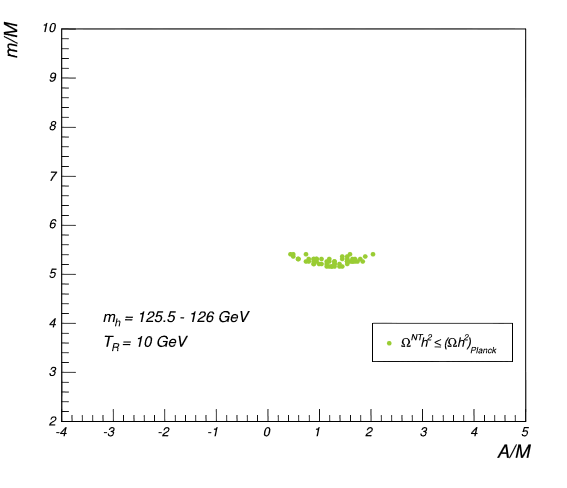}} 
\caption{Case with $\mu>0$: correlation between $a=m/M$ and $b=A/M$ after imposing LEP, LHC, Planck, Fermi and LUX bounds for $m_h = 125.5$ - $126$ GeV and $T_R = 2, 10$ GeV.}
\label{Fig9}
\end{figure}

Moreover, for $T_R=2$ GeV there are red points which saturate the observed DM content. In this case the neutralinos are becoming more pure higgsinos in order to enhance the annihilation cross section and the Fermi constraint becomes harder to avoid but there are still regions allowed by both direct and indirect detection searches. There are more green points for $T_R=10$ GeV since the annihilation cross section becomes smaller due to bino mixing which means a larger allowed region after using Fermi data but the constraint from the direct detection becomes more stringent (due to bino-higgsino mixing in the LSP) and so there are no red points which saturate the observed DM content.\footnote{The direct detection exclusion however depends on various uncertainties, e.g.~strange quark content of proton, form factor etc.~\cite{Accomando:1999eg}.} For the points shown in Fig.~\ref{Fig9}, the GUT values $c= B/M$ and $d=\mu/M$ are around $0.6$ and $1$ respectively.

\newpage
\item \textbf{Results for negative} $\mathbf{\mu}$.

The results for the negative $\mu$ case are shown in Figs.~\ref{Fig10} and~\ref{Fig11}.

\begin{figure}
\centering
\subfloat[]
{\includegraphics[width=.45\textwidth]{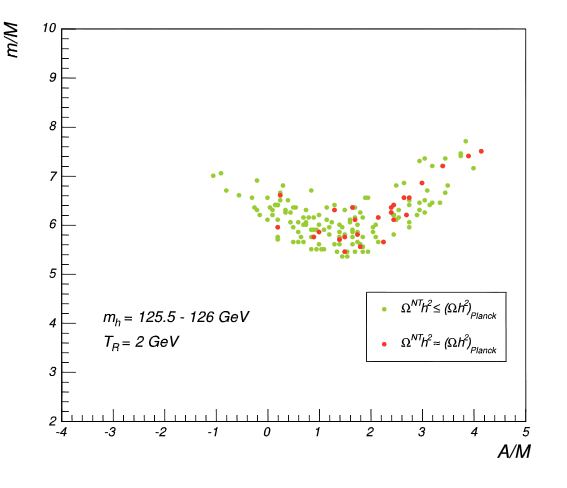}} \quad
\subfloat[]
{\includegraphics[width=.45\textwidth]{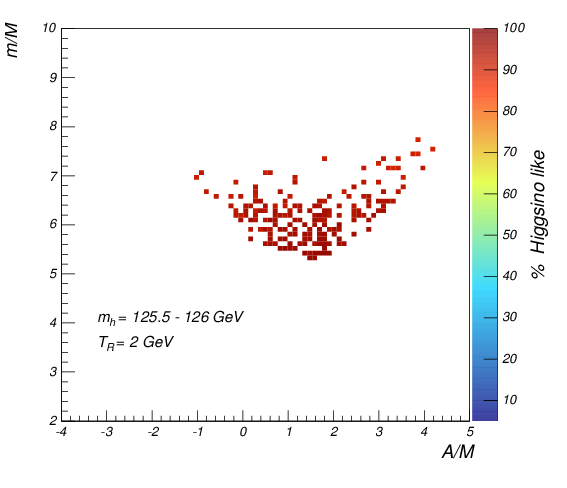}} \\
\subfloat[]
{\includegraphics[width=.45\textwidth]{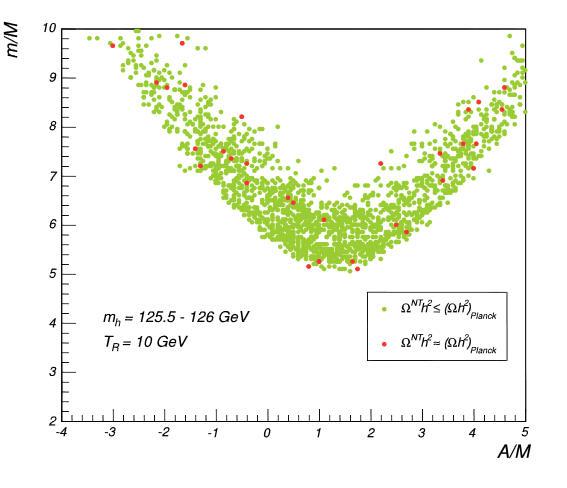}} \quad
\subfloat[]
{\includegraphics[width=.45\textwidth]{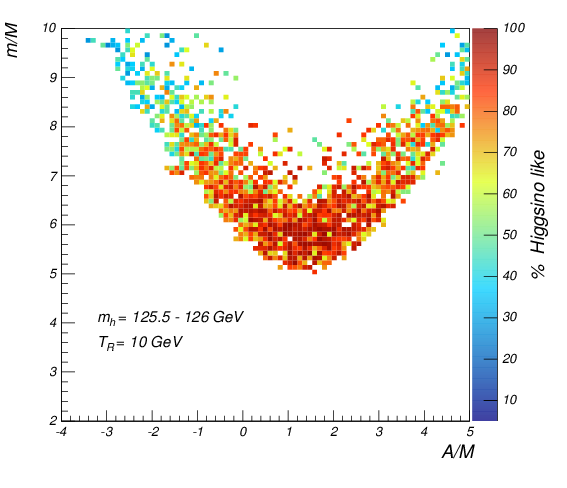}}
\caption{Case with $\mu<0$: $a$ versus $b$ after imposing LEP, LHC, Planck and Fermi (pass 8 limit) bounds (left) and LSP composition (right) for $m_h = 125.5$ - $126$ GeV and $T_R = 2, 10$ GeV.}
\label{Fig10}
\end{figure}

Note that under the same conditions the allowed parameter space for $\mu<0$ is larger than the one for $\mu>0$. This can be understood as follows~\cite{Arnowitt:2001yh, Ellis:2000jd}: $\sigma_{\tilde\chi^0_1-p}$ is dominated by the t-channel $h$, $H$ exchange diagrams which mostly arise from down type (s-quark) interaction:
\begin{align}
A^d\propto m_d\left({\frac{\cos\alpha}{\cos\beta}}{\frac{F_H}{m_H^2}}-{\frac{\sin\alpha}{\cos\beta}}{\frac{F_h}{m_h^2}}\right)\,,
\end{align}
where $\alpha$ is the higgs mixing angle, $F_h=(N_{12}-N_{11}\tan\theta_W)(N_{14}\cos\alpha+N_{13}\sin\alpha)$ and $F_H=(N_{12}-N_{11}\tan\theta_W)(N_{14}\sin\alpha-N_{13}\cos\alpha)$ using $\tilde\chi^0_1=N_{11}\tilde B+N_{12}\tilde W+N_{13}\tilde H_1+N_{14}\tilde H_2$. For $\mu<0$, the ratio $N_{14}/N_{13}$ is positive and this amplitude can become small due to cancellations if:
\begin{align}
\frac{N_{14}}{N_{13}}=- {\frac{\tan\alpha+m_h^2/m_H^2 \cot\alpha}{1+m_h^2/m_H^2}}\,,
\end{align}
is satisfied (for $\tan\alpha<0$). In Fig.~\ref{Fig11} there are more allowed points compared to Fig.~\ref{Fig9} even if there are still no points which saturate the observed DM content for $T_R=10$ GeV due to stringent direct detection constraints. For the points shown in Fig.~\ref{Fig11}, the GUT scale values of $B$ and $\mu$ are still both of order $M$.

\begin{figure}
\centering
\subfloat[]
{\includegraphics[width=.45\textwidth]{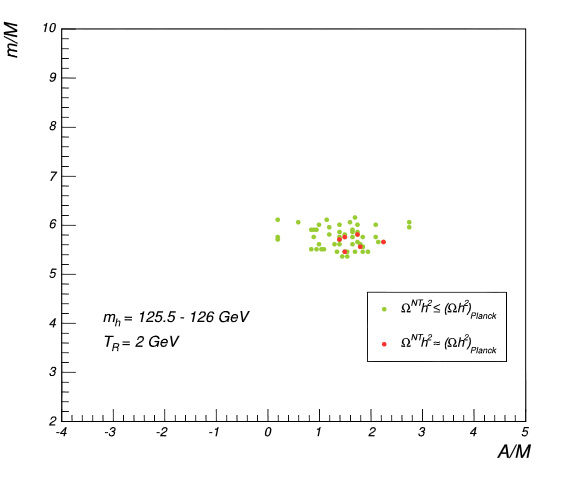}} \quad
\subfloat[]
{\includegraphics[width=.45\textwidth]{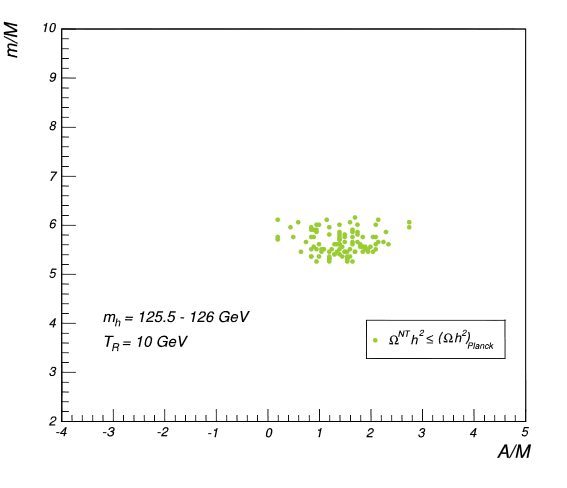}} 
\caption{Case with $\mu<0$: correlation between $a=m/M$ and $b=A/M$ after imposing LEP, LHC, Planck, Fermi (pass 8 limit)
and LUX for $m_h = 125.5$ - $126$ GeV and $T_R = 2, 10$ GeV.}
\label{Fig11}
\end{figure}
\end{itemize}

\section{Discussion of Results}\vspace{-0.1cm}
\label{ssec:DiscussionResults}

\subsection{Analysis of the Allowed Parameter Space}\vspace{-0.1cm}
We are now in a position to put all our results together and explain the effect of each experimental bound on our parameter space. In the end we shall analyze the spectrum of superpartners that appears for the points surviving all the phenomenological constraints. All the observables analyzed in this section have been computed numerically using \verb"micrOMEGAS" v3~\cite{Belanger:2013oya} linked to \verb"SPheno" v3.3.3~\cite{Porod:2003um, Porod:2011nf}.\\

\begin{figure}
\centering
\subfloat[]
{\includegraphics[width=.45\textwidth]{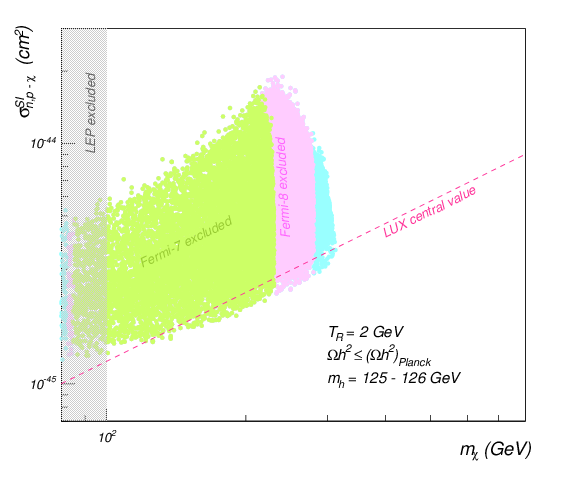}} \quad
\subfloat[]
{\includegraphics[width=.45\textwidth]{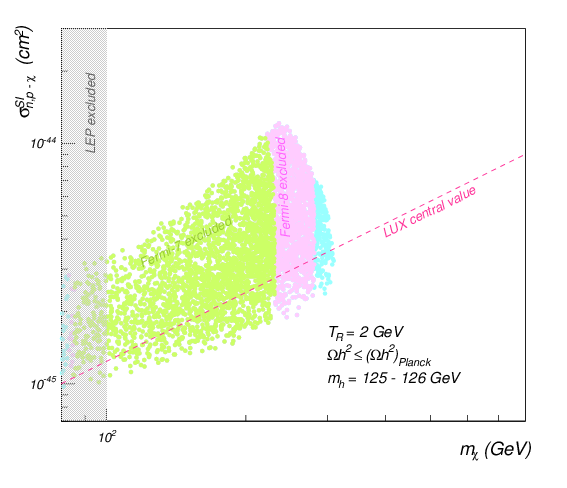}} \\
\subfloat[]
{\includegraphics[width=.45\textwidth]{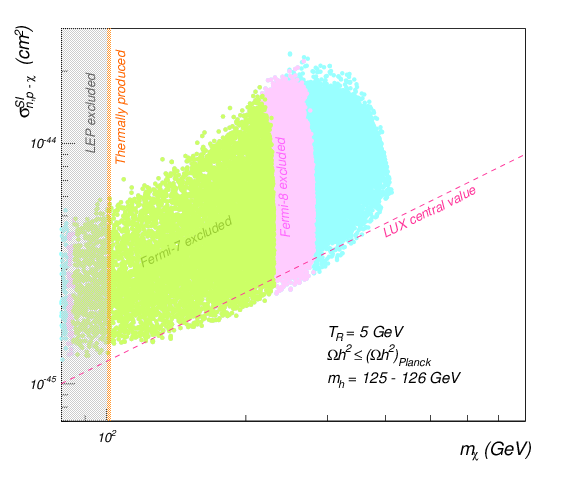}} \quad
\subfloat[]
{\includegraphics[width=.45\textwidth]{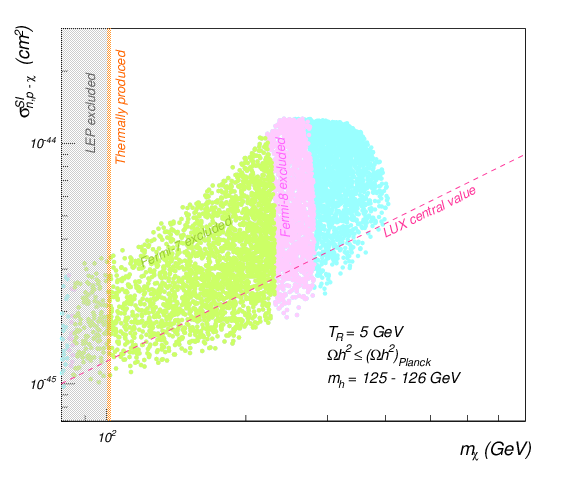}} \\
\subfloat[]
{\includegraphics[width=.45\textwidth]{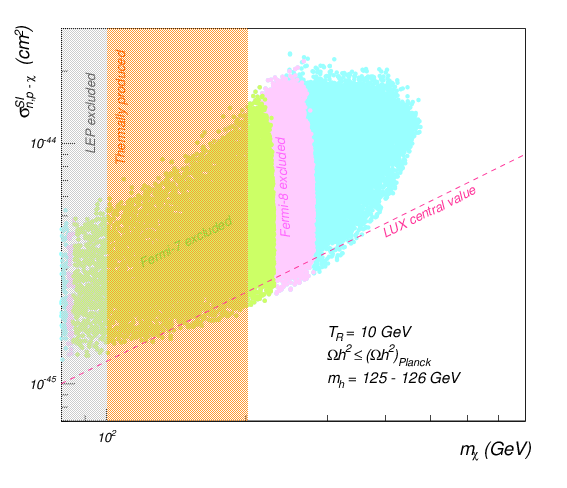}} \quad
\subfloat[]
{\includegraphics[width=.45\textwidth]{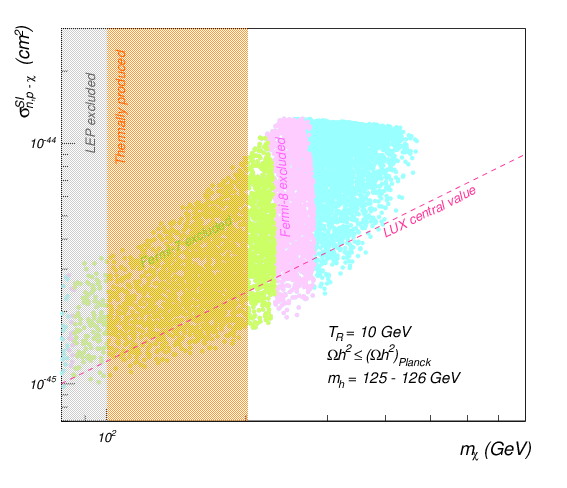}}
\caption{Spin-independent cross section versus LSP mass for $\mu>0$ (left) and $\mu<0$ (right). The light blue points are not ruled out by indirect detection experiments. We show $m_{\chi}$ up to 800 GeV.}
\label{Fig12}
\end{figure}

Fig.~\ref{Fig12} shows the relation between the spin-independent WIMP-nucleon cross section and the LSP mass. Depending on $T_R$ there is a different upper bound for neutralino masses which is given by the Planck constraint on DM. For larger values of $T_R$, the non-thermal relic density decreases, and so heavier neutralinos can pass the Planck constraint on the DM relic density. On the other hand, for larger $T_R$ the parameter space for standard thermal DM (orange band) becomes also larger. Note that LUX 2013 results exclude at 90\% most of the parameter space and the next round of results (LUX 300 days) will be able to probe the remaining regions (the light blue points below the LUX line).\\

In the scenario we considered, the gaugino masses are unified at the GUT scale and therefore the evolution of electroweakinos is totally dominated by the RG flow. This implies that the LSP can only be higgsino- or bino-like (or a mixed combination of them). The largest contributions to the thermal averaged annihilation rates are given by (see for example~\cite{ArkaniHamed:2006mb} and references therein):
\begin{align}
\langle \sigma_{\rm eff} v \rangle = \frac{g_2^4}{512 \pi \mu^2}\left(21 + 3 \tan^2\theta_W +  11 \tan^4\theta_W \right)\,,
\label{sigmamu}
\end{align}
for higgsino-like neutralinos (in the limit $M_W \ll \mu$) annihilating into vector bosons through chargino or neutralino interchange, and:
\begin{align}
\langle  \sigma_{\rm eff} v \rangle =  \sum_f \frac{g_2^4 \tan^2 \theta_W \left( T_{3_f} - Q_f \right)^4 r (1+r^2)}{2\pi m_{\tilde f}^2(1+r)^4}\,,
\label{sigmab}
\end{align}
for bino-like LSP annihilation into fermion-antifermion ($T_{3_f}$ and $Q_f$ are the third component of isospin and the fermion charge and $r = M_1^2 / m_{\tilde f}^2$). This process is driven at tree level by the t-channel exchange of a slepton ${\tilde f}$. In the case where the LSP is a mixed composition of higgsino and bino, the expression of the annihilation rate is an interpolation between (\ref{sigmamu}) and (\ref{sigmab}).
Fig.~\ref{Fig12} shows also the effect of Fermi bounds. As suggested by Fig. \eqref{sigmamu}) and Fig. \eqref{sigmab}, the most constrained regions are those with smaller LSP masses. The grey band corresponds to points excluded by LEP bounds on chargino direct production.\\

\begin{figure}
\centering
\subfloat[]
{\includegraphics[width=.45\textwidth]{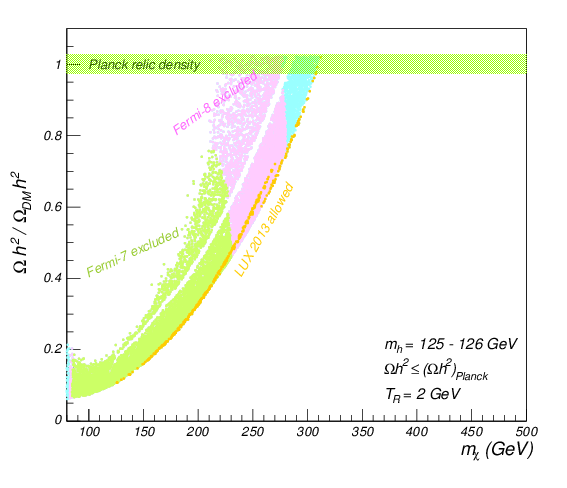}} \quad
\subfloat[]
{\includegraphics[width=.45\textwidth]{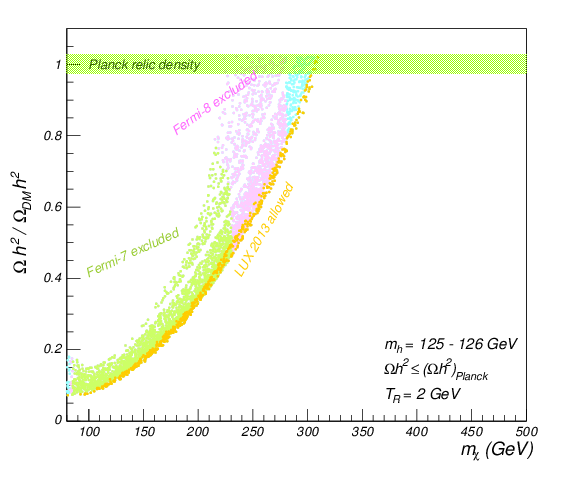}} \\
\subfloat[]
{\includegraphics[width=.45\textwidth]{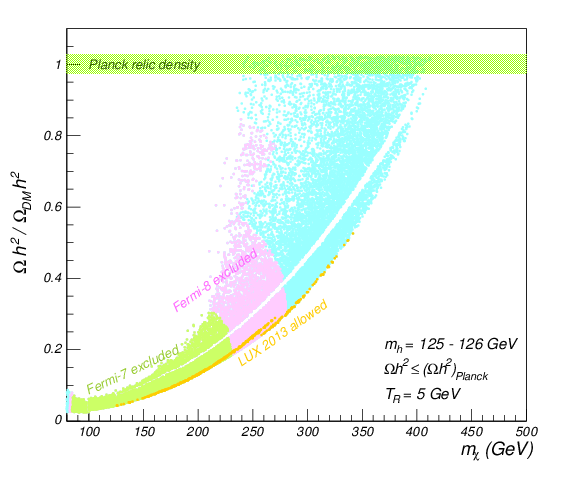}} \quad
\subfloat[]
{\includegraphics[width=.45\textwidth]{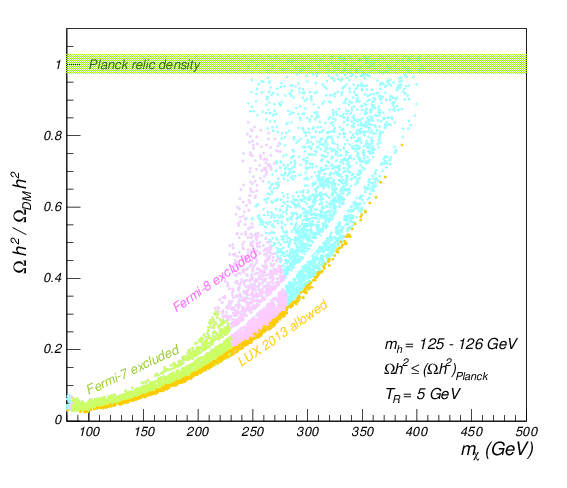}} \\
\subfloat[]
{\includegraphics[width=.45\textwidth]{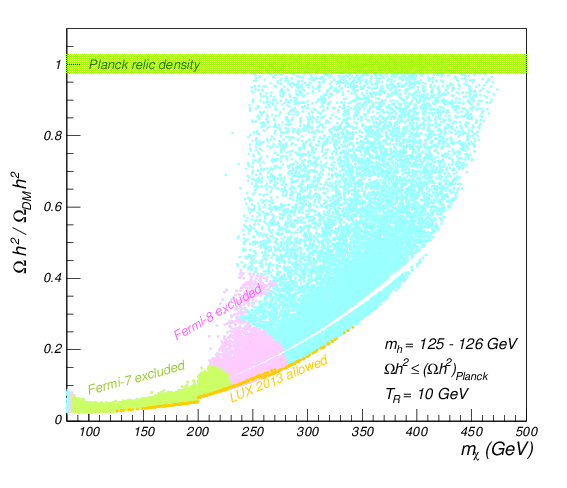}} \quad
\subfloat[]
{\includegraphics[width=.45\textwidth]{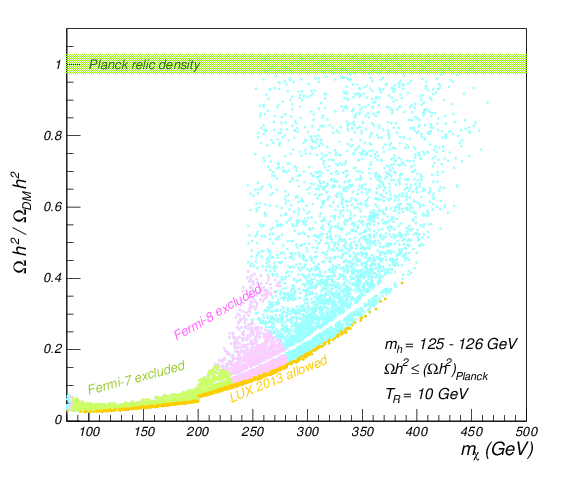}}
\caption{Non-thermal DM abundance predictions versus LSP mass for $\mu>0$ (left) and $\mu<0$ (right) and $T_R=2, 5, 10$ GeV. The light blue points are not ruled out by indirect detection experiments while only the yellow points are allowed by LUX 2013 results.}
\label{Fig13}
\end{figure}

Fig.~\ref{Fig13} shows the amount of non-thermal DM relic density provided by the LSP in terms of its mass, together with the bounds from indirect detection and LUX. The Planck value of the DM content can be saturated in the region which is not ruled out by direct detection bounds only for $T_R = 2$ GeV. Given that for larger $T_R$ the amount of LSP DM gets smaller, the cases with $T_R > 2$ GeV require multi-component DM. Combining Fig.~\ref{Fig12} and Fig.~\ref{Fig13}, we find  that the LUX allowed regions, indirect detection limits and the abundance of LSP DM are correlated for different $T_R$. The allowed regions, where the observed DM content is saturated, depend on $T_R$ but they are generically around $m_{\rm DM} \simeq 300$ GeV.\\

\begin{figure}
\centering
\subfloat[]
{\includegraphics[width=.45\textwidth]{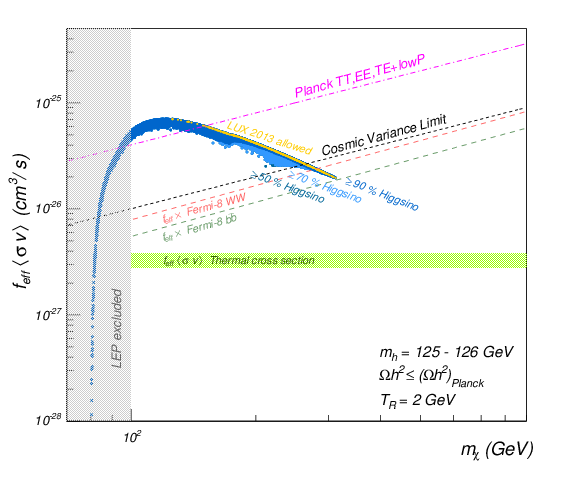}} \quad
\subfloat[]
{\includegraphics[width=.45\textwidth]{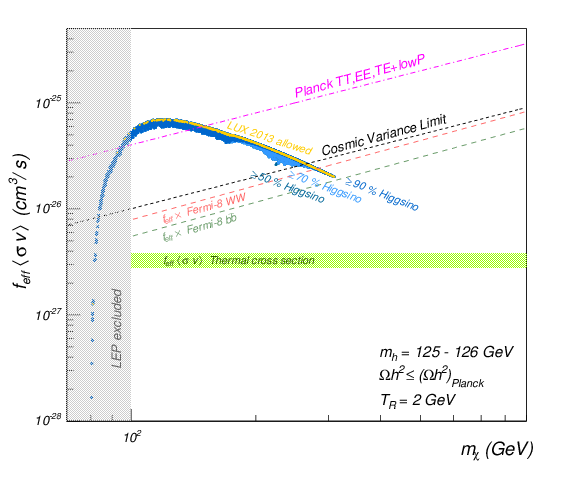}} \\
\subfloat[]
{\includegraphics[width=.45\textwidth]{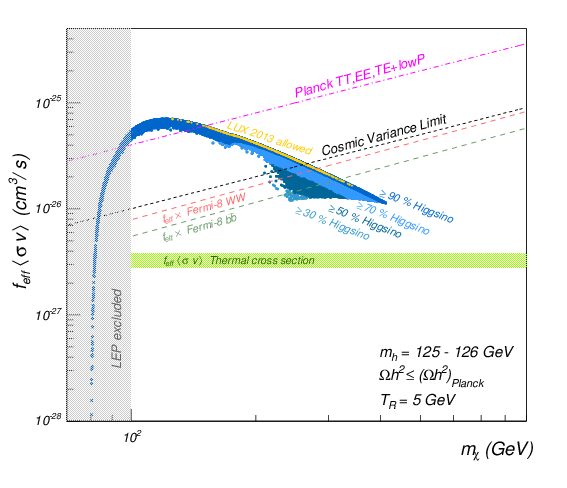}} \quad
\subfloat[]
{\includegraphics[width=.45\textwidth]{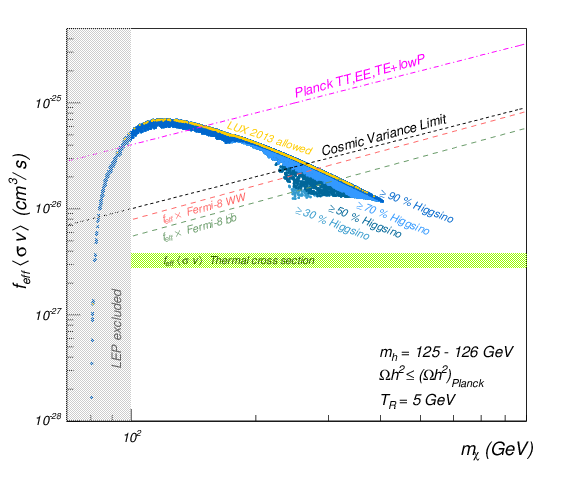}} \\
\subfloat[]
{\includegraphics[width=.45\textwidth]{Images/relicproj10_n.png}} \quad
\subfloat[]
{\includegraphics[width=.45\textwidth]{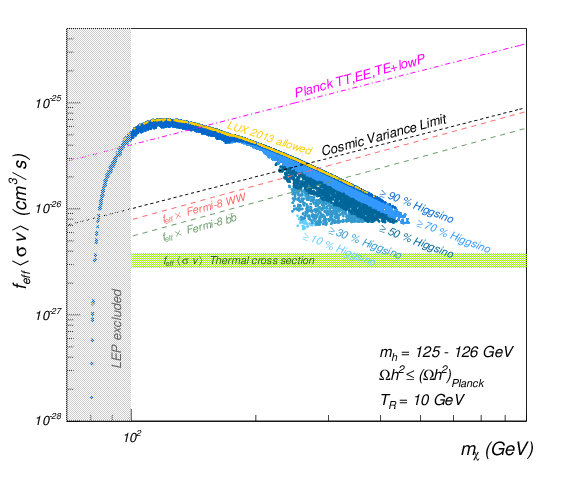}}
\caption{Comparison between detection constraints from Planck, Fermi and LUX for $\mu>0$ (left) and $\mu<0$ (right). We have set $f_{\rm eff}= 0.2$. We show $m_{{\rm DM}}$ up to 900 GeV.}
\label{Fig14}
\end{figure}

In Fig.~\ref{Fig14}, we show the Planck constraints on indirect detection through DM annihilation during the recombination epoch. We use $WW$ final states corresponding to an efficiency factor $f_{\rm eff}= 0.2$~\cite{Madhavacheril:2013cna}. Even going all the way to the cosmic variance bound, these constraints turn out to be less stringent than those coming from Fermi and LUX. On the other hand, Fig.~\ref{Fig14} shows the correlation between the LSP composition and the bounds coming from both direct and indirect detection. Concerning Fermi and Planck limits on DM annihilation, these bounds allow almost all possible combinations of bino/higgsino neutralinos. The restrictions coming from the $f \overline{f}$ and $W\overline{W}$ channels depend on the neutralino composition: for a higgsino-like LSP, the most stringent constraint is due to annihilation into vector bosons, while for a bino-like LSP the main constraint comes from annihilation into a fermion-antifermion pair.\\

The LUX constraints in Fig.~\ref{Fig14} reduce the parameter space to the region where the LSP is mostly higgsino-like. This could be a bit puzzling since the WIMP-nucleon cross section is dominated by the higgs exchange channel:
\begin{align}
\sigma_{\chi-p} \propto \frac{a_{\tilde H}^2 (g' a_{\tilde B} - g \ a_{\tilde W})^2}{m_h^4}\,,
\label{xip}
\end{align}
where $a_{\tilde H}$, $a_{\tilde B}$ and $a_{\tilde W}$ are respectively the higgsino, bino and wino LSP components. According to this expression, the cross section is enhanced when the higgsino component increases. However in Fig.~\ref{Fig14} direct detection bounds allow only points which are mainly higgsino-like. The reason of this effect is in the effective coupling $\tilde{\chi} \tilde{\chi}  h$ which for a bino-like LSP looks like:
\begin{align}
C_{\tilde{\chi}\tilde{\chi} h} \simeq \frac{m_Z \sin \theta_W \tan \theta_W}{M_1^2 -\mu^2}\left( M_1 + \mu \sin 2\beta \right)\,,
\label{Cb}
\end{align}
where for moderate to large $\tan \beta$ the second term is negligible and $\mu > M_1$. Hence this coupling is dominated by $M_1$. On the other hand, the coupling for a higgsino-like LSP is:
\begin{align}
C_{\tilde{\chi}\tilde{\chi} h} \simeq \frac{1}{2}\left( 1 \pm \sin 2\beta  \right)\left( \tan^2 \theta_W\frac{m_Z \cos \theta}{M_1 - |\mu|} +\frac{m_Z \cos \theta}{M_2 - |\mu|} \right)\,,
\label{Cn}
\end{align}
where $\pm$ is for the $H_u$ and $H_d$ components and $\mu < M_1$. Contrary to the bino-like case, this coupling is inversely proportional to $M_1$. Thus the WIMP-nucleon cross section grows in the regions where the LSP is higgsino-like and $M_1$ is small or where the LSP is bino-like and the gaugino mass is large. If we compare Fig.~\ref{Fig4} (right) which shows the distribution of gaugino masses along the V-shaped band, with Fig.~\ref{Fig6} (right) and~\ref{Fig10} (right), we realize that the region with a smaller cross section is the one at the bottom of the V-shaped band where gaugino masses are big and the LSP is very higgsino-like. The region where the LSP is more bino-like has smaller gaugino masses and the cross section is larger. Fermi constraints however become more stringent in the case with more higgsino content due to larger annihilation cross section. The competition between LUX and Fermi constraints produces the allowed parameter space where the Planck value of the DM content is saturated for $T_R=2$ GeV.\\

\begin{figure}
\centering
\subfloat[]
{\includegraphics[width=.45\textwidth]{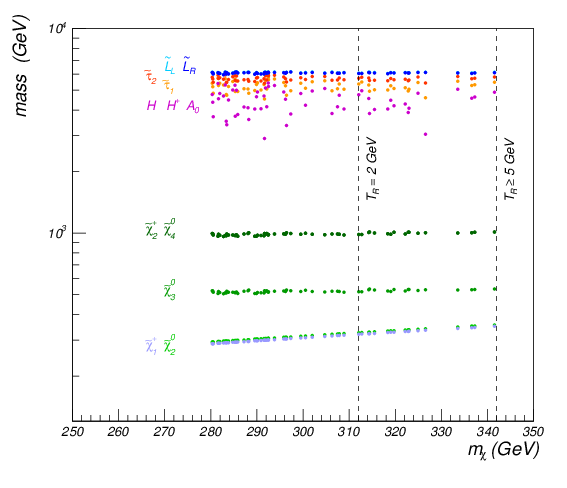}} \quad
\subfloat[]
{\includegraphics[width=.45\textwidth]{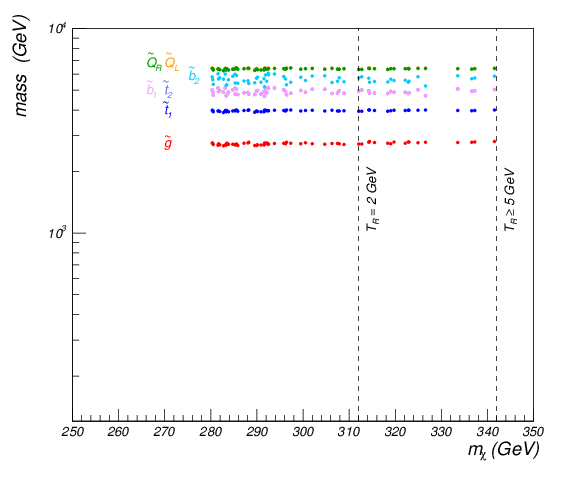}} 
\caption{The mass spectra of superpartners for allowed points shown in Fig.~\ref{Fig9} for different values of $T_R$.}
\label{Fig7a}
\end{figure}

Finally, in Fig.~\ref{Fig14} there is a change of behavior of the thermal averaged cross section for masses around $130$ GeV. The reason is the following: this region is (as can be shown in the plot) higgsino-like, but the masses are closer to $M_W$ and  $\langle \sigma v \rangle$ is no longer described by (\ref{sigmamu}) but by something like (with $x=\mu/m_W$):
\begin{align}
\langle \sigma_{\rm eff} v \rangle \sim \frac{9g^4}{16 \pi m_W^2} \frac{x^2}{(4x^2-1)^2}\,.
\end{align}
In Fig.~\ref{Fig7a}, we show the spectra of SUSY particles for the allowed regions of Fig.~\ref{Fig9} (blue points below the LUX line). We find that sleptons, staus, higgses, all other scalar masses and gluinos are rather heavy since they are between about $2$ and $7$ TeV. The lightest and second to lightest neutralino and the lightest chargino are around $280$-$340$ GeV while all other neutralinos and charginos are heavy. The allowed region for $T_R=2$ GeV is shown on the left side of the vertical line with the label $T_R=2$ GeV where the points situated exactly on the line satisfy all the constraints including the current DM content as measured by Planck. Similarly, the allowed region for $T_R\geq 5$ GeV is shown on the left side of the vertical line with the label $T_R\geq 5$ GeV even if there are no points in this region which saturate the current DM content. Notice that the spectrum is essentially independent of the reheating temperature $T_R$ and the hierarchy between the different sparticles is 
robust.

\subsection{Astrophysical Uncertainties}

The direct detection cross section can involve various uncertainties, e.g.~strange quark content of proton, form factor, local DM density and LSP contribution to the total amount of observed DM abundance. The local density can be 0.1-0.7 GeV/cm$^3$~\cite{Read:2014qva}. There could also be astrophysical uncertainties in the indirect detection results beyond what has been considered so far. Recently, it is  mentioned in~\cite{Baer:2015tva, Cahill-Rowley:2013dpa} that if the thermal neutralinos do not produce the entire amount of cold DM, the  direct and indirect detection cross sections should be reduced by $R$ and $R^2$ respectively with $R\equiv \Omega h^2/0.12$. Possible bounds arising from Fermi are now almost negligible since they are suppressed by $R^2$. Once the suppression factor $R$ is taken into account, Fermi, Planck and other indirect detection experiments have lower impacts. Concerning the effect on LUX and other direct detection bounds, the cross section is now reduced by $R$ which is equivalent to multiplying the effective couplings in eq. \eqref{Cb} and eq. \eqref{Cn} by $\sqrt{R}$. This clearly introduces a new parameter in the discussion performed in the previous section.\\

\begin{figure}
\centering
\subfloat[]
{\includegraphics[width=.45\textwidth]{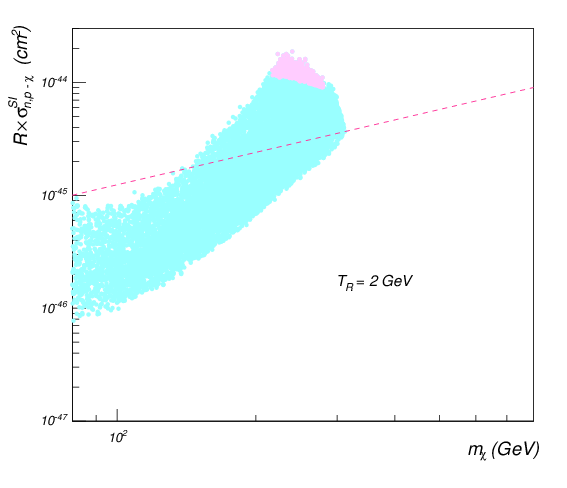}} \quad
\subfloat[]
{\includegraphics[width=.45\textwidth]{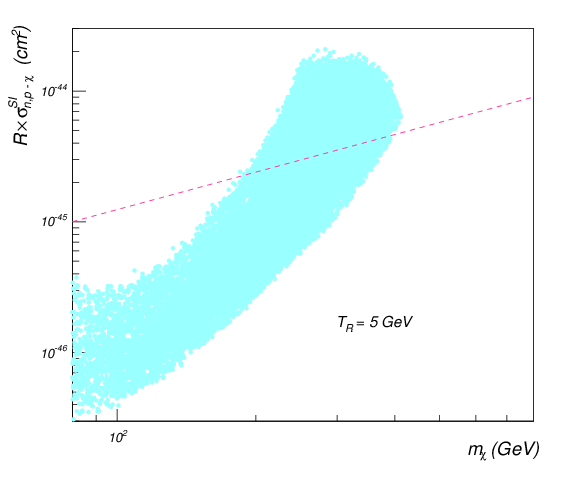}} 
\caption{$R\times\sigma^{SI}_{n,p-\tilde\chi^0_1}$ vs $m_{\tilde\chi^0_1}$ for $T_R=2, 5$ GeV. $R$ is defined in the text. More parameter space is allowed.}
\label{Fig15}
\end{figure}

If we assume such a reduction in the cases where $\Omega^{\rm NT}_{\rm DM} h^2\le 0.12$, more parameter space could be allowed for multi-component DM regions as shown in Fig.~\ref{Fig15}. The pink region is disallowed by Fermi data. The $T_R$ dependence of the region constrained by Fermi in Fig.~\ref{Fig15} is due to the fact that the  factor R is now a function of $T_R$
\begin{align}
R = \frac{\Omega^{\rm NT}_{\rm DM} h^2}{0.12} \simeq \frac{T_f}{T_R} \frac{\Omega^{\rm T}_{\rm DM} h^2}{0.12}\,.
\end{align}
$R$ becomes larger for smaller values of $T_R$ (=2 GeV) and the Fermi constraint becomes important. The region below the dashed line satisfied present LUX limits. In particular this implies that the region with lighter neutralinos is now unconstrained by LUX. This region typically corresponds to more bino component in the LSP as shown in Fig.~\ref{Fig6} (right) and~\ref{Fig10} (right). We have therefore a different situation compared to before, because now neutralinos with a larger bino component are allowed.\\

Let us stress, however, that the prediction for the region where $\Omega^{\rm NT}_{\rm DM} h^2$ saturates the DM content remains unchanged, i.e.~only the case $T_R=2$ GeV contains points which are still allowed by all data and saturate the DM content with an LSP mass around $300$ GeV. This new factor $R$ helps us to extract more parameter space for the multi-component DM scenarios. However, the DM simulations need to establish the validity of  the assumption that proportions of various DM components in the early universe is maintained even after the large scale structures are formed.

\section{Conclusions}

Non-thermal DM scenarios emerge in UV theories like string theory due to the presence of gravitationally coupled scalars which decay at late times when they are dominating the energy density of the universe. In such models the reheating temperature due to moduli decays is typically below the freeze-out temperature, $T_R<T_f$ when assuming an MSSM particle as main DM constituent.\\

In this chapter we have studied the non-thermal version of the CMSSM/mSUGRA and contrasted it with both particle physics and astrophysical experimental constraints. The experimental information available at this moment, including the well known value of the higgs mass, is enough to rule out the vast majority of the non-thermal CMSSM parameter space. There is still a small region which is consistent with all observations and is at the edge of detection by both astrophysics and particle physics experiments, resulting in a very interesting situation for beyond the SM physics.\\

In our determination of the allowed parameter space we have used constraints from collider experiments (LEP, LHC), cosmic microwave background observations (Planck) and direct and indirect DM searches (Fermi, XENON100, IceCube, LUX, CDMS). We have found that the most restrictive constraints come from Fermi and LUX which single out a small region of the non-thermal CMSSM parameter space corresponding to a higgsino-like LSP with a mass around $300$ GeV that can saturate the observed DM abundance for $T_R\simeq 2$~GeV while larger reheating temperatures require additional contributions to the present DM abundance. These results are summarized in Fig.~\ref{Fig13} which shows the comparison between the cases of positive and negative $\mu$.\\

This non-thermal scenario leads to a clear pattern of SUSY particles. In particular, the fact that the LSP is higgsino-like makes the lightest chargino, the lightest neutralino and next to lightest neutralino to be almost degenerate in mass. This kind of scenario can be probed at the LHC using monojet plus soft leptons plus missing energy~\cite{Baer:2014kya, Han:2014kaa}, monojet signal~\cite{Han:2013usa} and two Vector Boson Fusion jets and large missing transverse energy~\cite{Han:2013usa}. On the other hand, all the other superpartners are much heavier and beyond the LHC reach but accessible to potential future experiments such as a $100$ TeV machine.\\

It is worth mentioning that non-thermal CMSSM scenarios with TeV-scale soft-terms and reheating temperatures around $1$ - $10$ GeV can emerge in string models where the visible sector is sequestered from the sources of SUSY-breaking~\cite{Aparicio:2014wxa, Blumenhagen:2009gk}. In a subset of the parameter range these string scenarios lead to $M_{\rm soft}\sim M_{\rm P} \epsilon^2 \ll m_{\rm mod}\sim M_{\rm P} \epsilon^{3/2} \ll M_{\rm GUT} \sim M_{\rm P} \epsilon^{1/3}$ and $T_R\sim M_{\rm P} \epsilon^{9/4}$ where $\epsilon \simeq m_{3/2}/M_{\rm P}\ll 1$. For $\epsilon \sim 10^{-8}$, one obtains TeV-scale soft-terms, $M_{\rm GUT}\sim 10^{16}$ GeV, $T_R\sim 1 -10$ GeV and $m_{\rm mod}\sim 10^6$ GeV for $m_{3/2}\sim 10^{10}$ GeV.\\

We point out that our analysis is based on the CMSSM/mSUGRA for which all superpartners are expected to be at similar masses close to the TeV scale. In this sense we restricted ourselves to scalar masses lighter than $5$ TeV which is the range of validity of the codes we have used to perform our analysis.\\

There are several additional ways to generalize the analisys:
\begin{itemize}
\item Consider non-universal extensions of the CMSSM. Small departures from universality, even though strongly constrained by flavour changing neutral currents, allow more flexibility in the parameter space and will slightly enhance the allowed region.
\item Consider sfermions heavier than $5$ TeV as in the split-SUSY case. This is not only an interesting phenomenological possibility but is also the other class of scenarios that were derived in the string compactifications studied in~\cite{Aparicio:2014wxa, Blumenhagen:2009gk}.
\item Consider MSSM scenarios with R-parity violation and late decaying moduli fields. This class of models has not been studied in detail and given the fact that even with R-parity conservation there seems to be a need for other sources of DM such as axions or axinos, this should be a possibility worth studying.
\item Consider explicit D-branes models which tend to generate models beyond the standard MSSM (see for instance~\cite{Dolan:2011qu, Maharana:2012tu, Krippendorf:2010hj} for such models).
\end{itemize}

It is encouraging that new planned experiments such as upcoming LUX result and XENON1T will be enough to rule out the rest of the allowed parameter space, independent of the upcoming LHC run. Clearly also new LHC runs and future planned colliders will be crucial for this class of models. Combining astrophysical and collider measurements is probably the most efficient way to constrain beyond the SM physics and the analysis performed in this chapter is a clear illustration of this strategy.

\chapter{Dark Radiation in Sequestered Models}
\label{chap:DR}

\section{Introduction}
\label{sec:DRIntroduction}

According to the cosmological Standard Model (SM), neutrinos were in thermal equilibrium at early times and decoupled at temperatures of order $1$ MeV. This decoupling left behind a cosmic neutrino background which has been emitted much earlier than the analogous cosmic microwave background (CMB). Due to the weakness of the weak interactions, this cosmic neutrino background cannot be detected directly, and so goes under the name of `dark radiation'. Its contribution to the total energy density $\rho_{\rm tot}$ is parametrized in terms of the effective number of neutrino-like species $N_{\rm eff}$ as:
\begin{align}
\rho_{\rm tot} = \rho_{\gamma} \left( 1 + \frac{7}{8}\left( \frac{4}{11} \right)^{4/3} N_{\rm eff} \right)\,.
\end{align}
The SM predictions for $N_{\rm eff}$ are $N_{\rm eff}=3$ during Big Bang Nucleosynthesis (BBN) and $N_{\rm eff}=3.046$ at CMB times since neutrinos get slightly reheated when electrons and positrons annihilate. Any departure from these values would be a clear signal of physics beyond the SM due to the presence of extra dark radiation controlled by the parameter $\Delta N_{\rm eff}\equiv N_{\rm eff} - N_{\rm eff, SM}$.\\

Given that $N_{\rm eff}$ is positively correlated with the present value of the Hubble constant $H_0$, the comparison between indirect estimates of $H_0$ from CMB experiments and direct astrophysical measurements of $H_0$ could signal the need for extra dark radiation. The Planck 2013 value of the Hubble constant is $H_0 = (67.3\pm 1.2)$ km s$^{-1}$ Mpc$^{-1}$ ($68\%$ CL)~\cite{Ade:2013zuv} which is in tension at $2.5\sigma$ with the Hubble Space Telescope (HST) value $H_0 = (73.8\pm 2.4)$ km s$^{-1}$ Mpc$^{-1}$ ($68\%$ CL)~\cite{Riess:2011yx}. Hence the Planck 2013 estimate of $N_{\rm eff}$ with this HST `$H_0$ prior' is $N_{\rm eff} = 3.62^{+0.50}_{-0.48}$ ($95\%$ CL) which is more than $2\sigma$ away from the SM value and gives $\Delta N_{\rm eff} \leq 1.07$ at $2\sigma$.\\

However the HST Cepheid data have been reanalyzed by~\cite{Efstathiou:2013via} who found the different value $H_0 = (70.6\pm 3.3)$ km s$^{-1}$ Mpc$^{-1}$ ($68\%$ CL) which is within $1\sigma$ of the Planck 2015 estimate $H_0 = (67.3\pm 1.0)$ km s$^{-1}$ Mpc$^{-1}$ ($68\%$ CL)~\cite{Ade:2015xua}. Hence the Planck 2015 collaboration performed a new estimate of $N_{\rm eff}$ without using any `$H_0$ prior' and obtaining $N_{\rm eff} = 3.13 \pm 0.32$ ($68\%$ CL)~\cite{Ade:2013zuv} which is perfectly consistent with the SM value and gives $\Delta N_{\rm eff} \leq 0.72$ at around $2\sigma$. This result might seem to imply that extra dark radiation is ruled out but this naive interpretation can be misleading since larger $N_{\rm eff}$ corresponds to larger $H_0$ and there is still an unresolved controversy in the direct measurement of $H_0$. In fact the Planck 2015 paper~\cite{Ade:2015xua} analyzes also the case with the prior $\Delta N_{\rm eff}=0.39$ obtaining the result $H_0 = (70.6\pm 1.0)$ km s$^{-1}$ Mpc$^{-1}$ ($68\%$ CL) which is even in better agreement with the new HST estimate of $H_0$ performed in~\cite{Efstathiou:2013via}. Thus we stress that trustable direct astrophysical measurements of $H_0$ are crucial in order to obtain reliable bounds on $N_{\rm eff}$.\\

$N_{\rm eff}$ is also constrained by measurements of primordial light element abundances. The Planck 2015 estimate of $N_{\rm eff}$ based on the helium primordial abundance and combined with the measurements of~\cite{Aver:2013wba} is $N_{\rm eff}= 3.11^{+0.59}_{-0.57}$ ($95\%$ CL) giving $\Delta N_{\rm eff} \leq 0.65$ at $2\sigma$~\cite{Ade:2015xua}. However measurements of light element abundances are difficult and often affected by systematic errors, and so also in this case there is still some controversy in the literature since~\cite{Izotov:2014fga} reported a larger helium abundance that, in turn, leads to $N_{\rm eff}= 3.58 \pm 0.50$ ($99\%$ CL) which is $3\sigma$ away from the SM value and gives $\Delta N_{\rm eff} \leq 1.03$ at $3\sigma$. Due to all these experimental considerations, in the rest of this chapter we shall consider $\Delta N_{\rm eff}\lesssim 1$ as a reference upper bound for the presence of extra dark radiation.\\

Extra neutrino-like species can be produced in any beyond the SM theory which features hidden sectors with new relativistic degrees of freedom (\textit{dof}). In particular, extra dark radiation is naturally generated when reheating is driven by the decay of a gauge singlet since in this case there is no a priori good reason to suppress the branching ratio into hidden sector light particles~\cite{Cicoli:2012aq, Higaki:2012ar, Higaki:2013lra}. \\

This situation is reproduced in string models of the early universe due to the presence of gravitationally coupled moduli which get displaced from their minimum during inflation, start oscillating when the Hubble scale reaches their mass, quickly come to dominate the energy density of the universe since they redshift as matter and finally reheat the universe when they decay~\cite{Acharya:2008bk, Acharya:2009zt, Acharya:2010af, Allahverdi:2013noa}. In the presence of many moduli, the crucial one is the lightest since its decay dilutes any previous relic produced by the decay of heavier moduli.\\

Two important cosmological constraints have to be taken into account: (i) the lightest modulus has to decay before BBN in order to preserve the successful BBN predictions for the light element abundances~\cite{Coughlan:1983ci, Banks:1993en, deCarlos:1993wie}; (ii) the modulus decay to gravitinos should be suppressed in order to avoid problems of DM overproduction because of gravitinos annihilation or modifications of BBN predictions~\cite{Endo:2006zj, Nakamura:2006uc}. The first constraint sets a lower bound on the lightest modulus mass of order $m_{\rm mod} \gtrsim 100$ TeV, while a straightforward way to satisfy the second constraint is $m_{\rm mod}< 2 m_{3/2}$.\\

However in general in string compactifications the moduli develop a mass because of SUSY-breaking effects which make the gravitino massive via the super higgs mechanism and generate also soft-terms of order $M_{\rm soft}$. Because of their common origin, one has therefore $m_{\rm mod}\sim m_{3/2}\sim M_{\rm soft}$. The cosmological lower bound $m_{\rm mod}\gtrsim 100$ TeV then pushes the soft-terms well above the TeV-scale ruining the solution of the hierarchy problem based on low-energy SUSY. An intriguing way-out is given by Type IIB string compactifications where the visible sector is constructed via fractional D3-branes at singularities~\cite{Aldazabal:2000sa, Conlon:2008wa, Cicoli:2012vw, Cicoli:2013cha}. In this case the blow-up modulus resolving the singularity is fixed at zero size in a supersymmetric manner, resulting in the absence of local SUSY-breaking effects. SUSY is instead broken by bulk moduli far away from the visible sector singularity. Because of this geometric separation, the visible sector is said to be `sequestered' since the soft-terms can be suppressed with respect to the gravitino mass by $\epsilon = \frac{m_{3/2}}{M_{\rm P}}\ll 1$~\cite{Blumenhagen:2009gk}.\\

A concrete example of sequestered SUSY-breaking is given by the Type IIB LARGE Volume Scenario (LVS) with D3-branes at singularities, which is characterized by the following hierarchy of masses~\cite{Blumenhagen:2009gk}:
\begin{align}
M_{1/2}\sim m_{3/2} \epsilon \ll m_{\rm mod} \sim m_{3/2} \sqrt{\epsilon} \ll m_{3/2}\,.
\label{masses}
\end{align}
This mass spectrum guarantees the absence of moduli decays to gravitinos and allows for gaugino masses $M_{1/2}$ around the TeV-scale for $m_{\rm mod}\sim 10^7$ GeV and $m_{3/2}\sim 10^{10}$ GeV. On the other hand, SUSY scalar masses $m_0$ are more model-dependent since their exact $\epsilon$-dependence is determined by the form of the K\"ahler metric for visible sector matter fields and the mechanism responsible for achieving a dS vacuum. The general analysis of Chap. \ref{chap:Soft-Terms} ~\cite{Aparicio:2014wxa} found two possible $\epsilon$-scalings for scalar masses: (i) $m_0\sim M_{1/2}$ corresponding to a typical MSSM-like scenario and (ii) $m_0 \sim m_{\rm mod} \gg M_{1/2}$ resulting in a Split-SUSY-like case with heavy squarks and sleptons.\\

Following the cosmological evolution of these scenarios, reheating takes place due to the decay of the volume modulus which produces, together with visible sector particles, also hidden sector \textit{dof} which could behave as extra dark radiation~\cite{Cicoli:2012aq, Higaki:2012ar}. Some hidden sector \textit{dof} are model-dependent whereas others, like bulk closed string axions, are always present, and so give a non-zero contribution to $\Delta N_{\rm eff}$. In fact, as shown in~\cite{Allahverdi:2014ppa}, the production of axionic dark radiation is unavoidable in any string model where reheating is driven by the moduli decay and some of the moduli are stabilized by perturbative effects which keep the corresponding axions light. Note that light closed string axions can be removed from the low-energy spectrum via the St\"uckelberg mechanism only for cycles collapsed to zero size since in the case of cycles in the geometric regime the combination of axions eaten up by an anomalous $U(1)$ is mostly given by open string axions~\cite{Allahverdi:2014ppa}.\\

R-parity odd visible sector particles produced from the lightest modulus decay subsequently decay to the lightest SUSY particle, which is one of the main DM candidates. Due to their common origin, axionic dark radiation and neutralino DM have an interesting correlation~\cite{Allahverdi:2014ppa}. In fact, by combining present bounds on $N_{\rm eff}$ with lower bounds on the reheating temperature $T_{\rm rh}$ as a function of the DM mass $m_\DM$ from recent Fermi data, one can set interesting constraints on the $(N_{\rm eff}, m_\DM)$-plane.~\cite{Allahverdi:2014ppa} found that standard thermal DM is allowed only if $\Delta N_{\rm eff}\to 0$ while the vast majority of the allowed parameter space requires non-thermal scenarios with higgsino-like DM, in agreement with the results of~\cite{Aparicio:2015sda} for the MSSM-like case.\\

Dark radiation production for the MSSM-like case has been studied in~\cite{Cicoli:2012aq, Higaki:2012ar} which showed that the leading decay channels of the volume modulus are to visible sector higgses via the Giudice-Masiero (GM) term and to ultra-light bulk closed string axions. The simplest model with two higgs doublets and a shift-symmetric higgs sector yields $1.53 \lesssim \Delta N_{\rm eff}\lesssim 1.60$, where the window has been obtained by varying the reheating temperature between $500$ MeV and $5$ GeV, which are typical values for gravitationally coupled scalars with masses in the range $m_{\rm mod}\simeq (1 \div 5) \cdot 10^7$ GeV. These values of $\Delta N_{\rm eff}$ lead to dark radiation overproduction since they are in tension with current observational bounds.\footnote{Radiative corrections to the modulus coupling to higgs fields do not give rise to a significant change in the final prediction for $\Delta N_{\rm eff}$~\cite{Angus:2013zfa}.} Possible way-outs to reduce $\Delta N_{\rm eff}$ involve models with either a larger GM coupling or more than two higgs doublets.\\

Due to this tension with dark radiation overproduction, different models have been studied in the literature.~\cite{Angus:2014bia} showed how sequestered LVS models where the CY volume is controlled by more than one divisor are ruled out since they predict huge values of extra dark radiation of order $\Delta N_{\rm eff}\sim 10^4$. On the other hand,~\cite{Hebecker:2014gka} focused on non-sequestered LVS models where the visible sector is realized via D7-branes wrapping the large cycle controlling the CY volume.\footnote{Another option involves flavour D7-branes wrapped around the volume divisor and intersecting the visible sector D3-branes localized at a singularity~\cite{Hebecker:2014gka}.} In this way, the decay rate of the lightest modulus to visible sector gauge bosons becomes comparable to the decay to bulk axions, and so the prediction for $\Delta N_{\rm eff}$ can become smaller. In fact, the simplest model with a shift-symmetric higgs sector yields $\Delta N_{\rm eff}\simeq 0.5$~\cite{Hebecker:2014gka}. However this case necessarily requires high-scale SUSY since without sequestering $M_{\rm soft}\sim m_{3/2}$ (up to loop factors), and so from~\eqref{masses} we see that the cosmological bound $m_{\rm mod}\sim m_{3/2} \sqrt{\epsilon}\gtrsim 100$ TeV implies $M_{\rm soft} \sim m_{3/2} \gtrsim \left(100\,{\rm TeV}\right)^{2/3} M_{\rm P}^{1/3}\sim 3 \times 10^9$ GeV. Moreover in this case the visible sector gauge coupling is set by the CY volume $\vo$, $\alpha_\SM^{-1} \sim \vo^{2/3} \sim 25$, and so it is hard to achieve large values of $\vo$ without introducing a severe fine-tuning of some underlying parameters. A possible way-out could be to consider anisotropic compactifications where the CY volume is controlled by a large divisor and a small cycle which supports the visible sector~\cite{Cicoli:2011qg, Cicoli:2011yy, Angus:2012dd}.\\

In this chapter we take instead a different point of view and keep focusing on sequestered models of Chap.~\ref{chap:Soft-Terms}~\cite{Cicoli:2012aq, Higaki:2012ar}, since they are particularly promising for phenomenological applications: they are compatible with TeV-scale SUSY and gauge coupling unification without suffering from any cosmological moduli and gravitino problem, they can be embedded in globally consistent CY compactifications~\cite{Cicoli:2012vw, Cicoli:2013cha} and allow for successful inflationary models~\cite{Conlon:2005jm, Burgess:2013sla, Cicoli:2011zz} and neutralino non-thermal DM phenomenology~\cite{Aparicio:2015sda}. Following the general analysis of SUSY-breaking and its mediation to the visible sector performed in Chap.~\ref{chap:Soft-Terms}~\cite{Aparicio:2014wxa} for sequestered Type IIB LVS models with D3-branes at singularities, we focus on the Split-SUSY case where squarks and sleptons acquire a mass of order the lightest modulus mass: $m_0 = c\, m_{\rm mod}$ with $c\sim \mc{O}(1)$. We compute the exact value of the coefficient $c$ for different Split-SUSY cases depending on the form of the K\"ahler metric for visible sector matter fields and the mechanism responsible for achieving a dS vacuum. We find that the condition $c\leq 1/2$, which allows the new decay channel to SUSY scalars, can be satisfied only by including string loop corrections to the K\"ahler potential~\cite{Cicoli:2007xp, Berg:2007wt}. However this relation holds only at the string scale $M_s \sim 10^{15}$ GeV whereas the decay of the lightest modulus takes place at an energy of order its mass $m_{\rm mod}\sim 10^7$ GeV. Hence we consider the Renormalization Group (RG) running of the SUSY scalar masses from $M_s$ to $m_{\rm mod}$ and then compare their value to $m_{\rm mod}$ whose running is in practice negligible since moduli have only gravitational couplings. Given that also the RG running of SUSY scalar masses is a negligible effect in Split-SUSY-like models, we find that radiative corrections do not alter the parameter space region where the lightest modulus decay to SUSY scalars opens up.\\

We then compute the new predictions for $\Delta N_{\rm eff}$ which gets considerably reduced with respect to the MSSM-like case considered in~\cite{Cicoli:2012aq, Higaki:2012ar} since the branching ratio to visible sector particles increases due to the new decay to squarks and sleptons and the new contribution to the decay to higgses from their mass term. We find that the simplest model with a shift-symmetric higgs sector can suppress $\Delta N_{\rm eff}$ to values as small as $0.14$ in perfect agreement with current experimental bounds. Depending on the exact value of $m_0$ all values in the range $0.14 \lesssim \Delta N_{\rm eff} \lesssim 1.6$ are allowed. Interestingly $\Delta N_{\rm eff}$ can be within the allowed experimental window also in the case of vanishing GM coupling $Z = 0$ since the main suppression of $\Delta N_{\rm eff}$ comes from the lightest modulus decay to squarks and sleptons. Given that a correct realization of radiative EWSB in Split-SUSY-like models requires in general a large $\mu$-term of order $m_0$, the lightest modulus branching ratio into visible sector \textit{dof} is also slightly increased due to its decay to higgsinos. However this new decay channel yields just a negligible correction to the final prediction for dark radiation production.\\

In Sec.~\ref{drs} we analyze the predictions for axionic dark radiation. We present our conclusions in Sec.~\ref{Concl}. This chapter is based on~\cite{Cicoli:2015bpq}.

\section{Dark Radiation in Sequestered Models}
\label{drs}

As already argued in Sec. \ref{sec:DRIntroduction}, the production of dark radiation is a generic feature of string models where some of the moduli are fixed by perturbative effects~\cite{Allahverdi:2014ppa}. In fact, if perturbative corrections fix the real part of the modulus $T=\tau + {\rm i}\psi$, the axion $\psi$ remains exactly massless at this level of approximation due to its shift-symmetry, leading to $m_\tau \gg m_\psi$. Hence very light relativistic axions can be produced by the decay of $\tau$, giving rise to $\Delta N_{\rm eff} \neq 0$. 

\subsection{Dark Radiation from Moduli Decays}

Following the cosmological evolution of the Universe, during inflation the canonically normalized modulus $\Phi$ gets a displacement from its late-time minimum of order $M_{\rm P}$. After the end of inflation the value of the Hubble parameter $H$ decreases. When $H \sim m_\Phi$, $\Phi$ starts oscillating around its minimum and stores energy. During this stage $\Phi$ redshifts as matter, so that it quickly comes to dominate the energy density of the Universe. Afterwards reheating is caused by the decay of $\Phi$ which takes place when:
\begin{align}
3 H^2 \simeq \frac43 \,\Gamma_\Phi^2\,,
\label{HG}
\end{align}
where $\Gamma_\Phi$ is the total decay rate into visible and hidden \textit{dof}:
\begin{align}
\Gamma_\Phi = \Gamma_{\rm vis} + \Gamma_{\rm hid} = \left(c_{\rm vis} + c_{\rm vis}\right) \Gamma_0\,,\qquad\text{with} \qquad \Gamma_0~\equiv \frac{1}{48\pi}\frac{m_\Phi^3}{M_{\rm P}^2}\,.
\label{G}
\end{align}
The corresponding reheating temperature is given by:
\begin{align} 
T_{\rm rh} = \left(\frac{30\,\rho_{\rm vis}}{\pi^2 g_*(T_{\rm rh})}\right)^{\frac 14}\,,
\label{Trh}
\end{align}
where $\rho_{\rm vis} = \left(c_{\rm vis}/c_{\rm tot}\right) 3 H^2 M_{\rm P}^2$ with $c_{\rm tot}=c_{\rm vis} + c_{\rm hid}$. 
Using~\eqref{HG} and~\eqref{G} $T_{\rm rh}$ can be rewritten as:
\begin{align}
T_{\rm rh} \simeq \frac{1}{\pi} \left(\frac{5 c_{\rm vis} c_{\rm tot}}{288 g_*(T_{\rm rh})}\right)^{1/4} m_\Phi\,\sqrt{\frac{m_\Phi}{M_{\rm P}}}\,.
\label{rt}
\end{align}
This reheating temperature has to be larger than about $1$ MeV in order to preserve the successful BBN predictions.\\

In the presence of a non-zero branching ratio for $\Phi$ decays into hidden sector \textit{dof}, i.e. for $c_{\rm hid}\neq 0$, extra axionic dark radiation gets produced, leading to~\cite{Cicoli:2012aq, Higaki:2012ar}:
\begin{align}
\Delta N_{\rm eff} = \frac{43}{7} \frac{c_{\rm hid}}{c_{\rm vis}} \left(\frac{g_*(T_{\rm dec})}{g_*(T_{\rm rh})}\right)^{1/3}\,,
\label{dn}
\end{align}
where $T_{\rm dec} \simeq 1$ MeV is the temperature of the Universe at neutrino decoupling with $g_*(T_{\rm dec}) = 10.75$. 
The factor in brackets is due to the fact that axions are very weakly coupled (they are in practice only gravitationally coupled), and so they never reach thermal equilibrium. Therefore, given that the comoving entropy density $g_*(T) T^3 a^3$ is conserved, the thermal bath gets slightly reheated when some species drop out of thermal equilibrium. Note that the observational reference bound $\Delta N_{\rm eff}\lesssim 1$ implies:
\begin{align}
c_{\rm vis}\gtrsim 3\, c_{\rm hid}\qquad\text{for}\qquad T_{\rm rh}\gtrsim 0.2\,\text{GeV}\,,
\label{cbound}
\end{align}
where we have used the fact that $g_*(T_{\rm rh}) = 75.75$ in the window $0.2\,{\rm GeV} \lesssim T_{\rm rh} \lesssim 0.7\, {\rm GeV}$ while $g_*(T_{\rm rh}) = 86.25$ for $T_{\rm rh} \gtrsim 0.7\,\rm GeV$.

\subsection{Light Relativistic Axions in LVS Models}

Let us summarize the main reasons why axionic dark radiation production is a typical feature of sequestered LVS models:
\bi
\item Reheating is driven by the last modulus to decay which is $\tau_b$ since the moduli mass spectrum takes the form (the axion $\psi_\SM$ is eaten up by an anomalous $U(1)$):
\begin{align}
m_{\tau_b}\sim m_{3/2} \sqrt{\epsilon} \ll m_{\tau_s} \sim m_{\psi_s} \sim m_S \sim m_U \sim m_{3/2} \ll m_{\tau_\SM} \sim \frac{m_{3/2}}{\sqrt{\epsilon}}\sim M_s\,,
\end{align}
where $\epsilon = m_{3/2}/M_{\rm P} \sim W_0 / \vo \ll 1$. Given that gaugino masses scale as $M_{1/2}\sim m_{3/2} \epsilon$, TeV-scale SUSY fixes $m_{\tau_b}$ around $10^7$ GeV which in turn, using~\eqref{rt}, gives $T_{\rm rh}$ around $1$ GeV.\footnote{As in standard Split-SUSY models, we require TeV-scale gauginos for DM and gauge coupling unification. In MSSM-like models we focus on low-energy SUSY to address the hierarchy problem.} Note that $m_{\tau_b}\ll m_{3/2}$, and so sequestering addresses the gravitino problem since the decay of the volume modulus into gravitinos is kinematically forbidden.

\item Given that axions enjoy shift symmetries which are broken only by non-perturbative effects, the axionic partner $\psi_b$ of the volume mode $\tau_b$ is stabilized by non-perturbative contributions to the superpotential of the form $W \supset A_b\, e^{- a_b T_b} \sim e^{-\vo^{2/3}}\ll 1$. These tiny effects give rise to a vanishingly small mass $m_{\psi_b}^2 \sim e^{- \vo^{2/3}} \sim 0$. Hence these bulk closed string axions are in practice massless and can be produced from the decay of $\tau_b$~\cite{Cicoli:2012aq, Higaki:2012ar}. 

\item Some closed string axions can be removed from the 4D spectrum via the St\"uckelberg mechanism in the process of anomaly cancellation. However, the combination of bulk axions eaten up by an anomalous $U(1)$ is mostly given by an open string mode, and so $\psi_b$ survives in the low-energy theory (the situation is opposite for axions at local singularities)~\cite{Allahverdi:2014ppa}. 
\ei

\subsection{Volume Modulus Decay Channels}

The aim of this section is to compute the ratio $c_{\rm hid}/c_{\rm vis}$ which is needed to predict the effective number of extra neutrino-like species $\Delta N_{\rm eff}$ using (\ref{dn}).

\subsubsection{Decays into Hidden Sector Fields}
\label{vmds}

Some hidden sector \textit{dof} are model-dependent whereas others are generic features of LVS models. As pointed out above, bulk closed string axions are always a source of dark radiation. On top of them, there are local closed string axions which however tend to be eaten up by anomalous $U(1)$s (this is always the case for each del Pezzo singularity) and local open string axions (one of them could be the QCD axion~\cite{Cicoli:2012vw, Cicoli:2013cha}) whose production from $\tau_b$ decay is negligibly small~\cite{Cicoli:2012aq}. Moreover the decay of $\tau_b$ into bulk closed string $U(1)$s is also a sub-dominant effect~\cite{Cicoli:2012aq}. Model-dependent decay channels involve light \textit{dof} living on hidden D7-branes wrapping either $D_b$ or $D_s$ and hidden D3-branes at singularities which are geometrically separated from the one where the visible sector is localized. However, as explained in~\cite{Cicoli:2012aq}, the only decay channels which are not volume or loop suppressed are to light gauge bosons on the large cycle and to higgses living on sequestered D3s different from the visible sector. Given that the presence of these states is non-generic and can be avoided by suitable hidden sector model building, we shall focus here just on $\tau_b$ decays into 
bulk closed string axions.\\

The corresponding decay rate takes the form~\cite{Cicoli:2012aq,Higaki:2012ar}: 
\begin{align}
\Gamma_{\Phi \rightarrow a a} = \Gamma_0\qquad\Rightarrow\qquad c_{\rm hid}=1\,,
\label{ddr}
\end{align}
where $\Phi$ and $a$ are, respectively, the canonically normalized real and imaginary parts of the big modulus $T_b$.
This result can be derived from the tree-level K\"ahler potential:
\begin{align}
K \simeq - 3 \ln\left(\frac{T_b+\overline{T}_b}{2}\right)\,,
\end{align}
which gives a kinetic lagrangian of the form:
\begin{align}
\mc{L}_{\rm kin} = \frac{3}{4 \tau_b^2} \left(\partial_\mu \tau_b \partial^\mu \tau_b + \partial_\mu \psi_b \partial^\mu \psi_b\right)\,.
\label{lag1}
\end{align}
After canonical normalization of $\tau_b$ and $\psi_b$:
\begin{align}
\label{cn1}
\frac{\Phi}{M_{\rm P}} = \sqrt{\frac{3}{2}} \ln \tau_b\,, \qquad \qquad \frac{a}{M_{\rm P}} = \sqrt{\frac{3}{2}} \frac{\psi_b}{\langle \tau_b \rangle}\,,
\end{align}
and expanding $\Phi$ as $\Phi = \Phi_0 + \hat \Phi$, the kinetic lagrangian~\eqref{lag1} can be rewritten as:
\begin{align}
\mc{L}_{\rm kin} = \frac 12 \partial_\mu \hat\Phi \partial^\mu \hat\Phi + \frac 12 \partial_\mu a \partial^\mu a - \sqrt{\frac 23 } \frac{\hat \Phi}{M_{\rm P}} \partial_\mu a \partial^\mu a\,,
\end{align}
which encodes the coupling of the volume modulus to its axionic partner. Integrating by parts and using the equation of motion 
$\Box \hat \Phi = - m_\Phi^2 \hat \Phi$ we obtain the coupling:
\begin{align}
\mathcal{L}_{\Phi a a} = \frac{1}{\sqrt{6}}\frac{m_\Phi^2}{M_{\rm P}} \hat\Phi a a\,,
\label{lphiaa}
\end{align}
which yields the decay rate~\eqref{ddr}.

\subsubsection{Decays into Visible Sector Fields}

The dominant volume modulus decays into visible sector \textit{dof} are to higgses via the GM coupling $Z$. Additional leading order decay channels can be to SUSY scalars and higgsinos depending respectively on $m_0$ and $\hat\mu$. On the other hand, as explained in~\cite{Cicoli:2012aq,Higaki:2012ar}, $\tau_b$ decays into visible gauge bosons are loop suppressed, i.e. $c_{\Phi\to A A}\sim \alpha_\SM^2 \ll 1$, whereas decays into matter fermions and gauginos are chirality suppressed, i.e. $c_{\Phi\to f f}\sim \left(m_f/m_\Phi\right)^2 \ll 1$. The main goal of this section is to compute the cubic interaction lagrangian which gives rise to the decay of the volume modulus into higgses, higgsinos, squarks and sleptons.

\subsubsection*{Decay into Scalar Fields}

Let us first focus on the volume modulus decays into visible scalar fields which are induced by the $\tau_b$-dependence of both kinetic and mass terms in the total effective lagrangian $\mc{L} = \mc{L}_{\rm kin} - V$. $\mc{L}_{\rm kin}$ is determined by the leading order K\"ahler potential:
\begin{align}
K \simeq - 3 \ln\left(\frac{T_b+\overline{T}_b}{2}\right) + \frac{2}{T_b + \overline{T}_b} \left[f_\alpha(U,S) \overline{C}^{\overline{\alpha}} C^\alpha + \left(Z H_u H_d + \rm h.c.\right)\right]\,,
\end{align}
where we included only the leading term of the K\"ahler matter metric $\tilde{K}_\alpha$ in~\eqref{mattermetric}. 
Writing each complex scalar field as $C^\alpha = \frac{\rm{Re} C^\alpha + i \rm{Im} C^\alpha}{\sqrt{2}}$, 
the canonically normalized real scalar fields look like:
\begin{align}
h_1 = \lambda_u {\rm Re} H_u^+ \qquad h_2 = \lambda_d {\rm Re} H_d^-
\qquad h_3 = \lambda_d {\rm Re} H_d^0 \qquad h_4 = \lambda_u {\rm Re} H_u^0  \nn \\
h_5 = \lambda_d {\rm Im} H_d^0  \qquad h_6 = \lambda_u {\rm Im} H_u^0 
\qquad h_7 = \lambda_u {\rm Im} H_u^+  \qquad h_8 = \lambda_d {\rm Im} H_d^- \nn \\
\sigma_\alpha = \lambda_\alpha {\rm Re} C_\alpha \qquad \chi_\alpha = \lambda_\alpha {\rm Im} C_\alpha
\qquad\text{where}\qquad \lambda_i\equiv \sqrt{\frac{f_i(U,S)}{\langle \tau_b\rangle}}\,.
\label{fields}
\end{align}
Keeping only terms which are at most cubic in the fields and neglecting axion-scalar-scalar interactions, we can schematically 
write the kinetic lagrangian as $\mc{L}_{\rm kin} = \mc{L}_{\rm{kin, quad}} + \mc{L}_{\rm{kin, cubic}}$ where:
\begin{align}
\mc{L}_{\rm{kin, quad}} = \frac 12 \partial_\mu \hat\Phi \partial^\mu \hat\Phi + \frac 12 \partial_\mu a \partial^\mu a 
+ \frac 12 \partial_\mu h_i \partial^\mu h^i + \frac{1}{2} \partial_\mu \sigma_\alpha \partial^\mu \sigma^\alpha 
+ \frac{1}{2} \partial_\mu \chi_\alpha \partial^\mu \chi^\alpha\,, \nn
\end{align}
while the cubic part can be further decomposed as $\mc{L}_{\rm{kin, cubic}} = \mc{L}_{\Phi a a} + \mc{L}_{\Phi h h} + \mc{L}_{\Phi C C}$, 
with $\mc{L}_{\Phi aa}$ given in~\eqref{lphiaa} and:
\begin{align}
\mc{L}_{\Phi h h} =  - \frac{1}{M_{\rm P}\sqrt{6}} \left[\partial_\mu \hat\Phi h_i\partial^\mu h^i + \hat\Phi \partial_\mu h_i \partial^\mu h^i
+Z\partial_\mu \hat\Phi \sum_{i=1}^4 (-1)^{i+1}\left(h_{2i}\partial^\mu h_{2i-1} + h_{2i-1}\partial^\mu h_{2i} \right)\right], \nn
\end{align}
and:
\begin{align}
\mc{L}_{\Phi C C} = - \frac{1}{M_{\rm P}\sqrt{6}} \left(\sigma_\alpha \partial_\mu \sigma^\alpha \partial^\mu \hat\Phi + \chi_\alpha \partial_\mu \chi^\alpha \partial^\mu \hat\Phi + \hat\Phi \partial_\mu \sigma_\alpha \partial^\mu \sigma^\alpha + \hat\Phi \partial_\mu \chi_\alpha \partial^\mu \chi^\alpha\right)\,. \nn
\end{align}
In addition to the LVS part, the scalar potential contains also the following terms:
\begin{align}
\label{sp1}
V \supset \frac{1}{2}  m^2_0 \left( \sigma_\alpha\sigma^\alpha + \chi_\alpha\chi^\alpha \right) 
+ \frac 12 \left(\hat{\mu}^2+m_0^2\right) h_i h^i + B\hat{\mu} \sum_{i=1}^4 (-1)^{i+1}h_{2i-1}h_{2i}\,.
\end{align}
Since the soft-terms depend on the volume modulus, we can expand them as:
\begin{align}
\label{stexp}
\hat{\mu}^2\propto \tau_b^{-\alpha}\quad\Rightarrow\quad \hat{\mu}^2(\hat\Phi) = \hat{\mu}^2 \left(1-\alpha \sqrt{\frac 23}\frac{\hat\Phi}{M_{\rm P}}\right)\,, \nn \\
m_0^2 \propto \tau_b^{-\beta}\quad\Rightarrow\quad m_0^2(\hat\Phi) = m_0^2 \left(1-\beta \sqrt{\frac 23}\frac{\hat\Phi}{M_{\rm P}}\right)\,, \nn \\
B\hat{\mu} \propto \tau_b^{-\gamma}\quad\Rightarrow\quad B\hat{\mu} (\hat\Phi) = B\hat{\mu} \left(1-\gamma \sqrt{\frac 23}\frac{\hat\Phi}{M_{\rm P}}\right)\,,
\end{align}
where $\alpha$, $\beta$ and $\gamma$ depend on the specific scenario. This expansion leads to new cubic interactions coming from the terms of the scalar potential in~\eqref{sp1}:
\begin{align}
V\supset - \frac{1}{M_{\rm P}\sqrt{6}} &\left[\gamma m_0^2 \, \hat\Phi \left(\sigma_\alpha\sigma^\alpha  + \chi_\alpha\chi^\alpha\right) 
+\left(\alpha \hat{\mu}^2 + \beta m_0^2 \right) \, \hat \Phi h_i h^i + \right. \nonumber \\
+& \left. 2 \gamma B\hat{\mu} \, \hat \Phi \sum_{i=1}^4 (-1)^{i+1}h_{2i-1}h_{2i}\right]. \nn
\end{align}
Including the relevant cubic interactions coming from the kinetic lagrangian and integrating by parts, we obtain a total cubic lagrangian of the form:
\begin{align}
\mc{L}_{\rm cubic} = \frac{1}{M_{\rm P}\sqrt{6}} \left[ \hat \Phi h_i \Box h^i + \hat \Phi \left(\sigma_\alpha \Box \sigma^\alpha + \chi_\alpha \Box \chi^\alpha \right) + \left(\alpha \hat{\mu}^2 + \beta m_0^2 \right) \, \hat \Phi h_i h^i\right. + \nn \\
+ \left. \gamma m_0^2 \, \hat \Phi \left(\sigma_\alpha\sigma^\alpha + \chi_\alpha\chi^\alpha\right)+\left(Z \Box \hat \Phi + 2\gamma B\hat{\mu} \, \hat\Phi\right) \sum_{i=1}^4 (-1)^{i+1}h_{2i-1}h_{2i}\right]. \nn
\end{align}
The leading order expressions of the equations of motion are:
\begin{align}
\Box \sigma_\alpha = - m_0^2 \sigma_\alpha \qquad 
\Box h_{2i-1} =  - \left(\hat{\mu}^2 + m_0^2 \right) h_{2i-1} + (-1)^i B\hat{\mu} \, h_{2i} \quad i=1,\dots,4 \nn \\
\Box \chi_\alpha = - m_0^2 \chi_\alpha \qquad
\Box h_{2j} =  - \left(\hat{\mu}^2 + m_0^2 \right) h_{2j} + (-1)^j B\hat{\mu} \, h_{2j-1} \quad j=1,\dots,4 \,, \nn
\end{align}
which have to be supplemented with:
\begin{align}
\Box \hat \Phi = - m_{\Phi}^2 \hat \Phi \qquad \qquad \Box a = - m_a^2 a \simeq 0\,.
\end{align}
Plugging these equations of motion into $\mc{L}_{\rm cubic}$, the final result becomes:
\begin{align}
\mc{L}_{\rm cubic} = - \frac{1}{M_{\rm P}\sqrt{6}} &\left[\left(\hat{\mu}^2 \left(1 - \alpha\right) + m_0^2 \left(1 - \beta \right)\right) \, \hat\Phi h_i h^i 
+ (1 - \gamma) m_0^2 \, \hat \Phi \left(\sigma^\alpha\sigma_\alpha + \chi^\alpha\chi_\alpha\right)\right. \nn \\
+& \left. \left(2 B\hat{\mu} \left(1 - \gamma \right) + Z\, m_\Phi^2 \right)\, \hat\Phi \sum_{i=1}^4 (-1)^{i+1}h_{2i-1}h_{2i}\right],
\label{finallagr}
\end{align}
from which it is easy to find the corresponding decay rates using the fact that: 
\begin{align}
\mc{L}_1 = \lambda \frac{m^2}{M_{\rm P}}\Phi \phi \phi \quad\Rightarrow \quad
\Gamma_{\Phi \rightarrow \phi \phi} = \lambda_1 \Gamma_0 \,,\quad \lambda_1 = 6\lambda^2\left(\frac{m}{m_{\Phi}}\right)^4 \sqrt{1 - 4 \left(\frac{m_\phi}{m_{\Phi}}\right)^2}\,, \label{dw1} \\
\mc{L}_2 = \lambda \frac{m^2}{M_{\rm P}}\Phi \phi_1 \phi_2\,\,\Rightarrow \,\,
\Gamma_{\Phi \rightarrow \phi_1 \phi_2} = \lambda_2 \Gamma_0\,, \quad \lambda_2 = \frac{\lambda_1}{2}\,\,\text{for}\,\, m_{\phi_1} = m_{\phi_2} = m_\phi\,.
\label{dw2}
\end{align}

\subsubsection*{Decay into Higgsinos}

The decay of the volume modulus into higgsinos is determined by expanding the higgsino kinetic and mass terms around the VEV of $\tau_b$ and then working with canonically normalized fields. The relevant terms in the low-energy lagrangian are:
\begin{align}
\mc{L} \supset i \tilde{H}^{\dagger}_i \overline{\sigma}^\mu \partial_\mu \tilde{H}^{i} \left(1-\sqrt{\frac 23}\frac{\hat\Phi}{M_{\rm P}}\right)- \frac{\hat\mu}{2} \left(\tilde{H}_u^+ \tilde{H}_d^- - \tilde{H}_u^0 \tilde{H}_d^0\right)\left(1-\frac{\alpha}{\sqrt{6}} \frac{\hat\Phi}{M_{\rm P}}\right) +\text{h.c.} \,.
\end{align}
After imposing the equations of motion, we get the following cubic interaction lagrangian:
\begin{align}
\mc{L}_{\rm cubic} \supset \, \frac{\alpha}{2 \sqrt{6}} \, \frac{\hat\mu}{M_{\rm P}} \,\hat\Phi \left(\tilde{H}_u^+ \tilde{H}_d^- - \tilde{H}_u^0 \tilde{H}_d^0\right)+\text{h.c.}\,.
\label{higgsinocubic}
\end{align}
The corresponding decay rates take the form:
\begin{align}
\Gamma_{\Phi \rightarrow \tilde{H}_u^+ \tilde{H}_d^-} = \Gamma_{\Phi \rightarrow \tilde{H}_u^0 \tilde{H}_d^0} = \frac{\alpha^2}{4} \left(\frac{\hat\mu}{m_\Phi}\right)^2 \left(1 - 4 \left(\frac{\hat\mu}{m_\Phi}\right)^2\right)^{3/2} \Gamma_0 \,.
\label{higgsinosdr}
\end{align}

\subsection{Dark Radiation Predictions}

It is clear from~\eqref{finallagr} and~\eqref{higgsinosdr} that the volume modulus branching ratio into visible sector \textit{dof} depends on the size of the soft-terms. Hence the final prediction for dark radiation production has to be studied separately for each different visible sector construction. 

\subsubsection{MSSM-like Case}

Firstly we consider MSSM-like models arising from the ultra-local dS$_2$ case where all soft-terms are suppressed relative to the volume modulus mass:
\begin{align}
m_0^2 \simeq M_{1/2}^2 \simeq B\hat{\mu} \simeq \hat{\mu}^2 \sim \frac{M_{\rm P}^2}{\vo^4} \ll m_{\Phi}^2 \sim \frac{M_{\rm P}^2}{\V^3}\,.
\end{align}
Let us briefly review the results for dark radiation production which for this case have already been studied in~\cite{Cicoli:2012aq, Higaki:2012ar}. 

Given that all soft-terms are volume-suppressed with respect to $m_\Phi$, only the last term in~\eqref{finallagr} gives a non-negligible contribution to the volume modulus branching ratio into visible sector fields. Thus the leading $\Phi$ decay channel is to MSSM higgses via the GM coupling. Using~\eqref{dw2}, we find:  
\begin{align}
\label{phihh1}
\Gamma_{\Phi \rightarrow hh} = c_{\rm vis}\Gamma_0\qquad\text{with}\qquad c_{\rm vis}= 4\times\frac{Z^2}{2} \sqrt{1 - 4\,\frac{(\hat\mu^2+m_0^2)}{m_\Phi^2}} \simeq 2 Z^2\,.
\end{align}
Plugging this value of $c_{\rm vis}$ together with $c_{\rm hid}=1$ (see~\eqref{ddr}) into the general expression~\eqref{dn} for extra dark radiation, we obtain the window:
\begin{align}
\frac{1.53}{Z^2}\lesssim \Delta N_{\rm eff}\lesssim  \frac{1.60}{Z^2}\,,
\label{MSSMpred}
\end{align}
for $0.2 \, {\rm GeV} \lesssim T_{\rm rh} \lesssim 10 \, {\rm GeV}$. Clearly this gives values of $\Delta N_{\rm eff}$ larger than unity for $Z = 1$. Using the bound~\eqref{cbound}, we see that we need $c_{\rm vis}\gtrsim 3$ in order to be consistent with present observational data, implying $Z\gtrsim \sqrt{3/2}\simeq 1.22$.

\subsubsection{Split-SUSY-like Case}
\label{splitsusycase}

Let us now analyze dark radiation predictions for Split-SUSY-like scenarios arising in the dS$_1$ (both local and ultra-local) and local dS$_2$ cases.
In these scenarios the hierarchy among soft-terms is (considering $\mu$ and $B\mu$-terms generated by $K$):
\begin{align}
\label{hierarchyss}
M_{1/2}^2 \simeq \hat{\mu}^2 \sim \frac{M_{\rm P}^2}{\vo^4} \ll m_0^2 \simeq B\hat\mu \simeq m_{\Phi}^2 \sim \frac{M_{\rm P}^2}{\vo^3}\,.
\end{align}
The main difference with the MSSM-like case is that now $B\hat\mu$ and $m_0^2$ scale as $m_\Phi^2$. In order to understand if volume modulus decays into SUSY scalars are kinematically allowed, i.e. $R~\equiv m^2_0/m_\Phi^2 \leq 1/4$, we need therefore to compute the exact value of $m_\Phi$ and compare it with the results derived in Sec.~\ref{sssec:SoftTerms}. It turns out that $m_\Phi$ depends on the dS mechanism as follows:
\begin{align}
{\rm dS}_1: \,\, m_\Phi^2 = \frac{9}{8 a_s\tau_s} \frac{m_{3/2}^2\tau_s^{3/2}}{\vo} \qquad\qquad 
{\rm dS}_2: \,\, m_\Phi^2 = \frac{27}{4 a_s\tau_s} \frac{m_{3/2}^2\tau_s^{3/2}}{\vo}\,.
\label{vmm}
\end{align}

Let us analyze each case separately:
\bi
\item \emph{Local and ultra-local dS$_1$ cases}: Even if the F-terms contribution to scalar masses in eq. ~\eqref{localscalar} for the local case can be made small by appropriately tuning the coefficient $c_s$, the D-terms contribution to $m_0^2$ given by~\eqref{genscalD} cannot be tuned to small values once the requirement of a dS vacuum is imposed. Hence for both local and ultra-local cases, $m_0^2$ cannot be made smaller than the value of eq.~\eqref{genscalD} giving:
\begin{align}
m_0^2\geq 2 m_\Phi^2 \qquad\Rightarrow\qquad R \geq 2\,,
\end{align}
which is clearly in contradiction with the condition $R\leq 1/4$ that has to be satisfied to open up the decay channel of $\Phi$ into SUSY scalars. Therefore 
the decay of $\Phi$ into squarks and sleptons is kinematically forbidden. 

\textit{Decay to Higgses}

Similarly to the MSSM-like case, $\Phi$ can still decay to higgs bosons via the GM term in (\ref{finallagr}). Given that $m_\Phi< m_0$, when $\Phi$ decays at energies of order $m_\Phi$, EWSB has already taken place at the scale $m_0$.\footnote{As we shall show later on, RG flow effects do not modify these considerations qualitatively.} The gauge eigenstates $h_i$ $i=1,...,8$ given in~\eqref{fields} then get rotated into $8$ mass eigenstates. $4$ higgs \textit{dof} which we denote by $A^0, H^0, H^\pm$ remain heavy and acquire a mass of order $m_{H_d}^2 \simeq m_0^2$, and so the decay of $\Phi$ into these fields is kinematically forbidden. The remaining $4$ \textit{dof} are the $3$ would-be Goldstone bosons $G^0$ and $G^{\pm}$ which become the longitudinal components of $Z^0$ and $W^{\pm}$, and the ordinary SM higgs field $h^0$. The $\Phi$ decay rate into light higgs \textit{dof} can be obtained from the last term in~\eqref{finallagr} by writing the gauge eigenstates in terms of the mass eigenstates as~\cite{Martin:1997ns}:
\begin{align}
h_1 = {\rm Re} G^+ \sin\beta + {\rm Re} H^+ \cos\beta\,,&\qquad& h_2 = - {\rm Re} G^+ \cos\beta + {\rm Re} H^+ \sin\beta\,, \nn \\
h_3 = \sqrt{2} v_d + h^0 \sin\beta + H^0 \cos \beta \,, &\qquad& h_4 = \sqrt{2} v_u + h^0 \cos\beta - H^0 \sin\beta \,, \nn \\
h_5 = - G^0 \cos\beta\, + A^0 \sin \beta\,,&\qquad& h_6 = G^0 \sin\beta\, + A^0 \cos \beta\,, \nn \\
h_7 = {\rm Im} G^+ \sin\beta + {\rm Im} H^+ \cos \beta\,,&\qquad& h_8 = {\rm Im} G^+ \cos\beta - {\rm Im} H^+ \sin\beta\,,
\label{gauge-mass}
\end{align}
where $v_u \equiv \langle H^0_u \rangle$, $v_d \equiv \langle H^0_d \rangle$ and $\tan \beta~\equiv v_u/v_d$. Since in Split-SUSY-like models $\tan \beta \sim \mc{O}(1)$ in order to reproduce the correct higgs mass~\cite{Giudice:2011cg, Bagnaschi:2014rsa}, the interaction lagrangian simplifies to:
\begin{align}
\mc{L}_{\rm cubic} \supset &\frac{Z}{2\sqrt{6}}\, \frac{m_\Phi^2}{M_{\rm P}} \, 
\hat\Phi \left[\left(h^0\right)^2+\left(G^0\right)^2+\left({\rm Re}G^+\right)^2+\left({\rm Im} G^+\right)^2 - \right. \nn \\
&\left.- \left(A^0\right)^2 - \left(H^0\right)^2 - \left({\rm Re}H^+\right)^2 - \left({\rm Im}H^+\right)^2\right]\,.
\label{ewsbint}
\end{align}
Neglecting interaction terms involving heavy higgses, (\ref{ewsbint}) gives a decay rate of the form:
\begin{align}
\Gamma^{\rm (GM)}_{\Phi \rightarrow hh,GG}=d_1 \Gamma_0\qquad\text{with}\qquad d_1 \simeq 4\times\frac{Z^2}{4}\simeq Z^2\,.
\end{align}

\textit{Decay to Higgsinos}

$\Phi$ can also decay to higgsinos via the interaction lagrangian (\ref{higgsinocubic}). We need only to check if this decay is kinematically allowed. In Split-SUSY-like models EWSB takes place at the scale $m_0$ where the following relations hold:
\begin{align}
\hat\mu^2 = \frac{m_{H_d}^2 - m_{H_u}^2 \tan^2 \beta}{\tan^2 \beta - 1} - \frac{m_Z^2}{2}\,, \qquad \quad \sin\left(2 \beta\right) = \frac{2 |B \hat\mu|}{m_{H_d}^2 + m_{H_u}^2 + 2 \hat\mu^2}\,.
\label{ewsb}
\end{align}
In the case of universal boundary conditions for the higgs masses, i.e. $m_{H_u}=m_{H_d}=m_0$ at the GUT scale, the $\hat\mu$-term has necessarily to be of order the scalar masses $m_0$ since for $\tan \beta \sim \mathcal{O}\left(1\right)$ the first EWSB condition in (\ref{ewsb}) simplifies to:
\begin{align}
\hat\mu^2 \simeq \frac{m_{H_d}^2}{\tan^2 \beta - 1} \simeq m_0^2 \,,
\label{ewsb2}
\end{align}
given that $m_{H_u}^2$ runs down to values smaller than $m_{H_d}^2$ due to RG flow effects. On the other hand, $\hat\mu$ could be much smaller than $m_0$ for non-universal boundary conditions, i.e. if $m_{H_u} \neq m_{H_d}$ at the GUT scale. In fact, in Split-SUSY models $m_{H_u}^2$ is positive around the scale $m_0$, and so the first EWSB condition in (\ref{ewsb}) for $\hat\mu \ll m_0$ becomes:
\begin{align}
m_{H_d}^2 \simeq m_{H_u}^2 \tan^2 \beta \,.
\end{align}
This condition can be satisfied at the scale $m_0$ for a proper choice of boundary conditions at the GUT scale with $m_{H_u}>m_{H_d}$. Let us point out that, if $\hat\mu$ is determined by K\"ahler potential contributions (see Tab.~\ref{tab:PhenoScenarios}), $\hat\mu$ is suppressed with respect to $m_0$ but, if $\hat\mu$ is generated by non-perturbative effects in $W$, $\hat\mu$ can be of order $m_0$. In this case, the parametrization of the $\hat\mu$-term (\ref{stexp}) reproduces the correct $\tau_b$-dependence of the non-perturbatively induced $\hat\mu$-term if $\alpha = 3 n + 1$.\footnote{If the instanton number is $n = 1$, the $\hat\mu$-term can easily be of order $m_\Phi$ since $c_{\mu,W}$ is a flux-dependent tunable coefficient. For example, setting $\xi = a_s = 1$ and $g_s = 0.1$, the requirement $\hat{\mu} \simeq m_\Phi$ implies that $c_{\mu,W} \simeq W_0/(4\,\vo^{1/6})$. For $W_0 \simeq 10$ and $\vo \simeq 10^7$ we get $c_{\mu,W} \simeq 0.2$.} If we parametrize the ratio between $\hat\mu$ and $m_{\Phi}$ as $\tilde{c} \equiv \hat\mu/m_\Phi$, the decay of $\Phi$ into higgsinos is kinematically allowed only for $\tilde{c} \leq 1/2$. Using (\ref{higgsinosdr}), this decay rate takes the form (where we set $n = 1\,\Leftrightarrow\,\alpha = 4$):
\begin{align}
\Gamma_{\Phi \rightarrow \tilde{H} \tilde{H}} = d_2 \Gamma_0 \qquad\text{with}\qquad d_2 \simeq 8 \tilde{c}^2 \left(1 - 4 \tilde{c}^2\right)^{3/2}\,.
\label{higgsinogamma}
\end{align}

\textit{Dark radiation prediction}

Plugging $c_{\rm vis} = d_1 + d_2$ into~\eqref{dn} with $c_{\rm hid}=1$ we get the following general result:
\begin{align}
\frac{3.07}{Z^2 + d_2}\lesssim \Delta N_{\rm eff}\lesssim  \frac{3.20}{Z^2 + d_2}\,.
\label{dS1pred}
\end{align}
Considering $\tilde{c} = 1/\sqrt{10}$ which maximizes $d_2 \simeq 0.37$ we find that for $0.2 \, {\rm GeV} \lesssim T_{\rm rh} \lesssim 10 \, {\rm GeV}$ this prediction yields values of $\Delta N_{\rm eff}$ larger than unity for $Z = 1$. Consistency with present observational data, i.e. $\Delta N_{\rm eff}\lesssim 1$, requires $Z\gtrsim 1.68$.

\item \emph{Local dS$_2$ case}: The situation seems more promising in this case since in the local limit the D-terms contribution to $m_0^2$ is volume-suppressed with respect to $m_\Phi^2$ since it scales as $\left. m_0^2\right|_D \sim \mc{O}\left(\vo^{-4}\right)$~\cite{Aparicio:2014wxa}. In this case it is therefore possible to tune the coefficient $c_s$ to obtain $R\leq 1/4$. By comparing the second term in~\eqref{vmm} with~\eqref{localscalar}, this implies that $c_s$ has to be tuned so that $\left(c_s - \frac 13\right) \leq \frac{9}{10 \, a_s \tau_s}$, where $a_s \tau_s \sim 80$ in order to get TeV-scale gaugino masses~\cite{Aparicio:2014wxa}. However the condition $m_0^2 > 0$ to avoid tachyonic masses translates into $\left(c_s - \frac 13\right)>0$, giving rise to a very small window: 
\begin{align}
0 < \left( c_s- \frac 13 \right) \leq \frac{9}{10 \, a_s \tau_s} \simeq 0.01\,.
\end{align}
Given that $c_s$ should be extremely fine-tuned, it seems very unlikely to open this decay channel. However the total K\"ahler potential, 
on top of pure $\alpha'$-corrections, can also receive perturbative string loop corrections of the form~\cite{Berg:2007wt}:
\begin{align}
K_{\rm loop} = \frac{g_s C_{\rm loop}}{\vo^{2/3}}\left(1+ k_{\rm loop}\sqrt{\frac{\tau_s}{\tau_b}}\right)\,,
\label{Kloop}
\end{align}
where $C_{\rm loop}$ and $k_{\rm loop}$ are two $\mc{O}(1)$ coefficients which depend on the complex structure moduli. Due to the \textit{extended no-scale structure}~\cite{Cicoli:2007xp}, $g_s$ effects do not modify the leading order scalar potential, and so the mass of the volume modulus is still given by~\eqref{vmm}. However, in order to reproduce a correct ultra-local limit~\eqref{ul}, we need to change the parametrization of the K\"ahler matter metric from~\eqref{newKtilde} to:
\begin{align}
\tilde{K} = \frac{1}{\vo^{2/3}} \left(1 - c_s \frac{\hat{\xi}}{\vo} - c_{\rm loop} \frac{g_s C_{\rm loop}}{\vo^{2/3}} \right),
\label{KtildeNew}
\end{align}
where we introduced a new coefficient $c_{\rm loop}$ and we neglected $k_{\rm loop}$-dependent corrections in~\eqref{Kloop} since they are sub-dominant in the large volume limit $\tau_s\ll\tau_b$. The new ultra-local limit is now given by $c_s = c_{\rm loop} = 1/3$. 

\begin{figure}[!ht]
\begin{center}
\includegraphics[width=0.5\textwidth, angle=0]{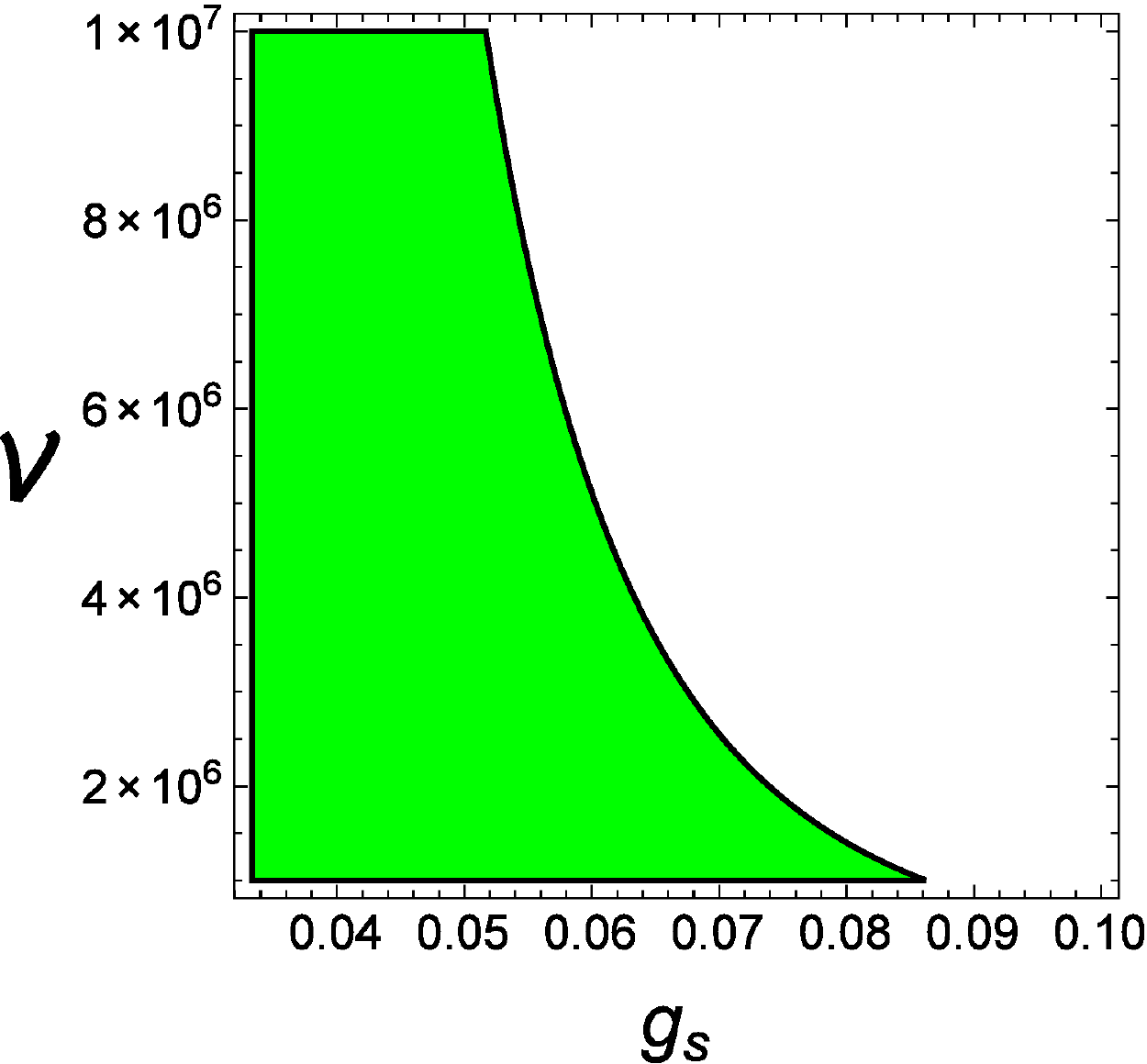}
\caption{The green region in the ($g_s$, $\vo$) parameter space gives $R \leq 1/4$, and so the decay channel of $\Phi$ into SUSY scalars is kinematically allowed.} 
\label{psf}
\end{center}
\end{figure}

These new $c_{\rm loop}$-dependent corrections in~\eqref{KtildeNew} affect the final result for scalar masses and can therefore open up the $\Phi$ decay channel to SUSY scalars. In fact, the result~\eqref{localscalar} for scalar masses in the local case gets modified to: 
\begin{align}
\left.m_0^2\right|_F = \frac{15}{2}\, \frac{m_{3/2}^2 \tau_s^{3/2}}{\vo}
\left[\left(c_s-\frac{1}{3}\right) - \frac{8 g_s C_{\rm loop}}{15} \left(c_{\rm loop} - \frac 13\right)
\frac{\vo^{1/3}}{\hat\xi}\right] \,.
\label{newlsm}
\end{align}
The two terms in square brackets are of the same order for $g_s\simeq 0.1$ and $\vo \sim 10^7$ which is needed to get TeV-scale gauginos, and so they can compete to get $R \leq \frac 14$. As an illustrative example, if we choose $c_s = 1/3$ and natural values of the other parameters: $C_{\rm loop} = a_s = 1$, $\xi = 2$ and $c_{\rm loop} = 0$ (non-tachyonic scalars require $c_{\rm loop} < 1/3$ for $c_s = 1/3$), the ratio between squared masses becomes:
\begin{align}
R = \frac{8}{81}\, g_s^{3/2}\,\vo^{1/3}\,.
\end{align}
As can be seen from Fig.~\ref{psf}, there is now a wide region of the parameter space where the $\Phi$ decay to SUSY scalars is allowed.

We finally point out that $g_s$ corrections to the K\"ahler matter metric affect the result for scalar masses only in the local case since in the ultra-local limit $m_0$ is generated by effects (D-terms for dS$_1$ and F-terms of $S$ and $U$-moduli for dS$_2$) which are sensitive only to the leading order expression of $\tilde{K}_\alpha$.
\ei
Let us now analyze the final prediction for dark radiation production for Split-SUSY-like models where the decay channel of $\Phi$ into SUSY scalars is kinematically allowed.

\subsubsection*{Dark Radiation Results}
\label{drresults}

We start by parametrizing the scalar mass $m_0$ in terms of the volume mode mass $m_\Phi$ as $m_0 = c \,m_\Phi$ and the $\hat\mu$-term as $\mu = \tilde{c}\,m_\Phi$ so that the corresponding kinematic constraints for $\Phi$ decays into SUSY scalars and higgsinos become $c \leq \frac 12$ and $\tilde{c}\leq \frac 12$.
Parametrizing also the $B\hat\mu$-term as in (\ref{eq:SummarySoftLocal}) and using the fact that for Split-SUSY-like models we have in (\ref{stexp}) $\beta = \gamma = 9/2$ and $\alpha = 4$,\footnote{We focus on the case where the $\hat\mu$-term is generated by non-perturbative effects in $W$ since when $\hat\mu$ is generated by $K$, it turns out to be very suppressed with respect to $m_0$, i.e. $\tilde{c}\ll 1$, and so it gives rise to a negligible contribution to the branching ratio of $\Phi$.} the leading order cubic lagrangian is given by the sum of (\ref{finallagr}) and (\ref{higgsinocubic}):
\begin{align}
\mc{L}_{\rm cubic} \simeq &\frac{7 c^2}{2\sqrt{6}}\frac{m_\Phi^2}{M_{\rm P}} \hat\Phi\left[  
\sigma^\alpha\sigma_\alpha + \chi^\alpha\chi_\alpha+\left(1+\frac{6\tilde{c}^2}{7 c^2}\right)h_i h^i+ \right.\nn \\
&\left.+ 2Z  \left(c_{B,K}  - \frac{1}{7 c^2} \right) \sum_{i=1}^4 (-1)^{i+1}h_{2i-1}h_{2i}\right] +\nn \\	
&+ \tilde{c}\sqrt{\frac 23} \, \frac{m_\Phi}{M_{\rm P}}\hat\Phi \left(\tilde{H}_u^+\tilde{H}_d^- -\tilde{H}_u^0\tilde{H}_d^0\right)+\text{h.c.}.
\label{intlagr}
\end{align}
Contrary to the MSSM-like case, now the decay of the volume modulus into squarks and sleptons through mass terms is kinematically allowed and also the decay rate into higgses is enhanced due to mass terms and $B\hat\mu$ couplings. Using~\eqref{dw1}, the total decay rate into squarks and sleptons reads:
\begin{align}
\Gamma_{\Phi \rightarrow \sigma\sigma,\chi\chi} = c_0 \Gamma_0\qquad\text{with}\qquad c_0 = N\times \frac{49 \, c^4}{4} \sqrt{1 - 4 c^2}\,, 
\end{align}
where $N=90$ is the number of real scalar \textit{dof} of the MSSM,\footnote{$12$ \textit{dof} for each left handed squark doublet ($3$ families), $6$ \textit{dof} for each right handed squark ($6$ squarks), $4$ \textit{dof} for each left handed slepton doublet ($3$ families), $2$ \textit{dof} for each right handed slepton ($3$ families).} except for the higgses.

\begin{figure}[!ht]
\begin{center}
\includegraphics[width=0.42\textwidth, angle=0]{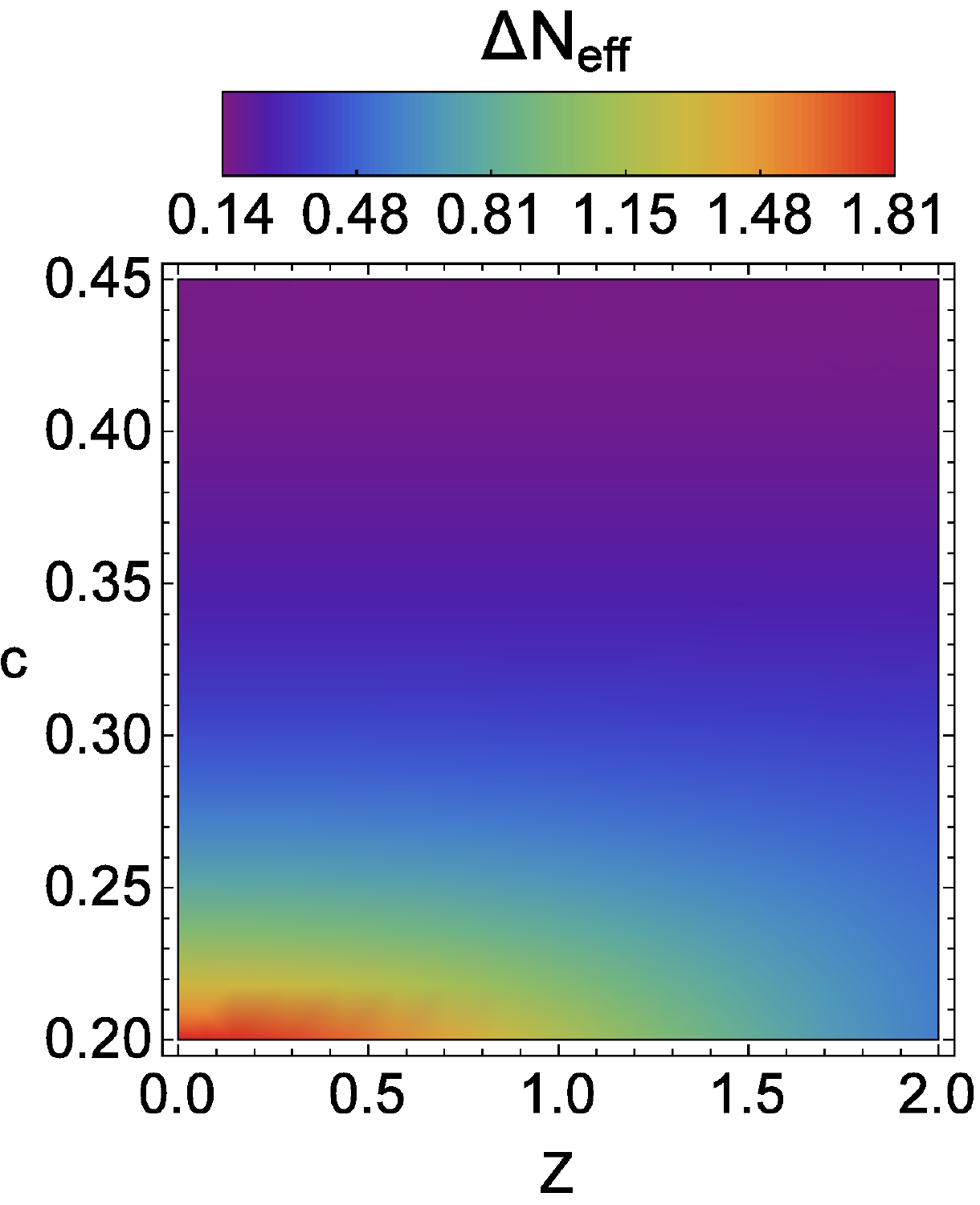}
\caption{Dark radiation production in Split-SUSY-like models for $T_{\rm rh} \gtrsim 0.7 \, {\rm GeV}$ and $\tilde{c}=0$.} 
\label{ris1}
\end{center}
\end{figure}

On the other hand, the decay rate into higgs bosons receives contributions from both mass and GM terms. Using~\eqref{dw1} we obtain:
\begin{align}
\Gamma^{\rm (mass)}_{\Phi \rightarrow hh} = c_1 \Gamma_0\qquad\text{with}\qquad c_1 = 8 \times \frac{49 \, c^4}{4} \left(1+\frac{2\tilde{c}^2}{7 c^2}\right)^2\sqrt{1 - 4 (c^2+\tilde{c}^2)}\,, 
\end{align}
where $8$ is the number of MSSM real higgs \textit{dof} while using (\ref{dw2}) we get:
\begin{align}
\Gamma^{\rm (GM)}_{\Phi \rightarrow hh} = c_2 \Gamma_0\qquad\text{with}\qquad c_2 = 4\times\frac{Z^2}{2} \left(7 c_{B,K} c^2  - 1 \right)^2 \sqrt{1 - 4\,c^2}\,.
\end{align}
The decay rate into higgsinos is given again by~\eqref{higgsinogamma} and thus the total $\Phi$ decay rate into visible sector fields becomes:
\begin{align}
\Gamma_{\rm vis} = \Gamma_{\Phi \rightarrow \sigma\sigma,\chi\chi}+\Gamma^{\rm (mass)}_{\Phi \rightarrow hh}+\Gamma^{\rm (GM)}_{\Phi \rightarrow hh}+\Gamma_{\Phi \rightarrow \tilde{H}\tilde{H}}= c_{\rm vis} \Gamma_0 \,,
\end{align}
\begin{align}
\text{where}\qquad c_{\rm vis} = c_0 + c_1 + c_2 + d_2 \,.
\end{align}
The final prediction for dark radiation production is then given by~\eqref{dn} with $c_{\rm hid}=1$, $g_*(T_{\rm dec}) = 10.75$ and $g_*(T_{\rm rh}) = 86.25$ for $T_{\rm rh} \gtrsim 0.7\,{\rm GeV}$.\footnote{The results do not change significantly for $g_*(T_{\rm rh}) = 75.75$ which is valid for $0.2\,{\rm GeV} \lesssim T_{\rm rh} \lesssim 0.7\, {\rm GeV}$.} The results are plotted in Fig.~\ref{ris1} where we have set $c_{B,K}=1$, $Z = 1$ and we are considering a conservative case in which the decay into higgsinos is negligible, i.e. $\tilde{c} = 0$. For $c>0.2$, the vast majority of the parameter space yields $\Delta N_{\rm eff} \lesssim 1$, in perfect agreement with present experimental bounds with a minimum value $\left.\Delta N_{\rm eff}\right|_{\rm min} \simeq 0.14$ at $c \simeq 1/\sqrt{5}$.\\

\begin{figure}[!ht]
\begin{center}
\includegraphics[width=0.7\textwidth, angle=0]{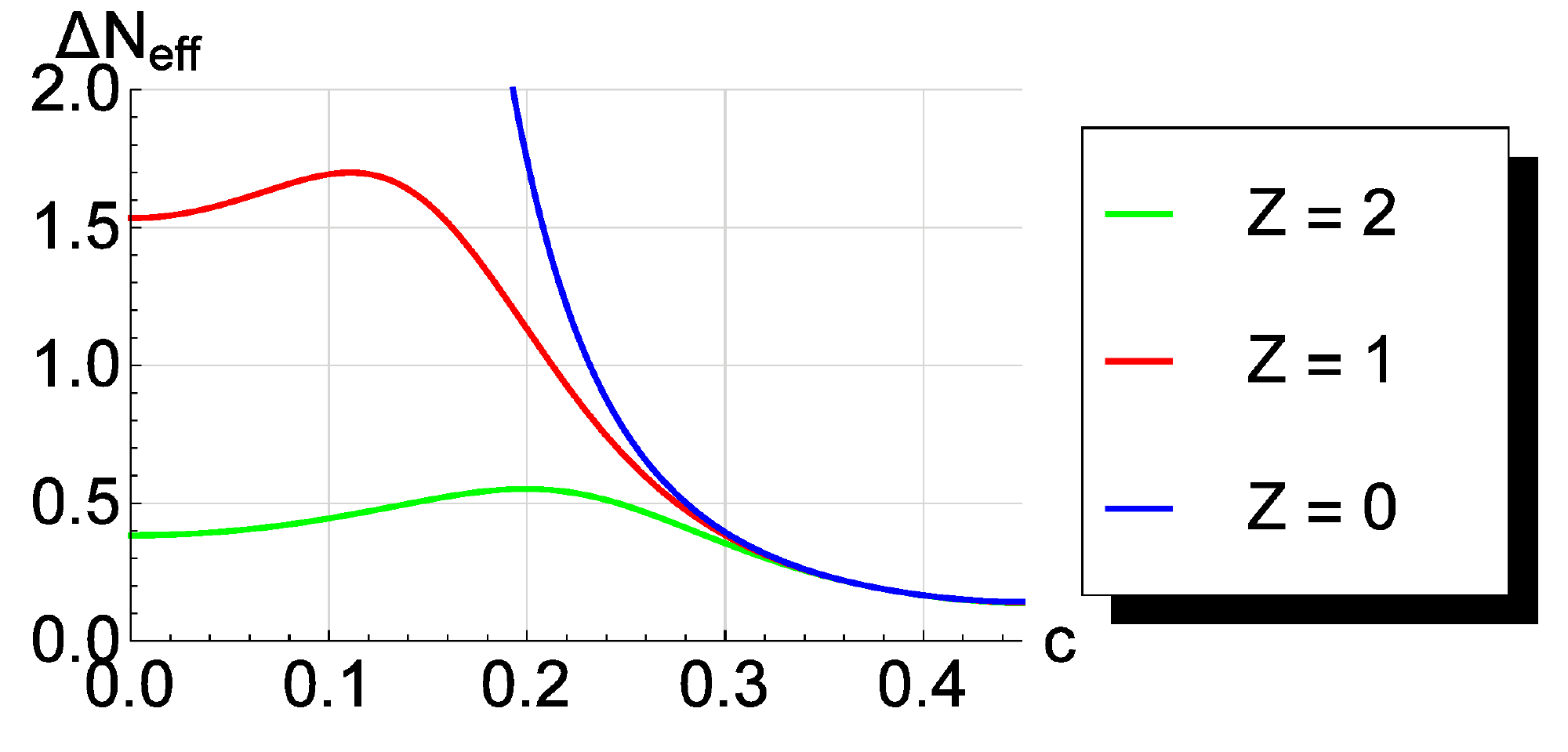}
\caption{Predictions for $\Delta N_{\rm eff}$ in Split-SUSY-like models with $\tilde{c}=0$ for $Z = 0$ (blue line), $Z = 1$ (red line) and $Z = 2$ (green line).} 
\label{z123}
\end{center}
\end{figure}

It is interesting to notice that, contrary to the MSSM-like case, dark radiation over-production can now be avoided if the GM term is absent or it is very suppressed. In fact even for $Z=0$, $\Delta N_{\rm eff} \lesssim 1$ if $c\gtrsim 0.23$, as a consequence of the fact that in this region of the parameter space almost the whole suppression of $\Delta N_{\rm eff}$ is due to the decay into scalar fields. The predictions for $\Delta N_{\rm eff}$ for different values of the GM coupling $Z = 0$ (blue line), $Z = 1$ (red line) and $Z = 2$ (green line) are shown in Fig.~\ref{z123}.\\

For $\tilde{c} \neq 0$ $\Delta N_{\rm eff}$ is even further suppressed than what is shown in Fig.~\ref{ris1} and~\ref{z123} but the correction is at the percent level in the interesting region where the decay into scalars dominates $\Gamma_{\rm vis}$. For example including the effect of decays into higgsinos and setting $\tilde{c} \simeq 1/\sqrt{10}$ to maximize the decay rate into higgsinos, the correction $\left.\delta \Delta N_{\rm eff}\right|_{\rm min}$ to $\left.\Delta N_{\rm eff}\right|_{\rm min}$ turns out to be:
\begin{align}
\frac{\left.\delta \Delta N_{\rm eff}\right|_{\rm min}}{\left.\Delta N_{\rm eff}\right|_{\rm min}} \simeq 0.03 \,.
\end{align}

Note that $\Delta N_{\rm eff}$ can be further suppressed by choosing $c_{B,K} > 1$, namely by enhancing the contribution due to the $B\hat\mu$-term. However 
the decay scale $m_\Phi$ is just slightly larger than the EWSB scale $m_0$ where the $4$ \textit{dof} of the two higgs doublets get rotated into the heavy higgs mass eigenstates $A^0, H^0, H^\pm$, the SM higgs $h^0$ and the longitudinal components $G^0, G^\pm$ of the vector bosons $Z, W^\pm$. Hence the $B\hat\mu$-term gets reabsorbed into the mass terms for $A^0, H^0, H^\pm$ and $h^0, G^0, G^\pm$, and so varying $c_{B,K}$ does not enhance $\Gamma_{\rm vis}$ which receives its main contributions from the decay into squarks, sleptons and heavy higgses through the mass term (assuming that is kinematically allowed) and into all higgs \textit{dof} via the GM term. These considerations will become more clear in the next section where we will take into account corrections due to RG flow effects. A possible way-out could be the separation between the EWSB scale and the volume modulus mass, which translates into requiring $c \ll 1$. However this choice would imply $m_0^2 \ll m_\Phi^2 \sim B\hat\mu$ which would be a quite unnatural situation from a top-down perspective since $m_\Phi$ and $m_0$ have the same volume scaling.

\subsubsection*{RG Flow Effects}
\label{rgflows}

The results obtained above have to be corrected due to RG flow effects from the string scale to the actual scale $m_\Phi$ where the modulus decay takes place~\cite{Angus:2013zfa}. However these corrections do not alter qualitatively our results since $\Phi$ interacts only gravitationally and the running of squarks and sleptons in Split-SUSY models is almost absent~\cite{Bagnaschi:2014rsa}. For sake of completeness, let us study these RG flow effects in detail.\\

The soft-terms $m_0$, $\hat\mu$ and $B\hat\mu$ entering in the interaction lagrangian~\eqref{finallagr} are just boundary conditions for the RG flow and should instead be evaluated at the scale $m_\Phi$ where the light modulus $\Phi$ decays. The RG equations for the first and second generation of squarks and sleptons are given by:
\begin{align}
m^2_\alpha = m_0^2 + \sum_{a=1}^3 c_{\alpha,a} K_a\,,
\end{align}
where $c_{\alpha,a}$ is the weak hypercharge squared for each SUSY scalar and the RG running contributions $K_a$ are proportional 
to gaugino masses~\cite{Martin:1997ns}. Given that in Split-SUSY-like models gaugino masses are hierarchically lighter than scalar masses, 
the RG running of first and second generation squarks and sleptons is a negligible effect. Thus we can consider their mass at the scale $m_\Phi$ as still given by $m_0$ to a high level of accuracy.\\

The situation for the third generation is slightly trickier since there are additional contributions from large Yukawa couplings. Using mSUGRA boundary conditions, the relevant RG equations become (ignoring contributions proportional to $M_{1/2}$)~\cite{Martin:1997ns}:
\begin{align}
16 \pi^2 \frac{d}{dt} m_{Q_3}^2 = X_t + X_b\,,\qquad 16 \pi^2 \frac{d}{dt} m_{\ov{u}_3}^2 = 2 X_t\,,
\qquad 16 \pi^2 \frac{d}{dt} m_{\ov{d}_3}^2 = 2 X_b\,,
\label{RG1}
\end{align}
\begin{align}
16 \pi^2 \frac{d}{dt} m_{L_3}^2 = X_\tau\,,\qquad 16 \pi^2 \frac{d}{dt} m_{\ov{e}_3}^2 = 2 X_\tau\,,
\label{RG2}
\end{align}
which are coupled to those involving higgs masses:
\begin{align}
16 \pi^2 \frac{d}{dt} m_{H_u}^2 = 3 X_t + X_b\,,\qquad 16 \pi^2 \frac{d}{dt} m_{H_d}^2 = 3 X_b + X_\tau\,.
\label{RG3}
\end{align}
The quantities $X_i$ look like:
\begin{align}
\label{xi1}
X_t = 2 |y_t|^2 \left(m_{H_u}^2 + m_{Q_3}^2 + m_{\ov{u}_3}^2\right) + 2 |a_t|^2\,, \\
\label{xi2}
X_b = 2 |y_b|^2 \left(m_{H_u}^2 + m_{Q_3}^2 + m_{\ov{d}_3}^2\right) + 2 |a_b|^2\,, \\
\label{xi3}
X_\tau = 2 |y_\tau|^2 \left(m_{H_u}^2 + m_{L_3}^2 + m_{\ov{e}_3}^2\right) + 2 |a_\tau|^2\,,
\end{align}
where $y_i$ are the Yukawa couplings and $a_i$ are the only sizable entries of the $A$-term couplings. Given that for sequestered scenarios the $A$-terms scale as $M_{1/2}$~\cite{Aparicio:2014wxa}, the contribution $2 |a_i|^2$ can be neglected with respect to the first term in each $X_i$. Moreover $X_t$, $X_b$ and $X_\tau$ are all positive, and so the RG equations~\eqref{RG1} and~\eqref{RG2} drive the scalar masses to smaller values at lower energies. This has a two-fold implication:
\bi
\item When $m_0 > m_\Phi$, RG running effects could lower $m_0$ to values smaller than $m_\Phi/2$ so that the decay channel to SUSY scalars opens up at the scale $m_\Phi$. However this never happens since RG effects are negligible.

\item When $m_0 \leq m_\Phi/2$, no one of the scalars becomes heavier than the volume modulus if $R < \frac 14$ at the boundary energy scale. On the other hand, RG running effects could still lower the scalar masses too much, suppressing the $\Phi$ decay rate to SUSY scalars. However this does not happen since RG effects are negligible.
\ei
In Split-SUSY-like models a correct radiative realization of EWSB requires a low value of $\tan \beta$~\cite{Bagnaschi:2014rsa}, which implies $y_b,\,y_\tau\ll y_t$. In turn, $X_b$ and $X_\tau$ give rise to a tiny effect, and so the running of $m_{H_d}^2$, $m_{\ov{d}_3}^2$, $m_{L_3}^2$ and $m_{\ov{e}_3}^2$ turns out to be negligible. In the end, the only relevant RG equations become:
\begin{align}
16 \pi^2 \frac{d}{dt} m_{Q_3}^2 \simeq X_t\,, \qquad 16 \pi^2 \frac{d}{dt} m_{\ov{u}_3}^2 \simeq 2 X_t\,, \qquad 16 \pi^2 \frac{d}{dt} m_{H_u}^2 = 3 X_t\,.
\end{align}

\begin{figure}[!ht]
\begin{center}
\includegraphics[width=0.8\textwidth, angle=0]{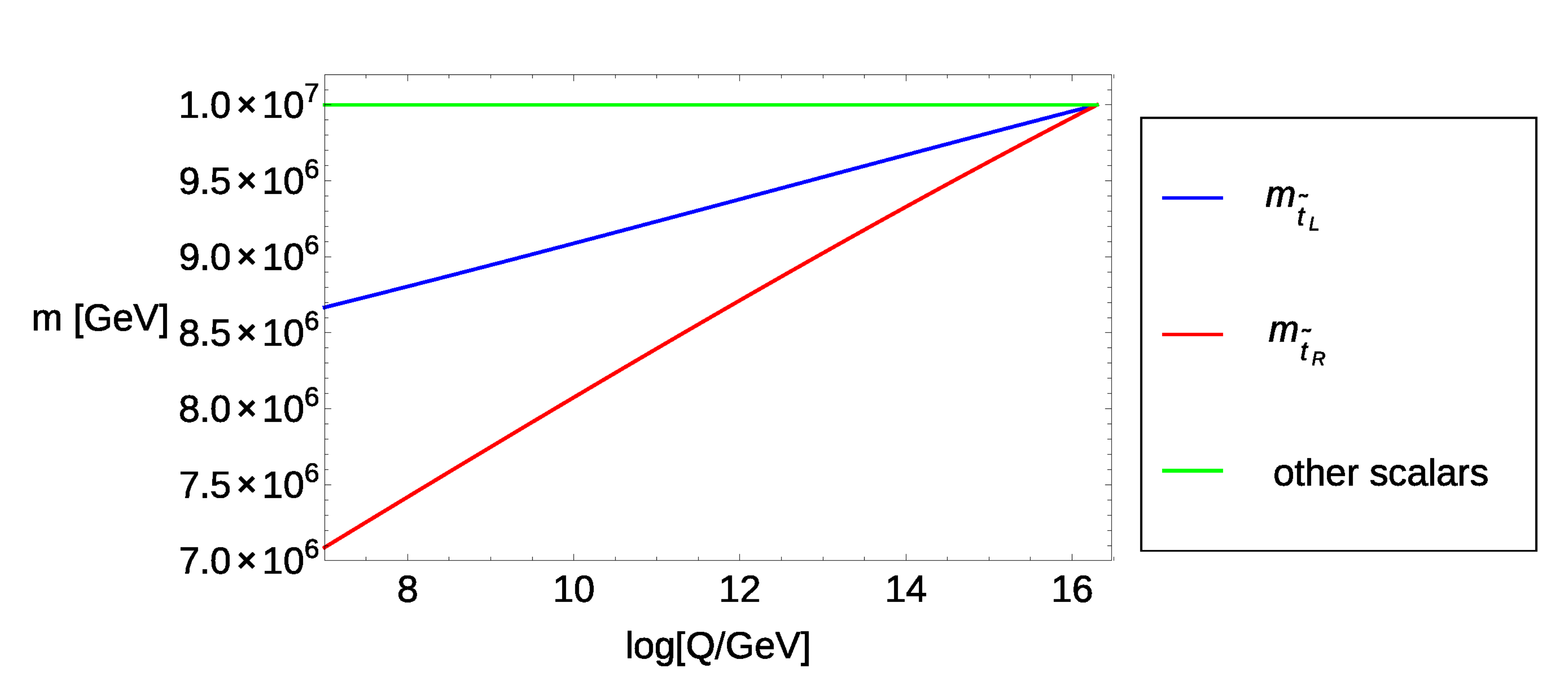}
\caption{Running of the scalar masses.\label{scr}}
\end{center}
\end{figure}

\begin{figure}[!ht]
\begin{center}
\includegraphics[width=0.8\textwidth, angle=0]{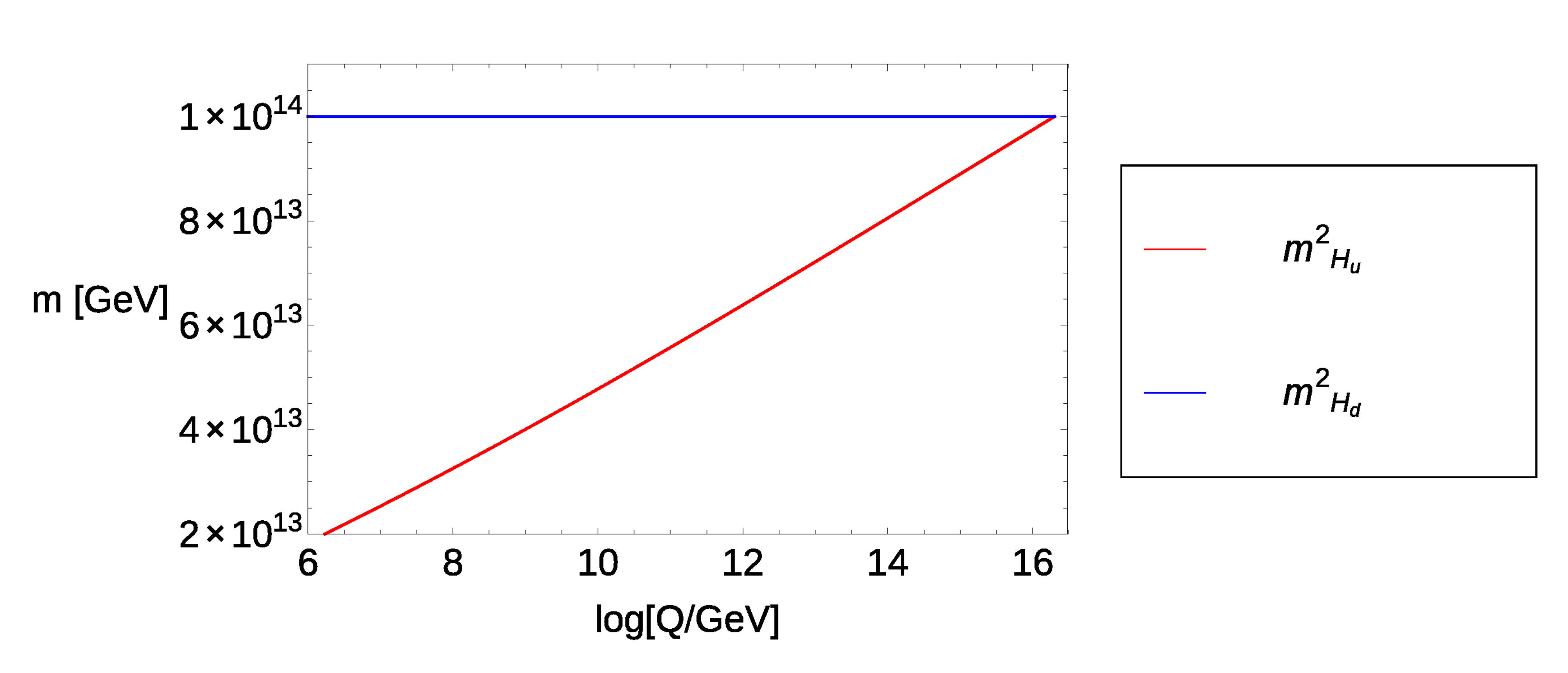}
\caption{Running of the higgs masses.}\label{hr}
\end{center}
\end{figure}

We performed a numerical computation of the RG running, using as boundary conditions $m_0 = B\hat\mu^{1/2} = \hat\mu = 10^7 \, \rm GeV$ and $M_a = 10^3 \, \rm GeV$ at the GUT scale $M_{\rm GUT} = 2 \times 10^{16}$ GeV and $\tan \beta \simeq 1.4$. We also used that the stop left-right mixing is given by $\chi_t = A_t - \hat\mu \cot \beta \simeq - m_0/{\tan \beta}$, being $A_t \simeq M_{1/2} \ll m_0$. We used SusyHD~\cite{Vega:2015fna} to run the Yukawa couplings from the top mass scale up to $m_0$ combined with SARAH~\cite{Staub:2008uz} to run them from $m_0$ up to the GUT scale.\footnote{We are grateful to J. P. Vega for useful discussions about this point.} These runnings have been computed at order one loop. Using the values of the Yukawa couplings obtained at the GUT scale, we have been able to compute the running of scalars, $\hat\mu$, $B\hat\mu$ and the GM coupling $Z$ down to the scale of the decay $m_\Phi$. Fig.~\ref{scr} shows the running of scalar masses while Fig.~\ref{hr} shows the running of $m_{H_u}^2$ and $m_{H_d}^2$. The running of $\hat\mu$ and $B\hat\mu$ is almost negligible. We clarify that our purpose here is not to study EWSB in full detail but to understand which kind of behavior we should expect for the running of soft-terms from the GUT scale to $m_\Phi$ using boundary conditions which are consistent with EWSB.

Due to RG running effects each of the scalars has a different mass at the scale $m_\Phi$, and so the exact prediction for $\Delta N_{\rm eff}$ becomes:
\begin{align}
\Delta N_{\rm eff} = \frac{43}{7} \mathcal{R} \left(\frac{g_* (T_{\rm decay})}{g_*(T_{\rm reheat})}\right)^{1/3} \,,
\end{align}
where:
\begin{align}
\mc{R} = \frac{\Gamma_0}{\Gamma_{\Phi \rightarrow \sigma\sigma,\chi\chi} + \Gamma_{\Phi \rightarrow hh} + \Gamma_{\Phi \rightarrow \tilde{H} \tilde{H}}}\,.
\end{align}
The decay rate into squarks and sleptons is given by:
\begin{align}
\label{gs}
\Gamma_{\Phi \rightarrow \sigma\sigma,\chi\chi} = \frac{49}{4} \sum_\alpha \, \kappa_\alpha \left(\frac{m_\alpha}{m_\Phi}\right)^4 \sqrt{1 - 4 \left(\frac{m_\alpha}{m_\Phi}\right)^2}\Gamma_0\,,
\end{align}
where the index $\alpha$ runs over all squarks and sleptons $m_\alpha = \left( m_{\tilde{Q}_{\rm L}}, m_{\tilde{u}_R}, m_{\tilde{d}_R}, m_{\tilde{L}}, m_{\tilde{e}_{\rm R}}\right)$ while $\kappa_\alpha$ is the number of \textit{dof} for each scalar.\footnote{It turns out that $\kappa_\alpha = \left(36, 18, 18, 12, 6\right)$.}\\

On the other hand the decay rate into higgs \textit{dof} is given by (we focus on the case where $\hat\mu\ll m_0$ since, as we have seen in the previous section, a large $\hat\mu$-term gives rise just to a negligible correction to the final dark radiation prediction):
\begin{align}
\label{gh}
\Gamma_{\Phi \rightarrow hh} = \left[\sum_{I \in \{A^0, H^0, H^\pm\}} \left(\frac{49}{4} \left(\frac{m_{I}}{m_{\Phi}} \right)^4 + Z^2 \right) \sqrt{1 - 4 \left(\frac{m_{I}}{m_{\Phi}}\right)^2} + Z^2\right]\Gamma_0\, .
\end{align}
All quantities in (\ref{gs}) and (\ref{gh}) have to be computed at the decay scale $m_\Phi$. As already explained in the previous section, $B\hat\mu$ does not explicitly contribute to (\ref{gh}) since it gets reabsorbed into the higgs masses due to EWSB. The decay of $\Phi$ into heavy higgses $A^0, H^0, H^\pm$ through the mass term can instead contribute to $\Gamma_{\rm Higgs}$, provided that $m_{A^0, H^0, H^\pm}/m_\Phi \leq 1/2$. The mass of the heavy higgses in the limit $m_Z, m_{W^\pm} \ll m_{A^0}$ can be written as~\cite{Martin:1997ns}:
\begin{align}
m_{H^0}^2 \simeq m_{H^\pm}^2 \simeq m_{A^0}^2 \simeq 2 |\hat\mu|^2 + m_{H_u}^2 + m_{H_d}^2 \,.
\end{align}
The decay rate into higgsinos instead reads:
\begin{align}
\Gamma_{\Phi \rightarrow \tilde{H} \tilde{H}} = 8 \left(\frac{\hat\mu}{m_\Phi}\right)^2 \left(1 - 4 \left(\frac{\hat\mu}{m_\Phi}\right)^2\right)^{3/2} \Gamma_0 \,,
\label{higgsinodecayrg}
\end{align}
where we added the two contributions in~\eqref{higgsinosdr} and we used $\alpha = 4$. In~\eqref{higgsinodecayrg} $\hat\mu/m_\Phi \neq \tilde{c}$, since $\hat\mu$ is computed at the decay scale $m_\Phi$.\\

We computed $\Delta N_{\rm eff}$ for different values of $m_\Phi$, keeping the boundary conditions fixed at $m_0 = B\hat\mu^{1/2} = \hat\mu = 10^7 \, \rm GeV$ and $M_a = 10^3$ GeV. The qualitative behavior of $\Delta N_{\rm eff}$ is the same as in the previous section where RG flow effects have been ignored. The results are shown in Tab.~\ref{tab1} which shows that the dominant contribution to $\Delta N_{\rm eff}$ is given by the decay into squarks and sleptons while the suppression coming from decay into higgsinos is always sub-dominant. If $m_\Phi \simeq 2.2 \times 10^7 \, \rm GeV$, corresponding to $m_0/m_\Phi \simeq 1/\sqrt{5}$, $\Delta N_{\rm eff} \simeq 0.15$ which is only slightly larger then $\left.\Delta N_{\rm eff}\right|_{\rm min} = 0.14$ computed in the previous section without taking into account RG flow effects. This is due to the fact that the running of the SUSY scalars is negligible as can be clearly seen from Fig.~\ref{scr}.

\begin{table}[ht!]
\begin{center}
\begin{tabular}{cccccc}
\hline
$m_\Phi$ & $\Gamma_{\rm scalars}/\Gamma_0$ & $\Gamma_{\rm Higgs}/\Gamma_0$ & $\Gamma_{\rm Higgsinos}/\Gamma_0$ & $\Delta N_{\rm eff}$  \\
\hline
$2.2 \times 10^7 \, \rm GeV$ & $18.53$ & $1.12 \,(*)$ & $0.08$  & $0.15$ \\
\hline
$3 \times 10^7 \, \rm GeV$ & $9.19$ & $1.12 \,(*)$ & $0.36$ & $0.29$ \\
\hline
$4 \times 10^7 \, \rm GeV$ & $3.36$ & $2.52$ & $0.33$ & $0.49$ \\
\hline
$5 \times 10^7 \, \rm GeV$ & $1.45$ & $2.45$ & $0.25$ & $0.74$ \\
\hline
\end{tabular}
\end{center}
\caption{Values of $\Delta N_{\rm eff}$ corresponding to different masses $m_\Phi$ of $\Phi$ for fixed boundary condition $m_0 = 10^7$ GeV at the GUT scale. We also indicate the relative importance of the various decay channels. In the case denoted by a $(*)$ the only non-vanishing contribution to $\Gamma_{\rm Higgs}$ is due to the decay into light higgs \textit{dof} through the GM coupling, since the decay into heavy higgs \textit{dof} turns out to be kinematically forbidden as a consequence of the RG flow: $2 m_{A^0, H^0, H^\pm} > m_\Phi$ at the decay scale $m_\Phi$. The decay into higgsinos is always a sub-leading effect.}
\label{tab1}
\end{table}

\section{Conclusions}
\label{Concl}

Extra dark radiation is a very promising window for new physics beyond the Standard Model. Its presence is a generic feature of string models where some of the moduli are stabilized by perturbative effects since the corresponding axionic partners remain very light and can behave as extra neutrino-like species~\cite{Allahverdi:2014ppa}. These light hidden sector \textit{dof} are produced by the decay of the lightest modulus~\cite{Acharya:2008bk, Acharya:2009zt, Acharya:2010af, Allahverdi:2013noa} leading to $\Delta N_{\rm eff}\neq 0$~\cite{Cicoli:2012aq, Higaki:2012ar, Higaki:2013lra}.\\

In this chapter we performed a general analysis of axionic dark radiation production in sequestered LVS models where the visible sector is localized on D3-branes at singularities~\cite{Aldazabal:2000sa, Conlon:2008wa, Cicoli:2012vw, Cicoli:2013cha}. These models yield a very interesting post-inflationary cosmological history where reheating is driven by the decay of the lightest modulus $\Phi$ with a mass of order $m_\Phi\sim 10^7$ GeV which leads to a reheating temperature of order $T_{\rm rh}\sim 1$ GeV. The gravitino mass is much larger than $m_\Phi$ ($m_{3/2}\sim 10^{10}$ GeV), so avoiding any gravitino problem~\cite{Endo:2006zj, Nakamura:2006uc}. Low-energy SUSY can still be achieved due to sequestering effects that keep the supersymmetric partners light. Gauginos are around the TeV-scale whereas squarks and sleptons can either be as light as the gauginos or as heavy as the lightest modulus $\Phi$ depending on the moduli dependence of the matter K\"ahler metric and the mechanism responsible for achieving a dS vacuum~\cite{Aparicio:2014wxa}.\\

The final prediction for dark radiation production due to the decay of the volume modulus into ultra-light bulk closed string axions depends on the details of the visible sector construction:
\ben
\item \textit{MSSM-like case}: \\
MSSM-like models arise from the ultra-local dS$_2$ case where the leading visible sector decay channel of $\Phi$ is to higgses via the GM coupling $Z$. The simplest model with two higgs doublets and $Z=1$ gives $1.53 \lesssim \Delta N_{\rm eff}\lesssim 1.60$ for $500\,{\rm GeV} \lesssim T_{\rm rh}\lesssim 5\,{\rm GeV}$~\cite{Cicoli:2012aq, Higaki:2012ar}. Values of $\Delta N_{\rm eff}$ smaller than unity require $Z\gtrsim 1.22$ or more than two higgs doublets.

\item \textit{Split-SUSY-like case with $m_0> m_\Phi/2$}: \\
Local and ultra-local dS$_1$ cases give rise to Split-SUSY-like scenarios where scalar masses $m_0$ receive a contribution from D-terms which cannot be made smaller than $m_\Phi/2$. Thus the decay of $\Phi$ into squarks and sleptons is kinematically forbidden. The leading visible sector decay channel of $\Phi$ is again to higgses via the GM coupling $Z$. However, given that EWSB takes place at the scale $m_0$ which in these cases is larger than the decay scale $m_\Phi$, the volume mode $\Phi$ can decay only to the 4 light higgs \textit{dof}. For a shift-symmetric higgs sector with $Z=1$, the final prediction for dark radiation production is $3.07 \lesssim \Delta N_{\rm eff}\lesssim 3.20$. Consistency with present experimental data, i.e. $\Delta N_{\rm eff}\lesssim 1$, requires $Z\gtrsim 1.68$. In most of the parameter space of Split-SUSY-like models a correct radiative EWSB can be achieved only if the $\hat\mu$-term is of order the scalar masses. Hence, depending on the exact value of $\hat\mu$, the decay of $\Phi$ into higgsinos could not be mass suppressed. However it gives rise just to a negligible contribution to $\Delta N_{\rm eff}$.

\item \textit{Split-SUSY-like case with $m_0\leq m_\Phi/2$}: \\
Given that in the local dS$_2$ case the D-terms contribution to scalar masses is negligible, the decay of $\Phi$ into SUSY scalars can become kinematically allowed. In fact, thanks to the inclusion of string loop corrections to the K\"ahler potential~\cite{Cicoli:2007xp, Berg:2007wt}, a large region of the underlying parameter space features $m_0\leq m_\Phi/2$. Hence the final prediction for $\Delta N_{\rm eff}$ gets considerably reduced with respect to the previous two cases since, in addition to decays into higgses via the GM term, leading order contributions to the branching ratio to visible sector particles involve decays into squarks and sleptons, decays into heavy higgses induced by mass terms and possible decays into higgsinos depending on the exact value of the $\hat\mu$-term. Depending on the exact value of $m_0$, the simplest model with $Z=1$ gives $0.14 \lesssim \Delta N_{\rm eff} \lesssim 1.6$. Hence these models feature values of $\Delta N_{\rm eff}$ in perfect agreement with present observational bounds. Note that dark radiation overproduction can be avoided even for $Z=0$ due to the new decay channels to squarks and sleptons.
\een

We finally studied corrections to these results due to RG flow effects from the string scale $M_s \sim 10^{15}$ GeV to the volume mode mass $m_\Phi\sim 10^7$ GeV where the actual decay takes place. However these corrections do not modify our predictions since the RG running of SUSY scalar masses is a negligible effect in Split-SUSY-like models and radiative corrections to $m_\Phi$ are tiny since moduli are only gravitational coupled.


\chapter{Volume Inflation}
\label{chap:VolumeInflation}

\section{Introduction}

Two crucial quantities of any string compactification are the Hubble scale during inflation $H$ and the gravitino mass $m_{3/2}$. The first sets the inflationary energy scale $M_{\rm inf}\sim \sqrt{H M_{\rm P}}$ which in turn is related to the tensor-to-scalar ratio $r$ as $M_{\rm inf}\sim M_{\rm GUT}\left(\frac{r}{0.1}\right)^{1/4}$, whereas the second gives the order of magnitude of the soft supersymmetry breaking terms $M_{\rm soft}\sim m_{3/2}$. Given that neither primordial tensor modes nor supersymmetric particles have been detected yet, experimental data yield just upper and lower bounds on these two quantities: 
\begin{align}
r\lesssim 0.1 \quad\,&&\Rightarrow \quad &&H \lesssim \frac{M_{\rm GUT}^2}{M_{\rm P}} \sim 10^{14}\,{\rm GeV} \label{rbound} \\
M_{\rm soft}\gtrsim 1\,{\rm TeV}\quad&&\Rightarrow\quad &&m_{3/2}\gtrsim 1\,{\rm TeV}\,. \label{Msoftbound}
\end{align}
However $H$ and $m_{3/2}$ are not two independent quantities since in any consistent string inflationary model the inflaton dynamics has not to destabilize the volume mode. This is guaranteed if the inflationary energy is smaller than the energy barrier to decompactification, i.e. $H^2 M_{\rm P}^2 \lesssim V_{\rm barrier}$, and the height of the barrier is generically set by the gravitino mass. In KKLT models $V_{\rm barrier}\sim m_{3/2}^2 M_{\rm P}^2$ which leads to $H\lesssim m_{3/2}$~\cite{Kallosh:2004yh} while in the LARGE Volume Scenario (LVS) $V_{\rm barrier}\sim m_{3/2}^3 M_{\rm P}$ giving $H\lesssim m_{3/2}\sqrt{\frac{m_{3/2}}{M_{\rm P}}}$~\cite{Conlon:2008cj}.\\

These theoretical bounds are not in contradiction with the experimental bounds (\ref{rbound}) and (\ref{Msoftbound}), in particular for the cases of high-scale supersymmetry and small field inflationary models with unobservable tensor modes. However the phenomenologically interesting cases of low-energy supersymmetry with $M_{\rm soft}\gtrsim \mc{O}(1)$ TeV and large field inflationary models with $r\gtrsim \mc{O}(0.01)$ would imply a high value of $H$ together with a small value of $m_{3/2}$, in clear tension with the theoretical bounds from volume destabilization problems.\\

Several ideas have been proposed in the literature to overcome this tension between TeV-scale supersymmetry and large field inflation. Here we briefly summarize them:
\ben
\item The relation $H^2 M_{\rm P}^2\lesssim V_{\rm barrier}$ would be independent of $m_{3/2}$ in models where the energy barrier and the gravitino mass are two uncorrelated quantities.  Racetrack superpotentials~\cite{Kallosh:2004yh}, models of natural inflation as in~\cite{Kappl:2015pxa} or models of D-terms inflation from field-dependent Fayet-Iliopoulos~\cite{Wieck:2014xxa} can provide viable models where $V_{\rm barrier}$ is decoupled from $m_{3/2}$.

\item The KKLT bound $H\lesssim m_{3/2}$ and the LVS bound $H\lesssim m_{3/2}\sqrt{\frac{m_{3/2}}{M_{\rm P}}}$ apply just to the inflationary era, and so in these expression $m_{3/2}$ is the gravitino mass during inflation which might be different from the present value of the gravitino mass. This is possible if $m_{3/2}= e^{K/2} |W| \simeq \frac{W_0}{\vo} M_{\rm P}$ evolves just after the end of inflation. Two viable realizations include inflationary models where the inflaton coincides with the volume mode $\vo$ so that $\vo$ relaxes from small to large values during inflation~\cite{Conlon:2008cj} or where the inflaton is a complex structure modulus or a matter field so that $W$ relaxes from large to small values during inflation~\cite{He:2010uk}. Another option is to consider models with two different volume stabilization mechanisms during and after inflation. If $\vo$ couples to the field $X$ whose F-term drives inflation, the volume's vacuum expectation value (VEV) during inflation would be determined by the F-terms potential of $X$ which however vanishes after the end of inflation when $\vo$ is fixed by a more standard KKLT or LVS mechanism. This kind of models with a K\"ahler potential coupling between $\vo$ and $X$ have been studied in~\cite{Antusch:2011wu} whereas superpotential interactions between $\vo$ and $X$ have been analyzed in~\cite{Yamada:2012tj}.

\item Another possible way-out to reconcile low-energy supersymmetry with high scale inflation is to consider models where the visible sector is sequestered from the sources of supersymmetry breaking so that the soft-terms are much smaller than the gravitino mass~\cite{Aparicio:2014wxa, Blumenhagen:2009gk}. This can be the case for compactifications where the visible sector is localized on fractional D3-branes at singularities which can lead to $M_{\rm soft}\sim m_{3/2}\frac{m_{3/2}}{M_{\rm P}}$.
\een 

Let us point out that all the solutions listed above require a high degree of tuning except for the sequestered case which might be however not enough to completely remove all the tension between observable tensor modes and TeV-scale supersymmetry. Moreover, it is technically rather complicated to provide consistent string models where these solutions are explicitly realized. Therefore they are at the moment at the level of string-inspired toy-models without a concrete string embedding where one can check if there is enough tuning freedom to achieve the amount of fine tuning needed to reproduce all the desired phenomenological details.\\

In this chapter we shall provide a first step towards an explicit stringy embedding of the case where the volume mode plays the r\^ole of the inflaton evolving from small to large values after the end of inflation~\cite{Conlon:2008cj}. We will describe a possible microscopic origin of the potential terms used in~\cite{Conlon:2008cj} to create an inflection point at small volumes around which slow-roll inflation can occur. Moreover, we shall also perform a deeper analysis of the relation between the position of the inflection point and the minimum with the tuning of the flux superpotential.\\

Before presenting the details of our analysis, let us stress some key-features of the model under study:
\bi
\item In order to have an evolving gravitino mass, we focus on models where the inflaton is the volume mode $\vo$. Given that we work in an effective supergravity theory where the K\"ahler potential has a logarithmic dependence on $\vo$, each term in the inflationary potential will be a negative exponential of the canonically normalized volume mode $\Phi\sim \ln\vo$, i.e. $V \supset e^{- k \Phi}$. This form of the potential is reminiscent of Starobinsky-like models which have a rather large inflationary scale since they are at the boarder between large and small field models~\cite{Starobinsky:1980te}. However Starobinsky-like potentials feature also an inflaton-independent constant which can never appear in cases where the inflaton is the volume mode $\vo$ since $\vo$ couples to all sources of energy because of the Weyl rescaling needed to obtain the correct effective action.\footnote{Starobinsky-like models with large $r$ and a constant inflaton-independent constant can instead be obtained in models where the inflaton is a K\"ahler modulus different from the volume mode~\cite{Cicoli:2008gp}.}

\item The best way to achieve volume inflation is therefore to consider a potential which has enough tuning freedom to create an inflection point and then realize inflation in the vicinity of the inflection point. The price to pay is that this inflationary scenario turns out to be rather fine tuned and it is necessarily a small field model with a sub-Planckian field range during inflation $\Delta\Phi\sim 0.4 \,M_{\rm P}$, unobservable tensor modes of order $r\sim 10^{-9}$ and low Hubble scale $H\sim 10^{10}$ GeV. Hence this approach cannot solve completely the tension between large tensor modes and TeV-scale supersymmetry. Still it provides a big step forward especially if combined with a sequestered visible sector so that low-energy supersymmetry can be safely reconciled with a high gravitino mass.

\item In LVS models where $H> m_{3/2} \sqrt{\frac{m_{3/2}}{M_{\rm P}}}$ and the inflaton is the volume mode, the destabilization problem of~\cite{Kallosh:2004yh} becomes an overshooting problem since the inflaton has an initial energy which is larger than the barrier to decompactification. The solution to this problem via radiation production after the end of inflation has already been discussed in~\cite{Conlon:2008cj}, and so we shall not dwell on this issue. 

\item In the models under study, the Hubble scale during inflation is set by the gravitino mass during inflation $m_{3/2}^{\rm inf}$ which is much larger than the gravitino mass today $m_{3/2}^{\rm today}$ due to the volume evolution. Hence the Hubble scale $H$ can be much larger than $m_{3/2}^{\rm today}$ since we have:
\begin{align}
H \sim m_{3/2}^{\rm inf} \sqrt{\frac{m_{3/2}^{\rm inf}}{W_0\,M_{\rm P}}} \gg m_{3/2}^{\rm today} \sqrt{\frac{m_{3/2}^{\rm today}}{W_0\,M_{\rm P}}}\,.
\end{align}
\ei
We shall analyze both the single modulus and the two moduli case focusing on three different visible sector realizations which lead to:
\ben
\item \textit{High-scale SUSY models}: in this case the requirement of low-energy supersymmetry is relaxed and the value of the gravitino mass both during and after inflation is huge. The volume mode evolves from values of order $100$ during inflation to values of order $200$ after the end of inflation. The flux superpotential has to be tuned to values of order $W_0\sim 10^{-5}$ in order to reproduce the correct amplitude of the density perturbations. Thus the order of magnitude of the gravitino mass during and after inflation is the same, $m_{3/2}^{\rm inf}\sim m_{3/2}^{\rm today}\sim 10^{11}$ GeV corresponding to $H\sim 10^{10}$ GeV. Due to the small value of $\vo$, in this case the validity of the effective field theory approach is not fully under control.

\item \textit{Non-sequestered models}: in these models during inflation the volume is of order $\vo \sim 10^5$ and $W_0\sim 1$ giving a gravitino mass of order $m_{3/2}^{\rm inf}\sim 10^{14}$ GeV which leads again to $H\sim 10^{10}$ GeV. After the end of inflation the volume evolves to $\vo\sim 10^{15}$ as required to get TeV-scale supersymmetry since the present value of the gravitino mass becomes $m_{3/2}^{\rm today}\sim 10$ TeV~\cite{Conlon:2005ki}.

\item \textit{Sequestered models}: in these models inflection point volume inflation takes place again for values of order $\vo \sim 10^5$ and $W_0\sim 1$ which yield $m_{3/2}^{\rm inf}\sim 10^{14}$ GeV and $H\sim 10^{10}$ GeV. After the end of inflation the volume evolves instead to $\vo\sim 10^7$ corresponding to $m_{3/2}^{\rm today}\sim 10^{11}$ GeV, as required to get low-energy supersymmetry gaugino masses in sequestered scenarios where $M_{1/2}\sim m_{3/2} \frac{m_{3/2}}{M_{\rm P}}\sim 10$ TeV~\cite{Aparicio:2014wxa, Blumenhagen:2009gk}. 
\een

This chapter is organized as follows. Sec.~\ref{SecIPInf} is a brief review of the basic concepts of inflection point inflation which will be used in the rest of the chapter while in Sec.~\ref{SecOrigin} we describe a possible microscopic origin of all the terms in the inflationary potential which are needed to develop an inflection point at small values of $\vo$ together with a dS minimum at larger values of the volume mode. In Sec.~\ref{SecSingle} we study the single modulus case presenting first an analytical qualitative description of the inflationary dynamics and then performing an exact numerical analysis. The two moduli case typical of LVS models is instead discussed in Sec.~\ref{SecTwo} before presenting our conclusions in Sec.~\ref{SecConcl}. This chapter is based on~\cite{Cicoli:2015wja}.

\section{Inflection Point Inflation}
\label{SecIPInf}

In this section we briefly review the generic features of inflection point inflation, closely following ~\cite{Baumann:2007ah} and~\cite{Linde:2007jn}. We summarize the main points and discuss the tuning involved in these models.\\

The basic assumption is that inflation takes place around an inflection point along some arbitrary direction in field space. The scalar potential around such an inflection point can always be expanded as:
\begin{align}
V = V_{\rm ip} \left(1+ \lambda_1 (\phi - \phi_{\rm ip})+ \frac{\lambda_3}{3!} (\phi - \phi_{\rm ip})^3+\frac{\lambda_4}{4!} (\phi - \phi_{\rm ip})^4+\dots\right),
\end{align}
where $\phi_{\rm ip}$ denotes the position of the inflection point. Given that we shall focus on cases where the field excursion during the inflationary period is small, i.e. $(\phi - \phi_{\rm ip})\ll 1$, the quartic term can be safely neglected. We therefore find that it suffices to analyze the following potential:
\begin{align}
V = V_{\rm ip} \left(1+ \lambda_1 (\phi - \phi_{\rm ip}) + \frac{\lambda_3}{3!} (\phi - \phi_{\rm ip})^3\right).
\label{appr}
\end{align}
The inflationary observables are determined by the slow-roll parameters:
\begin{align}
\epsilon &= \frac{1}{2} \left(\frac{V'}{V}\right)^2 = \frac{1}{2} \left(\lambda_1 + \frac{1}{2} \lambda_3 (\phi - \phi_{\rm ip})^2\right)^2\simeq \frac{1}{2}\lambda_1^2\,,
\label{eps} \\
\eta &= \frac{V''}{V} = \lambda_3 (\phi - \phi_{\rm ip})\,,
\label{eq:srParams}
\end{align}
which have to be evaluated at horizon exit where $\phi=\phi_*$. Note that in (\ref{eps}) and (\ref{eq:srParams}) we approximated $V \simeq V_{\rm ip}$ in the denominator, which is a good approximation for small field models where $(\phi - \phi_{\rm ip})\ll 1$. Notice that if $\lambda_3\gtrsim 1$ this small field condition has to be satisfied in order to obtain $\eta\ll 1$. If instead $\lambda_3$ is tuned such that $\lambda_3\ll 1$, the condition $\eta\ll 1$ could be satisfied also for large field values but then the approximation (\ref{appr}) would be under control only by tuning all the coefficients of the expansion. We shall therefore focus only on the case $(\phi - \phi_{\rm ip})\ll 1$. The number of e-foldings is given by:
\begin{align} 
N_e(\phi_*) = \int_{\phi_{\rm end}}^{\phi} \frac{d \phi}{\sqrt{2 \epsilon}} = \sqrt{\frac{2}{\lambda_1 \lambda_3}} 
\arctan \left.\left[\sqrt{\frac{\lambda_3}{2\lambda_1}}(\phi - \phi_{\rm ip})\right]\right|_{\phi_{\rm end}}^{\phi_*}\,.
\label{Ne}
\end{align}
In order to have enough e-foldings we need $\lambda_1\ll 1$, which is also needed to get $\epsilon\ll 1$, and $(\phi - \phi_{\rm ip})\gtrsim \sqrt{\lambda_1}$ so that the arc-tangent does not give a small number. Thus the slow-roll parameter $\eta$ turns out to be larger than $\epsilon$ since:
\begin{align}
\epsilon\sim \lambda_1^2 \ll \eta \sim (\phi - \phi_{\rm ip})\gtrsim \sqrt{\lambda_1}\ll 1\,.
\end{align}
Therefore the spectral index in these models is essentially given by $\eta$:
\begin{align}
n_s - 1 = 2 \eta(\phi_*) - 6 \epsilon(\phi_*) \simeq 2 \eta(\phi_*)\,,
\end{align}
By using (\ref{Ne}), it is possible to rewrite the spectral index as a function of the number of e-foldings as: 
\begin{align}
n_s - 1 \simeq - \frac{4}{N_e} + \frac 23 \lambda_1\lambda_3 N_e\,.
\label{eq:ns}
\end{align}
Since $1 - \frac{4}{N_e} \simeq 0.93$ for $N_e\simeq 60$, it is evident that for a very small $\lambda_1$ (or equivalently a very flat inflection point) the spectral index asymptotes to $0.93$. We have therefore to use (\ref{eq:ns}) to determine the value of $\lambda_1$ that gives a value of $n_s$ in agreement with recent Planck data, i.e. $n_s = 0.9655\pm 0.0062$ ($68\%$ CL)~\cite{Ade:2015lrj}. We find:
\begin{align}
n_s=0.965 \quad\Rightarrow\quad \lambda_1=7.92\cdot 10^{-4}\,\lambda_3^{-1}\qquad\text{for}\qquad N_e=60\,.
\label{correctns}
\end{align}
In these small field models, the tensor-to-scalar ratio $r$ turns out to be unobservable since:
\begin{align}
r= 16 \epsilon \simeq 8\lambda_1^2 \simeq 5.01\cdot 10^{-6} \lambda_3^{-2}\qquad\text{for}\qquad N_e=60.
\label{r}
\end{align}
Successful inflation not only gives rise to the correct spectral tilt and tensor fraction but does so at the right energy scale. The normalization of scalar density perturbations in these models can be written as: 
\begin{align}
\Delta^2 = \frac{1}{24 \pi^2} \left.\frac{V}{\epsilon}\right|_{\phi_*} \simeq \frac{1}{12 \pi^2} \frac{V_{\rm ip}}{\lambda_1^2} \simeq 2.4 \cdot 10^{-9}\,.
\label{cobe}
\end{align}
Once the parameter $\lambda_3$ is fixed, (\ref{correctns}) gives the value of $\lambda_1$ which produces the correct spectral index, $n_s=0.965$, and (\ref{cobe}) fixes the value of $V_{\rm ip}$ which reproduces the observed amplitude of the density perturbations. In turn, (\ref{r}) yields the prediction for the tensor-to-scalar ratio $r$.

\section{Microscopic Origin of the Inflationary Potential}
\label{SecOrigin}

In this section we will try to describe a possible microscopic origin of volume modulus inflation. The tree-level stabilization of the dilaton $S$ and of complex structure moduli proceeds as described in Sec.~\ref{ssec:ModuliStabilization}. In particular we will consider two explicit setups:
\bi
\item[a)] \textit{Single modulus case} \\
In this setup we have a single K\"ahler modulus $T_b = \tau_b + {\rm i} c_b$ where $c_b$ is an axion field and $\tau_b$ controls the overall volume: $\vo = \tau_b^{3/2}$. The canonical normalization of the inflaton field $\tau_b$ can be inferred from its kinetic terms: 
\begin{align}
\mc{L}_{\rm kin} = \frac{3}{4 \tau_b^2} \partial^\mu \tau_b \partial_\mu \tau_b\,,
\end{align}
so that the canonically normalized volume field can be written as:
\begin{align}
\Phi = \sqrt{\frac{3}{2}} \ln \tau_b \simeq \sqrt{\frac{2}{3}} \ln \vo\,.
\label{CanNorm}
\end{align}
In this setup the visible sector can be realized in two different ways:
\bi
\item[i)] The visible sector lives on a stack of D7-branes wrapping the 4-cycle associated to $\tau_b$. Given that the visible sector gauge coupling is set by $\tau_b$ as $\alpha_{\rm vis}^{-1}=\tau_b$, the late-time value of the volume has to be of order $100$ in order to reproduce a correct phenomenological value of $\alpha_{\rm vis}^{-1}\simeq 25$. This model is characterized by high scale SUSY.

\item[ii)] The visible sector is localized on D3-branes at singularities obtained by collapsing a blow-up mode to zero size due to D-terms stabilization~\cite{Aparicio:2014wxa, Blumenhagen:2009gk}. In this case the visible sector coupling is set by the dilaton, and so the volume can take much larger values of order $\vo \simeq 10^7$ which lead to TeV-scale supersymmetry via sequestering effects. In the presence of desequestering perturbative or non-perturbative effects~\cite{Conlon:2010ji}, low-energy SUSY requires larger values of $\vo$ of order $\vo\sim 10^{14}$.
\ei

\item[b)] \textit{Two moduli case}\\
In this setup we start with two K\"ahler moduli: $T_b = \tau_b + {\rm i} c_b$ and $T_s = \tau_s + {\rm i} c_s$ with $\tau_b \gg \tau_s$. The CY volume takes the swiss-Cheese form:
\begin{align}
\vo = \tau_b^{3/2} - \tau_s^{3/2}\,.
\end{align} 
The modulus which plays the role of the inflaton will still be $\tau_b$, and so its canonical normalization is given by (\ref{CanNorm}). 
However, in this setup the visible sector can be realized in three different ways:
\bi
\item[i)] The visible sector lives on a stack of D7-branes wrapping $\tau_b$ (for an explicit CY construction see~\cite{Cicoli:2011qg}). As explained above, in this case $\vo$ has to be of order $100$ for phenomenological reasons and the SUSY scale is very high.

\item[ii)] The visible sector lives on a stack of D7-branes wrapping $\tau_s$ (see again~\cite{Cicoli:2011qg} for an explicit model). Given that the visible sector gauge coupling is now independent on $\tau_b$ since $\alpha_{\rm vis}^{-1}=\tau_s$, the volume can take large values. A particularly interesting value is $\vo\simeq 10^{14}$ which leads to TeV-scale soft-terms, as in standard non-sequestered LVS models. 

\item[iii)] The visible sector is localized on D3-branes at singularities (see~\cite{Cicoli:2012vw, Cicoli:2013cha} for explicit global dS models) obtained by collapsing a blow-up mode to zero size due to D-terms stabilization. Since the gauge kinetic function of D3-branes is given by $S$, $\vo$ can take large values. Due to sequestering effects, in this case TeV-scale superpartners require $\vo\simeq 10^7$.
\ei
\ei

\subsubsection{$\alpha'$-corrections}

If $W_0\neq 0$, the no-scale structure is broken by $\alpha'^3$-corrections which show up in the K\"ahler potential as in eq. \eqref{eq:AlphaPrimeKahlerPotential}. The contribution of the $\alpha'$-correction to the scalar potential looks like: 
\begin{align}
V = e^K \left(K^{T_b \overline{T}_b} D_{T_b} W D_{\overline{T}_b} \overline{W} - 3 |W|^2 \right) \supset V_0 \,e^{-\sqrt{\frac{27}{2}} \Phi} \equiv V_{\alpha'}\,,
\label{Va}
\end{align}
where we have defined:
\begin{align}
V_0 = \frac{g_s}{8 \pi} \frac{3 \hat{\xi} |W_0|^2}{4}\,.
\label{eq:V0}
\end{align}
$V_0$ is a free parameter which can be tuned to get inflection point inflation with the right COBE normalization and a large volume minimum.

\subsubsection{Non-perturbative Effects}

Given that non-perturbative effects are exponentially suppressed, they tend to give rise to negligible contributions to the scalar potential. However, in the two moduli case, $T_s$-dependent non-perturbative effects could lead to potentially large contributions, as in the LVS setup, described in Sec.~\ref{sssec:LVS}. 
With the right normalization the LVS potential takes the form:
\begin{align}
V = \frac{g_s}{8 \pi} \left(\frac{8  a_s^2 A_s^2 \sqrt{\tau_s} e^{-2 a_s \tau_s}}{3 \V} - \frac{4 W_0 a_s A_s \tau_s e^{-a_s \tau_s}}{\V^2} + \frac{3 \hat{\xi} W_0^2}{4 \V^3}\right).
\label{eq:LVSScalarPotentialGs}
\end{align}
Integrating out $\tau_s$ from the scalar potential in eq. \eqref{eq:LVSScalarPotentialGs} we get an effective potential which depends only on the volume $\vo$:
\begin{align}
V = \frac{V_0}{\vo^3} \left[1 - \frac{2}{\hat\xi a_s^{3/2}} \left(\ln \V\right)^{3/2}\right]\,.
\label{effectivepotentialLVS}
\end{align}
The first term in (\ref{effectivepotentialLVS}) is just the $\alpha'$-correction (\ref{Va}) whereas the second term leads to a new non-perturbative contribution to the scalar potential:
\begin{align}
V_{\rm np} = - \kappa_{\rm np} V_0 \,\Phi^{3/2}\,e^{-\sqrt{\frac{27}{2}} \Phi} \qquad\text{with}\qquad 
\kappa_{\rm np}=\frac{2}{\xi}\left(\frac{3}{8\pi^2}\right)^{3/4}\left(g_s N\right)^{3/2}\,.
\label{Vnp}
\end{align}
Note that $\kappa_{\rm np}$ is another parameter that can be tuned to get inflection point inflation with the right COBE normalization and number of e-foldings.

\subsubsection{Higher Derivative $\alpha'$-corrections}
\label{sssec:HigherDerivativeAlphaCorrections}

Additional contributions to the scalar potential from higher derivative corrections to the 10D action have been computed in~\cite{Ciupke:2015msa}. These corrections have the same higher dimensional origin as the $\alpha'^3$-term (\ref{Va}). Both stem from the $\alpha'^3 \mc{R}^4$ contribution to the in 10D Type IIB action. The effect of these four derivative corrections is to modify the equations of motion of the auxiliary fields (F-terms) thereby giving rise to corrections to the kinetic terms, new quartic derivative couplings and more importantly new contributions to the scalar potential.  These $F^4$ corrections depend on the CY topology and take the generic form:
\begin{align}
V_{F^4} = - \frac{\hat{\lambda} |W_0|^4}{\V^4} \Pi_i t^i\,,
\label{v1}
\end{align}
where $\Pi_i$ are topological integers defined as:
\begin{align}
\Pi_i = \int_X c_2 \wedge \hat{D}_i\,,
\end{align}
with $c_2$ the second Chern class of the CY $X$ and $\hat{D}_i$ is a basis of harmonic $(1,1)$-forms allowing for the usual expansion of the K\"ahler form $J$ as:
\begin{align}
J = \sum_{i=1}^{h^{(1,1)}} t_i \hat{D}_i\,,
\end{align}
with $t^i$ being two-cycle volumes. The parameter $\hat{\lambda}$ is expected to be of order $\hat{\xi}/\chi(X)$, with $\chi(X)$ being the CY Euler number. For model building purposes, we will take it to be a real negative constant.\\

Depending on the details of the compactification space, these new terms can take different forms, yielding contributions to the scalar potential that scale differently with the overall volume. In the simplest single K\"ahler modulus case where $t_b = \sqrt{\tau_b} \simeq \V^{1/3}$, the correction \eqref{v1} takes the form:
\begin{align}
V_{F^4} =  \frac{\kappa_{F^4}\,V_0}{\vo^{11/3}} = \kappa_{F^4}\, V_0\,e^{-\frac{11}{\sqrt{6}} \Phi}\,,
\label{VF4}
\end{align}
where we defined:
\begin{align}
\kappa_{F^4} \equiv - \hat{\lambda}  \Pi_b \frac{|W_0|^4}{V_0}\,.
\end{align}
The form of these corrections for the two moduli case can be derived by focusing on the $\mathbb{C} \rm{P}^4_{[1,1,1,6,9]}$ case where the volume can be written in terms of the two-cycle volumes $t_1$ and $t_5$ as~\cite{Denef:2004dm}:
\begin{align}
\vo = \frac{1}{6} (3 t_1^2 t_5 + 18 t_1 t_5^2 + 26 t_5^3)\,,
\end{align}
implying that the four-cycles are given by:
\begin{align}
\tau_1 = \frac{\partial \V}{\partial t_1} = t_5 (t_1 + 3 t_5) \quad\text{and}\quad \tau_5 = \frac{\partial \V}{\partial t_5} = \frac{1}{2} (t_1 + 6 t_5)^2\,.
\end{align}
One can define $\tau_4$ as the linear combination:
\begin{align}
\tau_4 = \tau_5 - 6 \tau_1 = \frac{t_1^2}{2} \quad \Rightarrow \quad t_1 = \sqrt{2 \tau_4}\,.
\end{align}
Plugging $t_1$ back into the expression for $\tau_5$, solving the equation for $t_5$ and requiring that $t_5 > 0$ when $\tau_5 \gg \tau_4$ we get:
\begin{align}
t_5 = \frac{1}{3 \sqrt{2}} \left(\sqrt{\tau_5} - \sqrt{\tau_4}\right),
\end{align}
from which it can be inferred that the volume has the form:
\begin{align}
\vo = \frac{1}{9 \sqrt{2}} \left(\tau_b^{3/2} - \tau_s^{3/2}\right),
\end{align}
where we identified $\tau_5 \equiv \tau_b$ and $\tau_4 \equiv \tau_s$. Thus the $\alpha'$-correction of \eqref{v1} becomes:
\begin{align}
V_{F^4} \simeq V_0 \left(\frac{\kappa_{F^4_{(b)}}}{\vo^{11/3}} + \frac{\kappa_{F^4_{(s)}} \sqrt{\tau_s}}{\V^4}\right),
\label{deltav1}
\end{align}
where:
\begin{align}
\kappa_{F^4_{(b)}} = - \hat\lambda  \Pi_5 \frac{|W_0|^4}{6^{1/3} V_0} \qquad\text{and}\qquad \kappa_{F^4_{(s)}} = -\hat\lambda \sqrt{2} \left(\Pi_4 - \frac{\Pi_5}{6}\right) \frac{|W_0|^4}{V_0} \,.
\end{align}

\subsubsection{String Loop Corrections}

As we have observed in Sec.~\ref{sssec:CorrectionsToTreeLevel} the \Kahler potential receives string loop corrections as in eq. \eqref{eq:KKStringLoopCorrections} and eq. \eqref{eq:WindingStringLoopCorrections}. The final contribution to the scalar potential can be written as:
\begin{align}
V_{g_s} = \sum_{i = 1}^{h^{(1,1)}} \frac{|W_0|^2}{\vo^2} \left(g_s^2 \left(\mathcal{C}^{\rm KK}_i\right)^2 \frac{\partial^2 K_0}{\partial \tau_i^2} - 2 \delta K^{\rm W}_{(g_s)}\right).
\label{slccontributiontov}
\end{align}
Noting that $\partial^2 K_0/\partial\tau_b^2 = 3/(4 \tau_b^2)$, it turns out that in both scenarios the leading order string loop correction looks like:
\begin{align}
V_{g_s} = \frac{\kappa_{g_s}\,V_0}{\vo^{10/3}} = \kappa_{g_s}\,V_0\,e^{-\frac{10}{\sqrt{6}} \Phi}\,,
\label{Vgs}
\end{align}
where $\kappa_{g_s}$ is a real tunable number.

\subsubsection{Anti D3-branes}

Anti D3-branes yield a positive contribution to the scalar potential which in general provides a viable mechanism to realize a dS minimum. More precisely, the introduction of anti D3-branes in the compactification produces a term in the scalar potential of the form~\cite{Kachru:2003aw}:
\begin{align}
V_{\overline{D3}} = \frac{\kappa_{\overline{D3}}\,V_0}{\vo^2} = \kappa_{\overline{D3}}\,V_0\,e^{-\sqrt{6}\Phi}\,,
\label{VD3}
\end{align}
where $\kappa_{\overline{D3}}$ is a positive real number which can be tuned to realize inflection point inflation.

\subsubsection{Charged Hidden Matter Fields}

The possible presence on the big cycle of a hidden sector with matter fields $\phi$ charged under an anomalous $U(1)$ leads to the generation of moduli-dependent Fayet-Iliopoulos (FI) terms. The corresponding D-terms potential reads:
\begin{align}
V_D = \frac{1}{2 \text{Re}(f_b)} \left(q_{\phi} |\phi|^2 - \xi_b\right)^2\,,
\end{align} 
where $f_b = T_b$ and $q_\phi$ is the $U(1)$-charge of $\phi$ while the FI-term is given by:
\begin{align} 
\xi_b = - \frac{q_b}{4 \pi} \frac{\partial K_0}{\partial T_b} = \frac{3 q_b}{8 \pi} \frac{1}{\vo^{2/3}}\,,
\end{align} 
where $q_b$ the $U(1)$-charge of $T_b$. Since supersymmetry breaking effects generate a mass for $\phi$ of order the gravitino mass, the total scalar potential becomes:
\begin{align}
V = V_D + c m_{3/2}^2 |\phi|^2 + \mc{O}\left(\vo^{-3}\right),
\end{align}
where $c$ is an $\mc{O}(1)$ coefficient which can be positive or negative depending on hidden sector model building details.
Integrating out $\phi$ leads to a new contribution which has been used to obtain dS vacua and takes the form~\cite{Cicoli:2012vw, Cicoli:2013cha}:
\begin{align}
V_{\rm hid} = \frac{\kappa_{\rm hid}\,V_0}{\vo^{8/3}} = \kappa_{\rm hid}\,V_0\,e^{-\frac{8}{\sqrt{6}}\Phi}\,,
\label{vds2}
\end{align}
where $\kappa_{\rm hid} = \frac{3 c q_b W_0^2}{16 \pi q_\phi V_0}$ is a tunable coefficient.\footnote{Note that $q_b$ and $q_\phi$ must have the same sign otherwise the minimum for $|\phi|$ would be at zero.}

\subsubsection{Total Scalar Potential}

The total scalar potential that we shall consider can in general be written as:
\begin{align}
V_{\rm tot} = V_{\alpha'} + V_{\rm np} +  V_{F^4} + V_{g_s} + V_{\overline{D3}} + V_{\rm hid}\,,
\label{Vtot}
\end{align}
where $V_{\alpha'}$ is the universal $\alpha'$-correction (\ref{Va}), $V_{\rm np}$ is the non-perturbative generated potential (\ref{Vnp}) which is non-negligible only in the two moduli case, $V_{F^4}$ are the higher derivative effects (\ref{VF4}) and (\ref{deltav1}), $V_{g_s}$ is the string loop potential (\ref{Vgs}), $V_{\overline{D3}}$ is the contribution (\ref{VD3}) from anti D3-branes and $V_{\rm hid}$ is the potential (\ref{vds2}) generated by the F-terms of charged hidden matter fields.\\

Let us now add all these different contributions to the total scalar potential for the single modulus and the two moduli case separately: 
\bi
\item[a)] \textit{Single modulus case}\\
In this simple model with only a single K\"ahler modulus, the generic expression for the scalar potential is:
\begin{align}
V(\Phi) = V_0 \left(e^{-\sqrt{\frac{27}{2}} \Phi} + \kappa_{g_s} \,e^{- \frac{10}{\sqrt{6}} \Phi} + \kappa_{F^4} \,e^{- \frac{11}{\sqrt{6}} \Phi} 
+ \kappa_{\overline{D3}} \,e^{- {\sqrt{6}} \Phi} + \kappa_{\rm hid} \,e^{- \frac{8}{\sqrt{6}} \Phi}\right).
\label{V1}
\end{align}
In this setup the post-inflationary dS minimum is generated by the interplay between the universal $\alpha'^3$ term and the two terms proportional to $\kappa_{\overline{D3}}$ and $\kappa_{\rm hid}$, with the inflationary inflection point arising at smaller volumes in a region where the first three terms in \eqref{V1} are comparable in size. 

\item[b)] \textit{Two moduli case}\\
In the two moduli case, the total scalar potential (\ref{Vtot}) contains at least six terms. However, as we shall see in the next section, we need just five tunable parameters in order to get inflection point inflation. We shall therefore neglect the last term in (\ref{Vtot}) which might be removed by a model building choice. We stress that this choice does not affect our final results. In fact, if instead we neglected $V_{\overline{D3}}$ in (\ref{Vtot}), we would obtain qualitatively the same results. Thus in this case the total inflationary potential becomes:
\begin{align}
V(\Phi) = V_0 \left[\left(1 - \kappa_{\rm np} \Phi^{3/2}\right) e^{-\sqrt{\frac{27}{2}} \Phi} + \kappa_{g_s} \,e^{-\frac{10}{\sqrt{6}} \Phi} 
+ \kappa_{F^4_{(b)}} \,e^{-\frac{11}{\sqrt{6}} \Phi} + \kappa_{\overline{D3}}\, e^{- \sqrt{6} \Phi}\right].
\label{V2}
\end{align}
Here we are including only the leading term of the higher derivatives corrections \eqref{deltav1} which is proportional to $\kappa_{F^4_{(b)}}$. Hence we are assuming that the term proportional to $\kappa_{F^4_{(s)}}$ is either very suppressed (a natural possibility given its volume scaling) or exactly vanishing.
\ei

\section{Single Field Dynamics}
\label{SecSingle}

In this section we study the effective single field inflationary dynamics for both the single modulus and the two moduli case. A deeper analysis of the effect of the heavy field for the two moduli case will be performed in Sec.~\ref{SecTwo}. In order to obtain a phenomenologically viable model, we should require that:
\ben
\item There is an inflection point at $\Phi_{\rm ip}$.
\item The potential is such that the COBE normalization is satisfied and the number of e-foldings is $N_e \simeq 60$.
\item There is a large volume de Sitter minimum at $\Phi_{\rm min}$.
\een
As we will see below, these requirements translate into five conditions on the scalar potential, and so we need five tunable free parameters.

\subsection{Analytical Discussion}

As a first step, let us discuss the strategy used to determine the free parameters in \eqref{V1} and \eqref{V2} in order to get inflection point inflation. The position of the inflection point $\Phi_{\rm ip}$ and the minimum $\Phi_{\rm min}$ can be chosen independently with the only constraint (apart from $\Phi_{\rm min}>\Phi_{\rm ip}$) being:
\begin{align}
\vo_{\rm ip} = e^{\sqrt{\frac{3}{2}} \Phi_{\rm ip}} \gtrsim 10^3\qquad\Leftrightarrow\qquad\Phi_{\rm ip} \gtrsim 5\,,
\end{align}
in order to trust the effective field theory during inflation.\\

Once $\Phi_{\rm ip}$ and $\Phi_{\rm min}$ are chosen, we impose that the scalar potential actually produces an inflection point and the late time minimum at the desired positions. This can be done by scanning over flux parameters, intersections numbers and gauge groups so that the following constraints are satisfied:
\bi
\item \textit{Inflection point}
\begin{align}
&(1)\quad\left.V''\right|_{\Phi = \Phi_{\rm ip}} = 0 \label{eq:cond1}\\
&(2)\quad\left. V' \cdot V'''\right|_{\Phi = \Phi_{\rm ip}} = \left.\frac{2 \pi^2 V^2}{(170)^2}\right|_{\Phi = \Phi_{\rm ip}}\label{eq:cond2}
\end{align}
\item \textit{Late time minimum}
\begin{align} 
&(3) \quad \left.V'\right|_{\Phi = \Phi_{\rm min}} = 0 \label{eq:cond3}\\
&(4) \quad \left.V\right|_{\Phi = \Phi_{\rm min}} = 0 \label{eq:cond4}
\end{align}
\ei
The first two conditions produce an inflection point at $\Phi_{\rm ip}$ with the right slope to yield a scalar spectral tilt around $n_s = 0.96$, while the last two conditions imply the existence of a Minkowski minimum at $\Phi_{\rm min}$. These conditions are invariant under a rescaling of $V_0$ in \eqref{V1} and \eqref{V2} since it is just an overall multiplicative factor in the scalar potential. $V_0$ is instead fixed by the requirement of obtaining the right COBE normalization given by \eqref{cobe} which can also be rewritten as:
\begin{align}
(5) \quad \Delta^2 = \left.\frac{1}{24 \pi^2} \frac{V}{\epsilon}\right|_{\Phi = \Phi_*} \simeq \left.\frac{1}{12 \pi^2} \frac{V^3}{\left(V'\right)^2}\right|_{\Phi = \Phi_{\rm ip}} \simeq 2.4 \cdot 10^{-9}\,.
\end{align}
Given the definition of $V_0$ in \eqref{eq:V0}, condition (5) can be seen as a constraint on the magnitude of the flux superpotential $W_0$ which in Type IIB string compactifications naturally lies in the range $[0.1,\,100]$. We will show now how additional constraints on $\Phi_{\rm ip}$ arise from combining this naturalness criterion with the requirement of low-energy supersymmetry. In fact we shall carefully choose $\Phi_{\rm ip}$ so that the COBE normalization fixes $W_0$ in the natural range mentioned above. Fixing $W_0$ through the condition (5) sets also the energy scale of the soft-terms. Here we distinguish between two possibilities:
\bi
\item \textit{Non-sequestered models}: If the cycle supporting the visible sector is stabilized in geometric regime, the soft-terms are of order the gravitino mass:
\begin{align}
M_{\rm soft} \simeq m_{3/2} \simeq \sqrt{\frac{g_s}{8 \pi} }\frac{W_0}{\vo_{\rm min}}\,.
\label{eq:unseq}
\end{align}
This is usually referred to as the non-sequestered case which for $W_0\sim 1$ leads to TeV-scale supersymmetry only for values of the volume as large as $\vo\sim 10^{14}$. This is possible for the two moduli case where the value of the visible sector coupling does not depend on $\vo$. In the single modulus case, since $\alpha_{\rm vis}^{-1}=\vo^{2/3}$, the volume has to be of order $100$, resulting necessarily in a high-scale SUSY scenario.

\item \textit{Sequestered models}: If the visible sector modulus is fixed in the singular regime, the soft-terms can be very suppressed with respect to the gravitino mass: 
\begin{align}
M_{\rm soft} \simeq \frac{m_{3/2}}{\vo} \simeq \sqrt{\frac{g_s}{8 \pi}} \frac{W_0}{\vo_{\rm min}^2}\,.
\label{eq:seq}
\end{align}
This is usually referred to as the sequestered scenario which for $W_0\sim 1$ leads to low-energy supersymmetry only for $\vo\sim 10^7$. In these sequestered models supersymmetry is broken in the bulk of the extra-dimensions while the visible sector lives on branes localized at a singularity. Given that in this case the visible sector coupling is set by the dilaton, this scenario can be realized both in the single and in the two moduli case.
\ei
The magnitude of the flux superpotential that satisfies condition (5) for a generic inflection point $\Phi_{\rm ip}$ can be estimated by noting that at horizon exit $\epsilon\sim 10^{-10}$ and there is a percent level cancellation between the three dominant terms. This implies that at the inflection point $V\sim 0.01\ V_0 \ e^{-\sqrt{27/2}\Phi_{\rm ip}}$, and so the Hubble scale can be estimated as:
\begin{align}
\label{hubbleestimate}
H = \sqrt{\frac{V_{\rm ip}}{3 M_{\rm P}^2}} \simeq \frac{1}{10 \sqrt{3}} \frac{\sqrt{V_0 M_{\rm P}^2}}{\vo_{\rm ip}^{3/2}} = \sqrt{\frac{\xi}{2 \pi g_s^{1/2}}} \frac{W_0 M_{\rm P}}{40 \vo_{\rm ip}^{3/2}}\,,
\end{align}
where we used \eqref{eq:V0} and we restored the correct dependence on the Planck mass $M_{\rm P}$. Using the same argument we can also estimate the amplitude of the scalar perturbations in \eqref{cobe} as:
\begin{align}
\left.\frac{1}{24 \pi^2} \frac{V}{\epsilon}\right|_{\Phi = \Phi_*}\simeq \frac{1}{24 \pi^2} \frac{0.01\ V_0 \ e^{-\sqrt{27/2}\Phi_{\rm ip}}}{10^{-10}}
\simeq 2.4 \times 10^{-9}\,,
\end{align}
which yields:
\begin{align}
W_0^2\simeq 7.58 \times 10^{-15}\ 8 \pi \sqrt{g_s} \, \, e^{\sqrt{27/2}\ \Phi_{\rm ip}}\,.
\label{eq:w0}
\end{align}
Assuming $g_s\simeq 0.1$, we conclude that only inflection points in the range $\Phi_{\rm ip} \in [6,\,10]$ are compatible with natural values of $W_0$.\\

We can take these simple estimates further and for each $\Phi_{\rm ip}$ obeying \eqref{eq:w0} find the position of the late time minimum $\Phi_{\rm min}$ that gives rise to TeV-scale soft masses in both sequestered and non-sequestered scenarios. For the sequestered case we have:
\begin{align}
\frac{g_s}{8 \pi}\ W_0^2 \ e^{-2 \sqrt{6}\ \Phi_{\rm min}}\simeq 10^{-30}\,,
\end{align}
which by using \eqref{eq:w0} becomes:
\begin{align}
\Phi_{\rm min}\simeq 6.76 +\frac{3}{4}\Phi_{\rm ip}\,.
\label{eq:est_seq}
\end{align}
A similar estimate for the non-sequestered case yields:
\begin{align}
\Phi_{\rm min}\simeq 13.51 +\frac{3}{2}\Phi_{\rm ip}\,.
\label{eq:est_unseq}
\end{align}
Hence we see that the distance between the inflection point and the minimum in the non-sequestered case is exactly twice the corresponding distance in the sequestered setup. The factor of two descends directly from the extra volume suppression of \eqref{eq:seq} when compared with \eqref{eq:unseq}. In both cases the combination of the observational constraint on the amplitude of the density perturbations, the theoretical bias on natural values of $W_0$ and the requirement of TeV-scale soft-terms conspire to fix the distance between the inflationary inflection point and the late-time minimum.\\

Tables~\ref{tab1} and~\ref{tab2} illustrate some reference values obtained using \eqref{eq:w0} to fix the inflection point for different values of $W_0$, and then \eqref{eq:est_unseq} and \eqref{eq:est_seq} to get the late-time minimum in the non-sequestered (Tab.~\ref{tab1}) and sequestered (Tab.~\ref{tab2}) scenarios respectively.

\begin{table}[ht!]
\begin{center}
\begin{tabular}{cccccc}
\hline
$W_0$ & $\Phi_{\rm ip}$ & $\vo_{\rm ip}$ & $\Phi_{\rm min}$ & $\vo_{\rm min}$  \\
\hline
$0.1$ & $7.03$ & $5.5 \times 10^3$ & $24.06$  & $6.3 \times 10^{12}$ \\
\hline
$1$ & $8.29$ & $2.5 \times 10^4$ & $25.94$ & $6.3 \times 10^{13}$ \\
\hline
$10$ & $9.54$ & $1.2 \times 10^5$ & $27.82$ & $6.3 \times 10^{14}$ \\
\hline
$100$ & $10.79$ & $5.5 \times 10^5$ & $29.70$ & $6.3 \times 10^{15}$ \\
\hline
\end{tabular}
\end{center}
\caption{Positions of the inflection point and the late-time minimum for the non-sequestered case obtained by requiring a correct COBE normalization and low-energy supersymmetry for natural values of $W_0$ and setting $g_s=0.1$.}
\label{tab1}
\end{table}

\begin{table}[ht!]
\begin{center}
\begin{tabular}{cccccc}
\hline
$W_0$ & $\Phi_{\rm ip}$ & $\vo_{\rm ip}$ & $\Phi_{\rm min}$ & $\vo_{\rm min}$  \\
\hline
$0.1$ & $7.03$ & $5.5 \times 10^3$ & $12.03$  & $2.5 \times 10^{6}$ \\
\hline
$1$ & $8.29$ & $2.5 \times 10^4$ & $12.97$ & $7.9 \times 10^{6}$ \\
\hline
$10$ & $9.54$ & $1.2 \times 10^5$ & $13.91$ & $2.5 \times 10^{7}$ \\
\hline
$100$ & $10.79$ & $5.5 \times 10^5$ & $14.85$ & $7.9 \times 10^{7}$ \\
\hline
\end{tabular}
\end{center}
\caption{Positions of the inflection point and the late-time minimum for the sequestered case obtained by requiring a correct COBE normalization and low-energy supersymmetry for natural values of $W_0$ and setting $g_s=0.1$.}
\label{tab2}
\end{table}

Using the values listed in Tab.~\ref{tab1} and~\ref{tab2} to estimate the Hubble scale as in \eqref{hubbleestimate} we get $H \simeq 10^{10}$ GeV which corresponds to an inflationary scale of order $10^{14}$ GeV. This result can also be obtained numerically using the more precise values listed in the next section.\\

Let us comment on the consistency of our effective field theory approach. As derived in~\cite{Cicoli:2013swa}, the superspace derivative
expansion is under control if $m_{3/2}\ll M_{\rm KK}$ which translates into the bound:
\begin{align}
\delta \equiv \sqrt{\frac{g_s}{2}} \frac{W_0}{\vo^{1/3} } \ll 1\,.
\end{align}
This bound is satisfied in each case of Tab.~\ref{tab1} and~\ref{tab2} both around the inflection point and the late-time minimum. In fact, considering just the region around the inflection point ($\vo$ becomes larger around the minimum and so this bound is stronger during inflation) and setting $g_s=0.1$, we have: $\delta\simeq 10^{-3}$ for $W_0=0.1$, $\delta\simeq 10^{-2}$ for $W_0=1$, $\delta\simeq5\cdot 10^{-2}$ for $W_0=10$ and $\delta\simeq 0.1$ for $W_0=100$.
Thus the superspace derivative expansion is under control, and so higher derivative $\alpha'$-corrections should naturally be suppressed. Therefore, as we shall show in the next section, we have to tune the coefficient of the $F^4$ $\alpha'$ terms (\ref{VF4}) and (\ref{deltav1}) to large values. However the fact that the expansion parameter $\delta$ turns out to be small, allows us to neglect further higher derivative corrections in a consistent way.

\subsection{Numerical Results}

\begin{figure}[H]
\centering
\includegraphics[width=.6\textwidth]{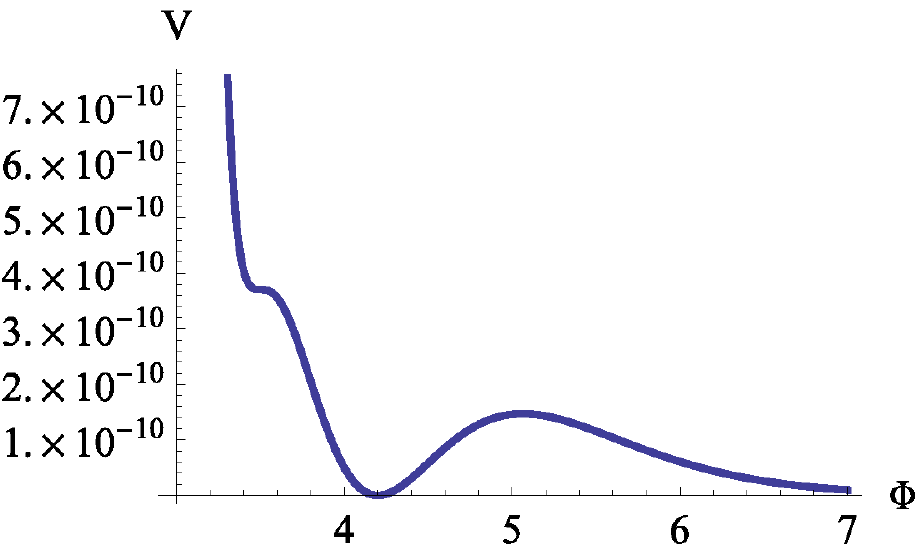}
\caption{Scalar potential obtained requiring $\Phi_{\rm ip} = 3.5$ and $\Phi_{\rm min} = 4.2$ in the single modulus case with the visible sector on D7-branes wrapping the volume cycle.}
\label{smallvol}
\end{figure}

In this section we present a detailed numerical study of inflection point volume inflation. We start by focusing on the non-sequestered single modulus case with high-scale SUSY. In this case the late-time minimum is bound to be of order $100$, so that the evolution of the canonically normalized field $\Phi$ is very limited. Nevertheless, as shown in Fig.~\ref{smallvol}, it is possible to get an inflection point and a late-time minimum at the desired values. The scalar potential is plotted in Fig.~\ref{smallvol}. We require $\Phi_{\rm ip} = 3.5$ and $\Phi_{\rm min} = 4.2$, corresponding respectively to values of the volume $\vo_{\rm ip} \simeq 72$ and $\vo_{\rm min} \simeq 171$. These values are clearly too small to trust the effective field theory approach. However we shall still present the numerical results for this case for illustrative purposes, and shall focus later on cases with low-energy supersymmetry where the volume during and after inflation is larger and the supergravity effective theory is under much better control.

In this example it is possible to reproduce the correct value of the spectral index, as a consequence of the condition \eqref{eq:cond2}. We list in Tab.~\ref{tab0} the numerical results obtained for different positions of the inflection point and the late-time minimum for which we get always $n_s \simeq 0.96$.
\begin{table}[H]
\begin{center}
\begin{tabular}{cccccccc}
\hline
$\Phi_{\rm ip}$ & $\Phi_{\rm min}$ & $W_0$ & $\kappa_{g_s}$ & $\kappa_{F^4}$ & $\kappa_{\overline{D3}}$ & $\kappa_{\rm hid}$ & $\Delta \Phi/M_{\rm P}$ \\
\hline
$2.5$ & $3.8$ & $3 \times 10^{-5}$ & $-2.55$ & $2.26$ & $1.05 \times 10^{-3}$ & $-0.14$  & $0.21$\\
\hline
$3$ & $4$ & $8 \times 10^{-5}$ & $-2.98$ & $3.11$  & $7.09 \times 10^{-4}$ & $-0.12$ & $0.17$ \\
\hline
$3.5$ & $4.2$ & $2 \times 10^{-4}$ & $-3.48$ & $4.28$ & $4.71 \times 10^{-4}$ & $-0.10$ & $0.12$ \\
\hline
\end{tabular}
\end{center}
\caption{Numerical results for the coefficients of the scalar potential for the single modulus case with the visible sector on D7-branes wrapping the volume cycle.}
\label{tab0}
\end{table}

$\Delta \Phi$ is the field excursion of the canonically normalized volume modulus $\Phi$ between horizon exit and the end of inflation. Since $\Delta \Phi\sim 0.1 M_{\rm P}$ we are clearly dealing with a small field inflationary model. Thus the tensor-to-scalar ratio is of order $r \simeq 10^{-10}$. The values of $W_0$ reported in Tab.~\ref{tab0} are the numerical results which satisfy the COBE normalization. The corresponding Hubble scale in each case is $H \simeq 10^9$ GeV which translates into an inflationary scale around $10^{14}$ GeV.\\

\begin{figure}
\centering
\subfloat[][\emph{Inflection point}.]
{\includegraphics[width=.45\textwidth]{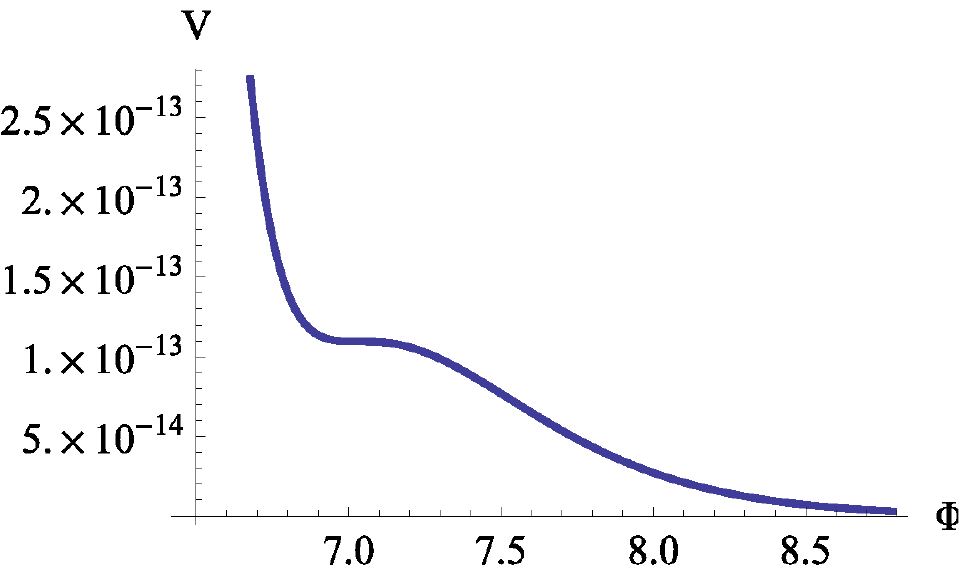}} \quad
\subfloat[][\emph{Late time minimum}.]
{\includegraphics[width=.45\textwidth]{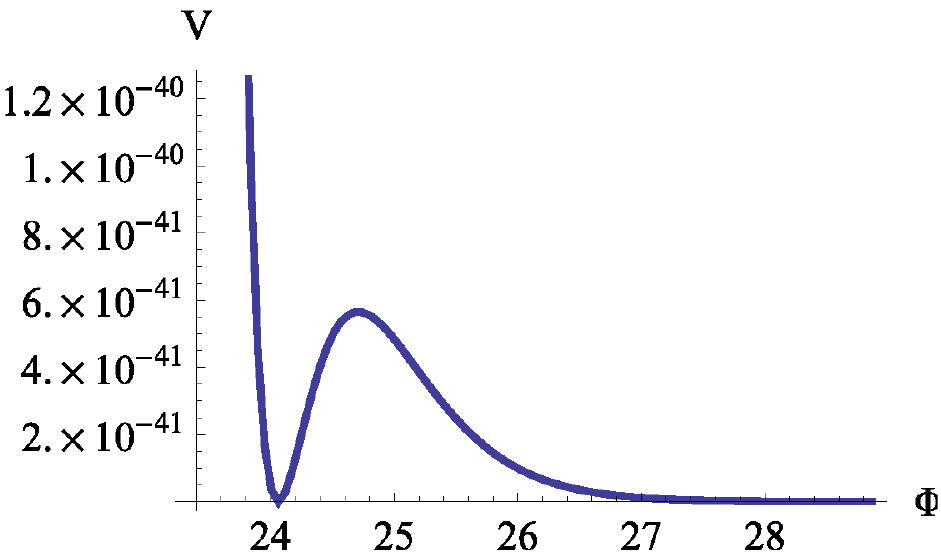}} 
\caption{Scalar potential for non-sequestered models with TeV-scale supersymmetry obtained requiring $\Phi_{\rm ip} = 7.03$ and $\Phi_{\rm min} = 24.06$.}
\label{SingleFieldNonSequestered1}
\end{figure}

We now turn to study the two more interesting sequestered and non-sequestered cases with larger values of the volume and TeV-scale supersymmetry. 
Since in both cases the shape of the scalar potential is always qualitatively the same around the inflection point and the late-time minimum, we plot it in Fig.~\ref{SingleFieldNonSequestered1} just for the non-sequestered case.\\ 

The numerical results listed below are obtained requiring that the inflection point and the late-time minimum are those given in Tab.~\ref{tab1} and~\ref{tab2} where we required natural values of $W_0$ and low-energy supersymmetry in both non-sequestered and sequestered cases. The tensor-to-scalar ratio turns out to be always of order $r \simeq 10^{-9}$. In the following tables the only inputs are $\Phi_{\rm ip}$ and $\Phi_{\rm min}$, while $W_0$, $\kappa_{g_s}$, $\kappa_{F^4}$ (or $\kappa_{F^4_{\rm (b)}}$), $\kappa_{\rm hid}$, $\kappa_{\overline{D3}}$ and $\kappa_{\rm np}$ are the numerical outputs obtained by solving \eqref{eq:cond1} and \eqref{eq:cond4}. As a consequence of the condition \eqref{eq:cond2}, the value of the spectral index is $n_s \simeq 0.967$ in each of the cases listed below.

\bi
\item \textit{Single modulus case}: 

Let us start with the simplest single modulus case of \eqref{V1}. The numerical results relative to the non-sequestered case with low-energy SUSY are listed in Tab.~\ref{tab3} while those relative to the sequestered case are listed in Tab.~\ref{tab4}:
\begin{table}[H]
\begin{center}
\begin{tabular}{cccccccc}
\hline
$\Phi_{\rm ip}$ & $\Phi_{\rm min}$ & $W_0$ & $\kappa_{g_s}$ & $\kappa_{F^4}$ & $\kappa_{\overline{D3}}$ & $\kappa_{\rm hid}$ & $\Delta \Phi/M_{\rm P}$ \\
\hline
$7.03$ & $24.06$ & $0.06$ & $-31.68$ & $253.85$ & $7.94 \times 10^{-14}$ & $-8.11 \times 10^{-5}$  & $0.42$\\
\hline
$8.29$ & $25.94$ & $0.6$ & $-53.02$ & $710.59$ & $7.93 \times 10^{-15}$ & $-3.76 \times 10^{-5}$ & $0.42$ \\
\hline
$9.54$ & $27.82$ & $6.2$ & $-88.35$ & $1972.74$ & $7.94 \times 10^{-16}$ & $-1.74 \times 10^{-5}$ & $0.42$ \\
\hline
$10.79$ & $29.70$ & $62.1$ & $-147.21$ & $5476.08$ & $7.94 \times 10^{-17}$ & $-8.12 \times 10^{-6}$ & $0.42$ \\
\hline
\end{tabular}
\end{center}
\caption{Numerical results for the coefficients of the scalar potential for the non-sequestered single modulus case with TeV-scale supersymmetry.}
\label{tab3}
\end{table}

\begin{table}[H]
\begin{center}
\begin{tabular}{cccccccc}
\hline
$\Phi_{\rm ip}$ & $\Phi_{\rm min}$ & $W_0$ & $\kappa_{g_s}$ & $\kappa_{F^4}$ & $\kappa_{\overline{D3}}$ & $\kappa_{\rm hid}$ & $\Delta \Phi/M_{\rm P}$ \\
\hline
$7.03$ & $12.03$ & $0.07$ & $-25.46$ & $187.38$ & $1.30 \times 10^{-7}$ & $-8.47 \times 10^{-3}$ & $0.39$ \\
\hline
$8.29$ & $12.97$ & $0.79$ & $-41.50$ & $504.98$ & $3.92 \times 10^{-8}$ & $-5.59 \times 10^{-3}$ & $0.38$ \\
\hline
$9.54$ & $13.91$ & $8.14$ & $-67.27$ & $1346.28$ & $1.17 \times 10^{-8}$ & $-3.68 \times 10^{-3}$ & $0.38$ \\
\hline
$10.79$ & $14.85$ & $83.6$ & $-108.80$ & $3575.02$ & $3.48 \times10^{-9}$ & $-2.42 \times 10^{-3}$ & $0.37$ \\
\hline
\end{tabular}
\end{center}
\caption{Numerical results for the coefficients of the scalar potential for the sequestered single modulus case with TeV-scale supersymmetry.}
\label{tab4}
\end{table}

\item \textit{Two moduli case}: 

Now we turn to the two moduli setup of \eqref{V2}. The results for the non-sequestered case are listed in Tab.~\ref{tab5} while the results for the sequestered case are presented in Tab.~\ref{tab6}.
\begin{table}[H]
\begin{center}
\begin{tabular}{cccccccc}
\hline
$\Phi_{\rm ip}$ & $\Phi_{\rm min}$ & $W_0$ & $\kappa_{g_s}$ & $\kappa_{F^4_{\rm (b)}}$ & $\kappa_{\overline{D3}}$ & $\kappa_{\rm np}$ & $\Delta \Phi/M_{\rm P}$ \\
\hline
$7.03$ & $24.06$ & $0.07$ & $-25.42$ & $193.01$ & $8.46 \times 10^{-15}$  & $8.91 \times 10^{-3}$ & $0.41$\\
\hline
$8.29$ & $25.94$ & $0.75$ & $-41.34$ & $525.04$ & $7.83 \times 10^{-16}$ & $7.93 \times 10^{-3}$ & $0.41$ \\
\hline
$9.54$ & $27.82$ & $7.55$ & $-67.16$ & $1421.51$ & $7.29 \times 10^{-17}$ & $7.12 \times 10^{-3}$ & $0.41$ \\
\hline
$10.79$ & $29.70$ & $75.8$ & $-109.33$ & $3858.01$ & $6.82 \times 10^{-18}$ & $6.43 \times 10^{-3}$ & $0.41$ \\
\hline
\end{tabular}
\end{center}
\caption{Numerical results for the coefficients of the scalar potential for the non-sequestered two moduli case with TeV-scale supersymmetry.}
\label{tab5}
\end{table}

\begin{table}[H]
\begin{center}
\begin{tabular}{cccccccc}
\hline
$\Phi_{\rm ip}$ & $\Phi_{\rm min}$ & $W_0$ & $\kappa_{g_s}$ & $\kappa_{F^4_{\rm (b)}}$ & $\kappa_{\overline{D3}}$ & $\kappa_{\rm np}$ & $\Delta \Phi/M_{\rm P}$ \\
\hline
$7.03$ & $12.03$ & $0.12$ & $-15.55$ & $97.00$ & $2.48 \times 10^{-8}$  & $2.28 \times 10^{-2}$ & $0.38$ \\
\hline
$8.29$ & $12.97$ & $1.27$ & $-23.17$ & $236.06$ & $6.86 \times 10^{-9}$ & $2.02 \times 10^{-3}$ & $0.37$ \\
\hline
$9.54$ & $13.91$ & $13.7$ & $-34.53$ & $572.48$ & $1.89 \times 10^{-9}$ & $1.80 \times 10^{-2}$ & $0.37$ \\
\hline
$10.79$ & $14.85$ & $147.9$ & $-51.58$ & $1390.83$ & $5.25 \times 10^{-10}$ & $1.62 \times 10^{-2}$ & $0.36$ \\
\hline
\end{tabular}
\end{center}
\caption{Numerical results for the coefficients of the scalar potential in the sequestered two moduli case with TeV-scale supersymmetry.}
\label{tab6}
\end{table}

Two important observations can be inferred from the values of the coefficients listed in Tab.~\ref{tab1} and~\ref{tab4}. The first one is that $\kappa_{g_s}$ is always required to be negative. In our models the negative sign can be obtained by the interplay of the two terms in \eqref{slccontributiontov}. Moreover the presence of the inflection point is highly sensitive to small variations of the coefficients in Tab.~\ref{tab3},~\ref{tab4},~\ref{tab5} and~\ref{tab6}. Thus in order to accurately reproduce the shape of the scalar potential in each case, it is necessary to tune the coefficients to a much higher level of precision than that reported in the Tables.
\ei

\section{Two Fields Dynamics}
\label{SecTwo}

Up to this point we have dealt exclusively with the single field limit, implicitly assuming that all other moduli, like the axio-dilaton, the complex structure moduli and additional K\"ahler moduli, are heavier than the Hubble scale during inflation. While this may be arranged for in the single K\"ahler modulus case, it is certainly not true for the model constructed within the LVS framework. In this section we comment on various aspects of the two field dynamics of this model.\\

\begin{figure}[ht!]
\centering
\subfloat[][\emph{Inflationary region}.]
{\includegraphics[width=.47\textwidth]{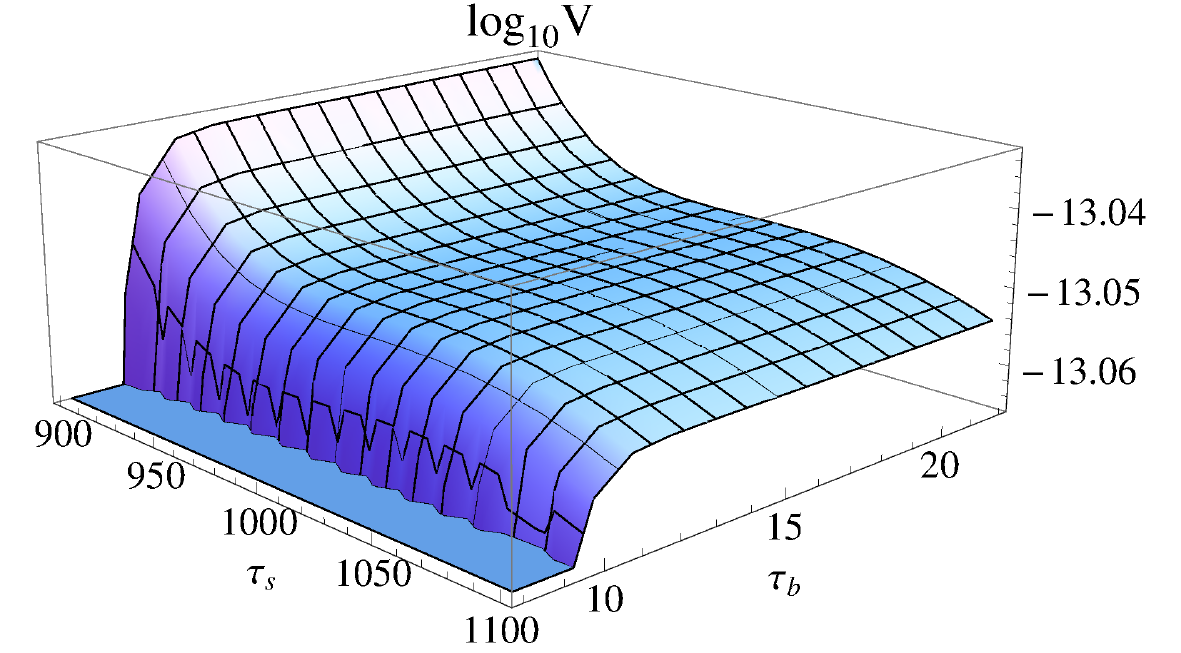}} \quad
\subfloat[][\emph{Large volume region}.]
{\includegraphics[width=.47\textwidth]{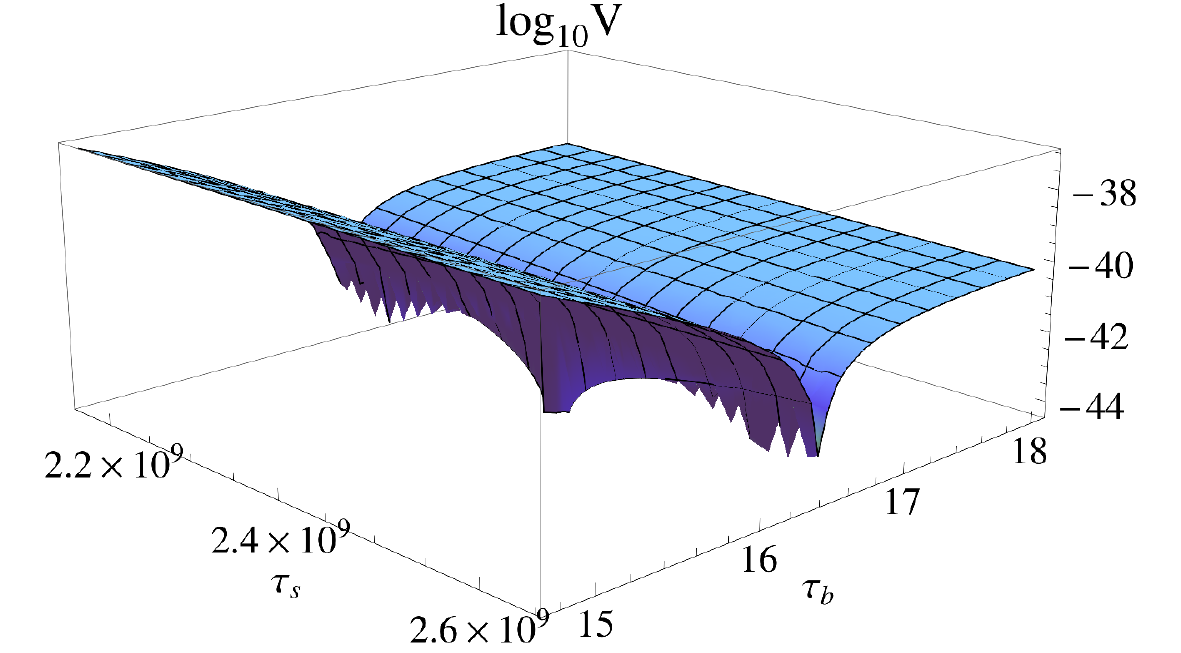}}
\caption{Scalar potential \eqref{V2} in the inflationary region, around the tuned point $(\tau_b,\tau_s)|_{\rm ip}=(1000,20)$ and in the large volume region $(\tau_b,\tau_s)|_{\rm min}=(2.4\times 10^9, 16.4)$.}
\label{fig:2FieldLVS_pot}
\end{figure}

As shown in Sec.~\ref{SecOrigin}, the effective single field potential of \eqref{V2} is obtained after integrating out the small blow-up modulus $\tau_s$. This procedure is valid in the vicinity of the LVS minimum where there is a clear mass hierarchy:
\begin{align}
m_{\tau_s}^2 \sim \frac{g_s}{8\pi}\frac{W_0^2}{\vo_{\rm min}^2} \gg \frac{g_s}{8\pi}\frac{W_0^2}{\vo_{\rm min}^3} \sim m_{\tau_b}^2\,,
\end{align}
however it fails around the inflection point where both fields are very light $m_{\tau_s}^2, m_{\tau_b}^2 \ll H^2 $, implying that both will be dynamical during inflation. This is illustrated in Fig.~\ref{fig:2FieldLVS_pot}.

Due to the flatness of the scalar potential along both directions, the correct way to analyze the system is by numerically solving the background field equations as done in the original work~\cite{Conlon:2008cj}. Here we will extend the aforementioned analysis by studying the sensitivity to the choice of initial conditions in a given potential and by clarifying to what extent the single field results constitute a valid approximation to the inflationary observables. The reader looking for more details of the setup is referred to~\cite{Conlon:2008cj} as we will focus only on the results obtained.\\

We proceed in the same spirit of the single field analysis by choosing the coefficients that induce an inflection point along the volume direction. For concreteness we choose $(\tau_b,\tau_s)|_{\rm ip}=(1000,20)$. We then consider a set of initial conditions around that point and numerically solve the equations of motion. In Fig.~\ref{fig:2FieldLVS} we plot the solutions for the different choices of initial positions for the system. We assume throughout that the fields are released with vanishing velocities and find that the sensitivity to the initial position, that is characteristic of inflection point models, is magnified in the two field setup as perturbing the initial conditions by a small amount can lead to drastically different outcomes. Starting uphill from the inflection point tends to lead to trajectories that produce insufficient expansion. In some cases, depending on the ratio $\tau_s^{\rm ip}/\tau_s^{\rm min}$, some of these trajectories can lead to the collapse of the compact manifold to vanishing volume. For trajectories that start downhill from the tuned point, it is easier to obtain a viable post inflationary evolution, with the system evolving towards the LVS minimum, but the number of e-foldings decreases drastically with the distance from the tuned point. One is therefore led to the conclusion that viable inflationary trajectories are obtained only in a narrow region around $\tau_b^{\rm ip}$ and $\tau_s\gtrsim \tau_s^{\rm ip}$.\\

\begin{figure}[ht!]
\centering
{\includegraphics[width=.47\textwidth]{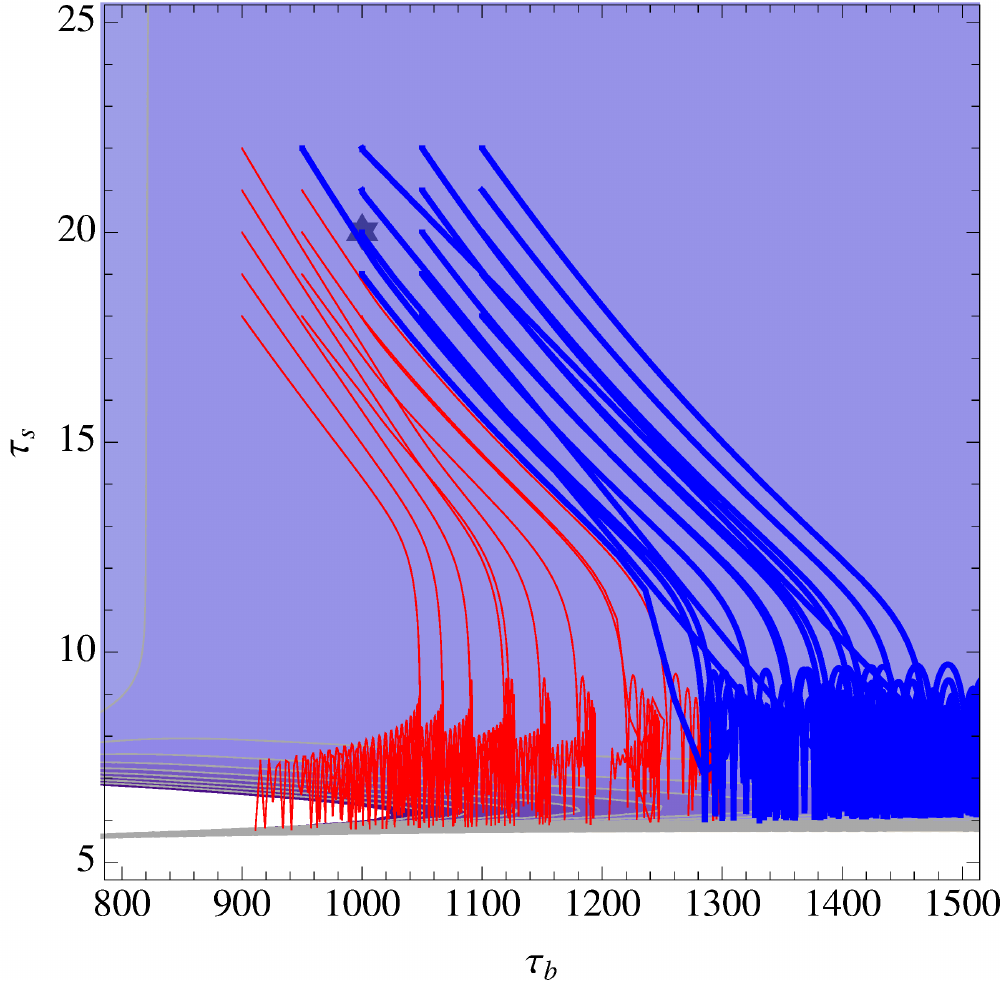}} \quad
{\includegraphics[width=.47\textwidth]{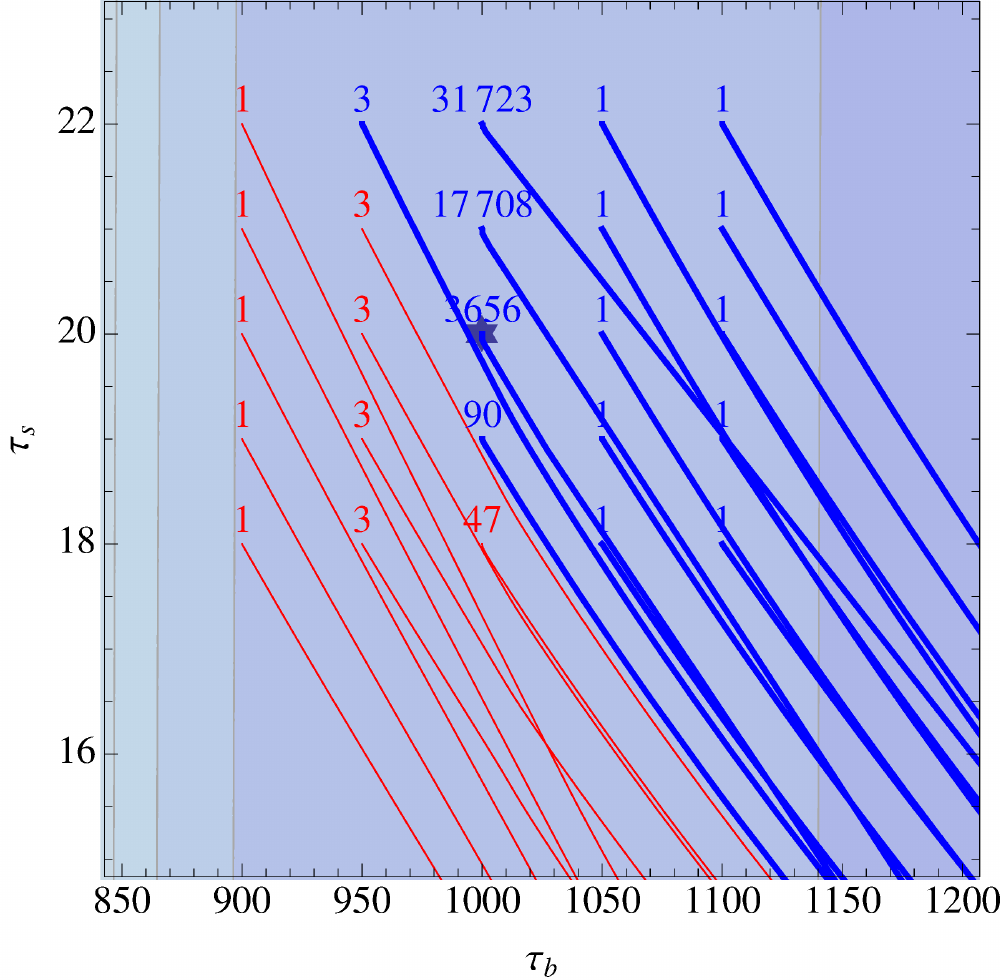}} 
\caption{Trajectories around the flat inflection point at $(\tau_b,\tau_s)= (1000 , 20)$ marked by the star. Blue (thicker) trajectories correspond to those evolving towards the post-inflationary large volume minimum, while the magenta (thinner) end with a collapsing volume modulus. Right: Magnification around the initial points. The numbers denote the total number of slow-roll e-foldings for each trajectory.}
\label{fig:2FieldLVS}
\end{figure}

In what concerns inflationary observables in the two field setup one expects the projection along the inflationary trajectory (and hence the single field estimates of Sec.~\ref{SecSingle}) to be a good approximation to the full result. This follows directly from the fact that the observable portion of the trajectories yielding $N_e\ge 60$, is rather straight, as can be seen by the smallness of the inverse curvature radius plotted in Fig.~\ref{fig:2FieldLVS}. This implies that curvature and isocurvature perturbations are essentially decoupled and that a straightforward generalization of the single field case leads to an accurate estimate of the cosmological observables.\footnote{For thorough discussion of this issue in the context of a local string inflation model see~\cite{Bielleman:2015lka}.}
\begin{figure}[ht!]
\centering
{\includegraphics[width=.5\textwidth]{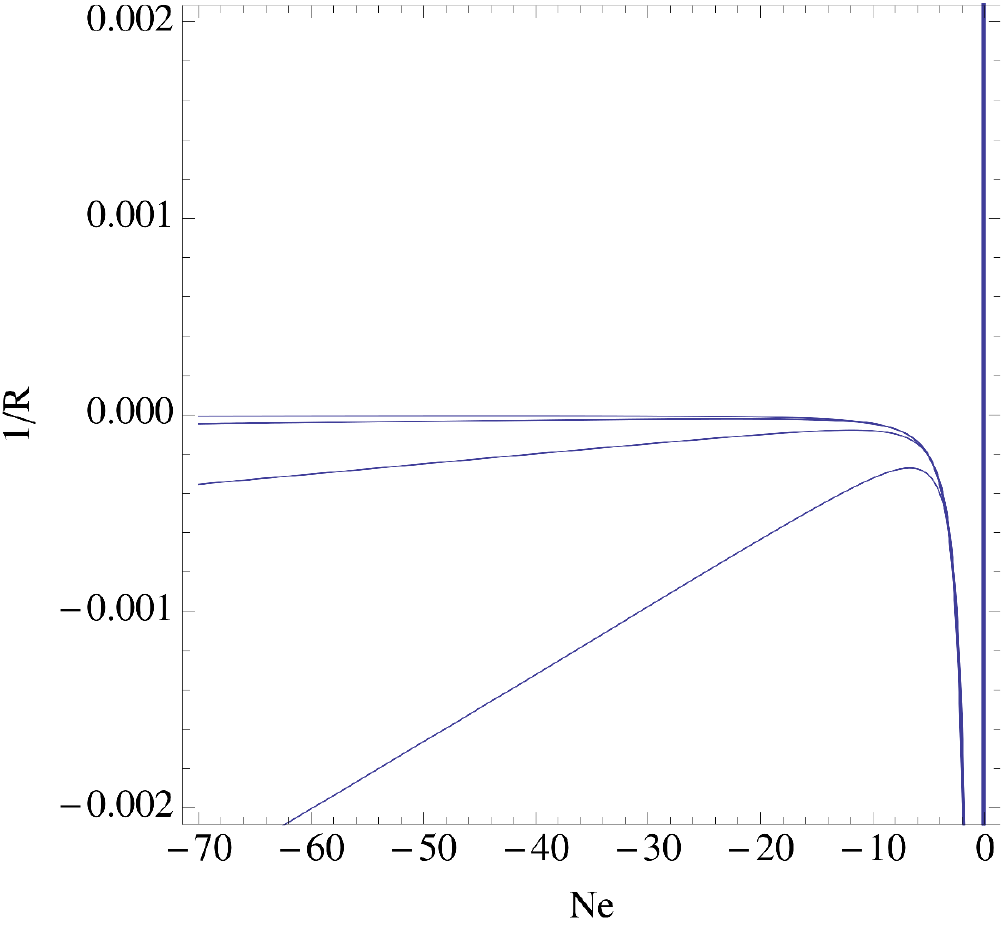}} 
\caption{Inverse curvature radius for trajectories leading to $N_e>60.$}
\label{fig:2FieldLVS}
\end{figure}

\section{Conclusions}
\label{SecConcl}

It has long been acknowledged that there is severe tension between having simultaneously low-scale supersymmetry and high scale inflation in supergravity models. This is due to the fact that the inflationary energy density contributes to the moduli potential and will in generally tend to destabilize them.\\

Several mechanisms to decrease or eliminate this tension have been proposed over the years and in this thesis we further develop the proposal of~\cite{Conlon:2008cj}. This particular model solves the tension between TeV-scale supersymmetry and the inflationary scale by having an evolving compactification volume between the inflationary epoch and today. Inflation is due to an inflection point in the volume direction of the scalar potential which also possesses a minimum at large volume where the modulus is supposed to sit at late-time after inflation. In order to prevent overshooting it is necessary to require that a small amount of radiation is generated after the end of inflation. The presence of this additional radiation is well justified in the two field model since it could be produced by the oscillations of the heavy modulus around its minimum while in the single field model particle production could be induced by a changing vacuum state between the end of inflation and today.\\

In this chapter we described a possible microscopic origin of the inflationary scalar potential that allows the gravitino mass to vary after the end of inflation and at the same time features a late-time dS minimum. In particular, we provided an explicit construction where the inflationary inflection point is generated by the interplay between string loops and higher derivative $\alpha'$-corrections to the scalar potential. Moreover, we supplemented the LVS construction with a new model that involves only one K\"ahler modulus. While in the LVS two moduli model non-perturbative effects play a crucial r\^ole in determining the presence of both the inflection point and the late-time minimum, in the single modulus model non-perturbative effects are absent and an additional contribution arising from the F-terms of charged hidden matter is needed. For both models we analyzed the relation between the value of the volume during inflation and at present with the size of the flux superpotential $W_0$. We found that if $W_0$ takes natural values, the distance between the inflection point and the late-time minimum is fixed.\\

We finally studied the full dynamics of the two field system in the LVS model and showed that, after tuning the potential such that it features the desired inflection point, there is a significant sensitivity to the choice of initial conditions. Perturbing the starting positions of the fields even by a small amount can lead to a radically different cosmological evolution. We also showed that, despite the presence of two dynamical fields, the predictions for the cosmological observables derived in the single field case are accurate since the field space trajectories are essentially straight over the last $60$ e-foldings of expansion.

\part{Conclusions}

\chapter{Summary and final remarks}

Let us conclude by providing a brief summary of the contents and of the results reported in this thesis.\\

In Chap.~\ref{chap:StateOfArt} we first presented a brief report of the current status of particle physics and cosmology. We also presented a quick overview of the possible alternatives for the physics beyond the Standard Model, focusing specially on supersymmetric theories. In the second part of the same chapter we presented string theory as the best candidate for a unified theory of all interactions, which describes also gravity at the quantum level in a consistent way. After having described its main features, we discussed the role of string phenomenology in the current search for new physics beyond the Standard Model.\\

In Chap.~\ref{chap:Compactifications&ModelBuilding} we presented an overview of the generic tools needed for the study of string compactifications. More in detail, in the first part we have shown that some phenomenological requirements along with our poor computational skills restrict the choice of the compact space to a very specific class, among the infinity that are available. We have also summarized how to get a $\mathcal{N} = 1$ supersymmetric effective field theory starting from the ten-dimensional action for massless string states. In general such an effective theory features the presence of many massless scalar fields, which are problematic from a phenomenological point of view. In the second part of the chapter we have shown how to connect this effective field theory with the real world. The main step is represented by moduli stabilization, which makes moduli massive and generally breaks supersymmetry. We used one of the most powerful techniques for moduli stabilization, which is called Large Volume Scenario. Finally we have described how to embed a chiral gauge theory using a stack of D3-branes on top of a singularity of the compact space. In the last part of the section we presented an explicit example of a string compactification in which the visible sector is consistently embedded in a global construction, and all the closed moduli are stabilized.\\

In Chap.~\ref{chap:Soft-Terms} we have computed the soft spectrum arising from sequestered compatifications~\cite{Aparicio:2014wxa}. We have shown that it is possible to get a MSSM-like spectrum only in a particular case in which the effective field theory is protected by an effective symmetry. In this case all the soft-terms have the same size, and they can be around $1 \,$ TeV without causing the cosmological moduli problem. In more generic situations the spectrum exhibits a hierarchy between gaugino masses and scalar masses, giving rise to Split-SUSY scenarios with gauginos around the TeV and scalars around $10^6-10^7 \,$ GeV. This is an interesting example of how string theory can act as a guidance among the myriads of beyond the Standard Model alternatives: Split-SUSY scenarios seem to be quite more natural than MSSM ones from a top-down perspective.\\

A model-independent feature of the sequestered models presented in Chap.~\ref{chap:Soft-Terms} is that the field whose VEV parametrizes the size of the compact extra-dimensions (\textit{volume modulus}) is always the lightest modulus. Due to its gravitational coupling the volume modulus decays late in the history of the universe and can modify significantly its cosmological evolution. In particular, if it decays after the freeze-out of thermally produced dark matter, it dilutes the previous abundance and produces a new amount of dark matter. In Chap.~\ref{chap:NonThermalDM} we performed a numerical study of the non-thermal dark matter production in the MSSM case of Chap.~\ref{chap:Soft-Terms} and we contrasted the results with the current experimental bounds~\cite{Aparicio:2015sda}. In the region of the parameter space which is not ruled out by experimental data, non-thermally produced dark matter saturates the observed relic abundance if the lightest supersymmetric particle is a higgsino-like neutralino with mass around $300$ GeV and the reheating temperature is about $2$ GeV.\\

Relativistic degrees of freedom in the hidden sector constitute the so-called dark radiation. The amount of dark radiation present in the universe can be constrained by CMB and BBN experiments, since the presence of additional relativistic degrees of freedom modifies the rate of expansion of the universe at early times. The existence of dark radiation is then a very promising window on beyond the Standard Model physics, and it is very interesting also from a string phenomenology point of view. Indeed, it provides a quite model-independent signature of string models, given that its existence is uniquely due to the presence of both hidden relativistic degrees of freedom and moduli, which are the four-dimensional manifestation of extra-dimensions. Hence it can be used either to rule out or to constrain entire classes of string compactifications. A generic feature of sequestered models is that the volume modulus decays into a massless axion-like field, which acts as dark radiation. In Chap.~\ref{chap:DR} we have shown that in Split-SUSY scenarios the dark radiation produced in sequestered models is within the current experimental bounds, provided that the decay of the volume modulus into scalars is kinematically allowed~\cite{Cicoli:2015bpq}.\\


In Chap.~\ref{chap:VolumeInflation} we studied the microscopic origin of a model of inflation in which the role of the inflaton is played by the volume modulus, called \textit{Volume Inflation}~\cite{Cicoli:2015wja}. This is a particularly interesting model because it provides a dynamical mechanism to overcome the well known tension between TeV soft-terms and high-scale inflation. It turns out that inflation is due to the interplay of quantum corrections to the scalar potential (both $\alpha'$ and string loop corrections) with contributions from anti-branes and/or from charged hidden matter. They give rise to an inflection point around which small-field inflation can take place. We have shown that requiring natural values for the parameters of the model and TeV scale soft-terms leads to a well defined relation between the value of the volume during inflation and its value in the late-time vacuum.\\


In general, the major open question is to exactly reproduce the Standard Model at low-energy from a class of string compactifications and to reconcile it with cosmological observations. This problem can be faced in vastly different ways depending on the desired pattern of supersymmetric particles, and hopefully in the near future new experimental data will help us in this task. In the present thesis we have observed that it is very hard to get a TeV MSSM spectrum (which could address the hierarchy problem) even in the most favored case in which the visible sector is placed on top of D3-branes at singularities. Hopefully in the near future we will understand whether it is actually worth addressing this problem or we should change our point of view on the naturalness issues.

\backmatter

\bibliographystyle{unsrt}
\bibliography{Thesis_PhD}

\end{document}